\documentclass[12pt]{elsart}
\usepackage{epsfig,graphics}
\usepackage{dsfont}               

\newcommand{\Cchi}{\hat C }
\newcommand{\barCchi}{\hat {\!\bar C}}
\newcommand{\ratVphi}{r }
\newcommand{\eA}{\hat e_A}
\newcommand{\eAbA}{\bar e_A}
\newcommand{\eAcptn}{\Big({\textstyle{1\over 3}}\,\hat e^{(K^*_+)}_A+{\textstyle{2\over 3}}\,\hat e^{(K^*_0)}_A \Big) } 
\newcommand{\eAcptnbA}{\Big({\textstyle{1\over 3}}\,\bar e^{(K^*_+)}_A+{\textstyle{2\over 3}}\,\bar e^{(K^*_0)}_A \Big) } 
\newcommand{\eAcmn}{\Big(\hat e^{(K^*_+)}_A-\hat e^{(K^*_0)}_A \Big) } 
\newcommand{\eAcmnbA}{\Big(\bar e^{(K^*_+)}_A-\bar e^{(K^*_0)}_A \Big) } 
\newcommand{\eAcmtn}{\Big(2\,\hat e^{(K^*_0)}_A-\hat e^{(K^*_+)}_A \Big)} 
\newcommand{\eAcmtnbA}{\Big(2\,\bar e^{(K^*_0)}_A-\bar e^{(K^*_+)}_A \Big)} 
\newcommand{\eAcmfn}{\Big({\textstyle{4\over 3}}\,\hat e^{(K^*_0)}_A-{\textstyle{1\over 3}}\,\hat e^{(K^*_+)}_A \Big)}  
\newcommand{\eAcmfnbA}{\Big({\textstyle{4\over 3}}\,\bar e^{(K^*_0)}_A-{\textstyle{1\over 3}}\,\bar e^{(K^*_+)}_A \Big)}  
\newcommand{\eAc}{\hat e_A^{(K^*_+)}}  
\newcommand{\eAcbA}{\bar e_A^{(K^*_+)}}  
\newcommand{\eAn}{\hat e_A^{(K^*_0)}}  
\newcommand{\eAnbA}{\bar e_A^{(K^*_0)}}  
\newcommand{\eAphi}{\hat e_A^{(\phi)}} 
\newcommand{\eAphibA}{\bar e_A^{(\phi)}} 

\newcommand{\tr}{{\rm tr} \,}

\def\rat#1#2{\renewcommand{\arraystretch}{0.5}
         \left(\!\begin{array}{c}#1 \\#2 \end{array}\!\right)}
\def\Re{{\rm Re\,}}

\def\prt{\partial}

\begin{document}\begin{frontmatter}
\title{On the radiative decays of \\ light vector and axial-vector mesons}
\author{M.F.M. Lutz and S. Leupold}
\address{GSI, Planckstrasse 1,  D-64291 Darmstadt, Germany}
\begin{abstract}
We study the  light vector and axial-vector mesons. According to the hadrogenesis conjecture
the nature of the two types of states is distinct. The axial-vector mesons are generated dynamically
by coupled-channel interactions based on the chiral Lagrangian written down in terms of the
Goldstone bosons and the light vector mesons. We propose a novel counting
scheme that arises if the chiral Lagrangian is supplemented by constraints from large-$N_c$
QCD in the context of the hadrogenesis conjecture. The counting scheme is successfully tested
by a systematic study of the properties of vector mesons. The spectrum of light axial-vector
mesons is derived relying on the leading order interaction of the Goldstone bosons with the vector
mesons supplemented by a phenomenology for correction terms. The
$f_1(1282),  b_1(1230), h_1(1386), a_1(1230)$ and $K_1(1272)$ mesons are recovered as molecular states.
Based on those results the one-loop contributions to the electromagnetic decay amplitudes of
axial-vector molecules into pseudo-scalar or vector mesons are evaluated systematically. In order to arrive at gauge invariant results in a transparent manner we choose to represent the vector particles by anti-symmetric tensor fields.
It is emphasized that there are no tree-level contributions
to a radiative decay amplitude of a given state if that state is generated by coupled-channel dynamics. The
inclusion of the latter would be double counting. At present we restrict ourselves to loops where a
vector and a pseudo-scalar meson couple to the axial-vector molecule.
We argue that final and predictive results require further
computations involving intermediate states
with two vector mesons. The relevance of the latter is predicted by our counting rules.

\end{abstract}
\end{frontmatter}

\newpage

\tableofcontents

\newpage

\section{Introduction}

The spectrum of light axial-vector mesons is one of the key objects to be studied in QCD.
The main objective is to unravel the effective degrees of freedom responsible for
the formation and decay properties of axial-vector states. Commonly the relevant degrees of freedom
for the vector and axial-vector state are assumed to be constituent quarks \cite{Godfrey:1985xj,Ishida:Yamada:Oda:1989}. This picture was
challenged by the hadrogenesis conjecture, which proposed the meson spectrum to be generated by coupled-channel
dynamics in terms of a preselected set of hadronic degrees of freedom \cite{Lutz:habil,Lutz-Kolomeitsev:2004},
which we would call 'quasi-fundamental' degrees of freedom.
Indeed in \cite{Lutz-Kolomeitsev:2004} it was demonstrated for the first time that the leading order chiral
interaction of the Goldstone boson octet with the nonet of light vector mesons gives rise to an
axial-vector spectrum that is amazingly compatible with the empirical one. This work hints at the possibility
that the effective degrees of freedom forming the vector and axial-vector states are distinct from each other.
A recent analysis of the spectral distribution of the $a_1$ meson as seen in the $\tau$
decay  \cite{Wagner:Leupold:2007,Wagner:2008gz} supports this picture.

Additional clues about the effective degrees of freedom were brought by analogous studies in the
open-charm sector of QCD. Based on the leading order chiral interaction of the Goldstone boson octet with the
open-charm triplets of pseudo-scalar and vector mesons a realistic spectrum of scalar and axial-vector molecules was
derived first in \cite{Kolomeitsev-Lutz:2004,Hofmann-Lutz:2004}. In this case the dynamics is constrained further by a
spin symmetry of QCD which arises in the limit of a large charm-quark mass.  The pseudo-scalar and
vector $D$-meson ground states degenerate as the charm-quark mass turns large. Thus it appears quite natural to
identify the pseudo-scalar and vector-meson ground states as quasi-fundamental degrees of freedom.

More recently the radiative decay amplitudes of the scalar and axial-vector open-charm mesons with strangeness were
studied in the hadrogenesis conjecture \cite{Lutz-Soyeur:2007}. The computation respects constraints from chiral and
heavy-quark symmetry. It was pointed out that in a coupled-channel framework it is advantageous to represent the
vector particles in terms of anti-symmetric tensor fields. This facilitates the derivation of gauge-invariant radiative
decay amplitudes significantly. Though in general the physics should be independent on the choice of fields, in a
non-perturbative approach like coupled-channel dynamics, certain field choices may be superior if they lead to a
more transparent realization of the electromagnetic gauge symmetry.

A first attempt to derive some radiative decays of the $b_1(1230)$
and the $a_1(1230)$  within a coupled-channel approach can be found
in \cite{Roca:Hosaka:Oset:2007}.  Though the numerical results are compatible with the empirical constraints the work suffers from conceptual inconsistencies.
The study allows for a gauge-dependent
transition of the neutral vector mesons $\rho_0, \omega$ and $\phi$
into a photon, i.e. the Lagrangian contains terms of the form $A_\mu
\,\omega^\mu$ etc. The latter play a  quantitatively significant
role via tree-level contributions to the only two
computed decays $b^+_1(1230) \to \gamma\,\pi^+ $ and $a^+_1(1230)
\to \gamma\,\pi^+$ \cite{Roca:Hosaka:Oset:2007}. However, such
couplings are consistent with the electromagnetic gauge symmetry and chiral constraints of QCD only if the vector mesons are treated in the hidden local-gauge symmetry
approach \cite{Bando} or the massive Yang-Mills approach (see e.g. the review \cite{Birse:1996}). The coupled-channel part of
\cite{Roca:Hosaka:Oset:2007} does not respect any
non-abelian gauge symmetry: it is performed starting from the
leading order chiral Lagrangian as suggested first in
\cite{Lutz-Kolomeitsev:2004}. As a consequence the realization of
electromagnetic gauge invariance in \cite{Roca:Hosaka:Oset:2007} is
obscure. A prescription, that appears incompatible with the
way the coupled-channel dynamics is set up, is used. Moreover, as
was emphasized in \cite{Lutz-Soyeur:2007} there are no tree-level
contributions to the radiative decay amplitudes of a given state if
that state is generated by coupled-channel dynamics. Thus in order
to arrive at solid results it is important to perform more
systematic computations based on the chiral Lagrangian and
rules of power counting.

The theoretical and experimental literature on the radiative decay
properties of axial-vector mesons is quite fragmentary. A quark
model computation by Rosner \cite{Rosner:1981} addresses the
$a^\pm_1(1230) \to \gamma\,\pi^\pm $ and $b^\pm_1(1230) \to
\gamma\,\pi^\pm $ processes only, a small subset of possible
radiative decays of the axial-vector states $f_1(1282), b_1(1230),
h_1(1386), a_1(1230)$ and $K_1(1272)$ considered in our work.
Similarly the hadronic tree-level model of Roca, Palomar and Oset
\cite{Roca:Palomar:Oset:2004} provides decay branchings of the
$a_1(1230)$ and $b_1(1285)$ into a photon and a pion only.
Radiative decay modes of axial-vector states with pseudo-scalar
mesons in the final states have been studied in the quark model by
Aznaurian and Oganesian \cite{Aznaurian:Oganesian:1988}. The study
recovers the empirical branching of the $b_1(1230)$ meson into the
$\gamma\,\pi$ final state and includes axial-vector mesons with
strangeness. A more systematic computation is due to Ishida,
Yamada and Oda \cite{Ishida:Yamada:Oda:1989} who apply the
covariant oscillator quark model to the radiative decays of
axial-vector mesons including decays with vector mesons in the
final state. However, like in the previous attempts by Rosner and
Roca et al. the empirical decay fraction of the $b_1(1230)$ meson
into a photon and a pion can not be reproduced at all. The
Particle Data Group \cite{PDG:2006} quotes firm branching ratios
for the three processes $b^+_1(1230) \to \,\gamma\,\pi^+$,
$f_1(1282)\,\to \gamma\,\rho,\, \gamma\,\phi$ only.

The purpose of the present work is to challenge further the hadrogenesis conjecture by a systematic study of the
radiative decay processes of light axial-vector molecules. Based on the chiral Lagrangian written in  terms of the
Goldstone boson octet and the nonet of light vector mesons we study the properties of vector and axial-vector
mesons systematically. Novel chiral power counting rules that rely on the hadrogenesis conjecture supplemented by
constraints from large-$N_c$ QCD are developed in this work. In a first step the counting rules are applied
successfully to the decay properties of the light vector mesons. We choose to represent the vector mesons in terms
of anti-symmetric tensor fields. This streamlines the realization of electromagnetic gauge invariance in a
non-perturbative context, like the coupled-channel approach we use to generate the axial-vector
molecules, significantly. We compute the radiative decays of the
axial-vector molecules with a Goldstone boson or a vector meson in the final state.
However, we confine ourselves to processes where the molecule
couples to a vector and a Goldstone boson. From the previous work \cite{Lutz-Soyeur:2007} and also from our novel power
counting rules we expect our study to be partial since contributions from two intermediate vector mesons will not be
considered in this work. Thus the main purpose is to establish the various contributions implied by the channels
involving a vector meson and a Goldstone boson.

The paper is organized in the following way: We demonstrate in section \ref{sec:generation}
how the axial-vector mesons can be generated from the coupled-channel dynamics of
Goldstone bosons and vector mesons. In section \ref{subsec:leading} the results of
\cite{Lutz-Kolomeitsev:2004} based on the leading order chiral Lagrangian are reviewed
and translated to the tensor representation of vector mesons. This leading-order
calculation is improved in section \ref{subsec:chiral-corr} by the inclusion of chiral
correction terms. Novel rules of power counting in the presence of vector mesons are
developed in section \ref{sec:counting}. The pertinent leading order Lagrangian is detailed
in section \ref{subsec:counting} and applied to the decay properties of vector mesons
in section \ref{subsec:dec-vec}. Section \ref{sec:rad-decay} is devoted to the
calculation of the radiative decay amplitudes of the axial-vector molecules. We present
the general formalism in sections \ref{subsec:dec-zerominus} and \ref{subsec:dec-oneminus}
for the radiative decay branches with a pseudo-scalar or a vector meson in the final state.
In sections \ref{subsec:f1}-\ref{subsec:K1} we discuss separately the results for the
various axial-vector mesons. Finally we summarize our results in section \ref{sec:summary}.
Technical issues and longer formulae are delegated to five appendices.

\newpage

\section{Axial-vector molecules from coupled-channel dynamics}
\label{sec:generation}

We set the stage for our study of radiative decays of axial-vector mesons. In a first step the axial-vector
mesons are formed by coupled-channel dynamics as derived from the chiral Lagrangian.
We follow \cite{Lutz-Soyeur:2007} and choose to represent the vector mesons
in terms of antisymmetric tensor fields \cite{Kyriakopoulos:1971,Kyriakopoulos:1972,Ecker:1989}. In
\cite{Lutz-Soyeur:2007} it was worked out how to adapt the coupled-channel formalism of
\cite{Kolomeitsev-Lutz:2004} to the case of tensor fields. We briefly recall the main motivation and results.
Representing the vector particles in terms of anti-symmetric tensor fields prevents scalar modes to mix with
longitudinal vector modes. This mixing phenomenon would cause severe complications when establishing gauge
invariant amplitudes  in a non-perturbative framework \cite{Lutz-Soyeur:2007}.

\subsection{Chiral coupled-channel dynamics at leading order}
\label{subsec:leading}

We construct an effective hadronic interaction, based on the  chiral Lagrangian
for the  Goldstone bosons and vector mesons. Consider the kinetic term for the vector mesons, written
in terms of a covariant derivative, $D_\mu$, as to respect the chiral SU(3) symmetry,
\begin{eqnarray}
{\mathcal L}_{kin} &=& -\frac{1}{4}\,{\tr }\, \Big\{(D^\mu\,V_{\mu \alpha})\,(D_\nu \,V^{\nu \alpha})\Big\}+
\frac{1}{8}\,m_{1^-}^2\,{\tr } \,\Big\{ V^{\mu \nu}\,V_{\mu \nu}\Big\}\,,
 \label{kinetic-term}
\end{eqnarray}
with the anti-symmetric field $V_{\mu \nu} = - V_{\nu \mu}$ and
\begin{eqnarray}
&& V_{\mu \nu} = \left(\begin{array}{ccc}
\rho^0_{\mu \nu}+\omega_{\mu \nu} &\sqrt{2}\,\rho_{\mu \nu}^+&\sqrt{2}\,K_{\mu \nu}^+\\
\sqrt{2}\,\rho_{\mu \nu}^-&-\rho_{\mu \nu}^0+\omega_{\mu \nu}&\sqrt{2}\,K_{\mu \nu}^0\\
\sqrt{2}\,K_{\mu \nu}^- &\sqrt{2}\,\bar{K}_{\mu \nu}^0&\sqrt{2}\,\phi_{\mu \nu}
\end{array}\right)\,.
\label{def-fields-vector}
\end{eqnarray}
The covariant derivative, $D_\mu$, in (\ref{kinetic-term}) involves the Goldstone boson fields in a manner that
the chiral Ward identities of QCD are transported into the effective field theory under consideration
(see e.g. \cite{Krause:1990}). We recall
\begin{eqnarray}
&&D_\mu \,V_{\alpha \beta}  = \partial_\mu\,V_{\alpha \beta} +\big[\Gamma_\mu ,V_{\alpha \beta}\big]_-
+i\,e\,A_\mu\,\big[Q, V_{\alpha \beta}\big]_- \,,
\nonumber\\
&&\Gamma_\mu = \frac{1}{2}\,\Big( u^\dagger \,\partial_\mu\,u +u
\,\partial_\mu\,u^\dagger \Big) \,, \qquad \qquad u = \exp \left( \frac{ i \,\Phi }{2\,f} \right)\,,
\label{eq:defcovder}
\end{eqnarray}
where $e = 0.303$ is the electromagnetic charge and $f \simeq f_\pi$ may be identified
with the pion-decay constant, $f_\pi =92.4$ MeV, at leading order. A precise determination of $f$
requires a chiral SU(3) extrapolation of some data set. In \cite{Lutz:Kolomeitsev:2002} the value
$f \simeq 90$ MeV was obtained from a detailed study of pion- and kaon-nucleon scattering data.
We will use $f = 90$ MeV throughout this work.

The photon field is denoted by
$A_\mu$. The Goldstone boson field, $\Phi$, and the charge matrix, $Q$, are normalized as follows:
\begin{eqnarray}
&&  \Phi =\left(\begin{array}{ccc}
\pi^0+\frac{1}{\sqrt{3}}\,\eta &\sqrt{2}\,\pi^+&\sqrt{2}\,K^+\\
\sqrt{2}\,\pi^-&-\pi^0+\frac{1}{\sqrt{3}}\,\eta&\sqrt{2}\,K^0\\
\sqrt{2}\,K^- &\sqrt{2}\,\bar{K}^0&-\frac{2}{\sqrt{3}}\,\eta
\end{array}\right) \,, \qquad Q = \left( \begin{array}{ccc} \frac{2}{3} & 0 & 0 \\
0 & -\frac{1}{3} & 0 \\
 0 & 0 & - \frac{1}{3}
\end{array}
\right)\,.
\label{def-fields-scalar}
\end{eqnarray}

We extract the leading order two-body interaction, the Weinberg-Tomozawa term, from (\ref{kinetic-term}). Expanding the
chiral connection, $\Gamma_\mu $, in powers of the Goldstone boson fields we identify
\begin{eqnarray}
{\mathcal L}_{WT}&=& -\frac{1}{16\,f^2}\,{\tr } \,\Big\{(\partial^\mu\,V_{\mu \alpha})\,\Big[[\Phi, \,(\prt_\nu \Phi )]_-,\,V^{\nu \alpha}\Big]_-\Big\} \,.
 \label{WT-term}
\end{eqnarray}
The result (\ref{WT-term}) is the starting point for a coupled-channel analysis of light axial-vector molecules
analogous to the study \cite{Lutz-Kolomeitsev:2004} based on the vector-field representation. As already pointed
out in \cite{Lutz-Soyeur:2007} one expects only minor differences. We focus here on the $J^P=1^+$ channels, where
the scattering amplitude takes the form
\begin{eqnarray}
&&T^{\rm on-shell}_{\mu \nu, \alpha \beta} =
\frac{3}{8}\,\Big\{\bar p_\nu\,\Big(\frac{w_\mu\,w_\alpha}{w^2}-g_{\mu \alpha}  \Big)\,p_\beta
-\bar p_\nu\,\Big(\frac{w_\mu\,w_\beta}{w^2}-g_{\mu \beta}  \Big)\,p_\alpha
\nonumber\\
&& \qquad \qquad -\bar p_\mu\,\Big(\frac{w_\nu\,w_\alpha}{w^2}-g_{\nu \alpha}  \Big)\,p_\beta
+\bar p_\mu\,\Big(\frac{w_\nu\,w_\beta}{w^2}-g_{\nu \beta}  \Big)\,p_\alpha\Big\}\,M(s) + \cdots\,,
\label{1plus-amplitude}
\end{eqnarray}
with $w_\mu=p_\mu+q_\mu = \bar p_\mu+ \bar q_\mu$. The dots in (\ref{1plus-amplitude}) represent additional contributions
from partial waves other than $1^+$. The momenta $p_\mu, q_\mu$ and $\bar p_\mu, \bar q_\mu$ denote the in- and out-going
momenta of the vector and pseudo-scalar mesons, respectively.

\begin{table}
\tabcolsep=0.mm
\renewcommand{\arraystretch}{1.4}
\fontsize{8}{2}
\begin{tabular}{||c|c|c|c||}
\hline\hline
$(0,+2)$ &
$(1,+2)$ &
$(\frac12,+1)$ &
$(\frac32, +1)$
\\\hline
$( {\textstyle{1\over \sqrt{2}}}\,K^t\,i\,\sigma_2 K_{\mu \nu})$ &
$( {\textstyle{1\over \sqrt{2}}}\,K^t\,i\,\sigma_2 \,\vec \sigma\,K_{\mu \nu} )$ &
$\left(\begin{array}{c}
({\textstyle{1\over \sqrt{3}}}\,\pi\cdot \sigma \, K_{\mu \nu})\\
({\textstyle{1\over \sqrt{3}}}\, \rho_{\mu \nu} \cdot \sigma\,K )\\
(K\, \omega_{\mu \nu})\\
(\eta \,K_{\mu \nu})\\
( K\, \phi_{\mu \nu})
\end{array}\right)$ &
$\left(\begin{array}{c}
(\pi\cdot T \,K_{\mu \nu})\\
(\rho_{\mu \nu} \cdot T\,K\, )
\end{array}\right)$
\\\hline\hline
$(0^+,0)$ &
 $(0^-,0)$ &
 $(1^+,0)$  &
 $(1^-,0)$
 \\ \hline
$\left({\textstyle{1\over 2}}\,( \overline{K}\,
K_{\mu \nu}-\overline{K}_{\mu \nu}\,K) \right)$ &
$\left(\!\!\begin{array}{c}
({\textstyle{1\over \sqrt{3}}}\, \pi\cdot \rho_{\mu \nu}) \\
(\eta \, \omega_{\mu \nu})\\
{\textstyle{1\over 2}}\,( \overline{K}\,
K_{\mu \nu}+\overline{K}_{\mu \nu}\,K) \\
(\eta\, \phi_{\mu \nu})
\end{array}\!\!\right)$&
$\left(\!\!\begin{array}{c}
(\pi\, \omega_{\mu \nu})\\
(\pi\, \phi_{\mu \nu})\\
(\eta \, \rho_{\mu \nu}) \\
{\textstyle{1\over 2}}\,( \overline{K}\,\sigma\,
K_{\mu \nu}+\overline{K}_{\mu \nu}\,\sigma\,K)
\end{array}\!\!\right)$ &
$\left(\!\!\begin{array}{c}
({\textstyle{1\over {i\sqrt{2}}}}\, \pi \times \rho_{\mu \nu}) \\
({\textstyle{1\over 2}}\,(
\overline{K}\,\sigma\, K_{\mu \nu}-\,\overline{K}_{\mu \nu} \,\sigma \,K)
\end{array}\!\!\right)$
\\
\hline\hline
\multicolumn{4}{||c|}{$(2,0)$}
\\ \hline
\multicolumn{4}{||c|}{$\left({\textstyle{1\over 2}}(\pi_i \,\rho^\mu_j
+\pi_j \,\rho^\mu_i) -{\textstyle{1\over 3}}\, \delta_{ij}\,\pi \cdot
\rho^\mu \right)$ }
 \\\hline\hline
\end{tabular}
\caption{
The coupled-channel states of the various sectors characterized  by isospin ($I$), G-parity ($G$)
and strangeness ($S$). We write $(I^G,S)$. The Pauli matrices
$\sigma_i$ act on isospin doublet fields $K,\,K_{\mu \nu}$ and $\bar K,\, \bar K_{\mu \nu}$ with
for instance $K_{\mu \nu} =(K_{\mu \nu}^{+},K^{0}_{\mu \nu})^t$. The 4$\times$2 matrices $T_j$  describe the
transition from isospin-$\frac{1}{2}$ to $\frac{3}{2}$ states. We use the normalization
implied by $\vec T \cdot \vec T^\dagger =1 $ and
$T^\dag_i T_j=\delta_{ij}-{\textstyle{1\over 3}}\,\sigma_i\,\sigma_j $.}
\label{tab:states}
\end{table}

The invariant amplitude $M(s)$ is computed in terms of an effective matrix interaction, $V^{}(s)$, and
a diagonal loop matrix, $J^{}(s)$, with
\begin{eqnarray}
&& M^{}(s) = \Big[ 1- V^{}(s)\,J^{}(s)\Big]^{-1}\,
V^{}(s)\,.
\label{M-generic}
\end{eqnarray}
The matrix structure of the scattering amplitude is defined with respect to states
of well defined isospin, where we apply the convention introduced in \cite{Lutz-Kolomeitsev:2004}. For the readers'
convenience the latter is recalled in Table \ref{tab:states}.
Each diagonal element of the loop matrix depends on the masses of the
pseudo-scalar  and vector meson, $m$ and $M$, that constitute the associated coupled-channel state. It holds
\begin{eqnarray}
&& J(s) = \Bigg(\frac{3}{2}\,M^2+
\frac{1}{2}\,p_{cm}^2\Bigg)\,\Big\{I(s)-I(\mu_M^2)\Big\}\,,
\nonumber\\
&& I(s)=\frac{1}{16\,\pi^2}
\left( \frac{p_{cm}}{\sqrt{s}}\,
\left( \ln \left(1-\frac{s-2\,p_{cm}\,\sqrt{s}}{m^2+M^2} \right)
-\ln \left(1-\frac{s+2\,p_{cm}\sqrt{s}}{m^2+M^2} \right)\right)
\right.
\nonumber\\
&&\qquad \qquad + \left.
\left(\frac{1}{2}\,\frac{m^2+M^2}{m^2-M^2}
-\frac{m^2-M^2}{2\,s}
\right)
\,\ln \left( \frac{m^2}{M^2}\right) +1 \right)+I(0)\;,
\label{i-def}
\end{eqnarray}
where $\sqrt{s}= \sqrt{M^2+p_{cm}^2}+ \sqrt{m^2+p_{cm}^2}$. The values of the matching scale, $\mu_M$, are taken from
\cite{Lutz-Kolomeitsev:2004}. The loop functions are folded with realistic spectral distributions whenever a $\rho$ or a
$K^*$ meson is involved as described in \cite{Lutz-Kolomeitsev:2004}.
It is left to recall the effective interaction $V^{}_{ij}(s)$ implied by (\ref{WT-term}). According to
\cite{Lutz-Soyeur:2007} it holds
\begin{eqnarray}
&&V^{(WT)}_{ij}(s) = \Bigg\{ \Big(\bar M^2+M^2 \Big)\,\Big[
3\,s-M^2-\bar M^2-m^2-\bar m^2
\nonumber\\
&& \quad  -\,\frac{M^2-m^2}{s}\,(\bar M^2-\bar m^2)\Big]
 -2\,(\bar M^2-M^2)(\bar m^2-m^2)\Bigg\}\,\frac{C^{(WT)}_{ij}}{24\,\bar M^2\,f^2\,M^2}\,,
\label{def-Vij-1plus}
\end{eqnarray}
where the matrix $C^{(WT)}_{ij}$ was already specified in \cite{Lutz-Kolomeitsev:2004} (see also Table 2). In
(\ref{def-Vij-1plus}) the parameters $M$ and $\bar M$ denote the masses of the vector
mesons in the initial and final state respectively. The parameters $m$ and
$\bar m$ stand for the masses of initial and final Goldstone bosons. Except for small
SU(3) breaking effects the expression (\ref{M-generic}) is equivalent to the
formal result of \cite{Lutz-Kolomeitsev:2004}. As a consequence the striking success of
\cite{Lutz-Kolomeitsev:2004} in generating the spectrum of low-lying axial vector mesons is
matched.

Since we aim at a computation of radiative decay widths of the
axial-vector mesons generated by coupled-channel dynamics it is important to consider correction terms as to obtain
more realistic resonance masses and hadronic coupling constants. At leading order where no free
parameters are encountered we observe typical discrepancies of (50-100) MeV only \cite{Lutz-Kolomeitsev:2004}.

\subsection{Phenomenology of chiral correction terms}
\label{subsec:chiral-corr}

In this section we construct a phenomenology how to model correction terms for the effective
coupled-channel interaction (\ref{def-Vij-1plus}). In a chiral expansion of the interaction
at subleading order one would consider contributions from the u- and t-channel exchange
processes of the Goldstone bosons and possibly also from the light vector mesons. However, it should be stressed
that  no chiral counting scheme is  established so far, that can be applied to the coupled-channel system of
section \ref{subsec:leading}. In particular the treatment of the u- and t-channel exchanges of the Goldstone bosons is unclear. This
is because the Goldstone bosons in such processes are not necessarily soft always. On the one hand, it is reasonable
to expect that the effect of the exchange of light vector mesons can be mimicked by appropriate local two-body
counter terms. On the other hand, the effect of u- and t-channel exchange processes of the Goldstone bosons
should be considered explicitly. Our request is motivated by the observation that the implied coupled-channel dynamics is
rather complicated with overlapping cut structures and anomalous threshold effects. Unfortunately, a straight forward
application of the on-shell reduction scheme \cite{Lutz:Kolomeitsev:2002} would fail due to close-by left-hand cut
structures.

Thus we follow here a more phenomenological approach where we assume the average strength implied by
the exchange processes of Goldstone bosons and light vector mesons to be modelled by effective
two-body counter terms. The effects of additional inelastic channels, like the vector-vector channels,
are also lumped into such effective interactions. As a guide we use chiral counter terms, which
will be constructed systematically in the following.

We develop chiral correction terms that are relevant for s-wave interactions.
Here we will rely on the standard counting only, which assigns the current quark masses to scale with $ Q_\chi^2$,
$V_{\mu \nu} \sim Q_\chi^0$ and $ \partial_\mu \Phi  \sim Q_\chi$ (see e.g. \cite{Jenkins:1995vb,Scherer:1,Scherer:2}).
In a covariant approach the form of local counter terms is not necessarily unique, i.e. one may choose from various
representations at a given order.  This holds if derivatives acting on the vector fields are counted as
$\partial_\mu  \sim Q_\chi^0$ in contrast to the order $Q_\chi$ they are counted if they act on Goldstone boson fields. The
difference of the possible realizations is, however, suppressed in chiral powers. Our choice will be to construct the terms
that involve the minimal number of derivatives acting on the vector fields.
A more thorough discussion of counting rules will be given in the next section, where we develop a novel expansion scheme
that combines chiral and large-$N_c$ properties of QCD in the context of the hadrogenesis conjecture. The parameter $N_c$
denotes the number of colors.

At order $Q^2_\chi$ there are two types of terms which one has to consider. The ones
that break the chiral SU(3) symmetry explicitly are proportional to the quark-mass matrix of QCD \cite{Cirigliano:2003yq}.
It holds
\begin{eqnarray}
&& {\mathcal L}^{(2)}_{\chi -SB}=
\frac{1}{8}\, b_0\,{\tr } \,\Big\{V^{\mu \nu}\,V_{\mu \nu}  \Big\}\,{ \tr }  \Big\{\chi_+ \Big\}
+\frac{1}{8}\, b_1\,{\tr } \,\Big\{V_{\mu \nu}  \Big\}\,{ \tr }  \Big\{V^{\mu \nu}\,\chi_+ \Big\}
\nonumber\\
&& \qquad + \frac{1}{8}\, b_2\,{\tr } \,\Big\{V_{\mu \nu}  \Big\}\,{ \tr }  \Big\{V^{\mu \nu} \Big\}
\,{ \tr }  \Big\{\chi_+ \Big\}
+ \frac{1}{8}\,b_D\,{\tr } \,\Big\{V^{\mu \nu}\,V_{\mu \nu} \, \chi_+ \Big\} \,,
\nonumber\\
&& \chi_\pm  = \frac{1}{2}\,u\,\chi_0 \,u \pm \frac{1}{2}\,u^\dagger\,\chi_0 \,u^\dagger
\,,
\label{def-chi-terms}
\end{eqnarray}
where the field $u$ was introduced in (\ref{eq:defcovder}) and
\begin{eqnarray}
\chi_0 &=& 2\,B_0 \, \left(
\begin{array}{ccc}
m_u & 0 & 0\\
0 & m_d & 0 \\
0 & 0 & m_s
\end{array}
\right) =
\left(
\begin{array}{ccc}
m_\pi^2 & 0 & 0\\
0 & m_\pi^2 & 0 \\
0 & 0 & 2 \,m_K^2 - m_\pi^2
\end{array}
\right)
\,.
\label{GMOR}
\end{eqnarray}
At leading order the diagonal matrix $\chi_0$ can be expressed in terms of the pion and kaon masses as
indicated in (\ref{GMOR}). Isospin breaking effects are neglected.
The parameters $b_{0,1,2}$ and $b_D$ can for instance be determined from  the
pion and kaon mass dependencies of the $\phi$- and $K^*$-meson masses as measured
in unquenched QCD lattice simulations. Unfortunately at present this is not possible due to a lack of
simulation data. Instead we consider further constraints
from QCD as they arise in the limit of a large number of
colors $N_c$ \cite{large-N_c-reference}. From large-$N_c$ arguments one would expect
the magnitude of $b_{0,1,2}$ to be somewhat smaller than
that of $b_D$ \cite{large-N_c-reference}. This follows, since $b_{0,1,2}$ are coefficients of an interaction
term that involves  a double trace in flavor space. The latter are suppressed by $1/N_c$ as compared
to single-flavor trace interactions, i.e. as is the case for the term proportional to $b_D$. Note that the
parameters $b_1$ and
$b_2$ imply a mixing of the $\omega$ and $\phi$ meson. The fact that such mixing effects are suppressed empirically
supports the above argument that the latter parameters should be small. We will assume $b_0=b_1=b_2=0$ in the following.

The parameter $b_D$ can be estimated from the mass splitting of the vector mesons. From
\begin{eqnarray}
&& m_\rho^2 \;\,\,\,\!=\, m^2_\omega =  m^2_{1^-}  +b_D\,m_\pi^2\,,
\nonumber\\
&& m^2_{K^*} = \,m^2_{1^-}  + b_D\,m_K^2 \,, \qquad
\nonumber\\
&& m_\phi^2 \;\,\,=\,m^2_{1^-}  + b_D\,(2\,m_K^2-m_\pi^2) \,,
\end{eqnarray}
it follows
\begin{eqnarray}
b_D = 0.92 \pm 0.05 \,.
\label{value-bD}
\end{eqnarray}
The fact that from the two possible mass differences $m^2_{K^*}- m_\rho^2$ and $m_\rho^2-m_\phi^2$ one
derives a universal parameter $b_D$ with a spread of about 5$\%$ only, is amazingly
consistent with the  assumption $b_1=b_2=0$ \cite{Cirigliano:2003yq}.

\begin{table}
\begin{tabular}{||c||c|c|c|c|c|c||}
\hline\hline
$(I^G,S)_{ij}$ & $ C^{(WT)}_{ij}  $  & $ C^{(D)}_{ij}  $ & $ C^{(F)}_{ij}  $ &  $ C^{(D)}_{\chi,ij} $\\
\hline\hline
$(1^+,0)_{ 1, 1}$ & $ 0$ &  $ 16$ & $ 0$ &  $ 8\,m_\pi^2 $ \\
$(1^+,0)_{ 2, 1}$ & $ 0$ &  $ 0$ & $ 0$ &  $ 0$ \\
$(1^+,0)_{ 2, 2}$ & $ 0$ &  $ 0$ & $ 0$ &  $ 0$ \\
$(1^+,0)_{ 3, 1}$ & $ 0$ &  $ 16/\sqrt{3}$ & $ 0$ &  $ (8\,m_\pi^2)/\sqrt{3} $ \\
$(1^+,0)_{ 3, 2}$ & $ 0$ &  $ 0$ & $ 0$ &  $ 0$ \\
$(1^+,0)_{ 3, 3}$ & $ 0$ &  $ 16/3$ & $ 0$ &  $ (8\,m_\pi^2)/3$ \\
$(1^+,0)_{ 4, 1}$ & $ 1$ &  $ 12$ & $ -4$ &  $ 2\,(m_K^2 + m_\pi^2)$ \\
$(1^+,0)_{ 4, 2}$ & $ -\sqrt{2}$ & $ 4\,\sqrt{2}$ & $ 4\,\sqrt{2}$ &  $ 2\,\sqrt{2}\,(m_K^2 + m_\pi^2)$ \\
$(1^+,0)_{ 4, 3}$ & $ \sqrt{3}$ &  $ 4/\sqrt{3}$ & $ -4\,\sqrt{3}$ &  $ (2\,(3\,m_\pi^2 - 5\,m_K^2))/\sqrt{3}$ \\
$(1^+,0)_{ 4, 4}$ & $ 1$ &  $ 12$ & $ -4$ &  $ 4\,m_K^2$ \\ \hline
$(1^-,0)_{ 1, 1}$ & $ 2$ &  $ -8$ & $ 24$ &  $ 8\,m_\pi^2 $ \\
$(1^-,0)_{ 2, 1}$ & $ \sqrt{2}$ &  $ -4\,\sqrt{2}$ & $ 12\,\sqrt{2}$ &  $ 2\,\sqrt{2}\,(m_K^2 + m_\pi^2) $ \\
$(1^-,0)_{ 2, 2}$ & $ 1$ &  $ -4$ & $ 12$ &  $ 4\,m_K^2 $ \\ \hline
$(2^{\phantom{+}},0)_{ 1, 1}  $ & $-2$ &  $ 8$ & $ 8$ &  $ 8\,m_\pi^2 $ \\ \hline
$(0^{\phantom{+}},2)_{1, 1}$ & $ 0$ &  $ 0$ & $ 0$ &  $ 0 $\\ \hline
$(1^{\phantom{+}},2)_{1, 1}$ & $ -2$ &  $ 8$ & $ 8$ &  $ 8\,m_K^2 $\\ \hline
\end{tabular}
\caption{The coefficients $C^{(WT)}_{ij}, C^{(D)}_{ij}, C^{(F)}_{ij}$ and $C^{(D)}_{\chi,ij}$
that characterize the  interaction of Goldstone bosons with vector mesons
as introduced in (\ref{WT-term}, \ref{def-chi-terms}, \ref{def-g-terms}) and (\ref{def-Vij-1plus}, \ref{def-correction-Vij}).
The coefficients are listed with respect to the channel labelling of Table \ref{tab:states}.}
\label{tab:Clebsches1}
\end{table}

\begin{table}
\begin{tabular}{||c||c|c|c|c|c|c||}
\hline\hline
$(I^G,S)_{ij}$ & $ C^{(WT)}_{ij}  $  & $ C^{(D)}_{ij}  $ & $ C^{(F)}_{ij}  $ &  $ C^{(D)}_{\chi,ij} $\\
\hline\hline
$(\frac{1}{2},1)_{ 1, 1}$ & $ 2$ & $4$ & $4$ &  $ 4\,m_\pi^2$ \\
$(\frac{1}{2},1)_{ 1, 2}$ & $ \frac{1}{2}$ & $ 10$ & $ -14$ &  $ -(m_K^2+m_\pi^2) $ \\
$(\frac{1}{2},1)_{ 2, 2}$ & $ 2$ &  $ 4$ & $ 4$ &  $ 4\,m_K^2 $ \\
$(\frac{1}{2},1)_{ 3, 1}$ & $ -\frac{\sqrt{3}}{2}$ &  $ 6\,\sqrt{3}$ & $ -2\,\sqrt{3}$ &  $ \sqrt{3}\,(m_K^2 + m_\pi^2) $ \\
$(\frac{1}{2},1)_{ 3, 2}$ & $ 0$ & $ 4\,\sqrt{3}$ & $ 4\,\sqrt{3}$ &  $ 4\,\sqrt{3}\,m_K^2 $ \\
$(\frac{1}{2},1)_{ 3, 3}$ & $ 0$ &  $ 4$ & $ 4$ &  $ 4\,m_K^2 $ \\
$(\frac{1}{2},1)_{ 4, 1}$ & $ 0$ &  $ -4$ & $ 12$ &  $ 4\,m_\pi^2 $ \\
$(\frac{1}{2},1)_{ 4, 2}$ & $ -\frac{3}{2}$ &  $ 2$ & $ -6$ &  $ 3\,m_\pi^2 - 5\,m_K^2 $ \\
$(\frac{1}{2},1)_{ 4, 3}$ & $ -\frac{\sqrt{3}}{2}$ &  $ 2/\sqrt{3}$ & $ -2\,\sqrt{3}$ &  $ (3\,m_\pi^2 - 5\,m_K^2)/\sqrt{3} $ \\
$(\frac{1}{2},1)_{ 4, 4}$ & $ 0$ &  $ 4/3$ & $ 12$ &  $  (32\,m_K^2)/3 - 4\,m_\pi^2 $ \\
$(\frac{1}{2},1)_{ 5, 1}$ & $ \sqrt{\frac{3}{2}}$ &  $ 2\,\sqrt{6}$ & $ 2\,\sqrt{6}$ &  $ \sqrt{6}\,(m_K^2 + m_\pi^2) $ \\
$(\frac{1}{2},1)_{ 5, 2}$ & $ 0$ &  $ 4\,\sqrt{6}$ & $ -4\,\sqrt{6}$ &  $ 0 $ \\
$(\frac{1}{2},1)_{ 5, 3}$ & $ 0$ &  $ 4\,\sqrt{2}$ & $ -4\,\sqrt{2}$ &  $ 0 $ \\
$(\frac{1}{2},1)_{ 5, 4}$ & $ \sqrt{\frac{3}{2}}$ &  $ -10\,\sqrt{\frac{2}{3}}$ & $ 2\,\sqrt{6}$ &  $ \sqrt{\frac{2}{3}}\,(3\,m_\pi^2 - 5\,m_K^2) $ \\
$(\frac{1}{2},1)_{ 5, 5}$ & $ 0$ &  $ 8$ & $ 8$ &  $ 8\,m_K^2 $ \\ \hline
$(\frac{3}{2},1)_{ 1, 1}$ & $ -1$ &  $ 4$ & $ 4$ &  $ 4\,m_\pi^2 $ \\
$(\frac{3}{2},1)_{ 2, 1}$ & $ -1$ &  $ 4$ & $ 4$ &  $ 2\,(m_K^2 + m_\pi^2) $ \\
$(\frac{3}{2},1)_{ 2, 2}$ & $ -1$ &  $ 4$ & $ 4$ &  $ 4\,m_K^2 $ \\ \hline
$(0^+,0)_{ 1, 1}$ & $ 3$ &  $ -12$ & $ 36$ &  $ 12\,m_K^2 $ \\ \hline
$(0^-,1)_{ 1, 1}$ & $ 4$ &  $ 32$ & $ -16$ &  $ 8\,m_\pi^2 $ \\
$(0^-,1)_{ 2, 1}$ & $ 0$ &  $ 16$ & $ 0$ &  $ 8\,m_\pi^2$ \\
$(0^-,1)_{ 2, 2}$ & $ 0$ &  $ 16/3$ & $ 0$ &  $ (8\,m_\pi^2)/3$ \\
$(0^-,1)_{ 3, 1}$ & $ \sqrt{3}$ &  $ 12\,\sqrt{3}$ & $ -4\,\sqrt{3}$ &  $ 2\,\sqrt{3}\,(m_K^2 + m_\pi^2)$ \\
$(0^-,1)_{ 3, 2}$ & $ \sqrt{3}$ &  $ 4/\sqrt{3}$ & $ -4\,\sqrt{3}$ &  $ (2\,(3\,m_\pi^2 - 5\,m_K^2))/\sqrt{3}$ \\
$(0^-,1)_{ 3, 3}$ & $ 3$ &  $ 36$ & $ -12$ &  $ 12\,m_K^2$ \\
$(0^-,1)_{ 4, 1}$ & $ 0$ &  $ 0$ & $ 0$ &  $ 0$ \\
$(0^-,1)_{ 4, 2}$ & $ 0$ &  $ 0$ & $ 0$ &  $ 0$ \\
$(0^-,1)_{ 4, 3}$ & $ -\sqrt{6}$ &  $ -20\,\sqrt{\frac{2}{3}}$ & $ 4\,\sqrt{6}$ &  $ 2\,\sqrt{\frac{2}{3}}\,(3\,m_\pi^2 - 5\,m_K^2)$ \\
$(0^-,1)_{ 4, 4}$ & $ 0$ &  $ 64/3$ & $ 0$ &  $ (32/3)\,(2\,m_K^2 - m_\pi^2)$ \\ \hline
\end{tabular}
\caption{Continuation of Table \ref{tab:Clebsches1}.}
\label{tab:Clebsches2}
\end{table}

The second class of terms we have to consider respects the chiral SU(3) symmetry. We detail
terms only that are relevant for s-wave interactions of vector mesons with Goldstone bosons.
The following form is used in this work
\begin{eqnarray}
&&{\mathcal L}^{(2)}_{s-wave}= \frac{1}{4}\,g_D\,{\tr } \,\Big\{V^{\mu \nu}\,\big[ V_{\mu \nu }\,,U_\alpha \big]_+\,U^\alpha \Big\}
+\frac{1}{4}\,g_F\,{\tr } \,\Big\{V^{\mu \nu}\,\big[ V_{\mu \nu }\,,U_\alpha \big]_-\,U^\alpha \Big\}
\nonumber\\
&& \qquad +\,\frac{1}{4}\,g_0\,\,{\tr } \,\Big\{V^{\mu \nu}\,V_{\mu \nu }\Big\}\,{\tr } \,\Big\{U_\alpha \,U^\alpha \Big\}
+\frac{1}{4}\,g_1\,\,{\tr } \,\Big\{V^{\mu \nu}\,U_\alpha\Big\}\,{\tr } \,\Big\{V_{\mu \nu } \,U^\alpha \Big\}
\nonumber\\
&& \qquad +\,\frac{1}{4}\,g_2\,\,{\tr } \,\Big\{V^{\mu \nu}\Big\}\,
{\tr } \,\Big\{V_{\mu \nu }\Big\}\,{\tr } \,\Big\{U_\alpha \,U^\alpha \Big\}
\nonumber\\
&& \qquad+\,\frac{1}{4}\,g_3\,\,{\tr } \,\Big\{V^{\mu \nu}\,U_\alpha\,U^\alpha\Big\}\,{\tr } \,\Big\{V_{\mu \nu }  \Big\}\,,
\label{def-g-terms}
\end{eqnarray}
where the field combination $U_\mu$ transforms like the vector meson field, $V_{\mu \nu}$, under
chiral transformations (see e.g. \cite{Krause:1990}), i.e.
\begin{eqnarray}
&&U_\mu = \frac{1}{2}\, u^\dagger \,\Big(\Big(\partial_\mu e^{i\,\frac{\Phi}{f}}\Big) + i\,e\,A_\mu\,
\Big[Q,e^{i\,\frac{\Phi}{f}}\Big]_- \Big)\,u^\dagger  \,,
\nonumber\\
&& D_\mu \,U_\nu =
\partial_\mu\,U_\nu +\Big[\Gamma_\mu ,\,U_\nu \Big]_-
+i\,e\,A_\mu\,\Big[Q,\, U_\nu \Big]_- \,,
\label{def-Umu}
\end{eqnarray}
with $\Gamma_\mu$ introduced in (\ref{eq:defcovder}).

Note that due to large-$N_c$ arguments, analogous to the ones applied above, one would expect the
magnitudes of the parameters $g_{0-3}$ to be smaller than those of $g_D$ and $g_F$. We point out that
owing to such arguments one would expect that any of u- or t-channel exchange has a flavor
SU(3) structure that can be linearly combined from the terms proportional to $g_D$ and $g_F$. Here we consider
three-point vertices that are leading in large-$N_c$ only. Recall that any such tree-level contribution from
meson exchange processes is leading in the large-$N_c$ world. This is an important observation since it justifies
the neglect of the terms $g_{0-3}$ even in our phenomenological approach in
which we do not resolve the detailed cut structure implied by the various meson exchange contributions. We argue
that the average strength can be represented by the counter interactions proportional to $g_D$ and $g_F$ at least
in an SU(3) symmetric world. Thus in this work we assume $g_{0}=g_{1}=g_{2}=g_{3}=0$ but use $g_D$ and $g_F$ to
tune some details of the axial-vector spectrum.

Utilizing the results of \cite{Lutz-Kolomeitsev:2004} it is a straightforward exercise to derive the
contribution of the terms (\ref{def-chi-terms}, \ref{def-g-terms}) to the effective interaction kernel. We obtain
\begin{eqnarray}
&& V_{ij}(s) = V^{(WT)}_{ij}(s)
+\frac{1}{12\,\bar M\,f^2\,M}\, b_D\,C^{(D)}_{\chi, ij}
\nonumber\\
&& \quad +\frac{(s-\bar M^2+\bar m^2)\,(s-M^2+m^2)}{48\,s\,\bar M\,f^2\,M}\,
\Big\{  g_D\,C^{(D)}_{ij} + g_F\,C^{(F)}_{ij} \Big\}\,,
\label{def-correction-Vij}
\end{eqnarray}
where the matrices $V^{(WT)}_{ij}$, $C^{(D,F)}_{\chi, ij}$ and $C^{(D,F)}_{ij} $ are specified in
Tables \ref{tab:Clebsches1}-\ref{tab:Clebsches2} with respect to the channel convention of Table \ref{tab:states}.

If in a given strangeness and isospin
sector a resonance of mass $M_{1^+}$ is formed, the scattering amplitude of (\ref{M-generic}), close to
the resonance structure, has the form:
\begin{eqnarray}
M_{ij}(s) \simeq - \frac{2}{3\,\bar M_i\,M_j}\,\frac{2\,g_i^*\,M^2_{1^+}g_j}{s-M_{1^+}^2+i\,\Gamma_{1^+}\,M_{1^+}}\,,
\label{def-gi-1plus}
\end{eqnarray}
with the dimensionless coupling constants $g_i$ and the hadronic width parameter $\Gamma_{1^+}$.
We provide the values for the resonance-coupling constants, $g_i$,
as extracted for axial-vector molecules. They determine the partial decay widths
\begin{eqnarray}
\Gamma_i = \frac{|g_i|^2}{4\,\pi}\,\left( 1+ \frac{p_{cm}^2}{3\,M_i^2}\right) p_{cm} \,,
\label{partial-decay-width}
\end{eqnarray}
and  play a central role in the determination of the radiative widths
of the axial-vector molecules.
As a consequence of the coupled-channel interaction (\ref{def-correction-Vij})
the $f_1(1282)$, $b_1(1230)$, $h_1(1386)$, $a_1(1230)$ and $K_1(1272)$ states are dynamically generated.

While we use a universal $b_D = 0.92$ as suggested by the mass splitting of the vector mesons,
the two parameters $g_D$ and $g_F$ are adjusted to reproduce the empirical masses and branching ratios.
There is no universal choice of the two parameters that reproduces all masses and widths simultaneously, though in
a given sector the two parameters suffice to reproduce the properties of a selected resonance.
We take this as a hint that one better treats the u- and t-channel processes of Goldstone bosons explicitly.
Once this is achieved we would expect universal values for the two parameters $g_D$ and $g_F$.
In Table \ref{tab:hadronic-decay} we collect the results obtained for the various sectors with isospin ($I$),
strangeness ($S$) and G-parity ($G$).  The parameters of Table \ref{tab:hadronic-decay} are obtained by Breit-Wigner
fits to the scattering amplitudes, where we quote the mass and width as extracted from the dominant channel. The coupling
constants of subleading channels are determined by an analysis of the transition amplitudes to the dominant channel.

It is pointed out that in some cases the properties of the axial-vector molecules are quite sensitive to
the spectral distributions of the $\rho$ and $K^*$ mesons. While in the present and in the original
work \cite{Lutz-Kolomeitsev:2004}
realistic spectral distributions of the $\rho$ and $K^*$ mesons are used, the
follow-up work \cite{Roca:Oset:Singh:2005}
relied on Breit-Wigner parameterizations only. Though the latter facilitates the determination of poles on higher
unphysical Riemann sheets, it does bring in further uncertainties. Since the search for poles on unphysical Riemann sheets
requires additional numerical effort  we refrain from doing so at a level where the uncertainties of the
coupled-channel dynamics are certainly larger than the ones extracting resonance properties from the physical
scattering amplitude as accessible in experiment.

\begin{table}
\begin{center}
\begin{tabular}{|c||c|c|c|c|c|}
\hline
$(I^G,S)$ & $(0^+,0)$ & $(1^+,0)$ & $(0^-,0)$ & $(1^-,0)$ & $(\frac{1}{2},1)$\\
  & $f_1(1282)$ & $b_1(1230)$ & $h_1(1386)$ & $a_1(1230)$  & $K_1(1272) $ \\
\hline \hline
M$_R$ [MeV]   & 1282      & 1230       & 1386      & 1260    &  1272       \\
\hline
$\Gamma$ [MeV]& 4         & 142        & 45        & 500     &  63    \\
\hline \hline
$g_1$         & $+4.6$    & $+2.1$     & $+0.1$    & $+3.8$  & $+0.4 $    \\
\hline
$g_2$         & -         & $-1.0$     & $+0.9$    & $+2.6$  & $+2.9$     \\
\hline
$g_3$         & -         & $+2.3$     &  $+2.5$   &-        & $+1.1$        \\
\hline
$g_4$         & -         & $+4.2$     & $-2.0$    &-        & $-2.5$        \\
\hline
$g_5$         & -         & -          & -         &-        &   $+0.1$       \\
\hline \hline
$g_D$         & $-0.3$    & $+0.7$     & $-0.8$    & 0       &   $+0.2$       \\
\hline
$g_F$         & $+0.4$    & $-2.8$     & $+0.8$    & 0       &   $-0.1$       \\
\hline
\end{tabular}
\vglue 0.3 truecm
\caption{Masses, widths and coupling constants for dynamically generated $1^+$ states. The coupling constants
$g_i$ are given in the isospin basis as specified in Table \ref{tab:states}.
We use $f=90$ MeV and $b_D=0.92$.}
\label{tab:hadronic-decay}
\end{center}
\end{table}

We discuss the various sectors.
The mass of the $f_1(1282)$ can be reproduced with the choice $g_D=-0.3$ and $g_F=0.4$. The resulting width
of about 4 MeV is somewhat larger than expected from \cite{PDG:2006}, which gives an upper bound of
$(9.0\pm 0.4)\,\%$ into the $K\,\bar K \,\pi$ channel and a total width of $(24.2 \pm 1.1)$ MeV.
Within the present scheme it is, however,  impossible to reduce the width. The properties of the $f_1(1282)$ molecule
depend on the parameter combination $3\,g_F-g_D \simeq 1.5$ only. The particular choice in Table \ref{tab:hadronic-decay}
is motivated by the requirement that the moduli of the two parameters $g_D$
and $g_F$ are minimal. The counter terms pull down the resonance mass by about 70 MeV to its empirical value.

The mass and total width of the $b_1(1230)$ molecule can be reproduced by the choice $g_D = 0.7$ and $g_F=-2.8$.
The rather large value for $g_F$ is needed to reduce the width of the $b_1(1230)$ down to the value of 142 MeV
quoted in Table \ref{tab:hadronic-decay}. According to the Particle Data Group \cite{PDG:2006} the dominant decay
mode is the $\pi \,\omega$ channel. Though the total width  is known quite accurately with $(142\pm 9)$ MeV, information
on partial decay widths are scarce. The effect of the counter terms is to pull down the mass by about 100 MeV
to its empirical value.

The mass of the $h_1(1380)$ molecule is reproduced with $g_F = -g_D=0.8$. The resulting width of 45 MeV is about half
of the total width $(91\pm 30)$ MeV quoted in \cite{PDG:2006}. The Particle Data Group \cite{PDG:2006} acknowledges
that the dominant decay product is $\bar K\,K^*+ K\,\bar K^*$, but leaves undetermined the branching fractions. The
parameter choice is not unique. The particular values were selected by the requirement that the moduli of the two
parameters $g_D$ and $g_F$ are minimal. The counter terms push up the resonance by about 75 MeV to its empirical value.

The properties of the $a_1$ molecule are most uncertain, due to its large width. Accordingly the
values quoted in Table \ref{tab:hadronic-decay} for the mass, width and coupling parameters should be taken
cautiously. As was pointed out
in \cite{Wagner:Leupold:2007} the  properties of the $a_1$ molecule
are best constrained by an analysis of the $\tau$ decay. Since already the leading order result of
\cite{Lutz-Kolomeitsev:2004} was shown to allow an amazingly accurate description of the
$a_1$ mass distribution as seen in the $\tau$ decay, we do not use non-zero values for $g_D$ and $g_F$ in this case.
The effect of the finite $b_D= 0.92$ on the $a_1$ properties is very minor.

The properties of the $K_1(1272)$ molecule can be reproduced by the choice $g_D = 0.2$ and $g_F=-0.1$.
Here we aim at a width of about $(69 \pm 14)\%$ of the total empirical width of $(90 \pm 20)$ MeV. According to
\cite{PDG:2006} the dominant decay  with a branching fraction of $(42\pm 6 )\%$ goes into the $K\,\rho$ channel.
This appears roughly compatible with the coupling constants listed in Table \ref{tab:hadronic-decay}.
The counter terms pull down the resonance by about 40 MeV to its empirical value.

\clearpage

\section{Power counting in the hadrogenesis conjecture}
\label{sec:counting}

In order to compute the radiative decays of the axial-vector molecules generated in the last
section we need to couple the photon field to the hadronic degrees of freedom, the molecules are made of. The kinetic term
of the vector mesons (\ref{kinetic-term}) involves the photon field $A_\mu$ (\ref{eq:defcovder}).
It contains a term providing the standard coupling of the photon to the charge of the vector mesons
\begin{eqnarray}
&&{\mathcal L}_{\rm e.m.} = i\,\frac{e}{2}\,A^\mu\, {\tr} \,
\Big\{(\partial_\mu \Phi)\,\Big[Q,\,\Phi\Big]_-\Big\} -\,i\,\frac{e}{2}\,A^\mu\,{\tr }\,
\Big\{\Big[Q,\,V_{\mu \nu}\Big]_-\, (\partial_\tau V^{\tau \nu} )\Big\}\,,
\label{em-kinetic}
\end{eqnarray}
where we include a corresponding term for the Goldstone
bosons. The latter is extracted from the kinetic term of the Goldstone bosons as expressed for convenience
in terms of the field combination, $U_\mu$, of (\ref{def-Umu}) with
\begin{eqnarray}
{\mathcal L}_{\rm kin} = f^2\, \tr \Big\{ U^\mu\,U^\dagger_\mu \Big\}\,,
\label{def-kin-Goldstone}
\end{eqnarray}
where upon expansion in the fields $\Phi$ and $A_\mu$ one recovers the first term on the r.h.s. of (\ref{em-kinetic}).

It is important to realize that given the interactions
(\ref{WT-term}, \ref{em-kinetic}) only, the electromagnetic decay amplitudes
$1^+ \to \, \gamma \,0^-, \gamma\,1^-$ vanish identically. These decay processes probe
3-point hadronic vertices, which will be developed in the following.

\subsection{Chiral and large-$N_c$ counting with vector mesons}
\label{subsec:counting}

We construct the relevant hadronic interaction vertices. As a starting point consider the following list
\begin{eqnarray}
{\mathcal L} &=& i\,\frac{m_V\,h_V}{4}\,{\rm tr}\,\Big\{
V_{\alpha \mu}\,V^{\mu \nu}\,V^{\alpha}_{\;\;\, \nu} \Big\}
+i\,\frac{\tilde h_V}{4\,m_V}\,{\rm tr}\Big\{
(D^\alpha V_{\alpha \mu})\,V^{\mu \nu}\,(D^\beta V_{\beta \nu}) \Big\}
\nonumber\\
&+& i\, \frac{h_{A}}{8}\,\epsilon^{\mu\nu\alpha\beta}\, {\rm tr}\,\Big\{
\Big(V_{\mu \nu}\,(D^\tau V_{\tau \alpha})+(D^\tau V_{\tau \alpha})\,V_{\mu \nu} \Big)\,U_\beta\Big\}
\nonumber\\
&+& i\,\frac{m_V\, h_{P}}{2}\,{\rm tr}\,\Big\{U_\mu\,V^{\mu\nu}\,U_\nu\Big\}
+ i\, \frac{b_A}{8} \, \epsilon^{\mu\nu\alpha\beta}\, {\rm tr}\,
\Big\{\Big[V_{\mu\nu},\,V_{\alpha \beta}\Big]_+ \, \chi_- \Big\} \,,
\label{interaction-tensor}
\end{eqnarray}
in terms of the dimensionless parameters $h_P, h_V, \tilde h_V, h_A$ and $b_A$ and the parameter $m_V$, which  carries
dimension mass.  The notation of the manuscript discriminates the mass of the light vector mesons $m_{1^-}$  (see (\ref{kinetic-term})) and the scale parameter $m_V$, to which no counting power will be assigned.

For the anti-symmetric tensor $\epsilon_{\mu \nu \alpha \beta}$ we use the
convention of \cite{Lutz-Soyeur:2007}.
Additional structures with four or more vector fields are not contributing in a one-loop computation of the
decay processes we are concerned about in this work. Therefore such terms are not included in (\ref{interaction-tensor}).
Moreover, terms similar to the ones introduced in (\ref{interaction-tensor})
which involve more than one flavor trace are neglected. The reason being that such terms
are suppressed in a large-$N_c$ world. The object $\chi_-$ was introduced in (\ref{def-chi-terms}).

We consider only terms in (\ref{interaction-tensor}) with up to two derivatives as they constitute the leading
elements in a systematic derivative expansion of the fields. If one assigned formal chiral powers by
\begin{eqnarray}
V_{\mu \nu} \sim Q_\chi^0\,, \qquad U_\mu \sim Q_\chi\,, \qquad    D_\mu \sim Q_\chi \,, \qquad \chi_\pm \sim Q_\chi^2 \,,
\label{power-counting}
\end{eqnarray}
the terms of (\ref{interaction-tensor})
are the ones that are relevant to chiral order $Q^2_\chi$. Note that (\ref{power-counting}) counts the covariant derivative
as order $Q_\chi$ irrespective whether it acts on Goldstone boson or vector-meson fields.
At first sight such a counting is troubled by the fact that
the empirical vector-meson masses are not really smaller than the chiral symmetry-breaking scale $4\,\pi f \simeq 1.13$ GeV.
This makes a scale separation difficult once decaying or virtual vector mesons are encountered.
However, we would argue that at least as we increase the number of colors $N_c$ away from the physical value
$N_c =3$ such a scale separation is possible. This follows since the chiral symmetry breaking scale
$4\, \pi f $ is proportional to the square root of $N_c$, whereas the pseudo-scalar  and vector-meson masses approach
finite values as $N_c$ turns large  \cite{large-N_c-reference,Harada:2003jx}. To this extent we would argue that the
power counting rules (\ref{power-counting}) do not reflect so much a clear scale separation, rather than at least in part
reflect a realization of a perturbative expansion in the small parameter $1/N_c$.

How about loop effects in the proposed scheme? Given a diagram with $L$ loops the counting
rules (\ref{power-counting}) suggest the formal power $Q^\nu$ with
\begin{eqnarray}
\nu = 4\,L-2\,I_V-2\,I_G + \sum_i\,V_i\,d_i \,,
\label{def-formal-power}
\end{eqnarray}
where $I_V$ and $I_G$ denote the number of internal vector and Goldstone boson lines. In (\ref{def-formal-power}) we sum over
the number of vertices $V_i$ of type $i$ with their associated number of derivatives $d_i$. Applying the well known
topological identity $L=I_G+I_V-\sum_i V_i+1$ the formal power can be rewritten as follows
\begin{eqnarray}
\nu = 2+2\,L + \sum_i\,V_i\,\Big\{d_i-2\Big\} \,.
\label{def-formal-power-rewrite}
\end{eqnarray}
The formula (\ref{def-formal-power-rewrite}) requires a detailed discussion. In a theory without explicit vector
mesons (\ref{def-formal-power-rewrite}) is nothing but the celebrated power-counting rule of standard chiral
perturbation theory. Since any vertex $V_i$ comes with at least two derivatives we have $d_i\geq 2$ and consequently
the larger $L$ is the larger the chiral index $\nu$. This implies that in a chiral expansion multiple-loop effects
are small. The situation is different once vector mesons are considered. In this case there is an infinite tower of
vertices with $d_i=0$. An example is given by the interaction of (\ref{interaction-tensor}) proportional to $h_V$.
As a consequence, to a given order an infinite number of diagrams would contribute. Fortunately there is an additional
element at our disposal that will remedy this problem.

According to large-$N_c$ QCD a vertex with $n$ mesons scales with $N_c^{1-n/2}$, i.e. the larger the number of
mesons involved in a vertex the more its size is suppressed as the number of colors
gets large \cite{large-N_c-reference}. From this we learn that besides the counting of derivatives we have to supply (\ref{power-counting}) with an additional
rule that counts the number of vector meson fields in a given vertex. We identify $Q_\chi \sim N_c^{-1/2}$ and
assign an extra power
\begin{eqnarray}
Q^{n_V-2}_\chi \qquad {\rm for} \qquad n_V \geq 2 \,,
\label{extra-power}
\end{eqnarray}
to an interaction involving $n_V$ vector fields\footnote{Such an
identification appears natural to the authors since the leading two-body interaction
of the Goldstone bosons scales with $1/f^2 \sim 1/N_c$ (see (\ref{def-kin-Goldstone})). According to chiral counting rules this terms carries
chiral power $Q^2_\chi$.}. Since for $n_V \geq 4$ the formal power turns larger or equal two it
follows that such vertices are not causing any trouble in the loop expansion (see (\ref{def-formal-power-rewrite})).
The cases $n_V =0,1,2,3$ are discussed separately.

A vertex with $n_V=0$ does not involve any vector mesons and
therfore no modification of (\ref{power-counting}) is required. In the absence of electromagnetic fields  a vertex
with $n_V=1$ involves at least two Goldstone boson fields which bring in two derivatives. An example is
the term proportional to $h_P$ in (\ref{interaction-tensor}). Again there is no reason for a modification of
(\ref{power-counting}). The same holds for terms with $n_V=2$. This follows since we count the vector meson mass
term as order $m_{1^-}^2\sim Q^2_\chi$.

The case $n_V=3$ is less transparent. In fact, there is a single term only
that causes complications and, therefore, requires a detailed discussion. It is the first term in
(\ref{interaction-tensor}) proportional to $h_V$. All other terms with $n_V=3$ have a chiral index equal
or larger than two and we do not see any reason not to apply (\ref{power-counting}, \ref{extra-power}).
Applying (\ref{extra-power}) to the troublesome term would give it a chiral index $\nu =1$. As a
consequence the 3-vector meson vertex would be non-perturbative. For instance the one-loop
vertex proportional to $h_V^3$ would receive the same chiral index $\nu =1$ as the tree-level contribution.
As a consequence the leading order  3-point vector vertex had to be
determined as a solution of an in-homogenous and non-linear integral equation: the driving term would be the
tree-level vertex to which its one-loop contribution had to  be added where the bare vertices in the
loop are replaced by the full vertex. At first sight it is unclear how to deal with
this challenge. However, in view of the fact that large-$N_c$ counting rules would predict loop corrections
to be perturbative always, an easy remedy would be to increase the chiral index of that troublesome term from one to two.
As a consequence the 3-point vector vertex turns perturbative as expected from large-$N_c$ arguments.

It may be useful to connect to a successful phenomenology that considers
the light vectors mesons as non-abelian gauge bosons of a hidden gauge symmetry \cite{Bando,Birse:1996}. The latter suggests a
correlation of the parameters $\tilde h_V$ and $h_P$, i.e.
\begin{eqnarray}
\tilde h_V = \frac{h_P\,m_V^2}{4\,f^2}   \,,
\label{value-hV}
\end{eqnarray}
but does not predict a vertex proportional to $h_V$. This observation supports the increase of the chiral index of
the $h_V$-vertex. As a consequence we will treat the first two terms
of (\ref{interaction-tensor}) on equal footing in this work.  Formally one may
justify this by promoting the second and
demoting the first term in (\ref{interaction-tensor}) by one power. Then both terms would carry the same formal power
$Q^2_\chi$ and a perturbative treatment would suffice.
The magnitudes of the coupling constants $h_V$ and
$\tilde h_V$ will be constrained by the magnetic and
quadrupole moments of the vector mesons.

Additional electromagnetic interaction terms involve the field strength tensor
$F_{\mu \nu}= \partial_\mu \,A_\nu - \partial_\nu\,A_\mu$. Such terms are a crucial element of effective
field theories. They bring in the physics of processes where the photon couples to the charged
degrees of freedom not considered explicitly. Chiral symmetry requests such terms to be constructed
with particular field combinations,
$f^\pm_{\mu \nu} \sim Q_\chi^2$, that transform under chiral transformations like the fields, $V_{\mu \nu}$ and $U_\mu$, identified
above. We recall from  \cite{Krause:1990}
\begin{eqnarray}
&& f^\pm_{\mu \nu} = \frac{1}{2}\,\Big(u\,Q\,u^\dagger \pm u^\dagger \,Q \,u \Big) \,F_{\mu \nu} \,.
\label{def-fmumnu}
\end{eqnarray}
At leading order $Q_\chi^2$ we identify two terms
\begin{eqnarray}
{\mathcal L}_{\rm e.m.}=&-& \frac{e_V\,m_V}{8}\,{\rm tr}\,\Big\{
V^{\mu\nu}\,f^+_{\mu \nu}\Big\}
- i\,\frac{e_M }{4} \,  {\rm tr}\,
\Big\{\Big[V_{\mu}^{ \;\; \alpha},\,V^{\mu \beta}\Big]_- \, f^+_{\alpha \beta} \Big\} \,,
\label{em-vertices:a}
\end{eqnarray}
where we neglect further structures that involve more than one flavor
trace\footnote{Note that the possible $Q^2_\chi$- term
$ \epsilon^{\mu\nu\alpha\beta} \,{\rm tr}\,
\big\{\big[V_{\mu\nu},\,V_\alpha^{\phantom{\alpha}\gamma}\big]_+ \, f^-_{\beta\gamma} \big\}
$ is zero identically. This follows from the identity
$g_{\sigma\tau}\epsilon_{\alpha\beta\gamma\delta}=
g_{\alpha\tau}\epsilon_{\sigma\beta\gamma\delta}
+g_{\beta\tau}\epsilon_{\alpha\sigma\gamma\delta}
+g_{\gamma\tau}\epsilon_{\alpha\beta\sigma\delta}
+g_{\delta\tau}\epsilon_{\alpha\beta\gamma\sigma}$
(see also \cite{Fearing:1994ga}). }. The latter are suppressed
in $1/N_c$.

In order to explore the usefulness of the power counting rules (\ref{power-counting}, \ref{extra-power}) we
construct  a selected term of order $Q_\chi^4$. We consider a single flavor-trace interaction that is linear in
$f^+_{\mu \nu}$,
\begin{eqnarray}
{\mathcal L}_{\rm e.m.}=i\,\frac{e_A}{4\,m_V}\,\,\epsilon^{\mu \nu \alpha \beta}\,{\rm tr}\,
\Big\{\Big(f^+_{\mu \nu}\,(D^\tau V_{\tau \alpha})+(D^\tau V_{\tau \alpha})\,f^+_{\mu \nu}\Big)\,U_\beta\Big\}\,,
\label{em-vertices:b}
\end{eqnarray}
in terms of the dimensionless parameter $e_A$. It is stressed that
we do not aim at a systematic study of order  $Q_\chi^4$ effects in this work. Such an enterprise would ask for
the construction of additional $Q^4_\chi$ terms as well as for a computation of various loop diagrams.
The particular term (\ref{em-vertices:b}) is instructive nevertheless
since the magnitude of the parameter $e_A$ can be studied under the assumption that
loop contributions can be absorbed roughly into effective counter terms. Given the values of such effective
counter terms one may estimate the size of the corrections implied in the radiative decay amplitudes of
the axial-vector molecules.

The readers are pointed to the fact that the success of a power counting scheme rises and falls with the effectiveness of
the naturalness assumption needed to truncate the infinite tower of operators in effective field theories. The latter
assumption can rarely be proven, in particular in a situation where the underlying fundamental theory is poorly known only.
It is argued that counting derivatives which act on vector fields as 'small' can be further justified by the following chain
of arguments. At leading order in the $1/N_c$ expansion the n-body vector vertex must be asymptotically bounded as a consequence
of the underlying quark-gluon dynamics
(see also \cite{RuizFemenia:2003hm} and references therein).
On the other hand there are no contributions yet from intermediate multi-meson states.
The latter occur at subleading order in the $1/N_c$ expansion only. Thus, owing to micro-causality formulated at the level
of hadronic degrees of freedom the n-body vector vertex must be point-like and involve only a minimal number of
derivatives, as not to spoil its asymptotic behavior. Micro-causality leads necessarily to dispersion-integral
representations  of the n-body vector vertex, their spectral weights being determined by physical intermediate states.
Only at subleading order  in the $1/N_c$ expansion the n-body vector vertices of our effective Lagrangian acquire
non-trivial structure, that
is represented by higher order derivatives on the vector fields. To this extent we deem it justified to take the power
counting rules (\ref{power-counting}, \ref{extra-power}) at least as a rough guide how to organize our expansion.

We should point out the crucial assumption our power counting relies on. It is conjectured that the infinite
tower of narrow meson states that arises for large-$N_c$ in QCD \cite{large-N_c-reference} exhibits a gap, i.e. the lowest
$0^+,1^+$ and $2^+$ states are significantly above the lowest $0^-$ and $1^-$ states. If the masses of the large-$N_c$ scalar,
axial-vector and tensor states are much larger than those of the first pseudo-scalar and vector states, the effect of the
former states as well as of the excited pseudo-scalar and vector states can be lumped into local counter terms that permit
systematic power counting rules. The latter conjecture is largely equivalent to the hadrogenesis conjecture that
all meson resonances except the pseudo-scalar and vector ground states are generated by coupled-channel dynamics
in terms of the $0^-,1^-$ and $\frac{1}{2}^+, \frac{3}{2}^+$ ground states.

\newpage

\subsection{Decay properties of vector mesons}
\label{subsec:dec-vec}

We continue with an estimate of the parameters introduced in the previous section.
This will shed first light on the usefulness of the proposed power counting rules (\ref{power-counting}, \ref{extra-power}).

The parameter $h_P$ in (\ref{interaction-tensor}) determines the $\rho\to \,\pi\, \pi$, $\phi \to \bar K\,K$ and
$K^* \to \,\pi\, K$ decay processes, with
\begin{eqnarray}
&& \Gamma_{\;\rho\;\;\to\, \pi \,\pi } \;= \frac{h_P^2}{6\,\pi} \left(\frac{m_V}{4\,f^2} \right)^2 \,q_{cm}^3\, , \qquad
\qquad q_{cm} \simeq 0.363 \,{\rm GeV } \,,
\nonumber\\
&& \Gamma_{K^* \to\, \pi\,K } = \frac{h_P^2}{8\,\pi} \left(\frac{m_V}{4\,f^2} \right)^2 \,q_{cm}^3\, , \qquad \qquad
q_{cm} \simeq 0.289\,{\rm GeV } \,,
\nonumber\\
&& \Gamma_{\phi \,\to\, \bar K\,K } \,= \frac{h_P^2}{6\,\pi} \left(\frac{m_V}{4\,f^2} \right)^2 \,q_{cm}^3\, , \qquad \qquad
q_{cm} \simeq 0.119\,{\rm GeV } \,.
\label{hadronic-decay}
\end{eqnarray}
Supplying (\ref{hadronic-decay}) by the empirical decay widths \cite{PDG:2006} we estimate
\begin{eqnarray}
h_P = 0.29 \pm 0.03 \,,
\label{value-hP}
\end{eqnarray}
where we use $f =90$ MeV and identify $m_V = 776$ MeV with the average of the $\rho$ and $\omega $ mass.
Note that the sign chosen in (\ref{value-hP}) defines our phase-convention for the vector meson field.
The uncertainty in (\ref{value-hP}) reflects the fact  which meson decay is used to determine the
parameter $h_P$. For instance, the choice $h_P\simeq 0.26$ reproduces the $\bar K K$-width of the $\phi$-meson and
the value $h_P\simeq 0.32$ the $\pi \pi$-width of the $\rho$-meson. A precise reproduction of the $K\,\pi$-width
of the $K^*$ meson requires $h_P \simeq 0.30$.

The magnitude of the parameter $e_V$ as introduced in (\ref{em-vertices:a}) is determined by the
decay of the neutral vector mesons $\rho_0, \omega$ and $\phi$ into di-electrons. It holds
\begin{eqnarray}
&&\Gamma_{\rho_0\,\rightarrow e^-\,e^+ } = \frac{e^2}{4 \pi}
\frac{m_V^2\,e^2_V}{48\,m_\rho^3}\,
\sqrt{1-\frac{4\,m_e^2}{m_\rho^2}}\left(2\,m_e^2+m_\rho^2\right) \,,
\nonumber\\
&&\Gamma_{\omega \,\rightarrow e^-\,e^+ } =\frac{e^2}{4 \pi}
\frac{m_V^2\,e^2_V}{9\times 48\,m_\omega^3}\,
\sqrt{1-\frac{4\,m_e^2}{m_\omega^2}}\left(2\,m_e^2+m_\omega^2\right) \,,
\nonumber\\
&&\Gamma_{\phi \,\rightarrow e^-\,e^+ } =\frac{e^2}{4 \pi}
\frac{2\,m_V^2\,e^2_V}{9\times 48\,m_\phi^3}\,
\sqrt{1-\frac{4\,m_e^2}{m_\phi^2}}\left(2\,m_e^2+m_\phi^2\right) \,,
\label{eq:dildecemass2}
\end{eqnarray}
with the electron mass $m_e$. From the empirical decay pattern \cite{PDG:2006} we obtain the estimate
\begin{eqnarray}
e_V = 0.22\pm 0.02 \,,
\label{value-eV}
\end{eqnarray}
where the phase was prejudiced from the assumption of universally coupled light vector mesons \cite{Bando,Birse:1996,Ecker:1989}.

The parameters $h_V, \tilde h_V$ and $e_M$ introduced in (\ref{interaction-tensor}, \ref{em-vertices:a}) relate to
the magnetic and quadrupole moment of the
charged vector mesons at leading order
in our expansion. Following the original works by Jones and Kyriakopoulos \cite{Kyriakopoulos:1971,Kyriakopoulos:1972,Jones:1962}
we derive the results
\begin{eqnarray}
&&\mu_{\rho_+}\, = - \mu_{\rho_-}\, =\frac{1}{2\,m_\rho}\, \Big\{ e +e_M + \frac{3\,h_V+\tilde h_V}{4}\,e_V  \Big\} \,, \qquad
\nonumber\\
&& Q_{\rho_+} =-Q_{\rho_-}= \;\frac{1}{\;m_\rho^2} \,\Big\{  e_M + \frac{3\,h_V-\tilde h_V}{4}\,e_V  \Big\}   \,,
\label{magnetic-moment-rho}
\end{eqnarray}
and
\begin{eqnarray}
&&\mu_{K^*_+} \,= -\mu_{K^*_-}\,= \frac{1}{2\,m_{K^*}}\, \Big\{ e +e_M +\Big( 2+\frac{m_V^2}{m_\phi^2}\Big)\, \frac{3\,h_V+\tilde h_V}{12}\,e_V  \Big\} \,, \qquad
\nonumber\\
&& Q_{K^*_+} = - Q_{K^*_+} =\;\frac{1}{\;m_{K^*}^2} \,\Big\{  e_M + \Big( 2+\frac{m_V^2}{m_\phi^2}\Big)\,\frac{3\,h_V-\tilde h_V}{12}\,e_V  \Big\}   \,,
\label{magnetic-moment-Ksplus}
\end{eqnarray}
and
\begin{eqnarray}
&&\mu_{K^*_0} \,= -\mu_{\bar K^*_0}\,=\frac{1}{2\,m_{K^*}}\, \Big( \frac{m_V^2}{m_\phi^2}-1\Big)\, \frac{3\,h_V+\tilde h_V}{12}\,e_V  \,, \qquad
\nonumber\\
&& Q_{K^*_0} = -Q_{\bar K^*_0}=\frac{1}{\;m_{K^*}^2} \,\Big( \frac{m_V^2}{m_\phi^2}-1\Big)\,\frac{3\,h_V-\tilde h_V}{12}\,e_V    \,,
\label{magnetic-moment-Kszero}
\end{eqnarray}
where we take into account the contributions implied by (\ref{interaction-tensor}, \ref{em-vertices:a}) and
identify again $m_V = 776$ MeV with the average of the $\rho$ and $\omega $ masses.
Neglecting effects proportional to $m_\rho-m_\omega $ the magnetic and quadrupole moments of the $\rho_0,\, \omega$ and $\phi$ mesons are
zero identically.

Here we count the $h_V$- and $\tilde h_V$-vertices as order $Q^2_\chi$.
Note that the two moments of the $\rho$ meson would be determined in leading order by one
parameter combination $e_M + 3\, h_V \, e_V /4$, if we counted the $\tilde h_V$-vertex as order
$Q_\chi^3$. On the other hand, if we counted the $h_V$-vertex as
order $Q_\chi$, its contribution to the magnetic moment would be dominant, in particular larger than the term
proportional to $e$. We would deem such a scenario unnatural.

We recall from \cite{Barua:1977gc,Brodsky:Hiller:1992} that if
the $\rho$ meson would be a point-like gauge boson it followed $\mu_{\rho^+}=e/m_\rho$ and
$Q_{\rho^+}=-e/m_\rho^2$. Empirically the magnetic and quadrupole moment of the charged vector mesons are
unknown, although a first quenched QCD lattice simulation is available \cite{lattice:Hedlich:2007}. There are various
theoretical studies performed \cite{Samsonov:2003hs,Aliev:2003ba,Choi:Ji:2004}. QCD sum rules suggest the range
$\mu_{\rho^+}=e\,(0.6-1.4)/m_\rho$ but don't provide so far an estimate for the quadrupole moment
\cite{Samsonov:2003hs,Aliev:2003ba}. Recent studies within the light-front quark model \cite{Choi:Ji:2004}
predicted the values
\begin{eqnarray}
\mu_{\rho^+}\simeq 0.96\,\frac{e}{m_\rho} \,, \qquad \qquad \qquad \quad \;\; Q_{\rho_+} \simeq -0.43 \,\frac{e}{m_\rho^2} \,.
\label{value-murho-LFQM}
\end{eqnarray}
Given the results (\ref{value-murho-LFQM}) we derive the estimates
\begin{eqnarray}
e_M + \frac{3}{4} \, h_V \, e_V  \simeq 0.245 \,e \,, \qquad \qquad
\tilde h_V \, e_V  \simeq 2.70 \,e\,,
\label{value-eM:b}
\end{eqnarray}
which may point towards the dominance of the $\tilde h_V$-vertex over the $e_M$-,$h_V$-vertices contrary
to what would be expected from the power-counting rules (\ref{power-counting}, \ref{extra-power}). Note, however,
the possibility of a cancellation mechanism amongst the two terms $e_M$ and $h_V\,e_V$ in (\ref{value-eM:b}).
The estimate (\ref{value-eM:b}) may be taken as a justification to demote the relevance of the $h_V$-vertex as
argued for above. We note that the estimate
\begin{eqnarray}
\tilde h_V \simeq 3.72 \,,
\label{value-tildehV}
\end{eqnarray}
suggested by (\ref{value-eM:b}) and (\ref{value-eV}) is somewhat
smaller than the phenomenological value $5.39$ implied by the relations (\ref{value-hV}, \ref{value-hP}).
The two parameters $e_M$ and $h_V$ would be determined by (\ref{value-eM:b}) together with an estimate of
the magnetic or quadrupole moment of the charged or neutral $K^*$ meson. It is interesting to explore the consequence
of the particular choice $e_M=0$, which implies
\begin{eqnarray}
h_V \simeq 0.45 \,, \qquad \quad &&
\mu_{K_+^*} \simeq +0.90\, \frac{e}{m_{K^*}} \, ,\qquad \,
\mu_{K_0^*} \simeq -0.07\, \frac{e}{m_{K^*}} \,,
\nonumber\\
&& Q_{K_+^*}\! \simeq -0.37\, \frac{e}{m^2_{K^*}} \, ,\qquad
Q_{K_0^*} \simeq +0.06\, \frac{e}{m^2_{K^*}} \,.
\end{eqnarray}

Now we estimate  the size of the parameters $h_A, b_A$ and $e_A$ introduced in (\ref{interaction-tensor}, \ref{em-vertices:b}).
They enter the computation of the radiative decay of
the light vector mesons. We recall that a non-relativistic constituent quark model gives $|h_A|\simeq 2$,
whereas a relativistic chiral quark model predicts $|h_A|\simeq 1.5$~\cite{Jenkins:1995vb}.
This leaves undetermined the relative phase of the coupling constants $h_P$ and $h_A$. We take over the
relative phase as suggested by any quark model, which predicts  identical signs for  the parameters $h_P$ and
$h_A$. Such a phase correlation is implied for instance by the analysis of \cite{Lutz-Soyeur:2007}, where the
consequence of the heavy-quark symmetry in the context of a phenomenological flavor SU(4) ansatz leads to a
correlation of the parameters $h_P,\tilde h_V$ and $h_A$. It is amusing to observe that a projection onto
the SU(3) subsector recovers (\ref{value-hV})
and suggests
\begin{eqnarray}
h_A = \frac{h_P\,m_V}{2\,f} \qquad \to \qquad h_A \simeq 1.4 \,,
\label{values-hA-SU4}
\end{eqnarray}
a value roughly compatible with the above estimates. We point out, however, that it is necessary to determine
the parameter $h_A$ within the given scheme, i.e. by  studying radiative decays of the light vector mesons.

Indeed, one gets a consistent description of the
decays $\rho \to\gamma\, \pi $, $\rho \to \, \gamma\,\eta $,
$\omega \to \, \gamma\,\pi $ and $\omega \to \, \gamma\,\eta $, if one uses
the vertices (\ref{interaction-tensor}, \ref{em-vertices:a}, \ref{em-vertices:b}). The relevant
parameter combination is readily established
\begin{eqnarray}
\tilde e_A = e_A + \frac{1}{4}\,h_A\,e_V - 2 \,b_A \, e_V \, \frac{m_\pi^2}{m_V^2} \,,
\label{relevant-combination}
\end{eqnarray}
where we identified $m_V \simeq m_\rho \simeq m_\omega$ for simplicity. It holds
\begin{eqnarray}
&&\Gamma_{\omega \,\to \, \gamma\,\pi^0 }  =  \frac{1}{24\times 16\,\pi}\,\frac{m_\omega^5}{(m_V\,f)^2} \,
\Big| \tilde e_A \Big|^2 \left(1- \frac{m_{\pi}^2}{m_{\omega}^2} \right)^3\,,
\nonumber \\
&&\Gamma_{\omega \,\to \,\gamma\,\eta \;}  =  \frac{1}{24\times 432\,\pi}\,\frac{m_\omega^5}{(m_V\,f)^2} \,
\Big| \tilde e_A \Big|^2 \left(1- \frac{m_{\eta}^2}{m_{\omega}^2} \right)^3\,,
\nonumber \\
&&\Gamma_{\rho^0 \to \, \gamma \,\pi^0 }  =  \frac{1}{24\times  144 \,\pi}\,\frac{m_\rho^5}{(m_V\,f)^2} \,
\Big| \tilde e_A \Big|^2 \left(1- \frac{m_{\pi}^2}{m_{\rho}^2} \right)^3\,,
\nonumber \\
&& \Gamma_{\rho^0 \to \,\gamma\,\eta \;}  =  \frac{1}{24 \times 48 \,\pi}\,\frac{m_\rho^5}{(m_V\,f)^2} \,
\Big| \tilde e_A \Big|^2 \left(1- \frac{m_{\eta}^2}{m_{\rho}^2} \right)^3 \,,
\label{vec-to-gam-decays-light}
\end{eqnarray}
which together with $f=90$ MeV and $m_V=776$ MeV implies
\begin{eqnarray}
\big| \tilde e_A \big| = 0.135 \pm 0.008 \,.
\label{value-eA}
\end{eqnarray}
According to the power counting rules we expect $e_A$ to be less important than the parameter combination
$h_A\,e_V$. The role of the parameter $b_A$ is suppressed by $m_\pi^2/m_V^2$ in the decays
of (\ref{vec-to-gam-decays-light}). Using the empirical value
$e_V \simeq 0.22$ from (\ref{value-eV}) together with the quark-model estimate for $h_A\simeq 2$
we would predict $e_A \simeq 0.025$, a value quite consistent with what we would expect from our power counting.

We turn to the  decays of $\phi$ and $K^*$. In contrast to the previously discussed
decays of the vector mesons into dileptons or two Goldstone bosons where the flavor breaking effects are
not very large, sizable flavor breaking effects are observed in the radiative decays. It is emphasized that
this is predicted by the power counting rules (\ref{power-counting}, \ref{extra-power}). Whereas the dilepton decays
are determined by a single parameter $e_V$, the radiative decays are determined by the two parameter combinations
$h_A\,e_V$ and $b_A\,e_V$ at leading order. Note that the size of flavor breaking
defines the $10 \%$ uncertainty of the parameter $h_P$ and $e_V $ (see also \cite{Leupold:2005ep}).
In order to work out the flavor breaking effects in the radiative decays we derive
\begin{eqnarray}
&& \Gamma_{K^*_\pm \to K_\pm \gamma}  =  \frac{1}{24\times 144\,\pi}\,\frac{m^5_{K^*_\pm}}{m^2_V\,f^2} \,
\Big| e^{(K^*_+)}_A \Big|^2 \left(1- \frac{m_{K_\pm}^2}{m_{K^*_\pm}^2} \right)^3\,,
\nonumber \\
&& \Gamma_{K^*_0 \to K_0 \gamma}  =  \frac{1}{24\times 36\,\pi}\,\frac{m^5_{K^*_0}}{m^2_V\,f^2}  \,
\Big| e^{(K^*_0)}_A \Big|^2 \left(1- \frac{m_{K_0}^2}{m_{K^*_0}^2} \right)^3\,,
\nonumber \\
&& \Gamma_{\phi \;\to\, \eta\, \gamma}  =  \frac{1}{24\times 54\,\pi}\,\frac{m^5_{\phi}}{m^2_V\,f^2}  \,
\Big| e^{(\phi)}_A \Big|^2 \left(1- \frac{m_{\eta}^2}{m_{\phi}^2} \right)^3\,,
\label{vec-to-gam-decays}
\end{eqnarray}
in terms of the effective decay parameters
\begin{eqnarray}
&&e_A^{(\phi)} \;\; =  e_A + e_V \, \frac{m_V^2}{m_\phi^2} \left(
\frac14 \, h_A \,  - 2 \,b_A \, \frac{2 \,m_K^2 -m_\pi^2}{m_\phi^2}  \right) \,,
\nonumber \\
&& e_A^{(K^*_0)\,}  =  e_A + e_V \,\left(1+\frac{m_V^2}{m_\phi^2} \right) \left( \frac18 \, h_A \,
- b_A \, \frac{m_K^2}{m_{K^*}^2} \right) \,,
\nonumber \\
&& e_A^{(K^*_+)}  =  e_A + e_V \, \left(2-\frac{m_V^2}{m_\phi^2} \right) \left( \frac14 \, h_A
- 2 \,b_A \, \frac{m_K^2}{m_{K^*}^2} \right) \,,
  \label{eq:deftildeqeA2}
\end{eqnarray}
and  $m_V \approx m_\rho \approx m_\omega $. The empirical data  \cite{PDG:2006} yield the values
\begin{eqnarray}
&& \Big|e_A^{(K^*_0)\,} \Big| = 0.094 \pm 0.004\,, \qquad \qquad \Big|e_A^{(\phi)} \Big| = 0.053 \pm 0.001 \,,
\nonumber\\
&& \Big|e_A^{(K^*_+)}  \Big| = 0.125  \pm 0.006 \,.
\label{value-eAeff}
\end{eqnarray}
The large variation of the parameters $\tilde e_A, e_A^{(K^*)}$ and $e_A^{(\phi)}$ reflect the sizable
flavor breaking in the radiative decays of the vector mesons. One may assign the value
$ \tilde e_A = 0.094 \pm 0.041$, with an error of about $43 \%$ much larger than the $10\%$ spread in $h_P$ and
$e_V$.

The parameters $h_A, b_A$ and $e_A$ are determined as follows. For a given value of $h_A$ we determine the range of
the parameter $b_A$ to be consistent with (\ref{eq:deftildeqeA2}, \ref{value-eAeff}). The parameter $e_A$ is
adjusted to remain compatible with (\ref{relevant-combination}, \ref{value-eA}). The optimal value
\begin{eqnarray}
h_A \simeq 2.1 \,,
\label{value-hA}
\end{eqnarray}
 is obtained by the requirement that the spread in $b_A$ is
minimal. This procedure leads to
\begin{eqnarray}
b_A = 0.27 \pm 0.05  \,, \qquad \qquad e_A = 0.023 \pm 0.001 \,,
\label{value-bA}
\end{eqnarray}
an amazingly consistent picture: at leading order the two parameters $h_A$ and $b_A$ describe the radiative
decay processes of the vector mesons reasonably well.

\newpage

\section{Radiative decay amplitudes of axial-vector molecules}
\label{sec:rad-decay}

We derive expressions for the radiative decay amplitudes of axial-vector mole\-cules. There are two types of processes
that are described by the generic decay amplitudes $M^{\mu, \alpha \beta}_{1^+ \to \, \gamma \,0^-}$ and
$M^{\bar \alpha \bar \beta, \mu, \alpha \beta}_{1^+ \to \, \gamma \,1^-}$. The first one is
characterized by a photon and a pseudo-scalar particle, the second by a photon and
a vector particle in the final state.

Given the transition amplitude,
$M^{\mu, \alpha \beta}_{1^+ \to \, \gamma \,0^-}$, we recall the partial-decay width $\Gamma_{1^+\to \, \gamma\,0^-}$. It is
determined  by one scalar number, $d_{1^+\to \,\gamma\, 0^-}$. We write:
\begin{eqnarray}
&&\epsilon_\mu^\dagger(q, \lambda_q)\, M^{\mu, \alpha \beta}_{1^+ \to \,\gamma \,0^-}\,\epsilon_{\alpha \beta}(p, \lambda_p)
\nonumber\\
&& \qquad = i\,d_{1^+ \to \,\gamma \,0^-} \,\epsilon_\mu^\dagger(q, \lambda_q)\,\Big\{ g^{\mu
\alpha}\,(\bar p\cdot q)- \bar p^\mu\,q^\alpha \Big\}\,\frac{p^\beta}{\sqrt{p^2}}\,
\epsilon_{\alpha \beta}(p, \lambda_p) \,,
\label{def-transition-tensor-axial}
\end{eqnarray}
which implies
\begin{eqnarray}
&&\Gamma_{1^+\to \,\gamma \,0^-}= \frac{|d_{1^+ \to \,\gamma\,0^-}|^2}{12\pi}\,\left(\frac{M_{1^+}^2-M_{0^-}^2}{2\,M_{1^+}} \right)^3\,,
\nonumber\\
&& d_{1^+ \to \,\gamma\,0^-} = \frac{-i\,}{(p \cdot q)\,M_{1^+}}\,\Bigg\{
g_{\mu \alpha}- \frac{p_\mu\,q_\alpha}{(p \cdot q)} \Bigg\}\,p_\beta\,M^{\mu, \alpha \beta} \,,
\label{result:width-1plus:a}
\end{eqnarray}
with $\bar p^2=(p-q)^2 =M_{0^-}^2$ and $p^2=M_{1^+}^2$. The projection formula of (\ref{result:width-1plus:a}) exploits
the fact that the decay amplitude is anti-symmetric in $\alpha \leftrightarrow \beta$ \cite{Lutz-Soyeur:2007}.
In (\ref{def-transition-tensor-axial}) $\epsilon^\dagger_\mu(q, \lambda_q)$ and
$\epsilon^{\alpha \beta}(p, \lambda_p)$ denote the wave functions of the emitted photon and the decaying axial-vector molecule \cite{Lutz-Soyeur:2007}.

We turn to the second process described by a rank-five
transition tensor, $M^{\bar \alpha \bar \beta, \mu, \alpha \beta}_{1^+ \to \, \gamma \,1^-}$.
It is characterized by two scalar decay parameters. The latter reflect the fact that the decay may go
via an s-wave or a d-wave transition. We write
\begin{eqnarray}
&& \epsilon_\mu^\dagger(q,\lambda_q)\,\epsilon^\dagger_{\bar \alpha \bar \beta}(\bar p, \lambda_{\bar p})\,
M^{\bar \alpha \bar \beta, \mu, \alpha \beta}_{1^+ \to \, \gamma \,1^-} \,\epsilon_{\alpha \beta}(p, \lambda_p)
 = - \epsilon_\mu^\dagger(q,\lambda_q)\, \frac{\bar p^{\bar \beta}\,\epsilon^\dagger_{\bar \alpha \bar \beta }(\bar p,
\lambda_{\bar p}) }{\sqrt{\bar p^2}} \,
\nonumber\\
&&   \times \,q_\tau \,\epsilon^{ \mu \tau \bar \alpha \sigma}\,\Bigg\{
d^{(1)}_{1^+\to \, \gamma \,1^-} \,\frac{q^\alpha\,p_\sigma}{(q \cdot p)}
+d^{(2)}_{1^+\to \, \gamma \,1^-} \left(g^\alpha_{\;\, \sigma}-\frac{q^\alpha\,p_\sigma}{(q \cdot p)}\right)
 \Bigg\}\,\frac{p^\beta\,\epsilon_{\alpha \beta}(p,\lambda_p) }{\sqrt{p^2}} \,,
\nonumber\\
\label{def-transition-amplitude-axial-b}
\end{eqnarray}
implying a width of
\begin{eqnarray}
&&\Gamma_{1^+ \to \, \gamma\,1^-} =\frac{1}{12\,\pi}\,
\Bigg\{
\Bigg|\frac{ d^{(1)}_{1^+\to \, \gamma \,1^-} }{M_{1^+} }\Bigg|^2
+\Bigg|\frac{ d^{(2)}_{1^+\to \, \gamma \,1^-} }{M_{1^-} }\Bigg|^2
 \Bigg\} \left(\frac{M_{1^+}^2-M_{1^-}^2}{2\,M_{1^+}} \right)^3\,,
\label{result:width-1plus:c}
\end{eqnarray}
with $\bar p^2 =M_{1^-}^2$ and $p^2=M_{1^+}^2$. The
extraction of the two decay parameters from a given amplitude is streamlined considerably
by the projection identities:
\begin{eqnarray}
&& d^{(1)}_{1^+\to \, \gamma \,1^-} =
\frac{-16\,M_{1^+}}{M_{1^-}\,(M_{1^+}^2-M_{1^-}^2)^3}\,p^{\sigma}\,q^{\tau}\,
\epsilon_{\sigma  \tau  \bar \alpha \mu }\,\bar p_{\bar \beta}\,
M^{\bar \alpha \bar \beta, \mu, \alpha \beta}_{1^+ \to \, \gamma\,1^-}\,q_{\alpha}\,p_\beta \,,
\nonumber\\
&& d^{(2)}_{1^+\to \, \gamma \,1^-} = \frac{-16\,M_{1^-}}{M_{1^+}\,(M_{1^+}^2-M_{1^-}^2)^3}\,p^{\sigma}\,q^{\tau}\,
\epsilon_{\sigma  \tau\mu \alpha }\,q_{\bar \alpha}\,\bar p_{\bar \beta}\,
M^{\bar \alpha \bar \beta, \mu, \alpha \beta}_{1^+ \to \, \gamma\,1^-} \,p_\beta\,,
\label{result-projection-1plus}
\end{eqnarray}
which are a consequence of the anti-symmetry and transversality of the decay amplitude \cite{Lutz-Soyeur:2007}.
Note that the two decay parameters can be measured separately
(see Appendix A).

\begin{figure}[t]
\begin{center}
\includegraphics[width=10cm,clip=true]{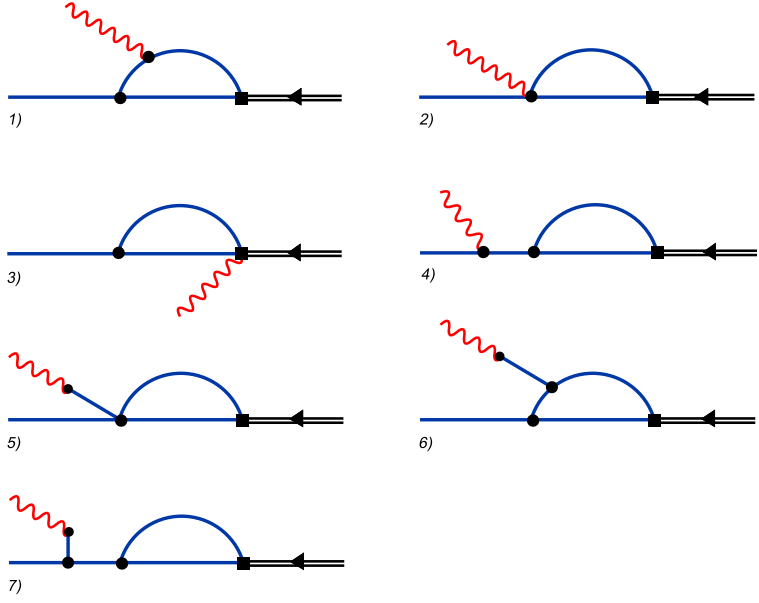}
\end{center}
\caption{Diagrams contributing to the decay amplitude of axial-vector molecules. Double lines
denote the axial-vector molecules, solid lines either pseudo-scalar or vector mesons  and the wavy lines
the photons.}
\label{fig:1}
\end{figure}

In Figure \ref{fig:1} a complete set of generic diagrams is depicted. Solid lines
represent the propagation of the pseudo-scalar or vector mesons. The wavy line is the photon.
We recall  the argument from \cite{Lutz-Soyeur:2007} that all contributions to the decay amplitude
must involve at least one-loop integral. Tree-level contributions like the ones studied in \cite{Roca:Palomar:Oset:2004}
within a vector-meson dominance picture are not present if a resonance is generated by coupled-channel dynamics.
The inclusion of such effects would be double counting. This issue is discussed in more detail in Appendix B.

Additional contributions involving an intermediate resonance line, that one may deem relevant at first sight, do
not affect the decay width. As discussed in \cite{Lutz-Soyeur:2007} this is a consequence of
the anti-symmetry of the tensor field as well as the particular form of the resonance vertex  implied
by (\ref{1plus-amplitude}, \ref{def-gi-1plus}). The scattering amplitude (\ref{def-gi-1plus}) may be reproduced by effective
resonance vertices of the form
\begin{eqnarray}
&&{\mathcal L} =  h_{\phi \,V}\,\Big\{ (\partial^\mu R^\dagger_{\mu \alpha})\,\phi
\,(\partial_\nu V^{\nu \alpha}) + h.c. \Big\} \,,
\label{def-hi}
\end{eqnarray}
close to the resonance mass. Here $R_{\mu \nu}$ denotes a generic resonance field and
$\phi V$ a final state carrying well defined isospin. It holds
\begin{eqnarray}
h_i = \sqrt{2}\, \frac{g_i}{M_i}\,,
\label{hi-gi}
\end{eqnarray}
with the vector-meson mass $M_i$ of the considered channel and the coupling constants $g_i$ introduced in
(\ref{def-gi-1plus}).
Within the applied scheme for the evaluation of the loop diagrams of Figure \ref{fig:1}
 the vertex (\ref{def-hi}) is basically equivalent to  the vertex,
$(\partial^\tau R^\dagger_{\tau \nu})(\partial_\mu \phi)\,V^{\mu \nu}$, as pointed out in \cite{Lutz-Soyeur:2007}.
The latter  form is manifestly compatible with chiral constraints. Note that as it should be, given the particular
form of the resonance
vertex (\ref{def-hi}) there are no tree-level contributions to the radiative decay of axial-vector mesons into
pseudo-scalar mesons. This follows since the vector-meson field  enters the resonance vertex in the form
$(\partial_\nu\,V^{\nu \alpha})$. In contrast the vertex
$(\partial^\tau R^\dagger_{\tau \nu})(\partial_\mu \phi)\,V^{\mu \nu}$ would allow a
vector-meson dominance type contribution. It is stressed that the equivalence of the two vertices holds only
given our scheme, for the evaluation of the loop diagrams in Figure \ref{fig:1}.
This is a consequence of the particular renormalization program introduced in \cite{Lutz-Soyeur:2007}, which we
will review briefly.

It is observed that the tensor integrals defined by Figure 1
 are ultraviolet diverging. Applying the Passarino-Veltman
reduction \cite{Passarino:Veltman:1979}, which is rigourously justified within dimensional regularization,
such tensors may be expressed in terms of a set of scalar integrals of the form
\begin{eqnarray}
&& I_a = -i\,\int
\frac{d^4l}{(2\pi)^4}\,S_a (l) \,, \qquad
 I_{ab}= -i\,\int
\frac{d^4l}{(2\pi)^4}\,S_a (l)\,S^{}_{b}(l+p)\,, \qquad
\nonumber\\
&&  \bar I_{ab}= -i\,\int
\frac{d^4l}{(2\pi)^4}\,S_a (l)\,S^{}_{b}(l+\bar p)\,,
\nonumber\\
&& J_{abc}= +i\,\int
\frac{d^4l}{(2\pi)^4}\,S_a (l)\,S^{}_{b}(l+q)\,S^{}_{c}(l+p)\,,
\nonumber\\
&& \bar J_{abc}= +i\,\int
\frac{d^4l}{(2\pi)^4}\,S_a (l)\,S^{}_{b}(l+\bar p)\,S^{}_{c}(l+p)\,,
\label{def-master-loops}
\end{eqnarray}
where we focus on the physical limit with space-time dimension four. The finite integrals $J_{abc}$
and $\bar J_{abc}$ may be evaluated applying Feynman's parametrization or
their dispersion-integral representation (see e.g. \cite{Lutz-Soyeur:2007}). Divergent structures
arise only from the tadpole integral $I_a$ and the two-propagator integrals $I_{ab}$ and $\bar I_{ab}$. Since the
combinations
\begin{eqnarray}
&& I_{ab}-\bar I_{ab} \,, \qquad \qquad
I_{ab} - \frac{I_b-I_a}{m_a^2-m_b^2} \,,
\label{finite-combinations}
\end{eqnarray}
are finite as the number of space-time dimensions approaches four,
one may take the viewpoint that all divergent structures are caused by the tadpole integral $I_a$.
As was emphasized in \cite{Lutz-Soyeur:2007} there is a consistency issue on how to treat the decay-amplitudes
in view of the coupled-channel approach (\ref{M-generic}). It was argued that tadpole contributions to
the decay amplitude must be dropped in a leading-order computation. In addition a finite renormalization should
be applied to the integral $I_{ab}$. It should be identified with the expression  introduced in (\ref{i-def}), with
\begin{eqnarray}
I_{ab}=I(p^2)-\Re I(\mu_M^2)\,, \qquad \qquad m_a=m \,, \qquad m_b = M \,,
\end{eqnarray}
and the matching scale $\mu_M$ \cite{Lutz-Kolomeitsev:2004}.

\clearpage

\subsection{The $1^+ \to \, \gamma \,0^-$ process}
\label{subsec:dec-zerominus}

The decay amplitudes can be expressed as linear combinations of generic loop tensors, which
are discussed in some  detail. We form gauge invariant combinations. Some of the tensors encountered
here have already been encountered in \cite{Lutz-Soyeur:2007}. Since the notation adopted in this work differs
from the one used in \cite{Lutz-Soyeur:2007} all loop tensors needed are defined here explicitly.
Appendix C  provides  results of the Passarino-Veltman reduction of the tensors integrals.

Begin with class A) and
class B) sets of diagrams which are defined by the processes where the photon is radiated from the charge of an intermediate
Goldstone boson or a vector meson, as depicted by the first diagram of Figure \ref{fig:1}. The contributions are
proportional to the parameter combination $e\,h_P$. To form gauge invariant tensors
we have to consider diagrams as labelled in Figure \ref{fig:1} by type 2)-3) in addition. The latter are required
since the resonance vertex as well as the hadronic 3-point vertex proportional to $h_P$
involve a covariant derivative that has a term proportional to the photon field.

We introduce the two tensor integrals,
$A^{\mu, \alpha \beta}_{ab}(\bar p,p)$ and $B^{\mu, \alpha \beta}_{ab}(\bar p,p)$, that describe the contributions
proportional to $e\,h_P$. We write
\begin{eqnarray}
&&A^{\mu, \alpha \beta}_{ab}(\bar p,p)=+i\,\int
\frac{d^4l}{(2\pi)^4}\,S_a(l)\,\Bigg\{ (l+\bar p)_{\sigma}\,S^{ \sigma \mu ,\,\beta}_{b}(p+l)\,p^\alpha
\nonumber\\
&& \qquad
 +\, \bar p_{\sigma }\, S^{\sigma  \beta }_{b}(\bar p+l)\,g^{\mu \alpha}
+S_a(l+q)\,\bar p_{\sigma }\,S^{\sigma  \beta}_{b}(p+l)\,(q+2\,l)^\mu\,p^\alpha
\Bigg\} \,,
\nonumber\\ \nonumber\\
&& B^{\mu, \alpha \beta}_{ab}(\bar p,p)=+i\,\int
\frac{d^4l}{(2\pi)^4}\,S_a (l)\,\Bigg\{
\bar p_{\sigma}\,g_{\tau \kappa}\,S_{b}^{\sigma \tau }(\bar p+l)\,S^{\mu \kappa,\,\beta}_{b}(p+l)\,p^\alpha
\nonumber\\
&& \qquad
+\, \bar p_{\sigma } \,g_{\tau \kappa}\,S_{b}^{ \sigma ,\, \mu \tau}(\bar p+l)\,
S^{\kappa \beta}_{b}(p+l)\, p^\alpha\, -
l_{\sigma }\,S^{  \mu \sigma ,\,\beta}_{b}(p+l)\,p^\alpha
\nonumber\\
&& \qquad
+\, \bar p_{\sigma }\,S^{ \sigma,\,\mu \beta}_{b}(\bar p+l)\,p^\alpha
+ \bar p_{\sigma }\, S^{ \sigma \beta}_{b}(\bar p+l)\,g^{\mu \alpha}
\Bigg\}\,,
\label{def-AB-tensor3}
\end{eqnarray}
where
\begin{eqnarray}
&&S_a(p) = \frac{1}{p^2-m_a^2}\,,\qquad \qquad  S^{\mu \nu}_a(p) =  p_\alpha \,S^{\alpha \mu, \beta \nu}_a(p)\,p_\beta \,,
\nonumber\\
&& S^{\mu \nu, \alpha \beta }_a(p)=-\frac{1}{m_a^2}\,\frac{1}{p^2-m_a^2+i\,\epsilon}\,
\Bigg[ (m_a^2-p^2)\,g_{\mu \alpha}\,g_{\nu \beta}
\nonumber\\
&& \qquad \qquad \qquad  + g_{\mu \alpha}\,p_\nu\,p_\beta-
g_{\mu \beta}\,p_\nu\,p_\alpha- (\mu \leftrightarrow \nu)\Bigg] \,,
\nonumber\\
&& S^{\mu ,\beta \nu}_a(p) =  p_\alpha \,S^{\alpha \mu, \beta \nu}_a(p)\,, \qquad \qquad
 S^{\alpha \mu ,\nu}_a(p) = S^{\alpha \mu, \beta \nu}_a(p)\,p_\beta \,.
 \label{def-S}
\end{eqnarray}

We continue with diagrams that are gauge invariant separately, where we consider first diagrams of type 6) in Figure
\ref{fig:1}. There are contributions proportional to $e_V\,h_A^2/f^2$,
$e_V\,h_A\,b_A/f^2$ and $e_V\,b_A^2/f^2$
described by the four tensors
\begin{eqnarray}
&& C^{\mu, \alpha \beta}_{abc}(\bar p,p)=+2\,i\,\int
\frac{d^4l}{(2\pi)^4}\,S_a (l)\,
q_\nu \,\epsilon^{\mu \nu }_{ \phantom{\mu \nu}\bar \alpha \bar \beta}\,l^{\bar \beta}\,
\epsilon_{\sigma \tau \kappa \lambda}\,\bar p^{\lambda}\,
\nonumber\\
&& \qquad \quad \,\times \;\Big\{
S_{b}^{\bar \alpha,\,\sigma \tau}(q+l)\,S^{\kappa \beta}_{c}(p+l) -
S_{b}^{\bar \alpha \kappa}(q+l)\,S^{\sigma
\tau ,\,\beta}_{c}(p+l) \Big\}\,p^\alpha
\nonumber\\ \nonumber\\
&& \bar C^{\mu, \alpha \beta  }_{abc}(\bar p,p)=
- \, i\,m_b^2 \,\int
\frac{d^4l}{(2\pi)^4}\,S_a (l)\,  q_\nu \,
\epsilon^{\mu \nu}_{\phantom{\mu\nu}\bar\alpha \bar\beta} \,
\epsilon_{\sigma \tau \kappa \lambda}\,\bar p^\lambda
\nonumber \\
&& \qquad \quad \times \;
\Big\{ S_b^{\bar\alpha \bar\beta, \sigma \tau}(q+l) \, S_c^{\kappa \beta}(p+l) -
S_b^{\bar\alpha \bar\beta, \kappa}(q+l) \, S_c^{\sigma \tau, \beta}(p+l)
\Big\} \, p^\alpha \,,
\label{def-C-tensor3}
\end{eqnarray}
and
\begin{eqnarray}
&& \Cchi^{\mu, \alpha \beta  }_{abc}(\bar p,p)=+2\, i\ \,\int
\frac{d^4l}{(2\pi)^4}\,S_a (l)\,q_{\nu} \,\epsilon^{\mu  \nu}_{\phantom{\mu \nu} \bar \alpha \bar \beta}\,l^{\bar \beta}  \,
 \epsilon_{\sigma \tau \kappa \lambda} \,
\nonumber \\
&& \qquad \quad \times \;
S_b^{\bar \alpha, \, \sigma \tau}(q+l) \,
S_c^{\kappa \lambda, \,\beta}(p+l) \, p^\alpha  \,,
\nonumber\\ \nonumber\\
&& \barCchi^{\mu, \alpha \beta  }_{abc}(\bar p,p)=
- i\,m_b^2 \,\int
\frac{d^4l}{(2\pi)^4}\,S_a (l) \,q_\nu \,\epsilon^{\mu \nu}_{\phantom{\mu\nu}\bar\alpha \bar\beta} \,
\epsilon_{\sigma \tau \kappa \lambda} \,
\nonumber \\
&& \qquad \quad \times \;
S_b^{\bar\alpha \bar\beta, \sigma \tau}(q+l) \,
S_c^{\kappa \lambda,\beta}(p+l) \, p^\alpha \,.
\label{def-Cchi-tensor3}
\end{eqnarray}
 In order to evaluate the various contributions
described by (\ref{def-C-tensor3}, \ref{def-Cchi-tensor3}) it is convenient to do so in terms of the effective vertices
\begin{eqnarray}
&& {\mathcal L} = - \frac{e_V\,h_{A}}{32\,f\,m_V}\, F_{\mu \nu}\,\epsilon^{\mu\nu\alpha\beta}\,{\rm tr}\,\Big\{
\Big[Q_{\rm eff},\, (\partial^\tau V_{\tau \alpha})\Big]_+\,(\partial_\beta \Phi)\Big\}
\nonumber\\
&& \qquad \,   -\frac{e_V\,b_A}{16\,f\,m_V} \,F_{\mu \nu}\, \epsilon^{\mu\nu\alpha\beta}\, {\rm tr}\,
\Big\{\Big[Q_{\rm eff},\,V_{\alpha \beta}\Big]_+ \, \Big[\Phi, \,\chi_0 \Big]_+ \Big\} \,,
\label{def-eff-eA}
\end{eqnarray}
which follow from the hadronic Lagrangian (\ref{interaction-tensor}) by means of the equation of motion
\begin{eqnarray}
V_{\mu \nu} =\frac{e_V}{2\,m_V}\,Q_{\rm eff} \,F_{\mu \nu}+ \cdots \,, \qquad
Q_{\rm eff} = \left(\begin{array}{ccc}
\frac{2}{3} & 0 & 0 \\
0 & -\frac{1}{3} & 0 \\
0 & 0 & -\frac{1}{3}\,\frac{m_V^2}{m_\phi^2}
\end{array} \right) \,.
\label{eom}
\end{eqnarray}
Here we assumed on-shell photons and introduced the effective charge matrix $Q_{\rm eff}$. Furthermore
we approximated $m_\omega \simeq m_\rho \simeq m_V= 778$ MeV.
The parameter $e_V$ was introduced in (\ref{em-vertices:a}). It describes in part the direct conversion of
a vector meson into a photon as illustrated in (\ref{eom}) and as encountered in Figure 1) by type 5)-7) diagrams.
Note that the tensors $C^{\mu, \alpha \beta}_{abc}(\bar p,p)$ and $\Cchi^{\mu, \alpha \beta}_{abc}(\bar p,p)$ of
(\ref{def-C-tensor3}, \ref{def-Cchi-tensor3}) arise also from the subleading
processes proportional to $e_A\,h_A/f$ and $e_A\,b_A/f$. The corresponding diagrams are
of type 1) in Figure 1.

There are further two tensors $D^{\mu, \alpha \beta}_{ab}(\bar p,p)$ and $\tilde D^{\mu, \alpha \beta}_{ab}(\bar p,p)$
of type 6) that probe the parameter combinations $e_V\,h_V\,h_P$ and  $e_V\,\tilde h_V\,h_P$ respectively,
\begin{eqnarray}
&& D^{\mu, \alpha \beta}_{ab}(\bar p,p)=-i\,\int
\frac{d^4l}{(2\pi)^4}\,S_a (l) \,
\Big\{q_\lambda \, g^{\mu}_{\phantom{\mu}\nu} - q_\nu \, g^{\mu}_{\phantom{\mu}\lambda }\Big\}\,
\nonumber \\
&& \qquad \quad \, \times\;
\bar p_{\tau} \,S_b^{\tau ,\, \rho \lambda }(\bar p+l)\,
g_{\kappa\rho} \, S_b^{\kappa \nu,\, \beta}(p+l)  \,p^\alpha \,,
\nonumber\\ \nonumber\\
&& \tilde D^{\mu, \alpha \beta}_{ab}(\bar p,p)=-i\,\int
\frac{d^4l}{(2\pi)^4}\,S_a (l)\,
\Big\{q_\lambda \, g^{\mu}_{\phantom{\mu}\nu} - q_\nu \, g^{\mu}_{\phantom{\mu}\lambda}\Big\}\,
\nonumber \\
&& \qquad \quad \, \times\;
\bar p_{\tau } \,S_{b}^{ \tau \lambda }(\bar p+l)\,
S^{\nu\beta}_{b}(p+l)\,p^\alpha \,.
\label{def-D-tensor3}
\end{eqnarray}
The derivation of  the corresponding contributions is streamlined by the application of the effective vertices
\begin{eqnarray}
&& {\mathcal L} = 3\,i\,\frac{e_V\,h_V}{8}\,F^{\mu \nu}\,{\rm tr}\,\Big\{
V_{\alpha \mu}\,Q_{\rm eff}\,V^{\alpha}_{\;\;\, \nu} \Big\}
\nonumber\\
&& \quad +\,i\,\frac{e_V\,\tilde h_V}{8\,m^2_V}\,F^{\mu \nu}\,{\rm tr}\,\Big\{
(\partial^\alpha V_{\alpha \mu})\,Q_{\rm eff}\,(\partial^\beta V_{\beta \nu}) \Big\}\,,
\label{def-eff-eM}
\end{eqnarray}
as implied by the hadronic interaction (\ref{interaction-tensor}) and the equation of motion (\ref{eom}).
Note that the tensor $D^{\mu, \alpha \beta}_{ab}(\bar p,p)$ arises also from a contribution proportional to
$e_M\,h_P$. The latter is of type 1) in Figure \ref{fig:1}.

We are left with diagrams of type 2), 4), 5) and 7) in Figure \ref{fig:1}.
There are contributions proportional to $e_V/f^2$, which follow upon expanding the first vertex of
(\ref{em-vertices:a}) to second order in the Goldstone boson fields, i.e.
\begin{eqnarray}
{\mathcal L} =- \frac{e_V\,m_V}{64\,f^2}\,\,F^{\mu \nu}\,{\rm tr}\,\Big\{ V^{\mu\nu}\,\Big[ \Phi, \,\Big[Q, \,\Phi\Big]_- \Big]_-\Big\} \,.
\nonumber
\end{eqnarray}
This gives rise to diagrams of type 2 as of
Figure \ref{fig:1}. The tensor $E^{\mu, \alpha \beta}_{ab}(\bar p,p)$ will describe such processes with
\begin{eqnarray}
&& E^{\mu, \alpha \beta}_{ab}(\bar p,p)=2\,i\,\int \frac{d^4l}{(2\pi)^4}\,S_a (l)\, q_\tau\,
S^{\tau \mu ,\,\beta}_{b}(p+l)\,p^\alpha \,.
\label{def-E-tensor3}
\end{eqnarray}
Further contributions are implied by the Weinberg-Tomozawa interaction where the final vector meson is converted directly
into a photon. Such terms are proportional to $e_V/f^2$ as well and are characterized  by
diagram 5) of Figure \ref{fig:1}. They are associated with the effective vertex
\begin{eqnarray}
{\mathcal L}= -\frac{e_V}{32\,f^2\,m_V}\,F^{\nu \alpha}{\tr } \,\Big\{(\partial^\mu\,V_{\mu \alpha})\,\Big[[\Phi, \,(\prt_\nu \Phi )]_-,\,Q_{\rm eff}\Big]_-\Big\} \,,
\end{eqnarray}
and the tensor
\begin{eqnarray}
&& F^{\mu, \alpha \beta}_{ab}(\bar p,p)= i\,\int \frac{d^4l}{(2\pi)^4}\, S_a (l)\,
S^{\nu \beta}_{b}(p+l) \,(l - \bar p)^\lambda \,
\Big\{q_\lambda \,g^{\mu \nu}-q^\nu \,g^{\mu}_{\phantom{\mu} \lambda}\Big\}\,p^\alpha\,.
\label{def-F-tensor3}
\end{eqnarray}
Contributions analogous to (\ref{def-F-tensor3}) where the Weinberg-Tomozawa interaction is replaced
by the chiral correction operators
introduced in (\ref{def-chi-terms}, \ref{def-g-terms}) are described by the tensors (\ref{def-E-tensor3}) and
\begin{eqnarray}
&& G^{\mu, \alpha \beta}_{ab}(\bar p,p)=
-2\, i \,\int \frac{d^4l}{(2\pi)^4}\, S_a (l) \,
q_{\tau}\,S^{\tau  \mu,\, \beta}_{b}(p+l) \,(  \bar p \cdot l) \, p^\alpha\,.
\label{def-G-tensor3}
\end{eqnarray}
The corresponding effective vertices read
\begin{eqnarray}
&&{\mathcal L} = - \frac{e_V\,b_D}{128\,f^2\,m_V}\,F^{\mu \nu}\,{\tr } \,\Big\{\Big[Q_{\rm eff}, \,V_{\mu \nu}\Big]_+ \,
\Big[\Phi,\,\Big[\Phi,\,\chi_0\Big]_+\Big]_+ \Big\}
\nonumber\\
&& \quad -\, \frac{e_V\,g_D}{32\,f^2\,m_V}\,F^{\mu \nu}\,{\tr } \,
\Big\{\Big[ V_{\mu \nu }\,,(\partial_\alpha \Phi) \Big]_+\,\Big[ (\partial^\alpha \Phi),\,Q_{\rm eff}\ \Big]_+\Big\}
\nonumber\\
&& \quad -\,\frac{e_V\,g_F}{32\,f^2\,m_V}\,F^{\mu \nu}\,{\tr } \,
\Big\{\Big[ V_{\mu \nu }\,,(\partial_\alpha \Phi) \Big]_-\,\Big[ (\partial^\alpha \Phi),\,Q_{\rm eff}\ \Big]_-\Big\} \,.
\end{eqnarray}
While the terms proportional to $e_V\,b_D/f^2$ probe the tensor $E^{\mu, \alpha \beta}_{ab}(\bar p,p)$, the contributions
proportional to $e_V\,g_D/f^2$ or $e_V\,g_F/f^2$ require the tensor $G^{\mu, \alpha \beta}_{ab}(\bar p,p)$.

We are left with diagrams of type 4) and 7).
Two diagrams of type 4) are proportional to $e_A\,h_A$ and $e_A\,b_A$. They vanish identically since the
vertex proportional to $e_A$ involves the structure $(\partial_\mu V^{\mu \nu})$, i.e. the intermediate
vector meson carrying 4-momentum $p$ has  no non-propagating $J^P=1^+$ component and therefore can not connect
to the initial axial-vector state. The remaining two diagrams of type 7 are proportional to
$e_V\,b_A\,h_A$ and $e_V\,b^2_A$ and described by the two tensors
\begin{eqnarray}
&& H^{\mu, \alpha \beta  }_{abc}(\bar p,p)=
+2 \, i \, \int \frac{d^4l}{(2\pi)^4}\, S_a (l)\,
q_\nu \,\epsilon^{\mu \nu}_{\phantom{\mu\nu} \bar\alpha \bar\beta} \,
\epsilon_{\sigma \tau \kappa \lambda} \,l^\lambda \,
\nonumber \\
&& \qquad \quad \times \;
S_b^{\bar\alpha \bar\beta, \,\sigma \tau}(p)\, S_c^{\kappa \beta}(p+l) \, p^\alpha
\,,
\nonumber \\ \nonumber\\
&& \bar H^{\mu, \alpha \beta  }_{abc}(\bar p,p)=
+2 \, i \, \int \frac{d^4l}{(2\pi)^4}\, S_a (l)\,
q_\nu \,\epsilon^{\mu \nu}_{\phantom{\mu\nu} \bar\alpha \bar\beta}\,
\epsilon_{\sigma \tau \kappa \lambda}  \,
\nonumber \\
&& \qquad \quad \times \;
S_b^{\bar\alpha \bar\beta,\,\sigma \tau}(p) \,S_c^{\kappa \lambda,\beta}(p+l) \,p^\alpha \,,
\label{def-H-tensor3}
\end{eqnarray}
where we note the relevance of the effective vertices (\ref{def-eff-eA}).

We point the readers to the fact that the tensors as given in
(\ref{def-AB-tensor3}, \ref{def-C-tensor3}, \ref{def-Cchi-tensor3}, \ref{def-D-tensor3}, \ref{def-E-tensor3}, \ref{def-F-tensor3}, \ref{def-G-tensor3}, \ref{def-H-tensor3})
are not anti-symmetric in the indices $\alpha \leftrightarrow \beta$ as of notational convenience. While
deriving the decay parameters
$d_{1^+\to \, \gamma \,0^-}$  the wave-functions project onto the relevant component,
the application of
the projection formulae (\ref{result:width-1plus:a}) requires
an explicit anti-symmetrization. Detailed results for the projected tensor integrals
(\ref{def-AB-tensor3}, \ref{def-C-tensor3}, \ref{def-Cchi-tensor3}, \ref{def-D-tensor3}, \ref{def-E-tensor3}, \ref{def-F-tensor3}, \ref{def-G-tensor3}, \ref{def-H-tensor3})
presented in terms of the master integrals (\ref{def-master-loops}) are delegated to Appendix C.

\clearpage

\subsection{The $1^+ \to \, \gamma \,1^-$ process}
\label{subsec:dec-oneminus}

Like for the $1^+ \to \,\gamma\,0^-$ process we specify all loop tensors relevant for  the
$1^+ \to \, \gamma \,1^-$ process explicitly in terms of gauge invariant combinations. Some of the
structures have already been defined in \cite{Lutz-Soyeur:2007} using a slightly different notation.
Appendix D offers results of the Passarino-Veltman reduction of the tensor integrals.

Class A) and class B) sets of diagrams are defined by the processes where the photon is radiated from the charge
of an intermediate Goldstone boson or a vector meson, as depicted by the first diagram of Figure \ref{fig:1}.
The contributions are
proportional either to the parameter combination  $e\,h_A$ or $e\,b_A$. To form gauge invariant tensors
we have to consider diagrams as labelled in Figure \ref{fig:1} by type 2)-3) in addition. The latter are required
since the resonance vertex as well as the hadronic 3-point vertex proportional to $h_A$
involves  a covariant derivative that has a term proportional to the photon field. If the final state
is a charged vector meson a gauge invariant combination probes type 4) of
Figure \ref{fig:1} as well. Whereas the tensors
$A^{\bar \alpha \bar \beta ,\mu, \alpha \beta}_{ab}(\bar p, p) $ and
$B^{\bar \alpha \bar \beta ,\mu, \alpha \beta}_{ab}(\bar p, p)$
probe the parameter combination $e\,h_A$, the tensors
$A^{\bar \alpha \bar \beta ,\mu, \alpha \beta}_{\chi, ab}(\bar p, p) $ and
$B^{\bar \alpha \bar \beta ,\mu, \alpha \beta}_{\chi, ab}(\bar p, p)$ select the parameter combination $e\,b_A$.
We introduce

\begin{eqnarray}
&& A^{\bar \alpha \bar \beta ,\mu, \alpha \beta}_{ab}(\bar p, p)=- i\, \int \frac{d^4l}{(2\pi)^4}\,S_a(l)\,
\Bigg\{ \epsilon^{\bar \alpha \bar \beta}_{\;\;\;\;\; \sigma \tau} \,\Big[
-g^{\mu \sigma}\,S_{b}^{\tau \beta }(p+l)\, p^\alpha
\nonumber\\
&& \qquad
+\, l^\sigma\, S^{ \tau \beta}_{b}(\bar p+l)\,g^{\mu \alpha } \,
+(l+q)^\sigma\,S_a(l+q)\,S^{ \tau \beta}_{b}(p+l)\,(q+2\,l)^\mu\,p^\alpha
\Big]
\nonumber\\
&& \qquad  +\,
  \epsilon_{\kappa \rho\, \sigma \tau}\,
  \bar p^{\bar \alpha} \,S^{\mu \bar \beta, \,\kappa \rho}_{\bar p }(p)
  \,l^\sigma \,S_{b}^{\tau \beta }(p+l)\, p^\alpha
- \epsilon_{ \sigma \tau}^{\;\;\;\;\; \rho \bar \beta  }\,\Big[
l_\rho\,g^{\mu \bar \alpha }\,S_{b}^{ \sigma \tau,\,\beta }(p+l)\,  p^\alpha
\nonumber\\
&& \qquad-\,g^{\mu}_{\;\, \rho}\,\bar p^{\bar \alpha}\, S_{b}^{\sigma \tau,\,\beta }(p+l)\,p^\alpha
+  \bar p^{\bar \alpha}\,l_\rho\,S^{\sigma \tau,\,\beta }_{b}(\bar p+l)\,
 \,g^{\mu \alpha }
\nonumber\\
&& \qquad  +\,\bar p^{\bar \alpha }\,(l+q)_\rho\,
S_a(l+q)\,S^{\sigma \tau,\, \beta }_{b}(p+l)\,(q+2\,l)^\mu\,
p^\alpha
\Big] \Bigg\} \,,
\nonumber\\ \nonumber\\
&& A^{\bar \alpha \bar \beta ,\mu, \alpha \beta}_{\chi, \,ab}  (\bar p,p)= +i\, \int
  \frac{d^4l}{(2\pi)^4}\,S_a (l)\, \Bigg\{
    \epsilon_{\phantom{\bar\mu \bar\nu}\sigma \tau }^{ \bar \alpha \bar\beta}\,
    S_b^{\sigma \tau , \,\beta}(\bar p+l) \, g^{\mu \alpha}
\nonumber \\
&& \qquad   +\,\epsilon_{\phantom{\bar\mu \bar\nu}\sigma \tau }^{ \bar\alpha \bar\beta} \, (2\,l+q)^\mu\,
   S_a(q+l) \, S_b^{\sigma \tau , \,\beta }(p+l) \, p^\alpha
\nonumber \\
&& \qquad  +\, \epsilon_{\sigma \tau \kappa \rho} \,
   \Big\{ g^{\mu \bar\alpha} S_{\bar p}^{\bar\beta,\,\kappa \rho}(p)
     + \bar p^{\bar\alpha} \, S_{\bar p}^{\mu \bar\beta,\,\kappa \rho}(p) \Big\}\,S_b^{\sigma \tau, \,\beta}(p +l)\, p^\alpha
   \Bigg\} \,,
   \label{def-A-tensor5}
\end{eqnarray}
and
\begin{eqnarray}
&& B^{\bar \alpha \bar \beta ,\mu, \alpha \beta}_{ab}  (\bar p,p)=-i\, \int
\frac{d^4l}{(2\pi)^4}\,S_a (l)\,\Bigg\{ \epsilon^{\bar \alpha \bar \beta}_{\;\;\;\;\;  \sigma \tau}\,\Big[
+l_\sigma\,S_{b}^{ \mu \tau ,\,\beta}(p+l)\,p^\alpha
\nonumber\\
&& \qquad
+\, l^\sigma\,g_{\kappa \rho}\,\Big\{
S_{b}^{ \tau ,\,\mu \kappa}(\bar p+l)\,
S^{\rho \beta }_{b}(p+l) + S_{b}^{ \tau \kappa}(\bar p+l)\,S^{\mu \rho ,\,\beta }_{b}(p+l)\Big\}\,
p^\alpha
\nonumber\\
&& \qquad
+\, l^\sigma \,S^{ \tau \beta}_{b}(\bar p+l)\, g^{\mu \alpha }
+ l^\sigma\,S_{b}^{  \tau,\,\mu \beta }(\bar p+l)\, p^\alpha\,
\Big]
\nonumber\\
&& \qquad  +\,
  \epsilon_{\kappa \rho\, \sigma \tau}\,
\bar p^{\bar \alpha} \,S_{\bar p}^{\mu \bar \beta, \,\kappa \rho}(p)
  \,l^\sigma \,S_{b}^{\tau \beta }(p+l)\, p^\alpha
 - \epsilon_{ \sigma \tau}^{\;\;\;\;  \rho \bar \beta}\,\Big[
l_\rho\,g^{\mu\bar \alpha }\,S_{b}^{ \sigma \tau ,\,\beta}(p+l)\, p^\alpha\,
\nonumber\\
&& \qquad  +\,l_\rho\,\bar p^{\bar \alpha}\,g_{\kappa \rho}\,
\Big\{ S_{b}^{\sigma \tau,\,\kappa  }(\bar p+l)\,S^{\mu \rho, \,\beta }_{b}(p+l)
+ S_{b}^{ \sigma \tau ,\,\mu \kappa }(\bar p+l)\,S^{ \rho \beta}_{b}(p+l)\Big\}\,p^\alpha
\nonumber\\
&& \qquad   +\, l_\rho\,\bar p^{\bar \alpha} \,S^{\sigma \tau ,\,\beta}_{b}(\bar p+l)\,
g^{\mu \alpha } +l_\rho\,\bar p^{\bar \alpha }\,
S_{b}^{ \sigma \tau ,\,\mu \beta}(\bar p+l)\, p^\alpha\,
\Big]\Bigg\}\,,
\nonumber\\ \nonumber\\
&& B^{\bar \alpha \bar \beta ,\mu, \alpha \beta}_{\chi, \,ab}  (\bar p,p)= +i\, \int
  \frac{d^4l}{(2\pi)^4}\,S_a (l)\, \Bigg\{ \epsilon_{\phantom{\bar\mu \bar\nu}\sigma \tau}^{\bar\alpha \bar\beta}\,
   S_b^{\sigma \tau , \,\beta}(\bar p +l)\,g^{\mu \alpha}
 \nonumber \\
&& \qquad +\, \epsilon_{\phantom{\bar\mu \bar\nu}\sigma \tau}^{\bar\alpha \bar\beta}\,
  \Big[ S_b^{\sigma \tau,\,\mu \beta}(\bar p +l) +
     S_b^{\sigma \tau,\,\mu \kappa }(\bar p +l)\,g_{\kappa \rho}\, S_{b}^{\rho \beta} (p+l)
\nonumber \\
&& \qquad +\,  S_b^{\sigma \tau ,\,\kappa}(\bar p +l)\,g_{\kappa \rho} \,S_{b}^{\mu \rho,\,\beta }(p+l)
 \Big]\,p^\alpha
\nonumber \\
&& \qquad  +\, \epsilon_{\sigma \tau \kappa \rho} \,
   \Big\{ g^{\mu \bar\alpha} S_{\bar p}^{\bar\beta,\,\kappa \rho}(p)
     + \bar p^{\bar\alpha} \, S_{\bar p}^{\mu \bar\beta,\,\kappa \rho}(p) \Big\}\,S_b^{\sigma \tau, \,\beta}(p +l)\, p^\alpha
   \Bigg\} \,.
\label{def-B-tensor5}
\end{eqnarray}
Note the type 4) terms proportional to $S^{\bar \alpha \bar \beta,\alpha \beta}_{\bar p} (p)$
in (\ref{def-A-tensor5}, \ref{def-B-tensor5}).
Here we identify $m_{\bar p}^2 =\bar p^2$. Such contributions are not at odds with parity conservation.
This follows since the tensor field carries spin one quanta with both parities.  Analogous terms where the
photon couples to the initial axial-vector meson do not arise due to parity conservation. Since the resonance field
couples always with $(\partial_\tau R^{\tau \alpha })$, only the parity $P=+1 $ component is accessed.

We continue with class C)  processes where the Goldstone boson emitted from the molecule is
converted into a vector meson while radiating a photon. The contributions are either of type 1) or 6) in Figure 1
and described by four generic tensors.
There are terms proportional to $e_V\,h_A\,h_V/f$ and $e_V\,h_A\,\tilde h_V/f$ associated with the first
vertex of (\ref{def-eff-eA}) and the following two loop integrals
\begin{eqnarray}
&& C_{abc}^{\bar \alpha \bar \beta ,\mu, \alpha \beta}(\bar p, p)
=+2\,i\, \int
\frac{d^4l}{(2\pi)^4}\,S_a (l) \,q_\nu \,\epsilon^{\mu \nu }_{\quad \sigma \tau}\,  l^\tau\,
\nonumber\\
&& \qquad \;\times \;  S_b^{\bar \beta \kappa, \sigma}(l+q)\,g_{\kappa \rho } \,
S_c^{ \bar \alpha \rho, \beta}(l+p)\,p^\alpha \,,
\nonumber\\ \label{def-C-tensor5}\\
&& \tilde C_{abc}^{\bar \alpha \bar \beta ,\mu, \alpha \beta}(\bar p, p)
=+2\,i\, \int
\frac{d^4l}{(2\pi)^4}\,S_a (l)\,q_\nu\,\epsilon^{\mu \nu }_{\quad \sigma \tau}\,l^\tau\,
\Bigg\{S_b^{\bar \alpha \sigma}(l+q)\,S_c^{\bar \beta \beta}(l+p)\,
\nonumber\\
&& \qquad +\, \bar p^{\bar \alpha }\,g_{\kappa \rho }\Big\{ S_b^{ \kappa \bar \beta,\, \sigma}(l+q)\,S_c^{\rho \beta}(l+p)\,
+ S_b^{\kappa \sigma}(l+q)\,S_c^{\rho \bar \beta, \,\beta}(l+p)\Big\} \,
\Bigg\}\,p^\alpha \,.\nonumber
\end{eqnarray}
The tensors (\ref{def-C-tensor5}) arise also from subleading
processes of type 1) in Figure \ref{fig:1}. In this case they are proportional to
$e_A\, h_V$ (tensor $C$), $e_A\,\tilde h_V$ (tensor $\tilde C$).
Additional terms of type 6) in Figure 1 are implied by the second vertex of (\ref{def-eff-eA}).
They are proportional to either $e_V\,b_A\,h_V/f$  or $e_V\,b_A\,\tilde h_V/f$ and define the two structures
\begin{eqnarray}
&& C_{\chi, \, abc}^{\bar \alpha \bar \beta ,\mu, \alpha \beta}(\bar p, p)
= - i \,m_b^2\, \int
\frac{d^4l}{(2\pi)^4}\,S_a (l)\, q_\nu \,\epsilon^{\mu \nu}_{\phantom{\mu\nu}\sigma \tau} \,
\nonumber\\
&& \qquad \times \,
 S_b^{\bar\beta \kappa,\,\sigma \tau}(q+l) \,g_{\kappa \rho } \,
S_c^{\bar\alpha \rho,\,\beta}(p+l) \, p^\alpha  \,,
\nonumber\\ \label{def-Cchi-tensor5}\\
&& \tilde C_{\chi, \, abc}^{\bar \alpha \bar \beta ,\mu, \alpha \beta}(\bar p, p)
= - i \,m_b^2\, \int
\frac{d^4l}{(2\pi)^4}\,S_a (l)\,q_\nu \, \epsilon^{\mu \nu}_{\phantom{\mu\nu}\sigma \tau}  \,
\Bigg\{ S_b^{\bar\alpha, \,\sigma \tau}(q+l) \, S_c^{\bar\beta \beta}(p+l)
\nonumber\\
&& \qquad {}
+ \bar p^{\bar\alpha} \, g_{\kappa \rho} \,\Big\{S_b^{\kappa \bar\beta,\,\sigma \tau}(q+l) \, S_c^{\rho\beta}(p+l)
+ S_b^{\kappa, \,\sigma \tau}(q+l) \, S_c^{\rho \bar\beta,\,\beta}(p+l) \Big\}
\Bigg\} \,  p^\alpha  \,. \nonumber
\end{eqnarray}

Consider class D) processes where an intermediate  vector meson emits a photon via magnetic or quadrupole-like
couplings. They are of  type 1) or 6) in Figure 1. Four structures are induced by the effective
vertices (\ref{def-eff-eM}). Contributions proportional to $e_V\,h_V\,h_A/f$ and $e_V\,\tilde h_V\,h_A/f$
are introduced with
\begin{eqnarray}
&& D_{ab}^{\bar \alpha \bar \beta ,\mu, \alpha \beta}(\bar p, p)
= -i \, \int
\frac{d^4l}{(2\pi)^4}\,S_a (l)\,
\Big\{q_\lambda \, g^{\mu}_{\phantom{\mu}\nu}- q_\nu\, g^{\mu}_{\phantom{\mu}\lambda}\Big\}
\nonumber \\
&& \qquad \times\; \Bigg\{
\epsilon^{\bar \alpha \bar \beta}_{\phantom{\alpha \beta}\sigma  \tau}\,\, l^{\tau }\,
S_b^{\sigma,\,\kappa \lambda  }(\bar p +l) \,g_{\kappa\rho}\,S_b^{\rho \nu, \,\beta}(p+l)
\nonumber \\
&& \qquad  \quad +\,
\bar p^{\bar \alpha} \,\epsilon^{\rho \bar \beta}_{\phantom{\alpha \beta}\sigma  \tau}\,l_{\rho} \,
S_b^{\sigma \tau,\,\kappa \lambda}(\bar p +l) \,g_{\kappa\rho}\,S_b^{\rho \nu, \,\beta}(p+l)
 \Bigg\}\,p^\alpha  \,,
\nonumber\\ \nonumber\\
&& \tilde D_{ab}^{\bar \alpha \bar \beta ,\mu, \alpha \beta}(\bar p, p)
=-i\, \int
\frac{d^4l}{(2\pi)^4}\,S_a (l)\, \Big\{q_\lambda \, g^{\mu}_{\phantom{\mu}\nu}- q_\nu \, g^{\mu}_{\phantom{\mu}\lambda}\Big\}
\nonumber\\
&& \qquad \times \; \Bigg\{
\epsilon^{\bar \alpha \bar \beta}_{\;\;\;\;\;  \sigma \tau}\, l^\tau\,
S_{b}^{ \sigma \lambda}(\bar p+l)\,S^{\nu \beta }_{b}(p+l)
\nonumber\\
&& \qquad \quad  +\,\bar p^{\bar \alpha}\, \epsilon^{\rho \bar \beta}_{\phantom{\alpha \beta } \sigma \tau}\,l_\rho\,
S_{b}^{\sigma \tau,\,\lambda  }(\bar p+l)\,S^{\nu  \beta }_{b}(p+l)
\Bigg\}\,p^\alpha \,.
\label{def-D-tensor5}
\end{eqnarray}
Additional two terms  proportional to
$e_V\,h_V\,b_A/f$ and $e_V\,\tilde h_V\,b_A/f$ are associated with the loop tensors
\begin{eqnarray}
&& D_{\chi, \, ab}^{\bar \alpha \bar \beta ,\mu, \alpha \beta}(\bar p, p)
= -i \, \int \frac{d^4l}{(2\pi)^4}\,S_a (l)\,
\Big\{ q_{\lambda} \, g^\mu_{\phantom{\mu}\nu} -q_{\nu} \, g^{\mu }_{\phantom{\mu} \lambda} \Big\}
\nonumber\\
&& \qquad  \times \;\epsilon^{\bar\alpha \bar\beta}_{\phantom{\alpha\beta}\sigma \tau} \,
S_b^{\sigma \tau,\,\kappa \lambda }(\bar p+l)\,g_{\kappa \rho} \, S_b^{\rho \nu,\,\beta}(p+l) \, p^\alpha \,,
\nonumber\\ \nonumber\\
&& \tilde D_{\chi, \, ab}^{\bar \alpha \bar \beta ,\mu, \alpha \beta}(\bar p, p)
= -i \, \int \frac{d^4l}{(2\pi)^4}\,S_a (l)\,
\Big\{q_\lambda \, g^\mu_{\phantom{\mu}\nu} - q_\nu \, g^\mu_{\phantom{\mu}\lambda} \Big\} \,
\nonumber\\
&& \qquad  \times \;\epsilon^{\bar\alpha \bar\beta}_{\phantom{\alpha\beta}\sigma \tau} \,S_b^{\sigma \tau,\, \lambda}(\bar p+l) \,
S_b^{\nu \beta}(p+l) \, p^\alpha \,.
\label{def-D-chi-tensor5}
\end{eqnarray}
We continue with class E)  processes where the vector meson emitted from the molecule is
converted into a Goldstone boson while radiating a photon. Again type 1) and 6) diagrams in Figure 1 contribute.
Two terms proportional to $e_V\,h_A\,h_P/f$ and $e_V\,b_A\,h_P/f$ as implied by the vertices (\ref{def-eff-eA}) are
required. The corresponding  tensors are
\begin{eqnarray}
&& E_{abc}^{\bar \alpha \bar \beta ,\mu, \alpha \beta}(\bar p, p)
=+2\,i\, \int \frac{d^4l}{(2\pi)^4}\,S_a (l)\,
l^{\bar \alpha} \,\bar p^{\bar \beta}\,q_\nu\,\epsilon^{\mu \nu}_{\phantom{\mu \nu}\sigma \tau}\,
(l+\bar p)^{\tau}\,
\nonumber\\
&& \qquad  \times \;S_b(l+\bar p)\,S^{\sigma \beta}_c(l+p)\,p^{\alpha}\,,
\nonumber\\ \nonumber\\
&& \bar E_{abc}^{\bar \alpha \bar \beta ,\mu, \alpha \beta}(\bar p, p)
= - i \,m_c^2\, \int \frac{d^4l}{(2\pi)^4}\,S_a (l)\,
 l^{\bar\alpha} \,  \bar p^{\bar\beta}  \,
q_\nu \,\epsilon^{\mu \nu}_{\phantom{\mu\nu}\sigma \tau} \,
\nonumber\\
&& \qquad  \times \,
S_b(l+\bar p)\,S_c^{\sigma \tau,\beta}(p+l) \, p^\alpha  \,.
\label{def-E-tensor5}
\end{eqnarray}

We are left with class F) processes characterized by type 4) or 7) diagrams
in Figure 1. The photon is radiated of the final vector meson via magnetic
couplings (see (\ref{em-vertices:a}, \ref{def-eff-eM})).
Type 7) contributions are proportional to $e_V\,h_V\,h_A$ and
$e_V\,h_V\,b_A$ with the corresponding tensors
\begin{eqnarray}
&& F_{ab}^{\bar \alpha \bar \beta ,\mu, \alpha \beta}(\bar p, p)
=-i\, \int \frac{d^4l}{(2\pi)^4}\,S_a (l)\,
\Big\{q^{\bar \beta} \, g_\nu^{\phantom{\nu}\mu} - q_\nu \, g^{\bar \beta \mu} \Big\}\,
\epsilon_{\sigma \tau \kappa \rho} \, l^\rho \,
\nonumber \\
&& \qquad  \times\;S_{\bar p}^{ \bar\alpha \nu,\,\sigma \tau}(p) \,S_b^{\kappa \beta}(p+l) \, p^\alpha \,
 \,,
\nonumber\\ \label{def-F-tensor5}\\
&& F_{\chi, \, ab}^{\bar \alpha \bar \beta ,\mu, \alpha \beta}(\bar p, p)
= -i\, \int \frac{d^4l}{(2\pi)^4}\,S_a (l)\,
\Big\{q^{\bar\beta} \, g_\nu^{\phantom{\nu}\mu} - q_\nu \, g^{\bar\beta \mu} \Big\}
\, \epsilon_{\sigma \tau\kappa \rho} \,
\nonumber \\ && \qquad  \times\;
S_{\bar p}^{\bar\alpha \nu,\,\sigma \tau}(p) \, S_b^{\kappa \rho,\,\beta}(p+l)\,p^\alpha  \,. \nonumber
\end{eqnarray}
There are further two diagrams of type 7) proportional to $e_V\,\tilde h_V\,h_A$ and
$e_V\,\tilde h_V\,b_A$, however,  they vanish identically.
Note that the tensors of (\ref{def-F-tensor5}) describe also the process proportional to $e_M\,h_A$ and
$e_M \,b_A$, which is of type 4) in Figure 1.

We point the readers to the fact that the tensors as given in
(\ref{def-A-tensor5}-\ref{def-F-tensor5})
are not anti-symmetric in the indices $\alpha \leftrightarrow \beta$ and
$\bar \alpha \leftrightarrow \bar \beta$ as of notational convenience. While
deriving the decay parameters $d^{(1)}_{1^+\to \, \gamma\, 1^-}$ and
$d^{(2)}_{1^+\to \, \gamma\, 1^-}$ the wave-functions project onto the relevant component,
the application of the projection formulae (\ref{result-projection-1plus}) requires
an explicit anti-symmetrization. Explicit results for the projected tensor integrals
(\ref{def-A-tensor5}-\ref{def-F-tensor5})
presented in terms of the master integrals (\ref{def-master-loops}) are delegated to Appendix D.

\clearpage

\subsection{Radiative decays of the $f_1(1282) $}
\label{subsec:f1}

We confront the results predicted by the effective field theory developed in this work
with the empirical results on the radiative decay properties of the $f_1(1282)$ molecule.
There are three measured quantities to be compared with. From the three possible decays,
$f_1(1282) \to \,\gamma \,\rho_0,\, \gamma\,\omega, \, \gamma\,\phi$, two are studied so far experimentally.
Note that decays into a photon and a $\pi_0$ or $\eta$ meson are suppressed by charge conjugation considerations.

The process studied best involves the neutral $\rho$ meson in the final state. According to the Particle Data Group
\cite{PDG:2006} the branching ratio is
\begin{eqnarray}
\frac{\Gamma_{f_1(1282) \,\to \, \gamma\,\rho^0 }}{\Gamma_{f_1(1282),\;{\rm tot}}} =0.055 \pm 0.013 \,.
\label{f1:branching}
\end{eqnarray}
Given the total width of $(24.2 \pm 1.1)$ MeV quoted in \cite{PDG:2006} we derive the following condition on the two
decay constants
\begin{eqnarray}
\Big| d^{\,(1)}_{f_1(1282) \,\to \,\gamma \,\rho }\Big|^2 + 2.73\,\Big| d^{\,(2)}_{f_1(1282) \,\to \,\gamma \,\rho }\Big|^2
= 1.25 \pm 0.35 \,,
\label{f1-empirical-constraint:a}
\end{eqnarray}
as introduced in (\ref{def-transition-amplitude-axial-b}). The number $2.73\simeq m_{1^+}^2/m_{1^-}^2$ in
(\ref{f1-empirical-constraint:a}) follows upon the obvious identification of the vector and axial-vector masses
$m_{1^-}$ and $m_{1^+}$ (see
 (\ref{result:width-1plus:c})). Additional information on the decay parameters stems
from an analysis of the subsequent two-pion decay of the $\rho$ meson, i.e.
\begin{eqnarray}
f_1(1282)\to \rho^0 \gamma \to \pi^+ \pi^- \gamma \,,
\end{eqnarray}
which determines  the angular distribution of the primary decay process.
According to \cite{Amelin:1994ii}  and Appendix A the latter provides the constraint
\begin{eqnarray}
5.46\,\Bigg| \frac{d^{\,(2)}_{f_1(1282) \,\to \,\gamma \,\rho }}{d^{\,(1)}_{f_1(1282) \,\to \,\gamma \,\rho }}\Bigg|^2
= 3.9 \pm 0.9 ({\rm stat}) \pm 1.0 ({\rm syst}) \,,
\label{f1-empirical-constraint:b}
\label{}
\end{eqnarray}
with the identification $5.46 \simeq 2\,m_{1^+}^2/m_{1^-}^2$ (see Appendix A).

A further empirical result is available for the $f_1(1282) \to \gamma \,\phi$ process. According to the
Particle Data Group \cite{PDG:2006} the branching ratio is
\begin{eqnarray}
\frac{\Gamma_{f_1(1282)\, \to \, \gamma\,\phi } }{\Gamma_{f_1(1282),\;{\rm tot}}} =  (7.4 \pm 2.6) \,10^{-4}\,,
\end{eqnarray}
which implies the relation
\begin{eqnarray}
\Big| d^{\,(1)}_{f_1(1282) \,\to \,\gamma \,\phi }\Big|^2 + 1.58\,\Big| d^{\,(2)}_{f_1(1282) \,\to \,\gamma \,\phi }\Big|^2
= 0.087 \pm 0.035 \,.
\label{f1-empirical-constraint:c}
\end{eqnarray}

For the corresponding decay amplitude we establish our results in terms of the coupling constants and
loop tensors introduced in sections \ref{subsec:chiral-corr}, \ref{subsec:counting} and \ref{subsec:dec-oneminus}. It is emphasized that all parameters
were determined already with the exception of the parameters $e_M$ and $h_V$, for which only
a linear combination is determined  by the electromagnetic properties of the
$\rho$ meson (see (\ref{value-eM:b})). The knowledge of the magnetic moment of the $K^*$ meson would provide
an additional condition from which the parameters $e_M$ and $h_V$ could be determined unambiguously.

Before presenting explicit numerical results for the decay amplitudes expressed in terms of the loop tensors introduced in section \ref{subsec:dec-oneminus}
we introduce some notation as to streamline the presentation. Since the tensors $A,F$
(see (\ref{def-A-tensor5}, \ref{def-F-tensor5})) and also $B,D, \tilde D, F$
(see (\ref{def-B-tensor5}, \ref{def-D-tensor5}, \ref{def-F-tensor5})) will occur always in
the same combination it is useful to introduce the  objects
\begin{eqnarray}
&& e\,\hat A^{\bar \alpha \bar \beta, \mu, \alpha \beta}_{ab,x} =
e\,A^{\bar \alpha \bar \beta, \mu, \alpha \beta}_{ab}
+ \left(e_M + \frac{3}{4\,(1+x)}\,\left(1+ x\,\frac{m_V^2}{m_\phi^2} \right) e_V \, h_V\right)
F^{\bar \alpha \bar \beta, \mu, \alpha \beta}_{ab}  \,,
\nonumber\\
&& e\,\hat B^{\bar \alpha \bar \beta, \mu, \alpha \beta}_{ab,x} =
e\,B^{\bar \alpha \bar \beta, \mu, \alpha \beta}_{ab}
+ \frac{1}{4\,(1+x)\,m_V^2}\left(1+ x\,\frac{m_V^2}{m_\phi^2} \right)
e_V \,\tilde h_V\,\tilde D^{\bar \alpha \bar \beta, \mu, \alpha \beta}_{ab}
\nonumber\\
&& \quad  \;\, +\, \left(e_M + \frac{3}{4\,(1+x)}\,\left(1+ x\,\frac{m_V^2}{m_\phi^2} \right) e_V \, h_V\right)
\Big( D^{\bar \alpha \bar \beta, \mu, \alpha \beta}_{ab}+F^{\bar \alpha \bar \beta, \mu, \alpha \beta}_{ab} \Big) \,,
\label{def-AB-hat-tensor3}
\end{eqnarray}
where we identified $m_V=776$ MeV with the averaged mass of the $\rho$ and $\omega $ meson. Small effects proportional to
$m_\rho- m_\omega$ are neglected.
Correspondingly the tensors $A_\chi,F_\chi$ (see (\ref{def-A-tensor5}, \ref{def-F-tensor5})) and also
$B_\chi,D_\chi, \tilde D_\chi, F_\chi$ (see (\ref{def-B-tensor5}, \ref{def-D-chi-tensor5}, \ref{def-F-tensor5})) enter the result
in always the same combination
\begin{eqnarray}
&& e\,\bar A^{\bar \alpha \bar \beta, \mu, \alpha \beta}_{ab,x} =
e\,A^{\bar \alpha \bar \beta, \mu, \alpha \beta}_{\chi,\,ab}
+ \left(e_M + \frac{3}{4\,(1+x)} \left(1+ x\,\frac{m_V^2}{m_\phi^2} \right) e_V \, h_V\right)
F^{\bar \alpha \bar \beta, \mu, \alpha \beta}_{\chi, \,ab}  \,,
\nonumber\\
&& e\,\bar B^{\bar \alpha \bar \beta, \mu, \alpha \beta}_{ab,x} =
e\,B^{\bar \alpha \bar \beta, \mu, \alpha \beta}_{\chi, \,ab}
+ \frac{1}{4\,(1+x)\,m_V^2} \left(1+ x\,\frac{m_V^2}{m_\phi^2} \right) e_V \,\tilde h_V\,
\tilde D^{\bar \alpha \bar \beta, \mu, \alpha \beta}_{\chi,\,ab}
\nonumber\\
&& \quad \;\, +\,
\left(e_M + \frac{3}{4\,(1+x)} \left(1+ x\,\frac{m_V^2}{m_\phi^2} \right) e_V \, h_V\right)
\Big( D^{\bar \alpha \bar \beta, \mu, \alpha \beta}_{\chi,\,ab}+F^{\bar \alpha \bar \beta, \mu, \alpha \beta}_{\chi,\,ab} \Big) \,.
\label{def-AB-bar-tensor3}
\end{eqnarray}
We recall that while the tensors $A$ and $A_\chi$ describe the processes where the photon couples to the charge of an
intermediate $K$ meson, the tensors $B$ and $B_\chi$ describe the ones where the photon couples to the charge of an
intermediate $K^*$  meson.  Diagrams of type  1)-4) in  Figure 1 are involved. The index $\chi$ indicates that the
tensors $A_\chi$ and $B_\chi$ probe a 3-point vertex proportional to the parameter $b_A$ that breaks chiral symmetry
explicitly. As demonstrated in section \ref{subsec:dec-vec} the latter plays an important role in the description of the radiative decay
of vector mesons. The additional contributions combined in the tensors $ \hat B$ and $\bar B$ reflect
the magnetic and quadrupole moment of the $K^*$ mesons. They are described by diagrams of type 1),6) and 4),7).

Further diagrams of type 1) and 6) are described by four tensors $C, \,\tilde C$ and $C_\chi, \,\tilde C_\chi$, which
probe the parameters $h_V$ and $\tilde h_V$. For the explicit
definition of the tensors see (\ref{def-C-tensor5}, \ref{def-Cchi-tensor5}). Since they contribute in the same combination always we introduce
the two tensors
\begin{eqnarray}
&& \tilde h_V \,\bar C^{\bar \alpha \bar \beta, \mu, \alpha \beta}_{abc} =
 \tilde h_V\,\tilde C^{\bar \alpha \bar \beta, \mu, \alpha \beta}_{\chi,\,abc}
 -3\,m_V^2\,h_V\,C^{\bar \alpha \bar \beta, \mu, \alpha \beta}_{\chi,\,abc} \,,
\nonumber\\
&& \tilde h_V \,\hat C^{\bar \alpha \bar \beta, \mu, \alpha \beta}_{abc} =
 \tilde h_V\,\tilde C^{\bar \alpha \bar \beta, \mu, \alpha \beta}_{abc}
 -3\,m_V^2\,h_V\,C^{\bar \alpha \bar \beta, \mu, \alpha \beta}_{abc} \,.
\label{def-barC-tensor5}
\end{eqnarray}

We are now well prepared to present
our result for the $f_1(1282) \to \,\gamma \,\rho_0, \,\gamma \,\omega , \,\gamma\,\phi $ decay amplitudes
\begin{eqnarray}
\lefteqn{M^{\bar \alpha \bar \beta, \mu, \alpha \beta}_{f_1 \to \, \gamma \,\rho^0}
=\frac{e\,h_A}{8\,f}\,h^{(f_1)}_{K\,\bar K^*}\,
\Big\{ \hat A^{\bar \alpha \bar \beta, \mu, \alpha \beta}_{K K^*,0}(\bar p,p)
 -\hat B^{\bar \alpha \bar \beta, \mu, \alpha \beta}_{K K^*,0}(\bar p,p)\Big\}  }
\nonumber\\
&& {} +\frac{e\,b_A}{2\,f}\,m_K^2\,h^{(f_1)}_{K\,\bar K^*}\,
\Big\{ \bar A^{\bar \alpha \bar \beta, \mu, \alpha \beta}_{K K^*,0}(\bar p,p)
 -\bar B^{\bar \alpha \bar \beta, \mu, \alpha \beta}_{K K^*,0}(\bar p,p)\Big\}
\nonumber\\
&& {} - \frac{\tilde h_V}{8\,f\,m_V^2}\,  \eAcptn
\,h^{(f_1)}_{K\,\bar K^*}\,\hat C^{\bar \alpha \bar \beta, \mu, \alpha \beta}_{K K^* K^*}(\bar p,p)
\nonumber\\
&& {} - \frac{\tilde h_V}{8\,f\,m_V^2}\,  \eAcptnbA
\,h^{(f_1)}_{K\,\bar K^*}\,\bar C^{\bar \alpha \bar \beta, \mu, \alpha \beta}_{K K^* K^*}(\bar p,p)
\nonumber\\
&& {} + \frac{h_P}{16\,f^3}\, \eAcptn
\, h^{(f_1)}_{K\,\bar K^*}\,
E^{\bar \alpha \bar \beta, \mu, \alpha \beta}_{K K K^*}(\bar p,p)
\nonumber\\
&& {} + \frac{h_P}{16\,f^3}\, \eAcptnbA
\, h^{(f_1)}_{K\,\bar K^*}\,
\bar E^{\bar \alpha \bar \beta, \mu, \alpha \beta}_{K K K^*}(\bar p,p) \,,
\label{f1-rho-final-result}
\end{eqnarray}
and
\begin{eqnarray}
\lefteqn{M^{\bar \alpha \bar \beta, \mu, \alpha \beta}_{f_1 \to \, \gamma \,\omega}
=\frac{e\,h_A}{8\,f}\,h^{(f_1)}_{K\,\bar K^*}\,
\Big\{ \hat A^{\bar \alpha \bar \beta, \mu, \alpha \beta}_{K K^*,2}(\bar p,p)
 -\hat B^{\bar \alpha \bar \beta, \mu, \alpha \beta}_{K K^*,2}(\bar p,p)\Big\}   }
\nonumber\\
&& {} +\frac{e\,b_A}{2\,f}\,m_K^2\,h^{(f_1)}_{K\,\bar K^*}\,
\Big\{ \bar A^{\bar \alpha \bar \beta, \mu, \alpha \beta}_{K K^*,2}(\bar p,p)
 -\bar B^{\bar \alpha \bar \beta, \mu, \alpha \beta}_{K K^*,2}(\bar p,p)\Big\}
\nonumber\\
&& {} + \frac{\tilde h_V}{24\,f\,m_V^2}\, \eAcmtn
\,h^{(f_1)}_{K\,\bar K^*}\,\hat C^{\bar \alpha \bar \beta, \mu, \alpha \beta}_{K K^* K^*}(\bar p,p)
\nonumber\\
&& {} + \frac{\tilde h_V}{24\,f\,m_V^2}\, \eAcmtnbA
\,h^{(f_1)}_{K\,\bar K^*}\,\bar C^{\bar \alpha \bar \beta, \mu, \alpha \beta}_{K K^* K^*}(\bar p,p)
\nonumber\\
&& {} - \frac{h_P}{48\,f^3}\, \eAcmtn
\,h^{(f_1)}_{K\,\bar K^*}\,E^{\bar \alpha \bar \beta, \mu, \alpha \beta}_{K K K^*}(\bar p,p)
\nonumber\\
&& {} - \frac{h_P}{48\,f^3}\, \eAcmtnbA
\,h^{(f_1)}_{K\,\bar K^*}\,\bar E^{\bar \alpha \bar \beta, \mu, \alpha \beta}_{K K K^*}(\bar p,p)\,,
\label{f1-omega-final-result}
\end{eqnarray}
and
\begin{eqnarray}
\lefteqn{M^{\bar \alpha \bar \beta, \mu, \alpha \beta}_{f_1 \to \, \gamma\,\phi}
=\sqrt{2}\,\frac{e\,h_A}{8\,f}\,h^{(f_1)}_{K\,\bar K^*}\,
\Big\{ \hat A^{\bar \alpha \bar \beta, \mu, \alpha \beta}_{K K^*,2}(\bar p,p)
 -\hat B^{\bar \alpha \bar \beta, \mu, \alpha \beta}_{K K^*,2}(\bar p,p)\Big\}  }
\nonumber\\
&& {} +\sqrt{2}\,\frac{e\,b_A}{2\,f}\,m_K^2\,h^{(f_1)}_{K\,\bar K^*}\,
\Big\{ \bar A^{\bar \alpha \bar \beta, \mu, \alpha \beta}_{K K^*,2}(\bar p,p)
 -\bar B^{\bar \alpha \bar \beta, \mu, \alpha \beta}_{K K^*,2}(\bar p,p)\Big\}
\nonumber\\
&& {} - \sqrt{2}\,\frac{\tilde h_V}{24\,f\,m_V^2}\, \eAcmtn
\, h^{(f_1)}_{K\,\bar K^*}\,\hat C^{\bar \alpha \bar \beta, \mu, \alpha \beta}_{K K^* K^*}(\bar p,p)
\nonumber\\
&& {} - \sqrt{2}\,\frac{\tilde h_V}{24\,f\,m_V^2}\, \eAcmtnbA
\, h^{(f_1)}_{K\,\bar K^*}\,\bar C^{\bar \alpha \bar \beta, \mu, \alpha \beta}_{K K^* K^*}(\bar p,p)
\nonumber\\
&& {} + \sqrt{2}\,\frac{h_P}{48\,f^3}\, \eAcmtn
\,h^{(f_1)}_{K\,\bar K^*}\, E^{\bar \alpha \bar \beta, \mu, \alpha \beta}_{K K K^*}(\bar p,p)
\nonumber\\
&& {} + \sqrt{2}\,\frac{h_P}{48\,f^3}\, \eAcmtnbA
\,h^{(f_1)}_{K\,\bar K^*}\, \bar E^{\bar \alpha \bar \beta, \mu, \alpha \beta}_{K K K^*}(\bar p,p) \,.
\label{f1-phi-final-result}
\end{eqnarray}
in terms of the tensors (\ref{def-E-tensor5}, \ref{def-AB-hat-tensor3}, \ref{def-AB-bar-tensor3}, \ref{def-barC-tensor5}).
The resonance-coupling constant $h^{(f_1)}_{K \bar K^*} \simeq 7.3$ GeV$^{-1}$ was introduced in (\ref{def-hi}).
It is related to the
appropriate coupling constant $g_1$ of Table \ref{tab:hadronic-decay} by (\ref{hi-gi}).
It remains to specify the effective coupling constants $\hat e_A$ and $\bar e_A$ entering the results
(\ref{f1-rho-final-result}-\ref{f1-phi-final-result}).  They are related to the radiative decay constants of the light vector mesons
as introduced in (\ref{vec-to-gam-decays}, \ref{eq:deftildeqeA2}), which are
just the sum $e^{...}_A=\hat e^{...}_A + \bar e^{...}_A$ of the two terms. It holds
\begin{eqnarray}
&& \hat e_A^{(K^*_0)\,}  =  e_A + \frac18 \,\left(1+\frac{m_V^2}{m_\phi^2} \right)  e_V \, h_A \,,
\nonumber \\
&& \hat e_A^{(K^*_+)}  =  e_A + \frac14 \, \left(2-\frac{m_V^2}{m_\phi^2} \right) e_V \,  h_A \,,
  \label{eq:hateA}
\nonumber\\
&& \bar e_A^{(K^*_0)\,}  =  - \,  \left(1+\frac{m_V^2}{m_\phi^2} \right) \left(
\frac{m_K^2}{m_{K^*}^2} \right)e_V \, b_A \,,
\nonumber \\
&& \bar e_A^{(K^*_+)}  =   - 2  \, \left(2-\frac{m_V^2}{m_\phi^2} \right) \left(
 \frac{m_K^2}{m_{K^*}^2} \right)\, e_V \, b_A \,.
  \label{eq:bareA}
\end{eqnarray}

\begin{table}[t]
\tabcolsep=2.mm
\renewcommand{\arraystretch}{1.5}
\fontsize{10}{3.5}
\begin{center}
\begin{tabular}{|l||l |l |}
\hline
 &$10\,\times \,d^{(1)}_{f_1^0 \to \,\gamma\, \rho^0}$ & $10\,\times \,d^{(2)}_{f_1^0 \to \,\gamma\, \rho^0}$ \\
\hline
\hline
$ \bar K \,K^*- \bar K^* \,K$ \hfill  &$ -2.44\,e\,h_A +3.13\,e\,b_A $ & $-1.26\,e\,h_A+1.47\,e\,b_A $  \\
$\times \,h_A\,e_V$ & $+0.20\,h_P-0.07\,h_V+0.06\,\tilde h_V $  & $-1.99\,h_P+0.03\,h_V-0.00\,\tilde h_V $\\
$\times \,b_A\,e_V$ & $-0.49\,h_P+0.08\,h_V-0.29\,\tilde h_V $  & $+4.86\,h_P-0.17\,h_V-0.14\,\tilde h_V $\\
$\times \,e_M$ & $-0.10\,h_A+1.08\,b_A $  & $-0.06\,h_A+0.36\,b_A$\\
$\times \,e_A$ & $+0.80\,h_P +0.03\,h_V+0.17\,\tilde h_V$  & $-7.96\,h_P+0.29\,h_V-0.00\,\tilde h_V$\\
\hline
\hline
 &$10\,\times \,d^{(1)}_{f_1^0 \to \,\gamma\, \omega}$ & $10\,\times \,d^{(2)}_{f_1^0 \to \,\gamma\, \omega}$ \\
\hline
\hline
$\bar K \,K^*- \bar K^* \,K$ \hfill  &$ -2.49\,e\,h_A +3.15\,e\,b_A $ & $-1.31\,e\,h_A+1.52\,e\,b_A $  \\
$\times \,h_A\,e_V$ & $-0.01\,h_P-0.05\,h_V+0.01\,\tilde h_V $  & $+0.11\,h_P-0.04\,h_V-0.00\,\tilde h_V $\\
$\times \,b_A\,e_V$ & $+0.03\,h_P+0.61\,h_V-0.02\,\tilde h_V $  & $ -0.27\,h_P+0.22\,h_V+0.01\,\tilde h_V $\\
$\times \,e_M$ & $-0.10\,h_A+1.06\,b_A $  & $-0.06\,h_A+0.36\,b_A$\\
$\times \,e_A$ & $-0.27\,h_P -0.00\,h_V-0.06\,\tilde h_V$  & $+2.80\,h_P-0.10\,h_V-0.01\,\tilde h_V$\\
\hline
\hline
 &$10\,\times \,d^{(1)}_{f_1^0 \to \,\gamma\, \phi}$ & $10\,\times \,d^{(2)}_{f_1^0 \to \,\gamma\, \phi}$ \\
\hline
\hline
$ \bar K \,K^*- \bar K^* \,K$ \hfill  &$ -6.06\,e\,h_A +6.02\,e\,b_A $ & $-4.46\,e\,h_A+4.32\,e\,b_A $  \\
$\times \,h_A\,e_V$ & $\rat{+0.10}{-0.00}h_P -0.11\,h_V+0.02\,\tilde h_V $  & $\rat{-0.67}{-0.01}h_P-0.06\,h_V-0.01\,\tilde h_V $\\
$\times \,b_A\,e_V$ & $\rat{-0.23}{+0.01}h_P+0.60\,h_V-0.06\,\tilde h_V $  & $\rat{+1.64}{+0.03}h_P+0.28\,h_V+0.02\,\tilde h_V $\\
$\times \,e_M$ & $-0.16\,h_A+1.10\,b_A $  & $-0.14\,h_A+0.65\,b_A$\\
$\times \,e_A$ & $\rat{+2.47}{-0.10}h_P -0.51\,h_V+0.27\,\tilde h_V$  & $\rat{-16.9}{-0.28}h_P+0.42\,h_V-0.32\,\tilde h_V$\\
\hline
\hline
\end{tabular}
\caption{Decay constants that are implied by (\ref{f1-rho-final-result}, \ref{f1-omega-final-result}, \ref{f1-phi-final-result}).
The various contributions have to be multiplied by the appropriate resonance-coupling constants
$g_1 = 4.6$ of Table \ref{tab:hadronic-decay}.
We use  $f=90$ MeV, $m_V=776$ MeV. See also (\ref{collection-of-parameters}, \ref{unusual-convention}).}
\label{tab:decay-parameters:f1}
\end{center}
\end{table}

In Table \ref{tab:decay-parameters:f1} the various contributions
to the decay constants are listed. The values quoted in the second
and third column have to be multiplied by the $f_1(1282)$ coupling
constant $g_1 = 4.6$ of Table \ref{tab:hadronic-decay}. In
addition a multiplication of entries of most rows with specific
coupling constants as indicated in the first column is required.
To keep a transparent order in Table
\ref{tab:decay-parameters:f1} we use a somewhat unusual convention
for the representation of complex numbers
\begin{eqnarray}
\rat{x}{y} \equiv x+i\,y  \qquad \mbox{ for $x,y$  real} \,.
\label{unusual-convention}
\end{eqnarray}

All parameters with the exception of one parameter $e_M$ or $h_V$ are
determined in section \ref{sec:counting}.
Taking $e_M=0$ we collect the values of the parameters assumed below:
\begin{eqnarray}
\begin{array}{llll}
e = 0.303\,, \qquad \quad & e_V = 0.22\,,  \qquad \quad&  e_A =0.02\,,  \qquad \quad& e_M =0.00\,,   \\
h_P = 0.29\,,   & h_A= 2.10\,,  & b_A = 0.27\,,  & \\
h_V = 0.45\,,  & \tilde h_V = 3.72\,. & &
\end{array}
\label{collection-of-parameters}
\end{eqnarray}
We checked that none of our conclusions will change if we allow $e_M \neq 0$ within a reasonable range while keeping
the constraint $e_M + 3\,e_V\,h_V/4 \simeq 0.245\,e$.

Upon inspection of Table \ref{tab:decay-parameters:f1} it is clear that the empirical constraints
(\ref{f1-empirical-constraint:a}, \ref{f1-empirical-constraint:b}, \ref{f1-empirical-constraint:c}), in particular
the one for the decay into the $\phi$ meson (\ref{f1-empirical-constraint:c}), can not be met.
We discuss now the relevance of the various contributions. For all six decay parameters of Table
\ref{tab:decay-parameters:f1} the dominant contributions are proportional to $e\,h_A$. The latter are defined
by the processes where the photon couples to the charge of a pseudo-scalar or vector meson with strangeness
plus or minus one. Given the molecule-coupling constant $g_1 = 4.6$ this dominant term predicts the
following decay parameters
\begin{eqnarray}
&& d^{(1)}_{f_1(1282)\, \to\, \gamma \,\rho_0} \simeq -0.71
\,,\qquad \qquad \! d^{(2)}_{f_1(1282) \,\to \,\gamma \,\rho_0}
\simeq -0.37 \,,
\nonumber\\
&& d^{(1)}_{f_1(1282) \,\to \,\gamma \,\omega} \,\simeq
-0.73\,,\qquad \qquad d^{(2)}_{f_1(1282) \,\to \,\gamma \,\omega}
\,\simeq - 0.38\,,
\nonumber\\
&& d^{(1)}_{f_1(1282) \,\to \,\gamma \,\phi}\; \simeq
-1.77\,,\qquad \qquad d^{(2)}_{f_1(1282) \,\to \,\gamma \,\phi}\,
\simeq -1.31\,. \label{f1-leading-order}
\end{eqnarray}
On the other hand summing up all terms of Table \ref{tab:decay-parameters:f1} we arrive at the values
\begin{eqnarray}
&& d^{(1)}_{f_1(1282)\, \to\, \gamma \,\rho_0} \simeq -0.56
\,,\qquad \qquad \! d^{(2)}_{f_1(1282) \,\to \,\gamma \,\rho_0}
\simeq -0.44 \,,
\nonumber\\
&& d^{(1)}_{f_1(1282) \,\to \,\gamma \,\omega}\, \simeq
-0.60\,,\qquad \qquad d^{(2)}_{f_1(1282) \,\to \,\gamma \,\omega}
\,\simeq - 0.31\,,
\nonumber\\
&& d^{(1)}_{f_1(1282) \,\to \,\gamma \,\phi}\; \simeq
-1.52-i\,0.00\,,\;\, d^{(2)}_{f_1(1282) \,\to \,\gamma \,\phi}\,
\simeq -1.24-i\,0.00\,, \label{f1-full-order}
\end{eqnarray}
not much different from the ones in (\ref{f1-leading-order}).
While the decay constants (\ref{f1-leading-order},
\ref{f1-full-order}) are too small as to be consistent with the
empirical constraint (\ref{f1-empirical-constraint:a}) on the
radiative $f_1(1282)$ decay into the $\rho_0$, the decay
parameters are an order of magnitude too large to be consistent
with the observed decay width into the $\phi$ meson
(\ref{f1-empirical-constraint:c}). This result is easily
understood since the $f_1(1282)$ molecule is generated by the
$\bar K\, K^*- K\, \bar K^*$ channel. Since the $\phi$ meson
couples strongly to the latter channels, the contribution of such
channels to its radiative decay into the $\phi$  meson will be
large.

Our results point to the importance of additional channels not
considered yet. We emphasize that according to the power counting
rules developed in this work this is not surprising: a consistent
computation of the radiative decay pattern of the $f_1(1282)$
meson must consider additional contributions from  intermediate states
involving two vector mesons. Since the generalization
of the coupled-channel framework for the inclusion of such
channels is a formidable task this will be addressed in a separate
work. From the overestimate of the $f_1(1282)$ decay into the
$\phi$ meson we would anticipate that the $K^* \bar K^*$ channel
plays an important role in the latter decay. We point out that
since the final decay constant must be a result of a subtle
cancellation mechanism it is important to derive all contributions
implied by the effective Lagrangian developed in section
\ref{sec:counting} systematically. This illustrates the importance of
power-counting rules and motivates the detailed exposition of all
contributions to the decay parameters in Table \ref{tab:decay-parameters:f1}.

\clearpage

\subsection{Radiative decays of the $b_1(1230)$}
\label{subsec:b1}

We consider the radiative decays of the $b^\pm_1(1230)$ and $b^0_1(1230)$ molecules. While the charged states
decay  into the $\gamma \,\pi^\pm$ or the $\gamma\,\rho^\pm$ channels, the neutral one decays either into
$\gamma \,\pi^0$ or $\gamma\,\eta$. Experimental information is available only on the $b_1^\pm \to \, \gamma\,\pi^\pm$
process, for which the Particle Data Group \cite{PDG:2006,Collick:1984} reports the partial decay width
\begin{eqnarray}
\Gamma_{b_1^\pm (1230)\to \, \gamma\,\pi^\pm }= ( 230 \pm 60 )\,{\rm keV } \,.
\label{exp:b1-decay}
\end{eqnarray}
According to (\ref{result:width-1plus:a}) this implies the following decay parameter
\begin{eqnarray}
\Big| d_{b_1^\pm (1230)\to \, \gamma\,\pi^\pm} \Big|= (0.20 \pm 0.03)\, {\rm GeV}^{-1} \,.
\label{b1:decay-parameter}
\end{eqnarray}
We discuss here in  some detail the decays $b_1^\pm \to \, \gamma\,\pi^\pm $ and $b_1^0 \to \, \gamma\,\pi^0 $, for which
their respective decay amplitudes degenerate in the isospin limit. For the latter we provide explicit
expressions in terms of the loop tensors introduced in section \ref{subsec:dec-zerominus}. Analogous results for the decay amplitudes of the
$b_1^\pm \,\to \,\gamma \,\rho^\pm $ and $b_1^0\,\to \,\gamma \,\eta$ processes are delegated to Appendix E.

We remind the readers of the various contributions to the decay amplitude as depicted in Figure 1 generically.
Class A) and class B) sets of diagrams are associated with the tensors
$A^{\mu, \alpha \beta }_{ab}(\bar p, p)$ and $B^{\mu, \alpha \beta }_{ab}(\bar p, p)$ introduced in
(\ref{def-AB-tensor3}).  They are defined by the
processes where the photon is radiated from the charge of an
intermediate Goldstone boson or a vector meson, as depicted by the first diagram of Figure \ref{fig:1}.
The contributions are proportional to the parameter combination $e\,h_P$. To form gauge invariant objects the tensors
include the appropriate diagrams as labelled in Figure \ref{fig:1} by type 2)-3) in addition. Class D) diagrams
are implied by the coupling of the photon to the magnetic or quadrupole moment of an intermediate vector meson. The
gauge invariant tensors $D^{\mu, \alpha \beta}_{ab}(\bar p,p)$ and $\tilde D^{\mu, \alpha \beta}_{ab}(\bar p,p)$
introduced in (\ref{def-D-tensor3}) arise, being of type 1) and 6) in Figure 1. A detailed investigation reveals that the
above tensors contribute for the intermediate states $K\,\bar K^*$ or $\bar K\,K^*$ only. The tensors
$B^{\mu, \alpha \beta}_{ab}(\bar p,p)$, $D^{\mu, \alpha \beta}_{ab}(\bar p,p)$ and
$\tilde D^{\mu, \alpha \beta}_{ab}(\bar p,p)$ occur in the  particular combination
\begin{eqnarray}
&&e\,\hat B^{\mu, \alpha \beta}_{ab,x} = e\,B^{\mu, \alpha \beta}_{ab}
+ \left(e_M + \frac{3}{4\,(1+x)} \left(1+ x\,\frac{m_V^2}{m_\phi^2} \right) e_V \, h_V\right) D^{\mu, \alpha \beta}_{ab}
\nonumber\\
&& \qquad \quad +\, \frac{1}{4\,(1+x)\,m_V^2}\, \left(1+ x\,\frac{m_V^2}{m_\phi^2} \right)e_V \,\tilde h_V\, \tilde D^{\mu, \alpha \beta}_{ab}\,.
\end{eqnarray}

Further contributions of class C) associated with the four tensors $C^{\mu ,\alpha \beta}_{abc}(\bar p,p)$,
$\bar C^{\mu ,\alpha \beta}_{abc}(\bar p,p)$, $\Cchi^{\mu ,\alpha \beta}_{abc}(\bar p,p)$ and
$\barCchi^{\mu ,\alpha \beta}_{abc}(\bar p,p)$ as introduced in
(\ref{def-C-tensor3}, \ref{def-Cchi-tensor3}) arise from the anomalous processes where a Goldstone boson emitting
a photon (either directly or via a vector meson) is converted into a vector meson. Such contributions are proportional
to either $h_A$ or $b_A$ and the effective
decay constants $\bar e^{(K_*)}_A $ or $\hat e^{(K_*)}_A$ as introduced in (\ref{eq:hateA}). The relations
(\ref{eq:hateA}) are supplemented by
\begin{eqnarray}
&& \bar e_A = - 2\,\left( \frac{m_\pi^2}{m_V^2}\right) \,e_V\,b_A\,, \qquad \quad
\hat e_A^{}  =  e_A + \frac14 \,  e_V  \, h_A  \,,
\label{eq:bareA-light}
\end{eqnarray}
where we identify $m_V=776 $ MeV with the average of the $\rho$ and $\omega $ mass. Contributions proportional
to $m_\rho-m_\omega$ are dropped. Recall that the combination $\tilde e_A = \bar e_A+ \hat e_A$ determines the
radiative decays of the $\rho$ and $\omega $ mesons (see (\ref{vec-to-gam-decays-light})).

We point out the existence of a contribution
proportional to $e_V$, which follows upon expanding the first vertex of
(\ref{em-vertices:a}) to second order in the Goldstone boson fields. This gives rise to diagrams of type 2 as of
Figure \ref{fig:1} and the tensor $E^{\mu, \alpha \beta}_{ab}(\bar p,p)$. This term will play a  role
later in the numerical analysis. It is interesting to observe that it occurs only for intermediate $K\,\bar K^*$
or $\bar K\,K^*$ states.

The remaining contributions are proportional to $e_V$  being of type 5) and 7) in Figure 1. While type 7) diagrams
give rise to the tensors $H_{abc}^{\mu, \alpha\beta}(\bar p,p)$ and $\bar H_{abc}^{\mu, \alpha \beta}(\bar p,p)$
and are proportional to either $h_A\,b_A$ or $b_A^2$, type 5) diagrams imply the three tensors
$E_{ab}^{\mu, \alpha\beta}(\bar p,p)$, $F_{ab}^{\mu, \alpha\beta}(\bar p,p)$ and $G_{ab}^{\mu, \alpha\beta}(\bar p,p)$
as detailed in (\ref{def-E-tensor3}, \ref{def-F-tensor3}, \ref{def-G-tensor3}). The last tensor
$G_{ab}^{\mu, \alpha\beta}(\bar p,p)$ is required to describe the process where the photon is emitted by the chiral
counter terms proportional to $g_D$ or $g_F$ via an intermediate vector meson. A similar term where the photon is emitted
by the Weinberg-Tomozawa interaction is associated with the tensor $F_{ab}^{\mu, \alpha \beta}(\bar p, p)$. The tensor
$E_{ab}^{\mu, \alpha \beta}(\bar p, p)$ describes also contributions proportional to the symmetry-breaking
counter term $b_D$.

All together we establish the following decay tensor
\begin{eqnarray}
\lefteqn{i\,M^{\mu, \alpha \beta}_{b_1^+ \to \, \gamma \,\pi^+} =
i\,M^{\mu, \alpha \beta}_{b_1^0 \to \, \gamma \,\pi^0} =
-\frac{e\,h_P\,m_V}{4\,f^2}\,h^{(b_1)}_{K\,\bar K^*}\,\Big\{
A^{\mu, \alpha \beta}_{K K^*}(\bar p,p)
 -\hat B^{\mu, \alpha \beta}_{K K^*,2}(\bar p,p)\Big\}  }
\nonumber\\
&& {} - \frac{h_A}{96\,f^2\,m_V}\, \eAcmtn \, h^{(b_1)}_{K\,\bar K^*}\,C^{\mu, \alpha \beta}_{K K^* K^*}(\bar p,p)
\nonumber\\
&& {} - \frac{m_\pi^2\,b_A}{24\,f^2\,m_V}\,\eAcmtn \, h^{(b_1)}_{K\,\bar K^*}\,\Cchi^{\mu, \alpha \beta}_{K K^* K^*}(\bar p,p)
\nonumber\\
&& {} - \frac{h_A}{96\,f^2\,m_V}\, \eAcmtnbA \, h^{(b_1)}_{K\,\bar K^*}\,\bar C^{\mu, \alpha \beta}_{K K^* K^*}(\bar p,p)
\nonumber\\
&& {} - \frac{m_\pi^2\,b_A}{24\,f^2\,m_V}\,\eAcmtnbA \, h^{(b_1)}_{K\,\bar K^*}\,\barCchi^{\mu, \alpha \beta}_{K K^* K^*}(\bar p,p)
\nonumber\\
&& {} - \frac{e_V\,m_V}{32\,f^2}\,h^{(b_1)}_{K\,\bar K^*}\, E^{\mu, \alpha \beta}_{K K^*}(\bar p,p)
   + \frac{e_V}{48\,f^2\,m_V}\,\left(2\,\ratVphi+1 \right) h^{(b_1)}_{K\,\bar K^*}\, F^{\mu, \alpha \beta}_{K K^*}(\bar p,p)
\nonumber\\
&& {} -\frac{e_V\, b_D}{96\,f^2\,m_V} \, (m_\pi^2 + m_K^2) \,
\left(1-2\, \ratVphi \right)  \, h^{(b_1)}_{K\,\bar K^*}\,
E^{\mu, \alpha \beta}_{K K^*}(\bar p,p)
\nonumber\\
&& {} -\frac{e_V}{48\,f^2\,m_V} \, \Big( \left(3-2\,\ratVphi \right) g_D - \left(1+2 \,\ratVphi \right) g_F \Big) \, h^{(b_1)}_{K\,\bar K^*}\,
G^{\mu, \alpha \beta}_{K K^*}(\bar p,p)
\nonumber\\
&& {} + \frac{h_A \, b_A \, e_V}{48 \, f^2 \, m_V} \, m_\pi^2 \,  h^{(b_1)}_{K\,\bar K^*}\, H^{\mu, \alpha \beta}_{K \rho K^*}(\bar p,p)
 + \frac{(b_A)^2 \, e_V}{12 \, f^2 \, m_V} \, m_\pi^2 \, m_K^2 \, h^{(b_1)}_{K\,\bar K^*}\,
   \bar H^{\mu, \alpha \beta}_{K \rho K^*}(\bar p,p)
\nonumber\\ \nonumber\\
&& {} + \frac{\eA\,h_A}{48\,f^2\,m_V} \,\Big\{h^{(b_1)}_{\pi\,\omega}\,C^{\mu, \alpha \beta}_{\pi \rho \omega}(\bar p,p)
+ \frac{1}{\sqrt{3}}\, h^{(b_1)}_{\eta\,\rho}\,C^{\mu, \alpha \beta}_{\eta \omega \rho}(\bar p,p) \Big\}
\nonumber\\
&& {} + \frac{\eAbA\,h_A}{48\,f^2\,m_V} \,
\Big\{ h^{(b_1)}_{\pi\,\omega}\,\bar C^{\mu, \alpha \beta}_{\pi \rho \omega}(\bar p,p)
+ \frac{1}{\sqrt{3}}\, h^{(b_1)}_{\eta\,\rho}\,\bar C^{\mu, \alpha \beta}_{\eta \omega \rho}(\bar p,p) \Big\}
\nonumber\\
&& {} + \frac{\eA\,b_A}{12\,f^2\,m_V} \, m_\pi^2 \,\Big\{ h^{(b_1)}_{\pi\,\omega}\,\Cchi^{\mu, \alpha \beta}_{\pi \rho \omega}(\bar p,p)
 + \frac{1}{\sqrt{3}}\,h^{(b_1)}_{\eta\,\rho}\,\Cchi^{\mu, \alpha \beta}_{\eta \omega \rho}(\bar p,p) \Big\}
\nonumber\\
&& {} + \frac{\eAbA\,b_A}{12\,f^2\,m_V} \, m_\pi^2 \,\Big\{ h^{(b_1)}_{\pi\,\omega}\,\barCchi^{\mu, \alpha \beta}_{\pi \rho \omega}(\bar p,p)
 + \frac{1}{\sqrt{3}}\, h^{(b_1)}_{\eta\,\rho}\,\barCchi^{\mu, \alpha \beta}_{\eta \omega \rho}(\bar p,p) \Big\}
\nonumber\\
&& {} -\frac{e_V\, b_D}{24\,f^2\,m_V} \, m_\pi^2\,
\Big\{ h^{(b_1)}_{\pi\,\omega}\, E^{\mu, \alpha \beta}_{\pi \omega}(\bar p,p)
 +\frac{1}{\sqrt{3}} \, h^{(b_1)}_{\eta\,\rho}\, E^{\mu, \alpha \beta}_{\eta \rho}(\bar p,p) \Big\}
\nonumber\\
&& {} -\frac{e_V\,g_D}{12\,f^2\,m_V} \, \Big\{h^{(b_1)}_{\pi\,\omega}\, G^{\mu, \alpha \beta}_{\pi\omega}(\bar p,p)
 +\frac{1}{\sqrt{3}}\,
h^{(b_1)}_{\eta\,\rho}\, G^{\mu, \alpha \beta}_{\eta \rho}(\bar p,p) \Big\}
\nonumber\\
&& {} + \frac{h_A \, b_A \, e_V}{48 \, f^2 \, m_V} \, m_\pi^2 \, \Big\{
h^{(b_1)}_{\pi\,\omega}\, H^{\mu, \alpha \beta}_{\pi \rho \omega}(\bar p,p)
+ \frac{1}{\sqrt{3}} \, h^{(b_1)}_{\eta\,\rho}\,
H^{\mu, \alpha \beta}_{\eta \rho \rho}(\bar p,p) \Big\}
\nonumber\\
&& {} + \frac{(b_A)^2 \, e_V}{12 \, f^2 \, m_V} \, m_\pi^4 \, \Big\{
h^{(b_1)}_{\pi\,\omega}\,\bar H^{\mu, \alpha \beta}_{\pi \rho \omega}(\bar p,p)
+ \frac{1}{\sqrt{3}} \, h^{(b_1)}_{\eta\,\rho}\, \bar H^{\mu, \alpha \beta}_{\eta \rho \rho}(\bar p,p) \Big\}
\,,
\label{b1:gamma:pi}
\end{eqnarray}
where $r= m_V^2/m_\phi^2$. The resonance-coupling constants $h^{(b_1)}_{...}$ in (\ref{b1:gamma:pi})
are related to the $g_i$'s of Table \ref{tab:hadronic-decay} via (\ref{hi-gi}).
As already noted the explicit expressions for the other decays can be found in Appendix E.

\begin{table}[t]
\tabcolsep=2.mm
\renewcommand{\arraystretch}{1.5}
\fontsize{9}{3}
\begin{center}
\begin{tabular}{|c||l l |}
\hline
 &$10\,\times \,d_{b_1^+ \to\, \gamma\, \pi^+}$[GeV$^{-1}$]& \\
\hline
$ \pi \,\omega $  &$  $ & $+(\rat{-0.26}{-1.21}\,g_D+\rat{-0.01}{-0.05}\,b_D) \,e_V$ \\
& $(\rat{+0.92}{-0.05}h_A+\rat{-0.10}{+0.10}b_A)\,e_A $ & $ +(\rat{+0.23}{-0.01}h_A^2 +\rat{-0.09}{-0.02} b_A\,h_A +\rat{+0.02}{+0.04}\, b_A^2)\,e_V$ \\
\hline
$ \eta \,\rho $  &$$ & $+(- 0.31\,g_D-0.01\,b_D) \,e_V$ \\
& $(+0.28\,h_A-0.02\,b_A)\,e_A $ & $+ (+0.07\,h_A^2 -0.03\, b_A\,h_A +0.01\, b_A^2)\,e_V$ \\
\hline
$ \bar K \,K^* $  &$-4.46\,h_P\,e -0.84\,e_V  $ & $+(- 0.14\,g_D+0.16\,g_F+0.01\,b_D) \,e_V$ \\
& $-0.54\,h_P\,e_M$  & $+(-0.29\,h_V+0.55\,\tilde h_V)\,h_P\,e_V$ \\
& $(-0.33\,h_A+0.03\,b_A)\,e_A $ & $+ (-0.01\,h_A^2 +0.08\, b_A\,h_A +0.08\, b_A^2)\,e_V$ \\
\hline
\hline
 &$10\,\times \,d_{b_1^0 \to \,\gamma\, \eta}$[GeV$^{-1}$]& \\
\hline
\hline
$ \pi \,\omega $  &$  $ & $+(\rat{-0.51}{-2.38}\,g_D+\rat{-0.02}{-0.08}\,b_D) \,e_V$ \\
& $(\rat{+1.49}{-0.18}h_A+\rat{-0.16}{+0.16}b_A)\,e_A $ & $ +(\rat{+0.37}{-0.05}h_A^2 +\rat{-0.15}{-0.03} b_A\,h_A +\rat{+0.03}{+0.07}\, b_A^2)\,e_V$ \\
\hline
$ \eta \,\rho $  &$$ & $+(- 0.64\,g_D-0.02\,b_D) \,e_V$ \\
& $(+0.40\,h_A-0.03\,b_A)\,e_A $ & $+ (+0.10\,h_A^2 -0.06\, b_A\,h_A +0.02\, b_A^2)\,e_V$ \\
\hline
$ \bar K \,K^*$  &$-9.70\,h_P\,e -1.82\,e_V  $ & $+(- 0.16\,g_D+0.48\,g_F+0.34\,b_D) \,e_V$ \\
& $-0.86\,h_P\,e_M$  & $+(-0.65\,h_V-0.37\,\tilde h_V)\,h_P\,e_V$ \\
& $(-0.51\,h_A+2.03\,b_A)\,e_A $ & $+ (-0.13\,h_A^2 +1.21\, b_A\,h_A -3.73\, b_A^2)\,e_V$ \\
\hline
\end{tabular}
\caption{Decay constants that are implied by (\ref{b1:gamma:pi}, \ref{b1:gamma:eta}).
The various contributions have to be multiplied by the appropriate resonance-coupling constants
$g_i$ of Table \ref{tab:hadronic-decay}.
We use  $f=90$ MeV, $m_V=776$ MeV. See also (\ref{collection-of-parameters}, \ref{unusual-convention}).}
\label{tab:decay-parameters:b1:1}
\end{center}
\end{table}

In Tables \ref{tab:decay-parameters:b1:1}-\ref{tab:decay-parameters:b1:2} the various contributions to the
decay constants are listed.
Using the resonance-coupling constants $g_i$ of Table \ref{tab:hadronic-decay} together with the parameters
of (\ref{collection-of-parameters}) we arrive at the following values
\begin{eqnarray}
&& d_{\,b_1^\pm \,\to \, \gamma \,\pi^\pm} = d_{\,b_1^0 \,\to \, \gamma \,\pi^0} = (-0.20 -i\,0.04 )\,{\rm GeV}^{-1} \,, \qquad
\nonumber\\
&& d_{\,b_1^0 \;\,\to \, \gamma \;\eta} \;\,= (-0.62 -i\,0.09 )\,{\rm GeV}^{-1} \,,
\nonumber\\
&& d^{(1)}_{\,b_1^\pm \,\to \, \gamma \,\rho^\pm} = -0.21 +i\,0.25  \,, \qquad
d^{(2)}_{b_1^\pm \,\to \, \gamma \,\rho^\pm} = -0.26 +i\,0.11  \,.
\label{b1-full}
\end{eqnarray}
It appears that we recover the empirical
decay constant (\ref{b1:decay-parameter}) describing the $b_1^\pm \,\to \, \gamma \,\pi^\pm$ decay quite well.

We discuss the various contributions, first for the decays of the $b_1^0(1230)$ into a
$\pi$ or an $\eta$ meson.
The dominant effects result from the $\bar K \,K^*+\bar K^*\,K $ channel,
which provides terms proportional to the parameter combination $h_P\,e$. The latter are largest and describe the
processes where the photon is radiated from the charge of a strange pseudo-scalar or vector meson. These terms
alone would predict the following decay constants
\begin{eqnarray}
&& d_{\,b_1^0 \,\to \, \gamma \,\pi^0} = -0.17\,{\rm GeV}^{-1} \,, \qquad \quad
 d_{\,b_1^0 \,\to \, \gamma \;\eta} \;= -0.38\,{\rm GeV}^{-1} \,,
\label{b1-leading-term}
\end{eqnarray}
which are not too far from the values (\ref{b1-full}). We point out, however, that the contributions proportional to
the parameters $e_V$ and $\tilde h_V\,h_P\,e_V$ are almost as large separately. Since they enter with opposite signs
for the $\gamma \,\pi$ final state
there is a significant cancellation at work. Note that the term proportional to $e_V$ is the constructive sum of two
terms: the first implied by the Weinberg-Tomozawa interaction, the second directly by the term
${\tr} \,(V^{\mu \nu}\,f^+_{\mu \nu})$ expanded to second order in the Goldstone boson fields. The relative size
of the two contributions is about 5:3 for the decay into the pion, but about 5:2 for the decay into the $\eta$ meson.

To arrive at the final values (\ref{b1-full}) it is important to
consider also the contributions proportional to $g_D\,e_V$ and $g_F\,e_V$. Switching off such terms would reduce the
decay constant $d_{b_1^\pm \,\to \, \gamma \,\pi^\pm}$ by almost a factor two. Here we observe that the contributions
of the $\eta\, \rho$ channel are quite small: there is a cancellation of terms proportional to
$g_D\,e_V$ and $h_A^2\,e_V$. This holds for both decays. The small imaginary parts of the decay constants reflect
the contributions from the $\pi \,\omega$ channel. It is dominated by the terms proportional to $g_D\,e_V$.
Note that the $\pi \,\phi$ channel does not contribute to the decay parameters. All together the following decomposition
\begin{eqnarray}
&&  d_{\,b_1^0 \,\to \, \gamma \,\pi^0} = \Big\{(0.022-i\,0.021)\,g_1+ 0.003\,g_3 -0.059\,g_4 \Big\}\,{\rm GeV}^{-1}\,,
\nonumber\\
&& d_{\,b_1^0 \,\to \, \gamma \;\eta}\;
=\Big\{(0.033-i\,0.044)\,g_1+ 0.001\,g_3 -0.165\,g_4\Big\}\,{\rm
GeV}^{-1}\,, \label{b1-channel-split-1}
\end{eqnarray}
holds, with the coupling constants $g_i$ given in Table \ref{tab:hadronic-decay}.

\begin{table}[t]
\tabcolsep=6.mm
\renewcommand{\arraystretch}{1.5}
\fontsize{10}{3.5}
\begin{center}
\begin{tabular}{|l||l |l |}
\hline
 &$10\,\times \,d^{(1)}_{b_1^+ \to \,\gamma\, \rho^+}$ & $10\,\times \,d^{(2)}_{b_1^+ \to \,\gamma\, \rho^+}$ \\
\hline
\hline
$ \pi\,\omega $ \hfill  &$ \rat{+0.40}{+1.92}e\,h_A +\rat{-0.08}{-0.40}e\,b_A $ & $ \rat{+0.26}{+1.23}e\,h_A +\rat{-0.07}{-0.31}e\,b_A $  \\
$\times \,h_A\,e_V$ & $\rat{+0.38}{-1.18}h_P+\rat{+0.17}{+0.78}h_V $  & $\rat{-1.48}{-1.96}h_P+\rat{+0.07}{+0.31}h_V $\\
$\times \,b_A\,e_V$ & $\rat{-0.10}{+0.60}h_P+\rat{+0.08}{+0.37}h_V $  & $\rat{+0.38}{+0.80}h_P+\rat{+0.03}{+0.15}h_V $\\
$\times \,e_M$ & $\rat{+0.22}{+1.04}h_A+\rat{+0.11}{+0.49}b_A $  & $\rat{+0.09}{+0.41}h_A+\rat{+0.04}{+0.19}b_A$\\
$\times \,e_A$ & $\rat{+1.51}{-4.70}h_P $  & $\rat{-5.90}{-7.85}h_P$\\
\hline
$ \eta\,\rho $ \hfill  &$ -1.57\,e\,h_A +0.13\,e\,b_A $ & $-0.87\,e\,h_A+0.04\,e\,b_A $  \\
$\times \,h_A\,e_V$ & $+0.09\,h_V-0.01\,\tilde h_V $  & $-0.06\,h_V-0.00\,\tilde h_V $\\
$\times \,b_A\,e_V$ & $+0.26\,h_V+0.02\,\tilde h_V $  & $+0.11\,h_V+0.02\,\tilde h_V $\\
$\times \,e_M$ & $+0.14\,h_A+0.23\,b_A $  & $-0.01\,h_A+0.09\,b_A$\\
$\times \,e_A$ & $-0.05\,h_V-0.14\,\tilde h_V$  & $-0.20\,h_V+0.00\,\tilde h_V$\\
\hline
$ \bar K \,K^* $ \hfill  &$ -0.35\,e\,h_A -0.31\,e\,b_A $ & $-0.07\,e\,h_A-0.70\,e\,b_A $  \\
$\times \,h_A\,e_V$ & $+0.32\,h_P+0.17\,h_V-0.03\,\tilde h_V $  & $-2.02\,h_P-0.03\,h_V+0.00\,\tilde h_V $\\
$\times \,b_A\,e_V$ & $-0.78\,h_P+2.53\,h_V+0.19\,\tilde h_V $  & $+4.92\,h_P+1.17\,h_V+0.11\,\tilde h_V $\\
$\times \,e_M$ & $+0.22\,h_A+2.69\,b_A $  & $+0.08\,h_A+1.01\,b_A$\\
$\times \,e_A$ & $+1.28\,h_P +0.00\,h_V-0.18\,\tilde h_V$  & $-8.07\,h_P-0.35\,h_V+0.01\,\tilde h_V$\\
\hline
\end{tabular}
\caption{Decay constants that are implied by (\ref{b1:gamma:rho}).
The various contributions have to be multiplied by the appropriate resonance-coupling constants
$g_i$ of Table \ref{tab:hadronic-decay}.
We use  $f=90$ MeV, $m_V=776$ MeV. See also (\ref{collection-of-parameters}, \ref{unusual-convention}).}
\label{tab:decay-parameters:b1:2}
\end{center}
\end{table}

It is useful to contrast our result with the recent work by Roca, Hosaka and Oset \cite{Roca:Hosaka:Oset:2007}. As
discussed already in the introduction, we can not derive the gauge invariance of the results obtained in
\cite{Roca:Hosaka:Oset:2007} for the $b_1^\pm \,\to \,\gamma \,\pi^\pm$ process. Leaving this concern aside
we compare the contributions from the $\bar K\,K^*$  channel. The resonance-coupling constant
to that channel agrees within 20$\%$, the value of  \cite{Roca:Hosaka:Oset:2007} yields about
$|g_4|\simeq 3.5$ in our convention. Thus it is legitimate to directly compare the contributions to the decay constant.
Translating the partial decay rate of  \cite{Roca:Hosaka:Oset:2007} into a value of the decay constant we deduce the value
$|d_{b_1^\pm \,\to \, \gamma \,\pi^\pm}|= 0.08\,{\rm GeV}^{-1}$. This should be compared to (\ref{b1-leading-term}).
We observe a sizeable difference in the predictions. It is pointed out that this is not surprising since
Roca, Hosaka and Oset use a form of the $K^* \bar K \,\pi$ vertex that is forbidden in the chiral Lagrangian.
Chiral  symmetry requests the vector mesons to couple to a pair of Goldstone bosons with at least two derivatives
involved. The claim of \cite{Roca:Hosaka:Oset:2007} to recover the empirical branching is a consequence of
a tree-level term where the $b_1^+(1230)$ decays first into a charged pion and a neutral vector meson. The latter
converts directly into an on-shell photon via a gauge-dependent vertex of the form $A_\mu \,\phi^\mu $ or
$A_\mu \,\omega^\mu $. There is no room for such contributions in a systematic approach based on the
chiral Lagrangian\footnote{Note that some of the vertices used in \cite{Roca:Hosaka:Oset:2007} are
consistent with chiral constraints
and electromagnetic gauge symmetry only if used in the context of the model of Bando et al. \cite{Bando} or the massive Yang-Mills approach (see e.g. the review
\cite{Birse:1996}). However, this requires
that a non-abelian gauge symmetry is preserved. The latter is violated by the coupled-channel approach
of \cite{Roca:Hosaka:Oset:2007}.}. Moreover, as discussed in detail in \cite{Lutz-Soyeur:2007} there are no tree-level contributions
to a radiative decay amplitude of a given state if that state is generated by coupled-channel dynamics. The inclusion of the latter would be double counting.
For a more detailed discussion of this issue we refer to Appendix B.

We turn to a discussion of the $b_1^+(1230) \,\to \,\gamma\,\rho^+$ process. In contrast to the decay of the
$b_1$ into the pseudo-scalar mesons, in this case the dominant contributions arise from the
$\pi \,\omega$ and $\eta \,\rho$ channels rather than from the $K\,\bar K^*$ and $\bar K\, K^*$ channels.
The following decomposition arises
\begin{eqnarray}
&& d^{(1)}_{\,b_1^\pm \,\to \, \gamma \,\rho^\pm} = (+0.035 +i\,0.118 )\,g_1 -0.098\,g_3 -0.014\,g_4\,,
\nonumber\\
&&d^{(2)}_{\,b_1^\pm \,\to \, \gamma \,\rho^\pm} = (-0.006 -i\,0.052 )\,g_1 -0.056\,g_3 -0.029\,g_4\,,
\label{b1-channel-split-2}
\end{eqnarray}
with no contribution from the $\pi \,\phi$ channel.
It is interesting to observe that
while in the $\eta\, \rho$ channel ($\sim g_3$) the dominant terms are proportional to $e\,h_A$, in the
$\pi \,\omega $ channel ($\sim g_1$) the largest effects are defined by the terms proportional to $h_A\,h_P\,e_V$ and
$e\,h_A$.

We would like to emphasize that we consider the reproduction of
the empirical decay branching of the $b_1$ into a photon and a
pion as accidental. It is an important issue to evaluate
contributions implied by intermediate states with two vector
mesons. Since such contributions must be significant for the
radiative decays of the $f_1(1282)$ we do not see any reason a
priori why this is not the case for the radiative decay of the
$b_1(1230)$. To this extent one should not view the values for the
so far unobserved decay constants (\ref{b1-full}) as predictions
to be confronted with upcoming measurements. These are partial
results which await completion.

\clearpage
\newpage

\subsection{Radiative decays of the $h_1(1386)$}
\label{subsec:h1}

We study the radiative decays of the $h_1(1386)$ molecule. There are two decay channels possible with either the
$\gamma \, \pi_0$ or the $\gamma\,\eta$ in the final state. So far there is no empirical information on such decay
processes available. Explicit expressions for the corresponding decay amplitudes
can be found in Appendix E.

Using the parameter set (\ref{collection-of-parameters}) together with the resonance-coupling constants of
Table \ref{tab:hadronic-decay} we arrive at the following decay constants
\begin{eqnarray}
&& d_{\,h_1^0 \,\to \, \gamma \,\pi^0} = (+0.00+i\,0.02)\,{\rm
GeV}^{-1} \,, \qquad \quad
\nonumber\\
&& d_{\,h_1^0 \,\to \, \gamma \;\eta} \;= (-0.29+i\,0.01)\,{\rm
GeV}^{-1} \,. \label{h1-full}
\end{eqnarray}
The results (\ref{h1-full}) are a consequence of various cancellation mechanisms, in particular the smallness of
$d_{\,h_1^0 \,\to \, \gamma \,\pi^0} $. There are four channels considered in this work, the $\pi \,\rho$,
$\eta\,\omega$, $\bar K\,K^*+ \bar K^*K $ and the $\eta\,\phi$ channel. The corresponding coupling strengths $g_i$
of the $h_1(1386)$ to the latter are collected in Table \ref{tab:hadronic-decay}. We decompose the results
(\ref{h1-full}) into contributions from the four channels considered
\begin{eqnarray}
&&  d_{\,h_1^0 \,\to \, \gamma \,\pi^0} = \Big\{(0.143+i\,0.017)\,g_1+ (0.072+i\,0.025)\,g_2
\nonumber\\
&& \qquad \qquad \quad-\,0.029\,g_3 \Big\}\,{\rm GeV}^{-1}\,,
\nonumber\\
&& d_{\,h_1^0 \,\to \, \gamma \;\eta}\; =\Big\{(0.026+i\,0.047)\,g_1+ (0.014+i\,0.005)\,g_2
\nonumber\\
&& \qquad \qquad \quad -\, 0.135\,g_3-0.015\,g_4\Big\}\,{\rm GeV}^{-1}\,.
\label{h1-channel-split}
\end{eqnarray}
For the decay into the pion there are basically two channels contributing with opposite signs and almost equal
strength. The $\pi \,\rho$ channel is almost irrelevant with its small coupling constant $g_1=0.1$, even though the
corresponding coefficient in (\ref{h1-channel-split}) is largest. The contributions from the
$\eta \,\omega$ and $\bar K\,K^*+ \bar K^* K $ channels with $g_2=0.9$ and $g_3=2.5$ cancel to a large degree.
The decay constant $d_{\,h_1^0 \,\to \, \gamma \;\eta}$ is dominated by the $\bar K\,K^*+ \bar K^* K$ channel.

It is interesting to observe that the total contributions from the $\bar K\,K^*+ \bar K^* K $ channel
to the $h_1(1385)$ decay constants are subject to
a further cancellation mechanism. In Table \ref{tab:decay-parameters:h1} the decay constants are
split into their various contributions from the four channels and the various parameter combinations.
In the $\bar K\,K^*+ \bar K^* K $ channel the largest terms are defined by the contributions proportional to
$e_V$. Almost as large contributions are proportional to $h_P\,e$ and $g_D\,e_V, g_F\,e_V$, where the terms do not
add up constructively. Thus our partial result (\ref{h1-full}) is quite sensitive to the precise values of
the resonance-coupling constants, but also to the size of the counter terms $g_D$ and $g_F$.

We stress that the results presented for the radiative decay of the $h_1(1385)$ are partial, which
need to be completed by additional contributions from intermediate states involving two vector mesons.
In addition the values of the parameters $g_D$ and $g_F$ need to be established more systematically
in a coupled-channel computation that considers the effects of t- and u-channel processes explicitly.

\begin{table}[t]
\tabcolsep=2.mm
\renewcommand{\arraystretch}{1.5}
\fontsize{9}{3}
\begin{center}
\begin{tabular}{|c||l l |}
\hline
 &$10\,\times \,d_{h_1 \to\, \gamma\, \pi^0}$[GeV$^{-1}$]& \\
\hline
\hline
$ \pi \,\rho $  &$\rat{-11.2}{-3.19} h_P\,e + \rat{+0.01}{-10.5}e_V $ & $+(\rat{+0.00}{-8.03}\,g_D+\rat{-0.00}{+4.01}g_F+\rat{+0.00}{-0.10}\,b_D) \,e_V$ \\
& $\rat{-3.00}{+4.47}h_P\,e_M$  & $+(\rat{-2.25}{+3.35}h_V+\rat{+8.41}{+1.67}\tilde h_V)\,h_P\,e_V$ \\
& $(\rat{+1.75}{+0.35}h_A+\rat{-0.23}{+0.28}b_A)\,e_A $ & $ +(\rat{+0.44}{+0.09}h_A^2 +\rat{-0.17}{+0.29} b_A\,h_A +\rat{+0.01}{+0.18}\, b_A^2)\,e_V$ \\
\hline
$ \eta \,\omega $  &$$ & $+(\rat{-2.76}{-1.55}g_D+\rat{-0.06}{-0.04}b_D) \,e_V$ \\
& $(\rat{+0.86}{-0.06}\,h_A+\rat{+0.07}{+0.11}\,b_A)e_A $ & $+ (\rat{+0.22}{-0.02}h_A^2 +\rat{-0.08}{+0.01}b_A\,h_A +\rat{+0.06}{+0.03} b_A^2)\,e_V$ \\
\hline
$ \bar K \,K^*$  &$-5.66\,h_P\,e -3.19\,e_V  $ & $+(- 2.90\,g_D+0.97\,g_F-0.37\,b_D) \,e_V$ \\
& $+0.12\,h_P\,e_M$  & $+(+0.09\,h_V+1.12\,\tilde h_V)\,h_P\,e_V$ \\
& $(+0.87\,h_A+0.01\,b_A)\,e_A $ & $+ (+0.22\,h_A^2 -1.91\, b_A\,h_A +0.95\, b_A^2)\,e_V$ \\
\hline
\hline
 &$10\,\times \,d_{h_1^0 \to \,\gamma\, \eta}$[GeV$^{-1}$]& \\
\hline
\hline
$ \pi \,\rho $  &$$ & $+(\rat{+0.00}{-2.56}\,g_D+\rat{+0.00}{-0.06}b_D) \,e_V$ \\
& $(\rat{+0.93}{+0.09}h_A+\rat{-0.12}{+0.16}b_A)\,e_A $ & $ +(\rat{+0.23}{+0.02}h_A^2 +\rat{-0.09}{+0.01} b_A\,h_A +\rat{+0.01}{+0.05}\, b_A^2)\,e_V$ \\
\hline
$ \eta \,\omega $  &$$ & $+(\rat{-0.60}{-0.34}g_D+\rat{-0.01}{-0.01}b_D) e_V$ \\
& $(\rat{+0.14}{-0.03}\,h_A+\rat{+0.01}{+0.02}\,b_A)e_A $ & $+ (\rat{+0.03}{-0.01}h_A^2 +\rat{-0.02}{+0.00}b_A\,h_A +\rat{+0.01}{+0.01} b_A^2)\,e_V$ \\
\hline
$ \bar K \,K^*$  &$-12.2\,h_P\,e -4.39\,e_V  $ & $+(- 1.44\,g_D+1.38\,g_F-0.05\,b_D) \,e_V$ \\
& $+0.31\,h_P\,e_M$  & $+(+0.17\,h_V+0.03\,\tilde h_V)\,h_P\,e_V$ \\
& $(+0.14\,h_A+0.05\,b_A)\,e_A $ & $+ (+0.01\,h_A^2 +0.96\, b_A\,h_A +4.90\, b_A^2)\,e_V$ \\
\hline
$ \eta \,\phi$  &$ $ & $+(+ 0.51\,g_D+0.38\,b_D) \,e_V$ \\
& $(-1.35\,h_A+2.04\,b_A)\,e_A $ & $+ (-0.20\,h_A^2 +1.64\, b_A\,h_A -6.50\, b_A^2)\,e_V$ \\
\hline
\end{tabular}
\caption{Decay constants that are implied by (\ref{h1:gamma:pi}, \ref{h1:gamma:eta}).
The various contributions have to be multiplied by the appropriate resonance-coupling constants
$g_i$ of Table \ref{tab:hadronic-decay}. We use  $f=90$ MeV, $m_V=776$ MeV.
See also (\ref{collection-of-parameters}, \ref{unusual-convention}).}
\label{tab:decay-parameters:h1}
\end{center}
\end{table}

\clearpage

\subsection{Radiative decays of the $a_1(1230)$}
\label{subsec:a1}

We turn to the decays of the $a_1(1230)$ molecule. The charged states
may decay into either $\gamma\,\pi^\pm$ or $\gamma\,\rho^\pm$. Due to charge conjugation symmetry the neutral
state decays into
$\gamma\,\rho^0$, $\gamma\,\omega$ or $\gamma\,\phi$. The partial decay width of the charged state into a charged pion
is acknowledged by the Particle Data Group \cite{PDG:2006} as seen. In \cite{Zielinski:1984}
the following value
\begin{eqnarray}
\Gamma_{a_1^\pm(1230)\, \to \, \gamma\,\pi^\pm }= (640 \pm 246 )\,{\rm keV}\,,
\label{exp:a1-decay}
\end{eqnarray}
is claimed. Using the central value of the $a_1$ mass of 1230 MeV as advocated by the Particle Data
Group \cite{PDG:2006} the corresponding decay parameter as introduced in (\ref{result:width-1plus:a}) results
\begin{eqnarray}
\Big| d_{a_1^\pm (1230)\, \to \, \gamma\,\pi^\pm} \Big|= (0.33 \pm 0.09)\, {\rm GeV}^{-1}\,.
\label{exp:a1-decay-constant}
\end{eqnarray}
Since there appears to be only a single experiment having observed the process $a_1^\pm(1230) \to \, \gamma\,\pi^\pm $,
the decay constant (\ref{exp:a1-decay-constant}) has to be taken with a grain of salt.

Using the parameters (\ref{collection-of-parameters}) together with the $a_1(1230)$ coupling constants of Table
\ref{tab:hadronic-decay} we derive the following decay constants
\begin{eqnarray}
&& d_{\,a_1^\pm \,\to \, \gamma \,\pi^\pm} =  (-0.68 +i\,0.11 )\,{\rm GeV}^{-1} \,, \qquad
\nonumber\\
&& d^{(1)}_{\,a_1^0 \,\to \, \gamma \,\rho^0} = -0.24 -i\,0.06  \,, \qquad
d^{(2)}_{\,a_1^0 \,\to \, \gamma \,\rho^0} = -0.20 -i\,0.07  \,,
\nonumber\\
&& d^{(1)}_{\,a_1^0 \,\to \, \gamma \,\omega}\, = -1.59 -i\,1.16  \,, \qquad \,
d^{(2)}_{\,a_1^0 \,\to \, \gamma \,\omega}\, = -0.75 -i\,0.95  \,,
\nonumber\\
&& d^{(1)}_{\,a_1^0 \,\to \, \gamma \,\phi}\; = -1.10 +i\,0.00  \,, \qquad \,
d^{(2)}_{\,a_1^0 \,\to \, \gamma \,\phi}\, = -0.02 +i\,0.00  \,,
\label{a1-full}
\end{eqnarray}
where we recall that $G$-parity invariance requires
\begin{eqnarray}
d^{(i)}_{\,a_1^0 \,\to \, \gamma \,\rho^0} = d^{(i)}_{\,a_1^\pm \,\to \, \gamma \,\rho^\pm} \,.
\end{eqnarray}
We discuss first the process $a^\pm _1(1230)\to \, \gamma \,\pi^\pm$ in more detail.
The modulus of the decay constant obtained in (\ref{a1-full}) is significantly larger than the value
in (\ref{exp:a1-decay-constant}) as implied by \cite{Zielinski:1984}. It is useful to decompose
the decay constant into its contributions from the two channels, $\pi \,\rho$ and $\bar K\,K^*$, that are considered
in this work. We find
\begin{eqnarray}
&&  d_{\,a_1^\pm  \,\to \, \gamma \,\pi^\pm} = \Big\{(-0.149+i\,0.029)\,g_1 - 0.043\,g_2
 \Big\}\,{\rm GeV}^{-1}\,,
\label{a1-channel-split}
\end{eqnarray}
which illustrates that the $\pi \,\rho$ channel is dominant. We recall from Table \ref{tab:hadronic-decay} the
values $g_1= 3.8$ and $g_2=2.6$. A full decomposition of the decay constant into contributions proportional
to various products of coupling constants is offered in Table \ref{tab:decay-parameters:a1-1}.
The real part of the decay constant is subject to a cancellation amongst the term proportional to $h_P\,e$ and
$\tilde h_V\,h_P\,e_V$ in the $\pi \,\rho$ channel. As a consequence the result is quite sensitive to the
precise value used for the parameter $\tilde h_V=3.72$. Recall that the latter was determined by the requirement to
recover the magnetic and quadrupole moments of the $\rho^\pm$ meson as suggested by the light-front quark model
\cite{Choi:Ji:2004}. This is a model-dependent assumption and may be relaxed. If we used instead the value
$\tilde h_V=6.3$, we would arrive at a decay constant that is compatible with (\ref{exp:a1-decay-constant}).

\begin{table}[t]
\tabcolsep=2.mm
\renewcommand{\arraystretch}{1.5}
\fontsize{9}{3}
\begin{center}
\begin{tabular}{|c||l l |}
\hline
 &$10\,\times \,d_{a_1^+ \to \,\gamma\, \pi^+}$[GeV$^{-1}$]& \\
\hline
\hline
$ \pi \,\rho $  &$\rat{-23.1}{-2.44} h_P\,e + \rat{+0.24}{+1.41}e_V $ & $+(\rat{+0.24}{+1.35}\,g_D+\rat{-0.71}{-4.05}g_F+\rat{-0.02}{-0.10}\,b_D) \,e_V$ \\
& $\rat{-1.47}{+1.74}h_P\,e_M$  & $+(\rat{-1.10}{+1.31}h_V+\rat{+4.64}{+0.58}\tilde h_V)\,h_P\,e_V$ \\
& $(\rat{-2.10}{+0.07}h_A+\rat{+0.23}{-0.22}b_A)\,e_A $ & $ +(\rat{-0.52}{+0.02}h_A^2 +\rat{+0.23}{+0.13} b_A\,h_A +\rat{-0.02}{-0.04}\, b_A^2)\,e_V$ \\
\hline
$ \bar K \,K^*$  &$-5.46\,h_P\,e +0.18\,e_V  $ & $+(+ 0.23\,g_D-0.68\,g_F-0.14\,b_D) \,e_V$ \\
& $-0.54\,h_P\,e_M$  & $+(-0.40\,h_V+0.76\,\tilde h_V)\,h_P\,e_V$ \\
& $(-0.97\,h_A+0.08\,b_A)\,e_A $ & $+ (-0.24\,h_A^2 +1.21\, b_A\,h_A -0.14\, b_A^2)\,e_V$ \\
\hline
\end{tabular}
\caption{Decay constants that are implied by (\ref{a1:gamma:pi}).
The various contributions have to be multiplied by the appropriate resonance-coupling constants
$g_i$ of Table \ref{tab:hadronic-decay}.
We use  $f=90$ MeV, $m_V=776$ MeV. See also (\ref{collection-of-parameters}, \ref{unusual-convention}).}
\label{tab:decay-parameters:a1-1}
\end{center}
\end{table}

There is a further interesting point to be made. The contributions proportional to $e_V$
are a result of significant cancellations amongst two terms in both channels.
The ratio of the first term implied by the Weinberg-Tomozawa interaction and the second defined directly by the
interaction ${\tr} \,(V^{\mu \nu}\,f^+_{\mu \nu})$ is about 7:10 in the $\pi \,\rho$ and  about 8:10 in the
$\bar K \,K^*$ channel. As seen from Table \ref{tab:decay-parameters:a1-1} all together the terms proportional
to $e_V$ have little effect on the real part of the decay constant. However, for the imaginary part they stil play
an important role.

\begin{table}[t]
\tabcolsep=4.mm
\renewcommand{\arraystretch}{1.45}
\fontsize{10}{3.5}
\begin{center}
\begin{tabular}{|l||l |l |}
\hline
 &$10\,\times \,d^{(1)}_{a_1^+ \to \,\gamma\, \rho^+}$ & $10\,\times \,d^{(2)}_{a_1^+ \to \,\gamma\, \rho^+}$ \\
\hline
\hline
$ \pi\,\rho $ \hfill  &$ -$ & $- $  \\
$\times \,h_P\,e_V$ & $\rat{+0.19}{-0.55}h_A +\rat{-0.05}{+0.29}b_A$  & $\rat{-0.55}{-0.91}h_A+\rat{+0.14}{+0.38}b_A$\\
$\times \,h_A\,e_V$ & $\rat{-0.04}{-0.13}h_V+\rat{+0.09}{-0.01}\tilde h_V $  & $\rat{+0.17}{-0.07}h_V+\rat{-0.03}{-0.02}\tilde h_V $\\
$\times \,b_A\,e_V$ & $\rat{-0.02}{-0.14}h_V+ \rat{-0.03}{-0.02}\tilde h_V$  & $\rat{-0.06}{-0.05}h_V+\rat{+0.00}{-0.02} \tilde h_V  $\\
$\times \,e_A$ & $\rat{+0.74}{-2.21}h_P $  & $\rat{-2.19}{-3.65}h_P$\\
$\times \,e_A$ & $\rat{-0.16}{-0.52}h_V +\rat{+0.37}{-0.05}\tilde h_V $  & $\rat{+0.68}{-0.26}h_V +\rat{-0.11}{-0.06}\tilde h_V$\\
\hline
$ \bar K\,K^* $ \hfill  &$ -2.32\,e\,h_A +3.03\,e\,b_A $ & $-1.22\,e\,h_A+1.49\,e\,b_A $  \\
$\times \,h_A\,e_V$ & $-0.02\,h_P-0.05\,h_V+0.01\,\tilde h_V $  & $+0.11\,h_P-0.03\,h_V-0.00\,\tilde h_V $\\
$\times \,b_A\,e_V$ & $+0.04\,h_P+0.48\,h_V-0.01\,\tilde h_V $  & $-0.26\,h_P+0.17\,h_V+0.01\,\tilde h_V $\\
$\times \,e_M$ & $-0.09\,h_A+0.84\,b_A $  & $-0.04\,h_A+0.28\,b_A$\\
$\times \,e_A$ & $-0.43\,h_P+0.00\,h_V-0.06\,\tilde h_V$  & $+2.69\,h_P-0.12\,h_V+0.01\,\tilde h_V$\\
\hline
\hline
 &$10\,\times \,d^{(1)}_{a_1^0 \to \,\gamma\, \omega}$ & $10\,\times \,d^{(2)}_{a_1^0 \to \,\gamma\, \omega}$ \\
\hline
\hline
$ \pi\,\rho $ \hfill  &$ \rat{-5.26}{-4.87}e\,h_A +\rat{+0.64}{+0.41}e\,b_A $ & $ \rat{-2.07}{-3.83}e\,h_A +\rat{+0.23}{+0.27}e\,b_A $  \\
$\times \,h_A\,e_V$ & $\rat{-0.26}{-0.40}h_V+\rat{-0.02}{+0.06}\tilde h_V $  & $\rat{-0.09}{-0.51}h_V+\rat{+0.00}{+0.02}\tilde h_V $\\
$\times \,b_A\,e_V$ & $\rat{+0.16}{+0.43}h_V+\rat{-0.00}{-0.02}\tilde h_V $  & $\rat{+0.04}{+0.20}h_V+\rat{-0.00}{-0.00}\tilde h_V $\\
$\times \,e_M$ & $\rat{-0.34}{-0.53}h_A+\rat{+0.21}{+0.57}b_A $  & $\rat{-0.12}{-0.69}h_A+\rat{+0.05}{+0.27}b_A$\\
\hline
$ \bar K\,K^* $ \hfill  &$ -2.36\,e\,h_A +3.05\,e\,b_A $ & $-1.27\,e\,h_A+1.54\,e\,b_A $  \\
$\times \,h_A\,e_V$ & $+0.33\,h_P-0.07\,h_V+0.06\,\tilde h_V $  & $-2.15\,h_P+0.06\,h_V-0.01\,\tilde h_V $\\
$\times \,b_A\,e_V$ & $-0.81\,h_P+0.13\,h_V-0.26\,\tilde h_V $  & $+5.24\,h_P-0.21\,h_V-0.10\,\tilde h_V $\\
$\times \,e_M$ & $-0.09\,h_A+0.83\,b_A $  & $-0.04\,h_A+0.29\,b_A$\\
$\times \,e_A$ & $+1.32\,h_P-0.03\,h_V+0.19\,\tilde h_V$  & $-8.60\,h_P+0.36\,h_V-0.02\,\tilde h_V$\\
\hline
\hline
 &$10\,\times \,d^{(1)}_{a_1^0 \to \,\gamma\, \phi}$ & $10\,\times \,d^{(2)}_{a_1^0 \to \,\gamma\, \phi}$ \\
\hline
\hline
$ \bar K\,K^* $ \hfill  &$ -6.38\,e\,h_A +6.46\,e\,b_A $ & $-4.90\,e\,h_A+4.89\,e\,b_A $  \\
$\times \,h_A\,e_V$ & $\rat{-4.25}{+0.05} h_P+0.60\,h_V-0.25\,\tilde h_V $  & $\rat{+17.7}{+0.11} h_P-0.56\,h_V+0.39\,\tilde h_V $\\
$\times \,b_A\,e_V$ & $\rat{+10.2}{-0.20} h_P-0.50\,h_V+0.84\,\tilde h_V $  & $\rat{-43.3}{-0.35} h_P+1.94\,h_V-0.72\,\tilde h_V $\\
$\times \,e_M$ & $-0.13\,h_A+0.85\,b_A $  & $-0.10\,h_A+0.51\,b_A$\\
$\times \,e_A$ & $\rat{-17.0}{+0.20} h_P+2.77\,h_V-1.00\,\tilde h_V$  & $\rat{+70.7}{+0.44} h_P-1.93\,h_V+1.57\,\tilde h_V$\\
\hline
\end{tabular}
\caption{Decay constants that are implied by (\ref{a1:gamma:rho}, \ref{a1:gamma:omega}, \ref{a1:gamma:phi}).
The various contributions have to be multiplied by the appropriate resonance-coupling constants
$g_i$ of Table \ref{tab:hadronic-decay}.
We use  $f=90$ MeV, $m_V=776$ MeV. See also (\ref{collection-of-parameters}, \ref{unusual-convention}).}
\label{tab:decay-parameters:a1-2}
\end{center}
\end{table}

It may be desirable to confront our result with the recent work by Roca, Hosaka and Oset \cite{Roca:Hosaka:Oset:2007}.
As emphasized before, we can not reproduce the claim that the results obtained in
\cite{Roca:Hosaka:Oset:2007} for the $a_1^\pm \,\to \,\gamma \,\pi^\pm$ are gauge invariant.
Nevertheless we scrutinize their contributions from the $\pi \,\rho$ and  $\bar K\,K^*$ channels to the
decay constant  $d_{a_1^+ \to \,\gamma\, \pi^+}$, the only one considered in their work.
We note that the $a_1(1230)$ coupling constants used are somewhat smaller than the ones derived in our work:
given in our convention they read $|g_1| \simeq 2.6$ and $|g_2| \simeq 1.4$. We agree with
\cite{Roca:Hosaka:Oset:2007} to the extent that the decay is dominated by the $\pi\,\rho$ channel.
However, we observe a significant difference in the predictions. In particular, we do not confirm that the
loop contributions, that arise when the photon couples to the charge of one of the
two intermediate states, are small (see the term proportional to $h_P\,e$ in Table \ref{tab:decay-parameters:a1-1}).
We point the reader to the fact that
Roca, Hosaka and Oset use a form of the $\rho\, \pi \,\pi$ vertex that is forbidden in the chiral Lagrangian.
Chiral  symmetry requests the vector mesons to couple to a pair of Goldstone bosons with at least two derivatives
involved. The claim of \cite{Roca:Hosaka:Oset:2007} to recover the empirical branching is a consequence of an
additional tree-level contribution where the $a_1^+(1230)$ decays first into a charged pion and a neutral $\rho$ meson.
In a second step the vector meson is converted into the photon via a gauge-dependent vertex of the form
$A_\mu \,\rho_0^\mu $. According to \cite{Roca:Hosaka:Oset:2007} the decay is completely dominated by that term.
As repeatedly emphasized in our work, if the $a_1(1230)$ is considered to be a molecule, tree-level
contributions to its radiative decay amplitudes do not arise in a systematic approach.
For a more detailed discussion of this issue we refer to Appendix B.

We turn to a discussion of the radiative decay of the $a_1(1230)$ into vector mesons.
Again it is instructive to decompose the decay constants into their contributions from the
two channels
\begin{eqnarray}
&& d^{(1)}_{\,a_1^0 \,\to \, \gamma \,\rho^0} = (+0.021-i\,0.015)\,g_1-0.122\,g_2  \,, \qquad
\label{a1-channel-split-vector}\\
&&d^{(2)}_{\,a_1^0 \,\to \, \gamma \,\rho^0} = (-0.010-i\,0.019)\,g_1-0.063\,g_2  \,,
\nonumber\\
&& d^{(1)}_{\,a_1^0 \,\to \, \gamma \,\omega}\, = (-0.338-i\,0.305)\,g_1-0.117\,g_2  \,, \qquad \,
\nonumber\\
&&d^{(2)}_{\,a_1^0 \,\to \, \gamma \,\omega}\, = (-0.131-i\,0.249)\,g_1-0.097\,g_2  \,,
\nonumber\\
&& d^{(1)}_{\,a_1^0 \,\to \, \gamma \,\phi}\; = (-0.423+i\,0.000)\,g_2  \,, \qquad \, d^{(2)}_{\,a_1^0 \,\to \,
\gamma \,\phi}\, = (-0.006+i\,0.001)\,g_2  \,. \nonumber
\end{eqnarray}
An intricate picture emerges. The decay into the $\phi$ meson probes the second channel only. While the
decay into the  $\omega$ is dominated by the first channel, the decay into the $\rho^0$ probes both channels
with almost equal weight. In Table \ref{tab:decay-parameters:a1-2} a complete decomposition of the decay constants
into terms proportional to the various parameter combinations can be found.

The table explains the source of the large asymmetries in the decay constants, i.e.
$d^{(1)} \neq d^{(2)}$. Most strikingly this is realized for the decay of the $a_1(1230)$ into a $\phi$ meson.
This would imply a significant asymmetry in the angular distribution of the decay
process. The asymmetry is the result of a cancellation of terms proportional to $e\,h_A$ and $h_A\,h_P\,e_V$ in
$d^{(2)}_{\,a_1^0 \,\to \, \gamma \,\phi}$. The analogous contributions in $d^{(1)}_{\,a_1^0 \,\to \, \gamma \,\phi}$
add up constructively. It remains to be seen whether this phenomenon persists in a complete computation.

\clearpage

\subsection{Radiative decays of the $K_1(1272)$}
\label{subsec:K1}

We turn to the radiative decays of the $K_1(1272)$ molecule. The charged state
may decay into either $\gamma\,K^+$ or $\gamma\,K^+_*$. The neutral state decays into
$\gamma\,K^0 $ or $\gamma\,K^*_0$. The Particle Data
Group \cite{PDG:2006} quotes the $K^0_1(1272) \,\to \gamma\,K^0$ process as seen, but does not provide
any information on the other radiative decay processes. So far there is a single experiment studying the
radiative decay of the $K_1(1272)$ meson relying on the Primakoff effect only. For the partial
decay width of the neutral state the KTeV collaboration \cite{KTeV:2002}  obtained the following result
\begin{eqnarray}
\Gamma_{K_1^0(1272) \to \, \gamma\,K^0 }= (73.2 \pm 6.1 \pm 28.3 )\,{\rm keV}\,.
\label{exp:K1-decay}
\end{eqnarray}
Using the central value of the $K_1$ mass of 1272 MeV as claimed by the Particle Data Group \cite{PDG:2006} the
corresponding decay parameter as introduced in (\ref{result:width-1plus:a}) results
\begin{eqnarray}
\Big| d_{K_1^0(1272) \to \, \gamma\,K^0} \Big|= (0.13\pm 0.04) \, {\rm GeV}^{-1}\,.
\label{exp:K1-decay-constant}
\end{eqnarray}
Clearly, an independent  experimental confirmation is necessary.

\begin{table}[t]
\begin{center}
\tabcolsep=2.mm
\renewcommand{\arraystretch}{1.45}
\fontsize{9}{3}
\begin{tabular}{|c||l l |}
\hline
 &$10\,\times \,d_{K_1^+ \to \,\gamma\, K^+}$[GeV$^{-1}$]& \\
\hline
\hline
$ \pi \,K^* $  &$\rat{+3.13}{-0.54} h_P\,e + \rat{-0.61}{-1.74}e_V $ & $+(\rat{-0.50}{-1.41}\,g_D+\rat{+0.67}{+1.89}g_F+\rat{+0.04}{+0.12}\,b_D) \,e_V$ \\
& $\rat{+0.53}{-0.41}h_P\,e_M$  & $+(\rat{+0.23}{-0.18}h_V+\rat{+0.04}{-0.10}\tilde h_V)\,h_P\,e_V$ \\
& $(\rat{+1.48}{-0.29}h_A+\rat{-1.68}{+1.48}b_A)\,e_A $ & $ +(\rat{+0.37}{-0.07}h_A^2 +\rat{-0.42}{+0.65} b_A\,h_A +\rat{+0.32}{+0.51}\, b_A^2)\,e_V$ \\
\hline
$ K \,\rho $  &$\rat{-4.87}{-0.34} h_P\,e + \rat{-0.25}{-0.06}e_V $ & $+(\rat{-0.48}{-0.11}\,g_D+\rat{-1.81}{-0.42}g_F+\rat{-0.76}{-0.18}\,b_D) \,e_V$ \\
& $\rat{+0.19}{+0.28}h_P\,e_M$  & $+(\rat{+0.14}{+0.21}h_V+\rat{-0.08}{+0.07}\tilde h_V)\,h_P\,e_V$ \\
& $(\rat{-0.70}{+0.03}h_A+\rat{+0.38}{-0.18}b_A)\,e_A $ & $ +(\rat{-0.10}{+0.00}h_A^2 +\rat{+1.21}{+0.18} b_A\,h_A +\rat{+1.43}{+0.42}\, b_A^2)\,e_V$ \\
\hline
$ K \,\omega $  &$+1.95\,h_P\,e +1.43\,e_V  $ & $+(- 0.81\,g_D-1.46\,g_F-0.77\,b_D) \,e_V$ \\
& $(+0.40\,h_A-0.26\,b_A)\,e_A $ & $+ (+0.14\,h_A^2 -0.68\, b_A\,h_A +2.89\, b_A^2)\,e_V$ \\
\hline
$ \eta  \,K^* $  &$+6.33\,h_P\,e +0.55\,e_V  $ & $+(- 0.25\,g_D+0.39\,g_F+0.14\,b_D) \,e_V$ \\
& $+0.98\,h_P\,e_M$  & $+(+0.63\,h_V+0.41\,\tilde h_V)\,h_P\,e_V$ \\
& $(+1.61\,h_A-1.63\,b_A)\,e_A $ & $+ (+0.29\,h_A^2 -1.36\, b_A\,h_A +0.29\, b_A^2)\,e_V$ \\
\hline
$ K  \,\phi $  &$-2.12\,h_P\,e -0.60\,e_V  $ & $+(- 0.20\,g_D+0.36\,g_F+0.09\,b_D) \,e_V$ \\
& $(+0.45\,h_A-0.45\,b_A)\,e_A $ & $+ (+0.16\,h_A^2-0.83\, b_A\,h_A +1.51\, b_A^2)\,e_V$ \\
\hline
\hline
 &$10\,\times \,d_{K_1^0 \to \,\gamma\, K^0}$[GeV$^{-1}$]& \\
\hline
\hline
$ \pi \,K^* $  &$\rat{+7.56}{+3.02} h_P\,e + \rat{+0.98}{+2.78}e_V $ & $+(\rat{+0.23}{+0.65}\,g_D+\rat{-0.35}{-1.00}g_F+\rat{-0.03}{-0.09}\,b_D) \,e_V$ \\
& $\rat{+1.07}{-0.83}h_P\,e_M$  & $+(\rat{+0.63}{-0.49}h_V+\rat{+0.10}{-0.27}\tilde h_V)\,h_P\,e_V$ \\
& $(\rat{-0.03}{-0.00}h_A+\rat{+0.03}{+0.01}b_A)\,e_A $ & $ +(\rat{-0.01}{-0.00}h_A^2 +\rat{-0.17}{-0.49} b_A\,h_A +\rat{-0.09}{-0.24}\, b_A^2)\,e_V$ \\
\hline
$ K \,\rho $  &$\rat{+8.39}{+0.88} h_P\,e + \rat{+3.08}{+0.71}e_V $ & $+(\rat{+0.67}{+0.15}\,g_D+\rat{-0.67}{-0.15}g_F) \,e_V$ \\
& $\rat{-0.19}{-0.28}h_P\,e_M$  & $+(\rat{-0.14}{-0.21}h_V+\rat{+0.08}{-0.07}\tilde h_V)\,h_P\,e_V$ \\
& $ $ & $ +(\rat{+0.07}{-0.00}h_A^2 +\rat{-1.08}{-0.17} b_A\,h_A +\rat{-2.06}{-0.54}\, b_A^2)\,e_V$ \\
\hline
$ K \,\omega $  &$  $ & $+(+ 0.90\,g_D+0.24\,g_F+0.39\,b_D) \,e_V$ \\
& $(-0.80\,h_A+0.53\,b_A)\,e_A $ & $+ (-0.16\,h_A^2 +0.75\, b_A\,h_A -3.22\, b_A^2)\,e_V$ \\
\hline
$ \eta  \,K^* $  &$-0.16\,e_V  $ & $+(- 0.10\,g_D-0.06\,g_F-0.16\,b_D) \,e_V$ \\
& $$  & $+(-0.10\,h_V-0.07\,\tilde h_V)\,h_P\,e_V$ \\
& $(+0.86\,h_A-0.79\,b_A)\,e_A $ & $+ (+0.10\,h_A^2 -0.93\, b_A\,h_A +2.39\, b_A^2)\,e_V$ \\
\hline
$ K  \,\phi $  &$ $ & $+(+ 0.22\,g_D-0.06\,g_F+0.09\,b_D) \,e_V$ \\
& $(-0.90\,h_A+0.90\,b_A)\,e_A $ & $+ (-0.18\,h_A^2+0.92\, b_A\,h_A -1.68\, b_A^2)\,e_V$ \\
\hline
\end{tabular}
\caption{Decay constants that are implied by (\ref{K1:gamma:Kp}, \ref{K1:gamma:K0}).
The various contributions have to be multiplied by the appropriate resonance-coupling constants
$g_i$ of Table \ref{tab:hadronic-decay}.
We use  $f=90$ MeV, $m_V=776$ MeV. See also (\ref{collection-of-parameters}, \ref{unusual-convention}).}
\label{tab:decay-parameters:K1-1}
\end{center}
\end{table}

Using the resonance-coupling constants $g_i$ of Table \ref{tab:hadronic-decay} together with the parameters
of (\ref{collection-of-parameters}) we arrive at the following decay constants
\begin{eqnarray}
&& d_{\,K_1^\pm \,\to \, \gamma \,K^\pm} = (-0.35 -i\,0.03 )\,{\rm
GeV}^{-1} \,, \qquad
\nonumber\\
&& d_{\,K_1^0 \;\,\to \, \gamma \,K^0} \,= (+0.44 +i\,0.09 )\,{\rm GeV}^{-1} \,,
\nonumber\\
&& d^{(1)}_{\,K_1^\pm \,\to \, \gamma \,K_*^\pm} = -0.15 +i\,0.01  \,, \qquad
d^{(2)}_{\,K_1^\pm \,\to \, \gamma \,K_*^\pm} = -0.53 -i\,0.12  \,,
\nonumber\\
&& d^{(1)}_{\,K_1^0 \;\,\to \, \gamma \,K_*^0}\, = +0.69 +i\,0.06  \,, \qquad
d^{(2)}_{\,K_1^0 \;\,\to \, \gamma \,K_*^0} \,= +0.67 +i\,0.06  \,.
\label{K1-full}
\end{eqnarray}
We do not recover the decay constant (\ref{exp:K1-decay-constant}) of the KTeV collaboration.
As repeatedly pointed out in this work this may hint at the importance of additional channels involving
a pair of vector mesons. The latter is expected from \cite{Lutz-Soyeur:2007}
and also from the power-counting rules (\ref{power-counting}, \ref{extra-power}). The decay constants given in
(\ref{K1-full}) are partial to the extent that so far contributions from two-body channels involving a Goldstone
boson and a vector meson are considered only.

We discuss the physics behind the values (\ref{K1-full}). Consider first the
radiative decays with a pseudo-scalar particle in the final state. The two decay parameters are decomposed into
contributions from the five channels $\pi \,K^*, \,K\,\rho, \,K\,\omega,\,\eta \,K^*$ and $K\,\phi$ characterized
by  the molecule-coupling constants $g_i$ of Table \ref{tab:hadronic-decay}. We obtain
\begin{eqnarray}
&& d_{\,K_1^\pm \,\to \, \gamma \,K^\pm} = \Big\{(0.050 -i\,0.052 )\,g_1
-(0.059+i\,0.002)\,g_2
\nonumber\\
&& \qquad \qquad \quad\; +\,0.044\,g_3+0.098\,g_4 -0.022\,g_5 \Big\}\,{\rm GeV}^{-1} \,, \qquad
\nonumber\\
&& d_{\,K_1^0 \;\,\to \, \gamma \,K^0} \,=  \Big\{(0.090 +i\,0.077 )\,g_1
+(0.138+i\,0.019)\,g_2
\nonumber\\
&& \qquad \qquad \quad\; -\,0.003\,g_3-0.003\,g_4 -0.009\,g_5\Big\}\,{\rm GeV}^{-1} \,.
\label{K1-channel-split}
\end{eqnarray}
The decomposition (\ref{K1-channel-split}) reveals an interesting pattern. While the radiative decay of the
charged state is dominated by the $K\,\rho$ and $\eta \,K^*$ channels, the decay of the neutral state is
largely determined by the $K\,\rho$ channel only. Note that the coupling constants
$g_2=2.9$ and $g_4 = -2.5$ of Table \ref{tab:hadronic-decay} have opposite signs such that the two dominant terms
in the $K_1^\pm \,\to \, \gamma \,K^\pm$ decay add up constructively. The contributions of the first and fifth channels
are suppressed due to small coefficients in (\ref{K1-channel-split}), but also due to small coupling constants $g_1=0.4$ and
$g_5=0.1$.

In Table \ref{tab:decay-parameters:K1-1} we present a further decomposition of the two decay constants into terms
proportional to various products of the coupling constants (\ref{collection-of-parameters}).
We discuss first the $K_1^0 \;\,\to \, \gamma \,K^0$ decay in more detail. The large contribution from the second channel
is dominated by the terms proportional to $h_P\,e$ and $e_V$.
The term proportional to $e_V$ is the constructive sum of two
terms: the first implied by the Weinberg-Tomozawa interaction, the second directly by the term
${\tr} \,(V^{\mu \nu}\,f^+_{\mu \nu})$. The relative size
of the two contributions is about 7:3. Since the terms proportional to $h_P\,e$ and $e_V$ are of almost equal
magnitude the relative phase of the parameters $h_P$ and $e_V$ matters crucially. We recall that the magnetic moment
of the $\rho $ meson determines the relative phase of $e_V$ and $\tilde h_V$
(see (\ref{magnetic-moment-rho}, \ref{value-eM:b})).
Furthermore the relation (\ref{value-hV}) suggests a correlation of phases for the parameters $\tilde h_V$ and $h_P$.
All together this implies both parameters $h_P$ and $e_V$ to be positive. Incidentally, the same conclusion follows
from a study of the pion electromagnetic form factor \cite{Ecker:1989}.

Consider now the $K_1^\pm \;\,\to \, \gamma \,K^\pm$ decay. In this case  the second and
fourth channel are dominated by the terms proportional to $h_P\,e$. This is a consequence of a significant
cancellation amongst the two contributions proportional to $e_V$. In the $K \rho$ channel the ratio of the
two is 9:8, in the $\eta \,K^*$ channel it is 11:5. In the latter channel note the sizeable contribution from
the terms proportional to $h_A^2\,e_V$ and $h_A\,b_A\,e_V$. They are of almost equal size, however, they contribute
destructively, such that their combined impact is much reduced.

\begin{table}[t]
\tabcolsep=4.mm
\renewcommand{\arraystretch}{1.35}
\fontsize{10}{3.5}
\begin{center}
\begin{tabular}{|c||l |l |}
\hline
 &$10\,\times \,d^{(1)}_{K_1^+ \to \,\gamma\, K^+_*}$ & $10\,\times \,d^{(2)}_{K_1^+ \to \,\gamma\, K^+_*}$ \\
\hline
$ \pi\,K^* $ \hfill  &$\rat{-0.80}{+1.08} e\,h_A+\rat{+0.04}{-0.21}e \,b_A$ & $\rat{-0.25}{+0.69} e\,h_A+\rat{-0.01}{-0.18}e \,b_A $  \\
$\times \,h_P\,e_V$ & $\rat{-0.02}{-0.24}h_A+\rat{+0.04}{+0.81} b_A $  & $\rat{-0.75}{-0.41}h_A +\rat{+1.84}{+1.21} b_A$\\
$\times \,h_A\,e_V$ & $\rat{+0.25}{+0.38}h_V+\rat{-0.16}{+0.03}\tilde h_V $  & $\rat{-0.18}{+0.18}h_V+\rat{+0.10}{+0.03}\tilde h_V $\\
$\times \,b_A\,e_V$ & $\rat{+0.08}{+0.31}h_V+ \rat{+0.05}{+0.03}\tilde h_V$  & $\rat{+0.12}{+0.15}h_V+\rat{-0.01}{+0.03} \tilde h_V  $\\
$\times \,e_M$ & $\rat{+0.09}{+0.43}h_A+\rat{+0.12}{+0.30}b_A $  & $\rat{+0.06}{+0.16}h_A+\rat{+0.05}{+0.15}b_A$\\
$\times \,e_A$ & $\rat{-0.12}{-1.64}h_P $  & $\rat{-5.21 }{-2.79}h_P$\\
$\times \,e_A$ & $\rat{+0.71}{+0.39}h_V +\rat{-0.59}{+0.10}\tilde h_V $  & $\rat{-0.88}{+0.23}h_V +\rat{+0.38}{+0.09}\tilde h_V$\\
\hline
$ K\,\rho $ \hfill  &$ \rat{-1.51}{-0.06} e\,h_A+\rat{+0.20}{-0.35}e \,b_A$ & $\rat{-1.00}{-0.12} e\,h_A+\rat{-0.51}{-0.27}e \,b_A $  \\
$\times \,h_P\,e_V$ & $\rat{+0.15}{-0.34}h_A +\rat{-0.04}{+0.13} b_A$  & $\rat{-4.35}{-1.53}h_A+\rat{+1.12}{+0.44}b_A$\\
$\times \,h_A\,e_V$ & $\rat{+0.19}{+0.04}h_V+\rat{+0.01}{+0.01}\tilde h_V $  & $\rat{-0.11}{-0.01}h_V+\rat{+0.03}{+0.00}\tilde h_V $\\
$\times \,b_A\,e_V$ & $\rat{+4.00}{+0.88}h_V+ \rat{+0.07}{-0.01}\tilde h_V$  & $\rat{+2.23}{+0.47}h_V+\rat{+0.03}{+0.02} \tilde h_V  $\\
$\times \,e_M$ & $\rat{+0.26}{+0.07}h_A+\rat{+5.33}{+1.15}b_A $  & $\rat{-0.03}{-0.01}h_A+\rat{+2.72}{+0.63}b_A$\\
$\times \,e_A$ & $\rat{+0.62}{-1.36}h_P $  & $\rat{-17.4}{-6.12}h_P$\\
$\times \,e_A$ & $\rat{+0.35}{+0.03}h_V +\rat{-0.25}{+0.01}\tilde h_V $  & $\rat{-0.38}{+0.03}h_V +\rat{+0.17}{+0.01}\tilde h_V$\\
\hline
$ K\,\omega  $ \hfill  &$ +2.00\,e\,h_A -3.28\,e\,b_A $ & $+1.38\,e\,h_A-2.41\,e\,b_A $  \\
$\times \,h_A\,e_V$ & $\rat{+0.47}{-0.51}h_P+0.12\,h_V+0.05\,\tilde h_V $  & $\rat{-6.92}{-2.65}h_P+0.17\,h_V-0.03\,\tilde h_V $\\
$\times \,b_A\,e_V$ & $\rat{-0.12}{+0.21}\,h_P+0.58\,h_V-0.23\,\tilde h_V $  & $\rat{+1.78}{+0.76}h_P+0.02\,h_V-0.04\,\tilde h_V $\\
$\times \,e_M$ & $+0.29\,h_A+1.46\,b_A $  & $+0.14\,h_A+0.72\,b_A$\\
$\times \,e_A$ & $\rat{+1.88}{-2.04}h_P-0.18\,h_V+0.14\,\tilde h_V$  & $\rat{-27.7}{-10.6}h_P+0.22\,h_V-0.09\,\tilde h_V$\\
\hline
$ \eta\,K^*  $ \hfill  &$ +0.94\,e\,h_A -3.46\,e\,b_A $ & $+0.53\,e\,h_A-1.59\,e\,b_A $  \\
$\times \,h_A\,e_V$ & $-0.53\,h_P-0.09\,h_V+0.00\,\tilde h_V $  & $+3.06\,h_P+0.06\,h_V-0.03\,\tilde h_V $\\
$\times \,b_A\,e_V$ & $+1.30\,h_P-1.70\,h_V-0.18\,\tilde h_V $  & $-7.46\,h_P-1.24\,h_V+0.03\,\tilde h_V $\\
$\times \,e_M$ & $-0.01\,h_A-2.67\,b_A $  & $+0.00\,h_A-1.21\,b_A$\\
$\times \,e_A$ & $-1.50\,h_P-0.53\,h_V+0.15\,\tilde h_V$  & $+8.61\,h_P+0.51\,h_V-0.21\,\tilde h_V$\\
\hline
$ K\,\phi  $ \hfill  &$ +1.60\,e\,h_A -1.92\,e\,b_A $ & $+1.09\,e\,h_A-1.37\,e\,b_A $  \\
$\times \,h_A\,e_V$ & $+0.66\,h_P+0.08\,h_V-0.05\,\tilde h_V $  & $-3.00\,h_P-0.04\,h_V+0.01\,\tilde h_V $\\
$\times \,b_A\,e_V$ & $-2.38\,h_P+0.57\,h_V+0.19\,\tilde h_V $  & $+10.8\,h_P+0.46\,h_V+0.05\,\tilde h_V $\\
$\times \,e_M$ & $+0.09\,h_A+0.62\,b_A $  & $+0.05\,h_A+0.30\,b_A$\\
$\times \,e_A$ & $+4.56\,h_P+0.05\,h_V-0.13\,\tilde h_V$  & $-20.7\,h_P-0.18\,h_V+0.03\,\tilde h_V$\\
\hline
\end{tabular}
\caption{Decay constants that are implied by (\ref{K1:gamma:Kmup}).
The various contributions have to be multiplied by the appropriate resonance-coupling constants
$g_i$ of Table \ref{tab:hadronic-decay}.
We use  $f=90$ MeV, $m_V=776$ MeV. See also (\ref{collection-of-parameters}, \ref{unusual-convention}).}
\label{tab:decay-parameters:K1-2}
\end{center}
\end{table}

We turn to the decays with a vector particle in the final state. While the two decay constants presented in
(\ref{K1-full}) are almost degenerate for the neutral state, a significant asymmetry is observed for the charged
state. In order to trace the source of this striking effect the decay constants are decomposed into their contributions
from the five channels
\begin{eqnarray}
&& d^{(1)}_{\,K_1^\pm \,\to \, \gamma \,K_*^\pm} =
(-0.075 + i\,0.081)\,g_1 - (0.076 +i\, 0.006)\,g_2
\nonumber\\
&& \qquad \qquad \quad\;+\, (0.116-i\,0.008)\,g_3 + 0.017\,g_4 + 0.093\,g_5\,,
\nonumber\\
&& d^{(2)}_{\,K_1^\pm \,\to \, \gamma \,K_*^\pm} =
(-0.012+i\, 0.048)\,g_1 -(0.126 + i\,0.032)\,g_2
\nonumber\\
&& \qquad \qquad \quad\;-\,( 0.044+i\,0.041)\,g_3 + 0.047\,g_4 + 0.026\,g_5 \,,
\nonumber\\
&& d^{(1)}_{\,K_1^0 \;\,\to \, \gamma \,K_*^0}\, =
(-0.197 - i\,0.135)\,g_1+ (0.282 + i\,0.037)\,g_2
\nonumber\\
&& \qquad \qquad \quad\;-\, (0.013-i\,0.008)\,g_3 + 0.014\,g_4 + 0.013\,g_5\,
\nonumber\\
&& d^{(2)}_{\,K_1^0 \;\,\to \, \gamma \,K_*^0} \,= (-0.094- i\,0.108)\,g_1 +
   (0.172 + i\,0.021)\,g_2
   \nonumber\\
&& \qquad \qquad \quad\;+\, (0.090+i\,0.041)\,g_3 - 0.047\,g_4 - 0.039\,g_5 \,.
\label{K1-channel-split-vector}
\end{eqnarray}
As is evident from (\ref{K1-channel-split-vector}) the large asymmetry in the charged decay is a consequence
of a significant cancellation in $d^{(1)}_{\,K_1^\pm \,\to \, \gamma \,K_*^\pm}$ of contributions from the
second and third channel. The role of the $K\,\omega$ channel is minor even though the corresponding
molecule-coupling constant $g_4= -2.5$ is large.

In  Tables \ref{tab:decay-parameters:K1-2} and \ref{tab:decay-parameters:K1-3} a further split of the
decay constants into moments proportional to various products of coupling constants is presented.
Again various cancellation mechanisms are at work.

\begin{table}[t]
\tabcolsep=4.mm
\renewcommand{\arraystretch}{1.5}
\fontsize{10}{3.5}
\begin{center}
\begin{tabular}{|c||l |l |}
\hline
 &$10\,\times \,d^{(1)}_{K_1^0 \to \,\gamma\, K^0_*}$ & $10\,\times \,d^{(2)}_{K_0^+ \to \,\gamma\, K^0_*}$ \\
\hline
$ \pi\,K^* $ \hfill  &$\rat{-3.05}{-2.20} e\,h_A+\rat{+0.28}{+0.16}e \,b_A$ & $\rat{-1.55}{-1.73} e\,h_A+\rat{+0.13}{+0.11}e \,b_A $  \\
$\times \,h_P\,e_V$ & $\rat{+0.01}{+0.17}h_A+\rat{-0.03}{-0.59}b_A $  & $\rat{+0.54}{+0.29}h_A+\rat{-1.34}{-0.88}b_A$\\
$\times \,h_A\,e_V$ & $\rat{-0.12}{-0.12}h_V+\rat{-0.02}{+0.03}\tilde h_V $  & $\rat{-0.04}{-0.12}h_V+\rat{-0.00}{+0.01}\tilde h_V $\\
$\times \,b_A\,e_V$ & $\rat{+0.04}{+0.05}h_V+ \rat{+0.00}{-0.01}\tilde h_V$  & $\rat{+0.01}{+0.03}h_V+\rat{+0.00}{-0.01} \tilde h_V  $\\
$\times \,e_M$ & $\rat{-0.16}{-0.12}h_A+\rat{+0.08}{+0.13}b_A $  & $\rat{-0.06}{-0.16}h_A+\rat{+0.03}{+0.07}b_A$\\
$\times \,e_A$ & $\rat{-0.03}{+0.00}h_V +\rat{+0.01}{+0.00}\tilde h_V $  & $\rat{+0.03}{+0.00}h_V +\rat{-0.01}{+0.00}\tilde h_V$\\
\hline
$ K\,\rho $ \hfill  &$ \rat{+5.31}{+0.82} e\,h_A+\rat{-6.53}{-0.94}e \,b_A$ & $\rat{+3.61}{+0.65} e\,h_A+\rat{-4.16}{-0.70}e \,b_A $  \\
$\times \,h_P\,e_V$ & $\rat{+0.35}{-0.33}h_A +\rat{-0.09}{+0.13}b_A$  & $\rat{-1.41}{-0.89}h_A+\rat{+0.36}{+0.28}b_A$\\
$\times \,h_A\,e_V$ & $\rat{+0.14}{+0.03}h_V+\rat{-0.02}{-0.01}\tilde h_V $  & $\rat{+0.25}{+0.05}h_V+\rat{-0.02}{-0.00}\tilde h_V $\\
$\times \,b_A\,e_V$ & $\rat{-2.35}{-0.50}h_V+ \rat{-0.02}{+0.02}\tilde h_V$  & $\rat{-1.36}{-0.28}h_V+\rat{-0.02}{-0.01} \tilde h_V  $\\
$\times \,e_M$ & $\rat{+0.31}{+0.06}h_A+\rat{-2.43}{-0.48}b_A $  & $\rat{+0.31}{+0.07}h_A+\rat{-1.28}{-0.30}b_A$\\
$\times \,e_A$ & $\rat{+1.40}{-1.31}h_P $  & $\rat{-5.65}{-3.55}h_P$\\
\hline
$ K\,\omega  $ \hfill  &$- $ & $- $  \\
$\times \,h_A\,e_V$ & $\rat{-0.55}{+0.51}h_P+0.04\,h_V-0.06\,\tilde h_V $  & $\rat{+5.28}{+2.65}h_P-0.10\,h_V+0.04\,\tilde h_V $\\
$\times \,b_A\,e_V$ & $\rat{+0.14}{-0.21}h_P+0.25\,h_V+0.26\,\tilde h_V $  & $\rat{-1.36}{-0.76}h_P+0.42\,h_V+0.04\,\tilde h_V $\\
$\times \,e_A$ & $\rat{-2.20}{+2.04}h_P+0.36\,h_V-0.28\,\tilde h_V$  & $\rat{+21.1}{+10.6}h_P-0.44\,h_V+0.19\,\tilde h_V$\\
\hline
$ \eta\,K^*  $ \hfill  &$ - $ & $- $  \\
$\times \,h_A\,e_V$ & $+0.59\,h_P-0.06\,h_V+0.06\,\tilde h_V $  & $-3.40\,h_P+0.12\,h_V-0.03\,\tilde h_V $\\
$\times \,b_A\,e_V$ & $-1.45\,h_P+0.25\,h_V-0.25\,\tilde h_V $  & $+8.30\,h_P-0.37\,h_V+0.03\,\tilde h_V $\\
$\times \,e_A$ & $+3.00\,h_P-0.46\,h_V+0.37\,\tilde h_V$  & $-17.2\,h_P+0.76\,h_V-0.21\,\tilde h_V$\\
\hline
$ K\,\phi  $ \hfill  &$- $ & $- $  \\
$\times \,h_A\,e_V$ & $+0.66\,h_P-0.03\,h_V+0.05\,\tilde h_V $  & $-3.00\,h_P+0.07\,h_V-0.01\,\tilde h_V $\\
$\times \,b_A\,e_V$ & $-2.38\,h_P-0.26\,h_V-0.21\,\tilde h_V $  & $+10.8\,h_P-0.33\,h_V-0.06\,\tilde h_V $\\
$\times \,e_A$ & $+4.56\,h_P-0.11\,h_V+0.25\,\tilde h_V$  & $-20.7\,h_P+0.37\,h_V-0.07\,\tilde h_V$\\
\hline
\end{tabular}
\caption{Decay constants that are implied by (\ref{K1:gamma:Kmu0}).
The various contributions have to be multiplied by the appropriate resonance-coupling constants
$g_i$ of Table \ref{tab:hadronic-decay}.
We use  $f=90$ MeV, $m_V=776$ MeV. See also (\ref{collection-of-parameters}, \ref{unusual-convention}).}
\label{tab:decay-parameters:K1-3}
\end{center}
\end{table}

\clearpage

\section{Summary}
\label{sec:summary}

In this work we studied strong and electromagnetic properties of light vector and axial-vector mesons based
on the chiral Lagrangian constructed for the octet of Goldstone bosons and the nonet of light vector mesons.
The computations were guided by novel counting rules that put chiral and large-$N_c$ properties of QCD
into the context of the hadrogenesis conjecture. For technical reasons we represented the vector states by
anti-symmetric tensor fields.

It was demonstrated that the counting rules recover successfully the hadronic and electromagnetic properties
of the light vector mesons. In  particular the substantial flavor breaking in their radiative decay amplitudes
was obtained naturally. To further scrutinize their usefulness it would be important to establish
QCD lattice simulations for the magnetic and quadrupole moments of the light vector mesons.

The original computation for the spectrum of axial-vector mesons was refined deriving
the chiral correction operators of order $Q^2_\chi$ to the coupled-channel interaction that generates the spectrum.
It was found that in the absence of explicit u- and t-channel processes the local counter terms acquire non-universal
values, i.e. the two free parameters that enter in a combined chiral and large-$N_c$ approach take values that
depend on the molecule state. Though such a phenomenology is able to recover the properties of the axial-vector meson
spectrum quite reasonably, this asks for further developments. In particular the incorporation of explicit t- and
u-channel processes as well as additional channels with two vector mesons, as mandated by the counting rules,
is highly desirable.

We derived systematically the one-loop contributions to the
radiative decay amplitudes of the axial-vector molecules
$f_1(1282),\, b_1(1230), \,h_1(1386),\,a_1(1230)$ and $K_1(1272)$
with a Goldstone boson or a vector meson in the final state. It
was emphasized that if a state is generated by coupled-channel
dynamics there are no tree-level contributions like one may expect
from the vector-meson dominance picture. The inclusion of such
effects would be double counting. Detailed tables where the decay
amplitudes are decomposed into the various contributions from
different channels but also from different parameter combinations
were presented and discussed. All parameters with the exception of
one, which could be determined by precise values for the magnetic
moments of the strange vector mesons, are known reasonably well. We
unravelled various significant cancellation mechanisms that
illustrate the importance of systematic computations.

The established results for the radiative decays of the axial-vector molecules are still partial since so far only
contributions of channels involving a pseudo-scalar and a vector meson were considered. From previous computations in
the charm sector and also from the counting rules we expect the impact of channels with a pair of vector mesons to
be significant.

{\bfseries{Acknowledgments}}

We acknowledges useful discussions with J. Hofmann, E.E. Kolomeitsev and M. Wagner.

\newpage

\section*{Appendix A}
\label{appendix:A}

In the following we establish how the scalar decay parameters (\ref{result-projection-1plus}) define the angular
observable measured in the decay chain $1^+ \,\to \gamma \,1^- \, \to \gamma\,0^-\,0^-$ (see \cite{Amelin:1994ii}).
One determines the angle $\vartheta$ between the photon (momentum $q$)
and one of the decay products (momentum $k$) of the
$1^-$ vector meson in the rest frame of the $1^-$. In this frame the following relations
hold:
\begin{eqnarray}
p \cdot q = M_{1^-} E_\gamma
\,, & \qquad &
p^2 = M_{1^-}^2 + 2\,M_{1^-} \,E_\gamma
\,, \nonumber \\
k\cdot q = E_{k} \,E_\gamma - | \vec k\,|\, E_\gamma \,\cos\vartheta
\,, & \qquad &
k \cdot p = E_{k}\, M_{1^-} + k \cdot q \,.
  \label{eq:rfrho}
\end{eqnarray}
On account of the $h_P$ interaction
of (\ref{interaction-tensor}) the matrix element for the
decay $1^- \to 0^- 0^-$ comes with $k_\beta \, (\bar p - k)_{\beta'}$ which is contracted
with $M_{\beta \beta',\mu,\alpha \alpha'}$. Therefore the relevant quantity is given by
\begin{eqnarray}
  &&
  p^{\alpha'} k^\beta \, \bar p^{\beta'} M_{\beta \beta',\mu,\alpha \alpha'}(p,q) \;
  g^{\mu \nu} g^{\alpha \gamma} \;
  p^{\gamma'} k^\delta \bar p^{\delta'} M^*_{\delta \delta',\nu,\gamma \gamma'}(p,q)
  \nonumber \\ &&
  = \frac{1}{8}\, E_\gamma^2 \,M_{1^+}^2\, M_{1^-}^2 \,\vec k\,^2
  \nonumber \\ && \quad \times
  \left(
    \cos^2 \vartheta \, \vert d^{(2)}_{1^+\to \, \gamma \,1^-} \vert^2 +
    \frac{M_{1^-}}{2(M_{1^-}+2\,E_\gamma)} \, \sin^2\vartheta \, \vert d^{(1)}_{1^+\to \, \gamma \,1^-} \vert^2 \right) \,,
  \label{eq:combangl}
\end{eqnarray}
which has to be matched with $\rho_{00} \cos^2 \vartheta + \rho_{11} \sin^2 \vartheta$.
Thus one gets for the experimentally accessible ratio \cite{Amelin:1994ii}
\begin{equation}
  \label{eq:anglratio}
  \frac{\rho_{00}}{\rho_{11}} =
 2\, \frac{\vert d^{(2)}_{1^+\to \, \gamma \,1^-} \vert^2}{\vert d^{(1)}_{1^+\to \, \gamma \,1^-} \vert^2} \, \frac{M_{1^-}+2\,E_\gamma}{M_{1^-}}
  = 2\,\frac{\vert d^{(2)}_{1^+\to \, \gamma \,1^-} \vert^2}{\vert d^{(1)}_{1^+\to \, \gamma \,1^-} \vert^2} \, \frac{M_{1^+}^2}{M_{1^-}^2}  \,.
\end{equation}

\newpage

\section*{Appendix B}
\label{appendix:new}

This appendix consists of two parts. We first  point out
why the decay of a dynamically generated state always happens via a loop.
In the second part we elaborate on possible double-counting issues as for instance encountered in \cite{Roca:Hosaka:Oset:2007}.

Consider the interaction $K$ between two scattering particles $A$ and $B$.
Their two-particle propagator which emerges in the rescattering process is denoted by $G$. We can be very schematic here. The scattering amplitude $T$ is given by the geometric series
\begin{equation}
  \label{eq:schemel}
 T = K + K\,G\,K + \ldots = \frac{K}{1-G \, K} \to \Gamma^\dagger_R \, G_R \, \Gamma_R   \,.
\end{equation}
In the last step we have indicated that this series can be reinterpreted as
a resonant propagator, $G_R$, and the coupling vertex $\Gamma_R$
of this dynamically generated resonance
to the states $A$, $B$.
Now we consider the case that the two considered particles scatter into a different
final state --- for the case of interest into $\gamma + B$. The corresponding interaction
$K_\gamma$ is supposed to be very small such that rescattering can be neglected here.
The corresponding scattering amplitude is given by
\begin{eqnarray}
  \label{eq:scheminel}
  T_\gamma & = & K_\gamma + K_\gamma \, G \, K + K_\gamma \, G \, K \, G \, K + \ldots
  = K_\gamma + K_\gamma \, G \, \frac{K}{1-G \, K}
  \nonumber \\
  & \to & K_\gamma + K_\gamma \, G \, \Gamma^\dagger_R \, G_R \, \Gamma_R  \,.
\end{eqnarray}
The first contribution is non-resonant. If we want to describe the decay of the resonance
into $\gamma + B$ instead of the reaction $A + B \to \gamma + B$, this first
part provides non-resonant background. One can argue that close to the resonance mass
(and for not too large resonance width) this first term can be neglected since $G_R$
becomes large. This corresponds to the experimental procedure of background subtraction.
The resonance decay is just given by
$K_\gamma \, G \, \Gamma^\dagger_R$. Inevitably the loop $G$ appears here! This line of argument
is the basis of our approach which we follow in the main part of the present work.

We note, however, that there is one case conceivable where one might take an alternative point of view: for the very special case that $A$ directly converts
into a photon, one has $K_\gamma = g_\gamma\, G_A \, K$ with a strength of
direct conversion $g_\gamma $ and the
propagator $G_A$ of the state $A$. In this case one might include the non-resonant
piece in (\ref{eq:scheminel}) in the resummation process and obtain
\begin{eqnarray}
  \label{eq:schem3}
  T_\gamma = g_\gamma \, G_A \, T \to g_\gamma \, G_A \, \Gamma^\dagger_R \, G_R \, \Gamma_R  =K_\gamma + K_\gamma \, G \, \Gamma^\dagger_R \, G_R \, \Gamma_R  \,,
\end{eqnarray}
i.e.\ the decay of the resonance is given by $g_\gamma \, G_A \, \Gamma^\dagger_R$. This expression does
not contain a loop. We stress, however, that the difference between the general
formula $K_\gamma \, G \, \Gamma^\dagger_R$ and $g_\gamma \, G_A \, \Gamma^\dagger_R$ is just a non-resonant term,
i.e.\ of no concern for the resonance decay. If one aims at a description
of the whole reaction $A + B \to \gamma + B$ one has to be aware of the possible
non-resonant term. But this is a different issue. In the following we will always
start with the general formula $K_\gamma \, G \, \Gamma^\dagger_R$. In the graphical representation
of the relevant processes we will see where one might switch to the corresponding
tree level diagrams for the special case discussed above.

Next we will discuss pertinent diagrams that emerge from specific  Lagrangians.
Different representations of vector mesons and their
interrelations will be compared. This will set the stage for a thorough discussion of double counting which can occur, if one considers tree and loop diagrams for the radiative decay amplitude of a dynamically generated state. We will follow the discussion in \cite{Ecker:1989yg} concerning the construction of Lagrangians which obey chiral constraints and gauge invariance. In  \cite{Ecker:1989yg}  vector mesons are represented in terms of six-component antisymmetric tensor fields but alternatively also in terms of four-component vector fields. Note, however, that the conventions of \cite{Ecker:1989yg} differ from ours to some extent. To stay compatible with the rest of our work we translate the formulae of \cite{Ecker:1989yg}
into our conventions.

Following \cite{Roca:Hosaka:Oset:2007} we will restrict ourselves in this appendix to the channel $1^+ \to \gamma \,0^- $ and assume the vector particles to be represented by vector fields. As discussed before the  coupled-channel part of the work \cite{Roca:Hosaka:Oset:2007} considers the vector mesons as heavy-matter fields coupled to the Goldstone bosons. The corresponding effective chiral Lagrangian including
electromagnetic fields is well established  \cite{Ecker:1989yg}. In terms of the conventions and notations of our work the relevant terms read
\begin{eqnarray}
  \label{eq:lagrvec1}
&&  {\mathcal L}_{\rm vector,1} = f^2\, \tr \Big\{ U^\mu\,U^\dagger_\mu \Big\}
-\frac{1}{8}\,
  {\tr }\, \Big\{\hat V_{\mu \nu} \, \hat V^{\mu \nu} \Big\}+
  \frac{1}{4}\,m_{1^-}^2\,{\tr } \,\Big\{ \hat V^{\mu}\, \hat V_{\mu}\Big\}
  \nonumber  \\
  && \quad
  - \,i\,\frac{m_V\, h_{P}}{2 \, m_{1^-}}\,
  {\rm tr}\,\Big\{U_\mu\,\hat V^{\mu\nu}\,U_\nu\Big\}
  + \frac{e_V\,m_V}{8 \, m_{1^-}}\,{\rm tr}\,\Big\{ \hat V^{\mu\nu}\,f^+_{\mu \nu}\Big\}
  - {\rm tr}\,\Big\{j_{\mu\nu} \, j^{\mu \nu}\Big\}
\nonumber\\
&& \quad +\, {\mathcal O} \left( \hat V^2\right)
  \,,
\end{eqnarray}
with
\begin{eqnarray}
  \label{eq:defjmunu}
&& \hat V_{\mu\nu} := D_\mu \hat V_\nu -  D_\nu \hat V_\mu \,, \quad
\nonumber\\
&&  j_{\mu\nu} = - \frac{1}{2 \, \sqrt{2}} \left(
    \frac{e_V\,m_V}{2 \, m_{1^-}} \, f^+_{\mu \nu}
    +  i \, \frac{m_V\, h_{P}}{m_{1^-}}\, [U_\mu,U_\nu]_-   \right)   \,.
\end{eqnarray}
The vector field $\hat V_\mu$ has chiral transformation properties identical to the
tensor field $V_{\mu \nu}$ used throughout our work. The covariant derivative $D_\mu$ and the objects $U_\mu$ were specified in (\ref{eq:defcovder}, \ref{def-Umu}).

As pointed out first in \cite{Lutz-Kolomeitsev:2004} the Lagrangian (\ref{eq:lagrvec1}) provides a Weinberg-Tomozawa type coupled-channel interaction that may be used to generate axial-vector molecules dynamically. Since the follow-up work
\cite{Roca:Hosaka:Oset:2007} bases its molecule formation on the latter Weinberg-Tomozawa interaction one may expect the computation of the radiative decay
properties to be based on the same Lagrangian (\ref{eq:lagrvec1}). Unfortunately this is not the case: the use of the direct gauge-dependent vertex $A_\mu\,\rho^\mu_0$ in
\cite{Roca:Hosaka:Oset:2007} is a clear signal of this inconsistency. Note that such
vertices are not part of (\ref{eq:lagrvec1}).

It should be mentioned, however, that the radiative decay computation of
\cite{Roca:Hosaka:Oset:2007} may be viewed as being derived from a phenomenological
Lagrangian, where the vector mesons are considered  as non-abelian gauge bosons
\cite{Bando,Birse:1996}.  In this scheme the chiral Ward identities and
electromagnetic gauge invariance follow as a consequence of a
non-abelian gauge symmetry. We recall from  \cite{Ecker:1989yg} the relevant terms
\begin{eqnarray}
  \label{eq:lagrvec2}
&&  {\mathcal L}_{\rm vector,2} = f^2\, \tr \Big\{ U^\mu\,U^\dagger_\mu \Big\} -\frac{1}{8}\,
  {\tr }\, \Big\{\bar V_{\mu \nu} \, \bar V^{\mu \nu} \Big\}
  \nonumber  \\   && \qquad
  +\frac{1}{4}\,m_{1^-}^2\,{\tr } \,
  \left\{ \Big(\bar V^{\mu}\, - \frac{2 \, i}{g} \, \Gamma^\mu +
  \frac{e}{g} \, A^\mu (u^\dagger \, Q \, u + u \, Q \, u^\dagger)\Big)^2 \right\}\,,
\end{eqnarray}
with
\begin{eqnarray}
  \label{eq:fieldstrengths}
\bar V_{\mu\nu} :=
\partial_\mu \bar V_\nu -  \partial_\nu \bar V_\mu
- i \, \frac{g}{2} \, \Big[\bar V_\mu , \bar V_\nu \Big]_-  \,.
\end{eqnarray}
It has been shown in \cite{Ecker:1989yg} that under certain conditions the two Lagrangians (\ref{eq:lagrvec1}, \ref{eq:lagrvec2}) yield identical results in their description of low-energy physics in a tree-level computation\footnote{The matching requires additional terms in (\ref{eq:lagrvec1}) of the form
\begin{eqnarray*}
&&{\mathcal L}_{\rm vector,1} =\frac{1}{4}\,{\tr }  \Big\{
\Big[\hat V_\mu,\,\hat V_\nu\Big]_-\,\Big[U^\mu,\,U^\nu\Big]_-\Big\}
-i\,\frac{e}{4}\,{\tr }  \Big\{
\Big[\hat V_\mu,\,\hat V_\nu\Big]\,f_+^{\mu \nu}\Big\}
\nonumber\\
&& \qquad \quad \;\,\, + \,i\, \frac{g}{8}\,{\tr }  \Big\{
\Big[\hat V_\mu,\,\hat V_\nu\Big]_-\,\hat V^{\mu \nu}\Big\}
+ \frac{g^2}{32}\,
{\tr }  \Big\{
\Big[\hat V_\mu,\,\hat V_\nu\Big]_-\,\Big[\hat V^\mu,\,\hat V^\nu\Big]_-\Big\} \,.
\end{eqnarray*}}. The latter requires
the particular choice
\begin{eqnarray}
g = \frac{4 \, e \, m_{1^-}}{e_V \, m_V} =
\frac{2 \,m_{1^-}}{m_V \, h_P} \,.
\label{condition}
\end{eqnarray}
Since the condition (\ref{condition}) holds reasonably well empirically, it is
justified to consider the phenomenological Lagrangian (\ref{eq:lagrvec2}) as
a rough approximation to the systematic  chiral Lagrangian (\ref{eq:lagrvec1}).
Note that the Lagrangian (\ref{eq:lagrvec2}) contains a term $A_\mu\,\rho^\mu_0$ and
also a $\rho\,\pi \pi$ vertex that involves one derivative only, as used in \cite{Roca:Hosaka:Oset:2007}.

It is important to understand that according to the different chiral transformation
properties  of the vector fields $\hat V_\mu $ and $\bar V_\mu $ the field strength
tensors $\hat V_{\mu \nu}$ and $\bar V_{\mu \nu}$ are distinct (see (\ref{eq:defjmunu},  \ref{eq:fieldstrengths}) and \cite{Ecker:1989yg}).
This will have important consequences for our discussion below.

We discuss the two Lagrangians (\ref{eq:lagrvec1}, \ref{eq:lagrvec2}) as to be applied
for a computation of radiative decay processes. The contributions we consider are
selective as the purpose of the presentation is to unravel a potential double-counting issue. A complete discussion with all relevant terms as they result using tensor
fields is presented in the main part of this work.

\begin{figure}
  \centering
\includegraphics[keepaspectratio,width=0.8\textwidth]{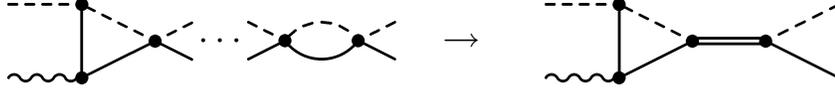}
  \caption{Feynman diagram emerging from the Lagrangian (\ref{eq:lagrvec1}). Full lines
  denote vector mesons, dashed lines pseudo-scalar mesons, wavy lines photons and the
  double line an axial-vector meson.}
  \label{fig:lagrstand}
\end{figure}

For simplicity we focus on processes where the intermediate vector meson carries
a charge but the intermediate Goldstone boson is charge neutral.
The Lagrangian (\ref{eq:lagrvec1}) contains
\begin{enumerate}
\item[1)] a coupling of a photon to two vector mesons from minimal substitution
\item[2)] a Weinberg-Tomozawa term, i.e.\ a
    four-point interaction with two vector and two pseud-scalar states
\item[3)] a three-point interaction $\sim h_P$ with one vector and two pseudo-scalar states which involves three derivatives
\item[4)] a four-point interaction $\sim h_P \, e$ of two pseudo-scalars, one vector and one photon emerging from minimal substitution
\item[5)] a term $\sim e_V$ for direct conversion of a vector meson into a photon
\item[6)] a second four-point interaction $\sim e_V$ of two pseudo-scalars, one vector and one photon.
\end{enumerate}
The Feynman diagram which emerges for the radiative decay of an axial-vector
state from the interactions of items 1) and 3) is depicted
in Figure \ref{fig:lagrstand} on the left hand side. The sum of bubbles just produces
the dynamically generated state in terms of the Weinberg-Tomozawa vertex 2). The corresponding diagram is shown on the right hand side of Figure \ref{fig:lagrstand}. Electromagnetic gauge invariance requires
the consideration of analogous contributions involving the interaction of item 4).
It is emphasized that even though one may draw additional contributions
where the final photon results from a direct conversion of a neutral vector meson, such terms are zero identically for on-shell photons.
We note in passing that the Feynman diagram implied by the four-point
interaction of item 6) is not considered in \cite{Roca:Hosaka:Oset:2007}.

The Lagrangian (\ref{eq:lagrvec2}) neither contains a coupling of a photon to two
vector mesons from minimal substitution nor a Weinberg-Tomozawa term. The reason is that
$\bar V_{\mu\nu}$ as given in (\ref{eq:fieldstrengths}) contains partial derivatives
and not covariant ones. Only the latter involve the photon field via minimal substitution and the coupling to Goldstone bosons (cf.\ the definition (\ref{eq:defcovder})).
However, the Lagrangian (\ref{eq:lagrvec2}) does contain the other interactions
described above in items 3) to 6)
with some remarkable differences: the interaction of one vector and
two pseudo-scalar states involves only one instead of three derivatives. The
direct conversion of a vector meson to a photon $\sim \bar V_\mu A^\mu$ does {\em not}
vanish for real photons. In addition, both Lagrangians (\ref{eq:lagrvec1}, \ref{eq:lagrvec2}) contain three- and four-point interactions of vector fields. The four-point interaction is of no concern for the
considered processes.

The axial-vector states are formed in the case of (\ref{eq:lagrvec2}) by an
iterated t-channel process: a vector meson is exchanged between a vector and a
pseudo-scalar state. As already noted there is a three-point interaction for one
vector and two pseudo scalars with one derivative and a three-point interaction for
three vector mesons --- also involving one derivative. For small enough energies
one might neglect the Mandelstam variable $t$ in the vector meson exchange and
effectively get back a Weinberg-Tomozawa type coupling with the right strength and form.
In this way the chiral low-energy theorems are satisfied by the
Lagrangian (\ref{eq:lagrvec2}). We will not judge here whether it is a reasonable
approximation to reduce the t-channel process to a point interaction in the
energy region of the axial-vector states. The important issue in the following is that
one should not forget that the original diagram is a t-channel process.

\begin{figure}
  \centering
\includegraphics[keepaspectratio,width=\textwidth]{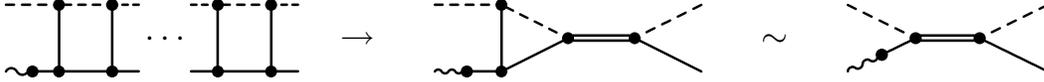}
  \caption{Feynman diagram emerging from the Lagrangian (\ref{eq:lagrvec2}). Full lines
  denote vector mesons, dashed lines pseudo-scalar mesons, wavy lines photons and
  double lines axial-vector mesons.}
  \label{fig:lagrhid}
\end{figure}

The Feynman diagram for the radiative decay of an axial-vector state which corresponds
to the one depicted in Figure \ref{fig:lagrstand} is shown in Figure \ref{fig:lagrhid}. Indeed, it is not complicated to make the diagram on the left hand side of Figure
\ref{fig:lagrhid} to resemble the corresponding one in Figure \ref{fig:lagrstand}.
One just has to remember our previous discussion: if one reduces
the vertical vector-meson lines to a point for all but the
interaction line most to the left, then the two diagrams already look very similar. The only difference
is that the photon is directly attached to the vector meson line in Figure
\ref{fig:lagrstand} while it converts first to a vector meson in Figure \ref{fig:lagrhid}.
But this is just how strict vector-meson dominance works:  electromagnetic
interactions of hadrons proceed via vector mesons. Passing from the left diagram of Figure \ref{fig:lagrhid} to the middle one indicates again the dynamical generation of the axial-vector state.

To summarize: the diagrams in Figure \ref{fig:lagrstand} emerge from the Lagrangian
(\ref{eq:lagrvec1}) whereas the ones in Figure \ref{fig:lagrhid} are implied by the
 Lagrangian (\ref{eq:lagrvec2}). In turn, neither the Lagrangian (\ref{eq:lagrvec1}) yields the diagrams in Figure \ref{fig:lagrhid}, nor does the Lagrangian (\ref{eq:lagrvec2})
lead to the diagrams of Figure \ref{fig:lagrstand}. In (\ref{eq:lagrvec1}) the direct
conversion of a photon to a vector meson vanishes for real photons. In (\ref{eq:lagrvec2}) there is no direct coupling of the photon to the charge of the vector meson --- it is generated by  the radiations of neutral vector mesons which  convert into a photon  via vector-meson dominance. The diagram on the left hand side of Figure \ref{fig:lagrhid} might be reinterpreted as a tree diagram with a dynamically
generated axial-vector meson. This is indicated in Figure \ref{fig:lagrhid} by the
last diagram. Such a reinterpretation is possible, albeit there
is no need to do so.

In \cite{Roca:Hosaka:Oset:2007} the tree diagram depicted on the right hand side of Figure \ref{fig:lagrhid} {\em and} the 
diagram shown on the right hand side of Figure \ref{fig:lagrstand} are studied. However, as worked out in depth above 
there is no effective Lagrangian that is consistent with the chiral constraints of QCD {\em and}  produces both diagrams. We have demonstrated that the two diagrams actually correspond to each other. Therefore, it is double counting to consider both diagrams simultaneously as done in \cite{Roca:Hosaka:Oset:2007}.

\newpage

\section*{Appendix C}
\label{appendix:B}

The  decay constant $d_{1^+\to \gamma 0^-}$ as of (\ref{result:width-1plus:a}) is
computed by contracting of the gauge invariant tensors
(\ref{def-AB-tensor3}, \ref{def-C-tensor3}, \ref{def-Cchi-tensor3}, \ref{def-D-tensor3}, \ref{def-E-tensor3}, \ref{def-F-tensor3},
\ref{def-G-tensor3}, \ref{def-H-tensor3})
with the anti-symmetric tensor
\begin{eqnarray}
&&  P^{(0-)}_{\mu, \alpha \beta} =
\frac{1}{2}\,
\Bigg( \Big\{g_{\mu \alpha}- \frac{p^\mu\,q^\alpha}{p\cdot q}\Big\}\,p_\beta
- \Big\{g_{\mu \beta }- \frac{p^\mu\,q^\beta}{p\cdot q}\Big\}\,p_\alpha \Bigg) \,.
\label{def-projections12}
\end{eqnarray}
We provide the results in terms of the master loop integrals $I_{ab}, \bar I_{ab}, J_{abc}$ and
$\bar J_{abc}$ introduced in (\ref{def-master-loops}). According to the
arguments of \cite{Lutz-Soyeur:2007} reduced tadpole contributions are dropped. Using $p^2=M_i^2$ and
$(p-q)^2=M_f^2$ we derive
\begin{eqnarray}
&& 16\,(p \cdot q)\,M_f^2\,P^{(0-)}_{\mu, \alpha \beta}\,A^{\mu, \alpha \beta }_{ab} =
2\, M_f^2\, \Big[M_i^6-\left(m_a^2-m_b^2\right)^2 M_i^2
\nonumber\\
&&   \qquad\,-\,\left(m_a^2+3\, m_b^2+M_i^2\right) \left(M_i^2-M_f^2\right)
   M_i^2
\nonumber\\
&&   \qquad\,+\,\left(M_f^2-M_i^2\right)^2 \left(-m_a^2+m_b^2+M_i^2\right)\Big]\,I_{ab}
\nonumber\\
&&   \,+\, \Big[  M_f^8-2 \left(m_a^2+m_b^2\right)
   M_f^6+\left(m_a^2-m_b^2\right)^2 M_f^4
\nonumber\\
&&   \qquad\,+\, 2 \left(\left(m_a^2-m_b^2\right)^2-M_f^4\right) M_i^2 M_f^2
\nonumber\\
&&   \qquad\,-\left(m_a^4-2
   \left(m_b^2+M_f^2\right) m_a^2+\left(m_b^2-M_f^2\right)^2\right) M_i^4 \Big]\,\bar I_{ab}
\nonumber\\
&&   \,+\, 4 \,m_a^2 \,M_f^2\,M_i^2 \left(m_a^2-m_b^2-M_f^2\right)
   \left(M_i^2-M_f^2\right) J_{aab}\,,
\nonumber\\ \nonumber\\
&& 16\,(p \cdot q)\,M_f^2\,P^{(0-)}_{\mu, \alpha \beta}\,B^{\mu, \alpha \beta }_{ab} =
M_f^2 \Big[-\left(m_a^2-m_b^2\right) M_f^4-\left(8\, m_b^2+M_f^2\right) M_i^4
\nonumber\\
&&   \qquad\,-\,\Big(M_f^4-\left(5\, m_a^2+3 \,m_b^2\right)
   M_f^2+2 \left(m_a^2-m_b^2\right)^2\Big) M_i^2\Big]\,I_{ab}
\nonumber\\
&&   \,+\,\Big[
   M_f^8-2 \left(m_a^2+m_b^2\right) M_f^6+\left(m_a^2-m_b^2\right)^2
   M_f^4
   \nonumber\\
&&   \qquad\,-\,\Big(M_f^4+\left(m_a^2-m_b^2\right) M_f^2
   -2 \left(m_a^2-m_b^2\right)^2\Big)\, M_i^2\, M_f^2
   \nonumber\\
&&   \qquad\,-\,\Big(-2\, M_f^4+\left(m_a^2-5\,
   m_b^2\right) M_f^2+\left(m_a^2-m_b^2\right)^2\Big)\, M_i^4 \Big]\,\bar I_{ab}
\nonumber\\
&&   \,-\,2 \,m_b^2\, M_f^2 \,M_i^2 \left(M_f^2-M_i^2\right) \left(-2\, m_a^2+2\,
   m_b^2-M_f^2+3\, M_i^2\right) \bar J_{abb} \,,
\end{eqnarray}
and

\begin{eqnarray}
&& 12\,(p \cdot q)\,M_f^2\,M_i^2\,P^{(0-)}_{\mu, \alpha \beta}\,C^{\mu, \alpha \beta }_{abc} =
M_f^2 \Big[6 \left(m^2_a-m^2_b\right) \left(m_b^2+m_c^2\right) M_i^6
\nonumber\\
&&\qquad \,-\,3 \left(m_b^2+m_c^2\right)
   \left(m_a^2-2 m_b^2+m_c^2-M_i^2\right) \left(M_i^2-M_f^2\right) M_i^4
   \nonumber\\
&&\qquad \,+\,\left(M_i^2-M_f^2\right)^3
   \Big(-M_i^4-\left(m_a^2+m_c^2\right) M_i^2+2 \left(m_a^2-m_c^2\right)^2\Big)
\nonumber\\
&&\qquad \,-\,\left(M_i^3-M_f^2\, M_i\right)^2 \Big(2\,
   M_i^4-\left(4 \,m_a^2+3\, m_b^2+7\, m_c^2\right) M_i^2
\nonumber\\
&&\qquad \, +\,\left(m_a^2-m_c^2\right) \left(2\, m_a^2+3\, m_b^2+m_c^2\right)\Big)\Big]\,I_{ac}
\nonumber\\
&& \,+\,3  \left(m_b^2+m_c^2\right) M_f^2\, M_i^4\, \Big[M_f^2 \left(-2\, m_a^2+3\, m_b^2-m_c^2+M_f^2\right)
\nonumber\\
&&\qquad \,-\,\left(m_b^2-m_c^2+M_f^2\right) M_i^2 \Big]\,\bar I_{bc}
\nonumber\\
&&\,+\, 6 \left(m_b^2+m_c^2\right) M_f^2\, M_i^4 \,\Big[-m_b^2\, M_f^4+\left(m_a^2-m_b^2\right)^2 M_f^2
\nonumber\\
&&\qquad \, +2\, m_b^2 \,M_i^2
   M_f^2-m_b^2\, M_i^4\Big] J_{abc}\,,
\end{eqnarray}
and
\begin{eqnarray}
&& 12\,\frac{M_i^2}{p\cdot q}\,P^{(0-)}_{\mu, \alpha \beta}\,\Bigg( \bar C^{\mu, \alpha \beta }_{abc}
-C^{\mu, \alpha \beta }_{abc} \Bigg) =
\Big[8 \,M_i^6-4 \left(m_a^2+13\, m_c^2+M_f^2\right) M_i^4
\nonumber\\
&&\qquad \, -\,4 \Big(\left(m_a^2-m_c^2\right)^2
+\left(m_a^2+m_c^2\right) M_f^2\Big)\,
   M_i^2+8 \left(m_a^2-m_c^2\right)^2 M_f^2 \Big]\,I_{ac}\,,
\nonumber\\ \nonumber\\
&& 3\,\frac{M_i^2}{p\cdot q}\,P^{(0-)}_{\mu, \alpha \beta}\,\Bigg( \frac{m_b^2+m_c^2}{2}\,\Cchi^{\mu, \alpha \beta }_{abc}
+C^{\mu, \alpha \beta }_{abc} \Bigg) =
\Big[-3\, M_i^6- 2   \left(m_a^2-m_c^2\right)^2 M_f^2
\nonumber\\
&&\qquad \,+\,\left(M_f^2+3 \left(m_a^2+m_c^2\right)\right) M_i^4+\left(m_a^2+m_c^2\right) M_f^2 \,M_i^2 \Big]\,I_{ac}\,,
\nonumber\\ \nonumber\\
&& 3\,\frac{M_i^2}{p\cdot q}\,P^{(0-)}_{\mu, \alpha \beta}\,\Bigg( \frac{m_b^2+m_c^2}{2}\,\barCchi^{\mu, \alpha \beta }_{abc}
+C^{\mu, \alpha \beta }_{abc} \Bigg) =
\Big[-3 \,M_i^6
\nonumber\\
&&\qquad \,+\,\left(3\, m_a^2+6 \,m_b^2 +9 \,m_c^2+M_f^2\right) M_i^4-2 \left(m_a^2-m_c^2\right)^2 M_f^2
\nonumber\\
&&\qquad \,+\,\left(\left(m_a^2+m_c^2\right) M_f^2-6 \left(m_a^2-m_c^2\right)
   \left(m_b^2+m_c^2\right)\right) M_i^2\Big]\,I_{ac}\,,
\end{eqnarray}
and
\begin{eqnarray}
&& 8\,M_f^2\,P^{(0-)}_{\mu, \alpha \beta}\,D^{\mu, \alpha \beta }_{ab} =
\Big[M_i^2 \,M_f^4+\left(m_a^2-m_b^2\right) M_f^4-2\, M_i^4\, M_f^2\Big]\,I_{ab}
\nonumber\\
&& \,+\,M_f^2 \left(-m_a^2+m_b^2+M_f^2\right) M_i^2\,\bar I_{ab}
+2\, m_b^2\, M_f^2\, M_i^2   \left(M_i^2-M_f^2\right) \bar J_{abb}\,,
\nonumber\\ \nonumber\\
&&24\,M_f^2\,P^{(0-)}_{\mu, \alpha \beta}\,\tilde D^{\mu, \alpha \beta }_{ab} =
M_f^2 \,\Big[-3 \,m_b^2 \left(-m_a^2+m_b^2+M_i^2\right) M_i^2
\nonumber\\
&&\qquad \,-\,\left(M_i^2-M_f^2\right) \left(m_a^4+\left(m_b^2-2\, M_i^2\right)
   m_a^2-2 \,m_b^4+M_i^4+m_b^2 M_i^2\right)\Big]\,I_{ab}
\nonumber\\
&&\,+\,  3\,\Big(m_a^4-2 \left(m_b^2+M_f^2\right) m_a^2+\left(m_b^2-M_f^2\right)^2 \Big)\, M_i^4
\nonumber\\
&&\qquad \,-\,3\, M_f^2 \left(m_a^4-\left(m_b^2+2\, M_f^2\right) m_a^2+M_f^4-3\, m_b^2 \,M_f^2\right) M_i^2\,\bar I_{ab}
\nonumber\\
&&\,+\,6\, m_b^4\, M_f^2\, M_i^2 \left(M_i^2-M_f^2\right) \bar J_{abb}\,,
\end{eqnarray}
and
\begin{eqnarray}
&& \frac{1}{p\cdot q}\,P^{(0-)}_{\mu, \alpha \beta}\,E^{\mu, \alpha \beta }_{ab} =
\Big[m_a^2-m_b^2-M_i^2 \Big]\,I_{ab}\,,
\nonumber\\ \nonumber\\
&& \frac{6}{p\cdot q}\,P^{(0-)}_{\mu, \alpha \beta}\,F^{\mu, \alpha \beta }_{ab} =
\Big[m_a^4+\left(m_b^2-2\, M_i^2\right) m_a^2-2\, m_b^4+M_i^4+7\, m_b^2\, M_i^2 \Big]\,I_{ab}\,,
\nonumber\\ \nonumber\\
&& \frac{6\,M_i^2}{p\cdot q}\,P^{(0-)}_{\mu, \alpha \beta}\,G^{\mu, \alpha \beta }_{ab} =
\Big[-3 M_i^6+3 \left(m_a^2-m_b^2\right)^2 M_i^2
\nonumber\\
&&\qquad \,+\,\left(M_i^2-M_f^2\right) \Big(M_i^4+\left(m_a^2+m_b^2\right) M_i^2-2
   \left(m_a^2-m_b^2\right)^2\Big) \Big]\,I_{ab}\,,
\nonumber\\ \nonumber\\
&& \frac{3\,m_b^2\,}{p\cdot q}\,P^{(0-)}_{\mu, \alpha
\beta}\,H^{\mu, \alpha \beta }_{abc} = -2  \Big[ m_a^4+\left(4\,
m_c^2-2\, M_i^2\right) m_a^2
\nonumber\\
&&\qquad \,-\,5\, m_c^4+M_i^4+4\, m_c^2\, M_i^2\Big]\,I_{ac}\,,
\nonumber\\ \nonumber\\
&& \frac{m_b^2\,}{p\cdot q}\,P^{(0-)}_{\mu, \alpha \beta}\,\bar
H^{\mu, \alpha \beta }_{abc} = \Big[ 8
\left(m_a^2-m_c^2-M_i^2\right) \Big]\,I_{ac}\,.
\end{eqnarray}

\newpage

\section*{Appendix D}
\label{appendix:C}

We detail the computation of the decay parameters $d^{(1)}_{1^+\,\to \,\gamma\,1^-}$ and
$d^{(2)}_{1^+\,\to \,\gamma\,1^-}$. According to (\ref{result-projection-1plus}) it suffices to compute
the projections of the tensor integrals
(\ref{def-A-tensor5}-\ref{def-F-tensor5})
 with two tensors
\begin{eqnarray}
&&P^{(1)}_{\bar \alpha \bar \beta, \mu, \alpha \beta } = \frac{1}{4}\,p^{\sigma}\,q^{\tau}\,\Big\{
 \epsilon_{\sigma  \tau  \bar \alpha \mu }\,\bar p_{\bar \beta}\,q_{\alpha}\,p_\beta
-\epsilon_{\sigma  \tau  \bar \alpha \mu }\,\bar p_{\bar \beta}\,q_{\beta}\,p_\alpha
\nonumber\\
&& \qquad \qquad \qquad \qquad \quad -\,\epsilon_{\sigma  \tau  \bar \beta \mu }\,\bar p_{\bar \alpha}\,q_{\alpha}\,p_\beta
+\epsilon_{\sigma  \tau  \bar \beta \mu }\,\bar p_{\bar \alpha}\,q_{\beta}\,p_\alpha
\Big\}\,,
\nonumber\\
&&P^{(2)}_{\bar \alpha \bar \beta, \mu, \alpha \beta } = \frac{1}{4}\,p^{\sigma}\,q^{\tau}\,\Big\{
 \epsilon_{\sigma  \tau\mu \alpha }\,q_{\bar \alpha}\,\bar p_{\bar \beta}\,p_\beta
-\epsilon_{\sigma  \tau\mu \beta }\,q_{\bar \alpha}\,\bar p_{\bar \beta}\,p_\alpha
\nonumber\\
&& \qquad \qquad \qquad \qquad \quad -\,\epsilon_{\sigma  \tau\mu \alpha }\,q_{\bar \beta}\,\bar p_{\bar \alpha}\,p_\beta
+\epsilon_{\sigma  \tau\mu \beta }\,q_{\bar \beta}\,\bar p_{\bar \alpha}\,p_\alpha
\Big\} \,.
\end{eqnarray}
We establish  the results

\begin{eqnarray}
&&48\,\frac{M_i^2}{(p \cdot q)}\,P^{(1)}_{\bar \alpha \bar \beta, \mu, \alpha \beta }\,
A^{\bar \alpha \bar \beta, \mu, \alpha \beta }_{ab} =
\Big[-12 \left(m_a^2-m_b^2+M_i^2\right) \left(m_b^2+M_i^2\right) M_i^4
\nonumber\\
&&   \qquad\,+\,6\left(M_i^2-M_f^2\right)
   \left(m_b^4+2 \,M_i^2 \,m_b^2+5 \,M_i^4+m_a^2 \left(M_i^2-m_b^2\right)\right) M_i^2
   \nonumber\\
&&   \qquad \,-\, \left(M_f^2-M_i^2\right)^2 \Big(19\,
   M_i^4-8 \left(m_a^2-2 \,m_b^2\right) M_i^2+\left(m_a^2-m_b^2\right)^2\Big)\Big]\,I_{ab}
\nonumber\\
&&   \,+\,   6 \left(m_a^2-m_b^2+M_f^2\right)
   \left(m_b^2+M_f^2\right) M_i^2 \left(M_f^2+M_i^2\right) \bar I_{ab}
\nonumber\\
&&   \, -\,12\, m_a^2 \left(m_b^2+M_f^2\right) M_i^2 \left(M_f^2-M_i^2\right)
   \left(M_f^2+M_i^2\right) J_{aab}\,,
\nonumber\\ \nonumber\\
&&48\,\frac{M_f^2\,M_i^2}{(p \cdot q)}\,P^{(2)}_{\bar \alpha \bar \beta, \mu, \alpha \beta }\,
A^{\bar \alpha \bar \beta, \mu, \alpha \beta }_{ab} =
 M_i^2\, \Big[-12 \left(m_a^2-m_b^2+M_i^2\right) \left(m_b^2+M_i^2\right) M_i^4
 \nonumber\\
&&   \qquad\,+\,12 \left(M_i^2-M_f^2\right)
   \left(m_a^2+m_b^2+3\, M_i^2\right) M_i^4
    \nonumber\\
&&   \qquad\,+\,6 \left(-m_a^2+m_b^2+M_i^2\right) \left(M_i^2-M_f^2\right)^3
 \nonumber\\
&&   \qquad \,-\,
   \left(M_f^2-M_i^2\right)^2 \Big(m_a^4-8 \left(m_b^2+M_i^2\right) m_a^2+7\, m_b^4
      \nonumber\\
&&   \qquad\,+\,31 \,M_i^4
   +22\, m_b^2\, M_i^2\Big)\Big]\,I_{ab}
 \nonumber\\
&&   \,+\, 12\, M_f^2
   \left(m_a^2-m_b^2+M_f^2\right) \left(m_b^2+M_f^2\right) M_i^4\,\bar I_{ab}
\nonumber\\
&&   \,+\,
   24 \,m_a^2\, M_f^2 \left(m_b^2+M_f^2\right) M_i^4
   \left(M_i^2-M_f^2\right) J_{aab}\,,
\end{eqnarray}
and
\begin{eqnarray}
&&96\,\frac{m_b^2}{M_f^2+ m_b^2}\,\frac{M_f^2\,M_i^2}{(p \cdot q)}\,
P^{(1)}_{\bar \alpha \bar \beta, \mu, \alpha \beta }\,B^{\bar \alpha \bar \beta, \mu, \alpha \beta }_{ab} =
M_f^2  \Big[24\, m_b^2 \left(M_i^2-m_a^2+m_b^2\right) M_i^4
      \nonumber\\
&&   \qquad \,-\,
   \left(M_i^2-M_f^2\right) \left(m_a^4+10 \,m_b^2 \,m_a^2-11\, m_b^4+M_i^4
   -2 \left(m_a^2+m_b^2\right) M_i^2\right) M_i^2
         \nonumber\\
&&   \qquad \,-\,
   \left(M_f^2-M_i^2\right)^2 \left(M_i^4-2 \left(m_a^2-5 \,m_b^2\right)
   M_i^2+\left(m_a^2-m_b^2\right)^2\right)\Big]\,I_{ab}
\nonumber\\
&&   \,-\, M_i^2 \,\Big[-\left(m_a^2+19 \,m_b^2\right)
   M_f^6-\left(m_a^4+10\, m_b^2 \,m_a^2-11\, m_b^4\right) M_f^4
         \nonumber\\
&&   \qquad \,+\,2\, M_f^8-3 \left(-m_a^4+6\, m_b^2 \,m_a^2-5 \,m_b^4
   +M_f^4-12 \,m_b^2\, M_f^2\right) M_i^2
   M_f^2
        \nonumber\\
&&   \qquad \, +\,\Big(M_f^4+\left(m_a^2+7\, m_b^2\right) M_f^2-2 \left(m_a^2-m_b^2\right)^2\Big)\, M_i^4\Big]\,\bar I_{ab}
\nonumber\\
&&   \,+\,   6 \,m_b^2\, M_f^2\, M_i^2 \left(M_f^2-M_i^2\right) \Big[-2 \,M_f^4+M_i^2\, M_f^2+M_i^4
         \nonumber\\
&&   \qquad \,+\,3\, m_a^2
   \left(M_f^2-M_i^2\right)+m_b^2 \left(M_f^2+7 \,M_i^2\right)\Big]\, \bar J_{abb}\,,
  \nonumber\\ \nonumber\\
&&48\,\frac{m_b^2}{m_b^2+ M_f^2}\,\frac{M_f^2}{(p \cdot q)}\,P^{(2)}_{\bar \alpha \bar \beta, \mu, \alpha \beta }\,
B^{\bar \alpha \bar \beta, \mu, \alpha \beta }_{ab} =
 \Big[12\, m_b^2 \left(-m_a^2+m_b^2+M_i^2\right) M_i^4
       \nonumber\\
&&   \qquad\,-\,\left(M_i^2-M_f^2\right) \left(M_i^4-2 \left(m_a^2-11
   m_b^2\right) M_i^2+\left(m_a^2-m_b^2\right)^2\right) M_i^2
          \nonumber\\
&&   \qquad\,+\,6\, m_b^2 \left(M_f^2-M_i^2\right)^2
   \left(m_a^2-m_b^2+M_i^2\right)\Big]\,I_{ab}
\nonumber\\
&&   \,+\,  \Big[  \Big(m_a^4-2 \left(m_b^2+M_f^2\right) m_a^2+\left(m_b^2-M_f^2\right)^2\Big)\, M_i^2
        \nonumber\\
&&   \qquad\,-\,M_f^2 \left(m_a^4-2 \left(7\, m_b^2+M_f^2\right) m_a^2+13\, m_b^4+M_f^4+10\, m_b^2\, M_f^2\right) \Big]
M_i^2\,\bar I_{ab}
\nonumber\\
&&   \,+\,    24 \,m_b^4\, M_f^2\, M_i^2  \left(M_f^2-M_i^2\right) \bar J_{abb}\,,
 \end{eqnarray}
and
\begin{eqnarray}
&& \frac{12\,M_i^2}{(p\cdot q)^3}\,
P^{(1)}_{\bar \alpha \bar \beta, \mu, \alpha \beta }\,\Bigg(A^{\bar \alpha \bar \beta, \mu, \alpha \beta }_{ab}
+\frac{m_b^2+M_f^2}{2}\,A^{\bar \alpha \bar \beta, \mu, \alpha \beta }_{\chi, ab} \Bigg)
\nonumber\\
\,= \, && \frac{12\,M_f^2}{(p\cdot q)^3}\,
P^{(2)}_{\bar \alpha \bar \beta, \mu, \alpha \beta }\,\Bigg(A^{\bar \alpha \bar \beta, \mu, \alpha \beta }_{ab}
+\frac{m_b^2+M_f^2}{2}\,A^{\bar \alpha \bar \beta, \mu, \alpha \beta }_{\chi, ab} \Bigg)
\nonumber\\
\,= \, && \frac{m_b^2}{m_b^2+M_f^2}\, \frac{12\,M_i^2}{(p\cdot q)^3}\,
P^{(1)}_{\bar \alpha \bar \beta, \mu, \alpha \beta }\,\Bigg(B^{\bar \alpha \bar \beta, \mu, \alpha \beta }_{ab}
+\frac{m_b^2+M_f^2}{2}\,B^{\bar \alpha \bar \beta, \mu, \alpha \beta }_{\chi, ab} \Bigg)
\nonumber\\
\,= \, && \frac{m_b^2}{m_b^2+M_f^2}\,\frac{12\,M_f^2}{(p\cdot q)^3}\,
P^{(2)}_{\bar \alpha \bar \beta, \mu, \alpha \beta }\,\Bigg(B^{\bar \alpha \bar \beta, \mu, \alpha \beta }_{ab}
+\frac{m_b^2+M_f^2}{2}\,B^{\bar \alpha \bar \beta, \mu, \alpha \beta }_{\chi, ab} \Bigg)
\nonumber\\
\,= \, &&
- \Big[M_i^4-2 \left(m_a^2-5\, m_b^2\right) M_i^2 +\left(m_a^2-m_b^2\right)^2\Big]\,I_{ab} \,,
\end{eqnarray}
and
\newpage

\begin{eqnarray}
&&192\,M_f^2\,M_i^2\,
P^{(1)}_{\bar \alpha \bar \beta, \mu, \alpha \beta }\,C^{\bar \alpha \bar \beta, \mu, \alpha \beta }_{abc} =
M_f^2 \,\Big[-12 \left(m_a^2-m_b^2\right)^2 M_i^6
\nonumber\\
&&   \qquad\,+\,12 \left(m^2_a-m^2_b\right)
   \left(m_a^2-m_b^2+M_i^2\right) \left(M_i^2-M_f^2\right) M_i^4
 \nonumber\\
&&   \qquad\, -\, \left(M_i^3-M_f^2\, M_i\right)^2 \Big(\left(m_a^2-m_c^2\right)^2-5\, M_i^4
 \nonumber\\
&&   \qquad\,+\,4  \left(4 \,m_a^2-3\, m_b^2+m_c^2\right) M_i^2\Big)
\nonumber\\
&&   \qquad\,  -\, \left(M_f^2-M_i^2\right)^3
   \Big(M_i^4-2 \left(m_a^2-5 \,m_c^2\right) M_i^2+\left(m_a^2-m_c^2\right)^2\Big)\Big]\, I_{ac}
\nonumber\\
&&\,+\,   M_i^2 \,\Big[2 \left(M_f^6+8\, M_i^2
   M_f^4-4\, M_i^4\, M_f^2+M_i^6\right) m_b^4 + 6\, m_a^4 \,M_f^4 \left(M_f^2+M_i^2\right)
  \nonumber\\
&&   \qquad\,  +\,\left(M_f^2-M_i^2\right) \Big(2\, M_f^6-21\, M_i^2\, M_f^4+7\, M_i^4 \,M_f^2
 \nonumber\\
&&   \qquad\,+\,m_c^2 \left(5\,
   M_f^4-9\, M_i^2\, M_f^2+4 \,M_i^4\right)\Big)\, m_b^2
    \nonumber\\
&&   \qquad\,-\,\left(m_c^2-M_f^2\right)
   \left(M_f^2-M_i^2\right)^2 \left(m_c^2 \left(M_f^2-2\, M_i^2\right)-M_f^2 \left(4\, M_f^2+M_i^2\right)\right)
  \nonumber\\
&&   \qquad\,  +\,3 \,m_a^2\, M_f^2
   \Big(\left(-3 \,M_f^4-6\, M_i^2\, M_f^2+M_i^4\right) m_b^2
   \nonumber\\
&&   \qquad\, +\,\left(M_f^2-M_i^2\right) \left(3 \,M_f^4-m_c^2\,
   M_f^2+\left(m_c^2+M_f^2\right) M_i^2\right)\Big)\Big]\, \bar I_{bc}
\nonumber\\
&&\,+\, 6\, M_f^2\, M_i^2 \Big[-2 \left(m_a^2-m_b^2\right)^3 M_i^4
\nonumber\\
&&   \qquad\,+\,\left(m_a^2-m_b^2\right)^2 \left(M_i^2-M_f^2\right) \left(3\, m_a^2-2\, m_b^2-m_c^2+3\, M_i^2\right)
   M_i^2
   \nonumber\\
&&   \qquad\,-\,\left(m^2_a-m^2_b\right) \left(M_f^2-M_i^2\right)^2 \Big(\left(5\, m_a^2-3\, m_b^2\right)
   M_i^2
    \nonumber\\
&&   \qquad\,+\,\left(m_a^2-m_b^2\right) \left(m_a^2-m_c^2\right)\Big)
\nonumber\\
&&   \qquad\, +\,\left(M_i^2-M_f^2\right)^3 \left(2\, m_a^4-m_b^2\, m_a^2-m_b^2
   \left(m_c^2+M_i^2\right)\right)\Big]\, J_{abc}\,,
\nonumber\\ \nonumber\\
&&96\,M_f^2\,
P^{(2)}_{\bar \alpha \bar \beta, \mu, \alpha \beta }\,C^{\bar \alpha \bar \beta, \mu, \alpha \beta }_{abc} =
   M_f^2 \,\Big[ 6 \left(m_a^2-m_b^2\right)^2 M_i^4
    \nonumber\\
&&   \qquad\,-\,3 \left(m^2_a-m^2_b\right) \left(m_a^2-m_c^2-M_i^2\right)
   \left(M_i^2-M_f^2\right) M_i^2
    \nonumber\\
&&   \qquad\,-\,3 \left(m_a^2-m_c^2-M_i^2\right) \left(M_i^2-M_f^2\right)^3
  \nonumber\\
&&   \qquad\,  -\,\left(M_f^2-M_i^2\right)^2 \Big(-2\,
   M_i^4+\left(m_a^2-6 \,m_b^2+m_c^2\right) M_i^2
    \nonumber\\
&&   \qquad\,  +\,\left(m_a^2-m_c^2\right)^2\Big)\Big]\, I_{ac}
\nonumber\\
&&\,+\,   M_i^2 \,\Big[\left(-2 \,M_f^4-5\, M_i^2
   M_f^2+M_i^4\right) m_b^4-6 \,m_a^4\, M_f^4
   \nonumber\\
&&   \qquad\,-\,\left(M_f^2-M_i^2\right) \left(8 \,M_f^4-5\, M_i^2\, M_f^2+m_c^2 \left(5 \,M_f^2-2\, M_i^2\right)\right)
   m_b^2
   \nonumber\\
&&   \qquad\,+\,\left(m_c^4+M_f^2\, m_c^2-2 \,M_f^4\right) \left(M_f^2-M_i^2\right)^2
 \nonumber\\
&&   \qquad\,+\,3\, m_a^2\, M_f^2 \left(\left(3\,
   M_f^2+M_i^2\right) m_b^2+\left(m_c^2+M_f^2\right) \left(M_f^2-M_i^2\right)\right)\Big]\,\bar I_{bc}
\nonumber\\
&&\,+\,6\, M_f^2\, M_i^2 \Big[ M_i^2
   \left(m_a^2-m_b^2\right)^3+\left(m^2_a-m^2_c\right) \left(M_f^2-M_i^2\right) \left(m_a^2-m_b^2\right)^2
   \nonumber\\
&&   \qquad\,+\,m_b^2
   \left(M_f^2-M_i^2\right)^3+\left(m^2_a-m^2_b\right) m_b^2  \left(M_f^2-M_i^2\right)^2\Big]\, J_{abc}\,,
\end{eqnarray}

and

\begin{eqnarray}
&& \frac{2}{(p\cdot q)^3}\,
P^{(1)}_{\bar \alpha \bar \beta, \mu, \alpha \beta }\,\Bigg(C^{\bar \alpha \bar \beta, \mu, \alpha \beta }_{abc}
 -C^{\bar \alpha \bar \beta, \mu, \alpha \beta }_{\chi, abc} \Bigg)
\nonumber\\
\,= \, && \frac{2}{(p\cdot q)^3}\,
P^{(2)}_{\bar \alpha \bar \beta, \mu, \alpha \beta }\,\Bigg(C^{\bar \alpha \bar \beta, \mu, \alpha \beta }_{abc}
-C^{\bar \alpha \bar \beta, \mu, \alpha \beta }_{\chi, abc} \Bigg)
=  \Big[-m_a^2+m_c^2+M_i^2\Big]\,I_{ac} \,,
\end{eqnarray}
and
\begin{eqnarray}
&&192\,M_f^2\,M_i^4\,
P^{(1)}_{\bar \alpha \bar \beta, \mu, \alpha \beta }\,\tilde C^{\bar \alpha \bar \beta, \mu, \alpha \beta }_{abc} =
M_f^2\, \Big[12 \,m_b^2\left(m_b^2-m_a^2\right)^2 M_i^8
 \nonumber\\
&&   \qquad \,-\,12\, m_b^2 \left(m_b^2-m_a^2\right) \left(-m_a^2+m_b^2-M_i^2\right)
   \left(M_i^2-M_f^2\right) M_i^6
    \nonumber\\
&&   \qquad \,+\,m_b^2 \left(M_f^2-M_i^2\right)^2 \Big(4 \left(m_a^2+m_c^2\right) M_i^2
    -5 \,M_i^4+\left(m_a^2-m_c^2\right)^2\Big)\, M_i^4
    \nonumber\\
&&   \qquad \,+\,\left(M_i^2-M_f^2\right)^3 \Big(3\, M_i^6-\left(7\, m_a^2-5 \,m_b^2+5\, m_c^2\right)
   M_i^4
 \nonumber\\
&&   \qquad \, +\, \left(5\, m_a^4+8 \,m_b^2 \,m_a^2+m_c^4-2 \left(3\, m_a^2+8\, m_b^2\right) m_c^2\right) M_i^2
 \nonumber\\
&&   \qquad \, -\, \left(m_a^2-m_c^2\right)^2 \left(m_a^2+m_b^2-m_c^2\right)\Big)\, M_i^2
-12 \left(m_a^2-m_b^2\right)^2 M_f^2\,M_i^6
    \nonumber\\
&&   \qquad \,-\,M_f^2 \Big(12 \left(m_b^2-m_a^2\right)
   \left(m_a^2+m_b^2-m_c^2\right) \left(M_i^2-M_f^2\right) M_i^4
   \nonumber\\
&&   \qquad \, -\, \left(M_f^2-M_i^2\right)^3 \Big(M_i^4-2
   \left(m_a^2-5 \,m_c^2\right) M_i^2+\left(m_a^2-m_c^2\right)^2\Big)
 \nonumber\\
&&   \qquad \,   +\, \left(M_i^3-M_f^2 M_i\right)^2
   \Big(m_a^4+\left(4 \,m_c^2-2\, M_i^2\right) m_a^2-5\, m_c^4+M_i^4
  \nonumber\\
&&   \qquad \, +\,4 \left(3\, m_b^2+m_c^2\right) M_i^2\Big)\Big)\,M_i^2
    \nonumber\\
&&   \qquad \,+\,\left(M_f^2-M_i^2\right)^4 \Big(-2 \,M_i^6+\left(5\, m_a^2+9\, m_c^2\right) M_i^4
  \nonumber\\
&&   \qquad \,  -\, 2 \left(2\, m_a^4-5\, m_c^2 \,m_a^2+3\,m_c^4\right) M_i^2+\left(m_a^2-m_c^2\right)^3\Big)\Big]\, I_{ac}
\nonumber\\
&&\,-\,M_i^4 \Big[2 \left(M_f^6+8\, M_i^2\, M_f^4-4 \,M_i^4\,M_f^2+M_i^6\right) m_b^6
+12\,m_b^4\,M_f^8
 \nonumber\\
&&   \qquad \, +\, \Big(3 \left(M_i^6-9\, M_i^2 \,M_f^4\right) M_f^2+m_c^2 \left(5\, M_f^2-4 \,M_i^2\right)
   \left(M_f^2-M_i^2\right)^2\Big)\, m_b^4
\nonumber\\
&&   \qquad \, -\,\left(M_f^2-M_i^2\right) \Big(12\, M_f^8+m_c^4
   \left(M_f^4-3\, M_i^2 \,M_f^2+2\, M_i^4\right)
     \nonumber\\
&&   \qquad \,-\,6\, M_i^2 \left(M_f^6+M_i^2\, M_f^4 \right)
     -3\, m_c^2 \left(4 \,M_f^6-13 \,M_i^2\, M_f^4+5\, M_i^4 \,M_f^2\right)\Big)\, m_b^2
\nonumber\\
&&   \qquad \, +\,6 \,m_a^4\, M_f^4
   \left(m_b^2-M_f^2\right) \left(M_f^2+M_i^2\right)
   \nonumber\\
&&   \qquad \,+\,\left(m_c^2-M_f^2\right) \left(M_f^2- M_i^2\right)^2 \left(2\, M_f^6-M_i^2\,
   M_f^4+m_c^2 \left(M_f^4+4 \,M_i^2\,M_f^2\right)\right)
   \nonumber\\
&&   \qquad \,+\,3\, m_a^2\, M_f^2 \Big(\left(-3 \,M_f^4-6\, M_i^2 M_f^2+M_i^4\right) m_b^4
  \nonumber\\
&&   \qquad \,-\,\Big(2
   \left(M_f^4-6\, M_i^2\, M_f^2+M_i^4\right) M_f^2+m_c^2 \left(M_f^2-M_i^2\right)^2\Big)\, m_b^2
   \nonumber\\
&&   \qquad \, +\, M_f^2 \left(M_f^2-M_i^2\right)
   \left(M_f^4-M_i^2 \,M_f^2+m_c^2 \left(M_f^2+3\, M_i^2\right)\right)\Big)\Big]\, \bar I_{bc}
\nonumber\\
&&\,+\, 6 \,M_f^2\, M_i^4 \Big[\Big(2    \left(m_a^2-m_b^2\right)^3 M_i^4
    -2\, m_a^2 \left(M_f^2-M_i^2\right)^4
    \nonumber\\
&&   \qquad \,+\,\left(m_a^2-m_b^2\right)^2 \left(-3\, m_a^2+2\, m_b^2+m_c^2-3 \,M_i^2\right)
   \left(M_i^2-M_f^2\right) M_i^2
    \nonumber\\
&&   \qquad \,-\,\left(M_i^2-M_f^2\right)^3 \left(\left(m_b^2+2\, m_c^2-2 \,M_i^2\right) m_a^2
   +m_b^2 \left(M_i^2-3\,m_c^2\right)\right)\Big)\, m_b^2
 \nonumber\\
&&   \qquad \,+\,\left(m^2_a-m^2_b\right)
   \left(M_f^2-M_i^2\right)^2 \Big( m_b^2 \left(m_a^2-m_b^2\right)   \left(m_a^2-m_c^2\right)
 \nonumber\\
&&   \qquad \,+\,m_b^2 \left(3\, m_a^2-m_b^2\right) M_i^2-\left(m_a^2+3\, m_b^2\right) M_f^2\,M_i^2
 \nonumber\\
&&   \qquad \,-\,\left(m_a^2+m_b^2\right) \left(m_a^2-m_c^2\right)M_f^2 \Big)
 \nonumber\\
&&   \qquad \, +\,M_f^2 \,\Big(-2 \left(m_a^2-m_b^2\right)^3 M_i^4
  + m_b^2 \left(m_a^2+m_c^2+M_i^2\right)
   \left(M_i^2-M_f^2\right)^3
  \nonumber\\
&&   \qquad \, +\,\left(m_a^2-m_b^2\right)^2 \left(3\, m_a^2+2\, m_b^2-3\, m_c^2+M_i^2\right) \left(M_i^2-M_f^2\right) M_i^2
   \Big)\Big]\,J_{abc}\,
\nonumber\\ \nonumber\\
&&192\,M_f^2\,M_i^2\,
P^{(2)}_{\bar \alpha \bar \beta, \mu, \alpha \beta }\,\tilde C^{\bar \alpha \bar \beta, \mu, \alpha \beta }_{abc} =
M_f^2\,  \Big[12 \left(m_a^2-m_b^2\right)^2 M_i^8
 \nonumber\\
&&   \qquad \,-\,6 \left(m^2_a-m^2_b\right)  \left(3\,
   m_a^2+m_c^2+M_i^2\right) \left(M_i^2-M_f^2\right) M_i^6
    \nonumber\\
&&   \qquad \,-\,6\, m_c^2 \left(-m_a^2+m_c^2+M_i^2\right)
   \left(M_i^2-M_f^2\right)^3 M_i^2
    \nonumber\\
&&   \qquad \,-\,2\left(M_i^2-M_f^2 \right)^2 \Big(M_i^4-\left(5 \,m_a^2-m_c^2-2\,m_b^2\right) M_i^2
  \nonumber\\
&&   \qquad \,-\,2 \left(m_a^4+m_c^4+\left(m_a^2-3\, m_b^2\right) m_c^2\right)
-m_b^2 \left(m_a^2-6 \,m_b^2+m_c^2\right)\Big)\, M_i^4
      \nonumber\\
&&   \qquad \,+\,2 \,m_b^2 \left(M_f^2-M_i^2\right)^2 \left(m_a^2-m_c^2\right)^2\, M_i^2
- 12 \,m_b^2\left(m_b^2-m_a^2 \right)^2 M_i^6
   \nonumber\\
&&   \qquad \,+\,6 \left(m^2_a-m^2_b\right) m_b^2 \left(m_a^2-m_c^2-M_i^2\right) \left(M_i^2-M_f^2\right) M_i^4
     \nonumber\\
&&   \qquad \,-\,2 \left(M_f^2-M_i^2\right)^4 \Big(-M_i^4-\left(m_a^2+m_c^2\right) M_i^2+2
   \left(m_a^2-m_c^2\right)^2\Big)
     \nonumber\\
&&   \qquad \,-\,\left(M_i^2-M_f^2\right)^3 \Big(-3\, M_i^6+\left(7\, m_a^2+6\, m_b^2+5\, m_c^2\right)
   M_i^4
   \nonumber\\
&&   \qquad \,-\,\left(m_a^2-m_c^2\right) \left(5\, m_a^2+6\, m_b^2-m_c^2\right) M_i^2
   +\left(m_a^2-m_c^2\right)^3\Big)\Big]\,I_{ac}
\nonumber\\
&& \,+\, 2\,M_i^4 \Big[\left(2 \,M_f^4+5\, M_i^2 \,M_f^2-M_i^4\right) m_b^6 +6\, m_a^4\,
   M_f^4 \left(m_b^2-M_f^2\right)
 \nonumber\\
&&   \qquad \,  +\,\Big(\left(5\, M_f^4-7\, M_i^2\, M_f^2+2\, M_i^4\right) m_c^2+3\, M_f^2
   \Big(3 \,M_f^4-6 \,M_i^2\, M_f^2
   \nonumber\\
&&   \qquad \,+\,M_i^4\Big)\Big)\, m_b^4
 +\left(M_f^2-M_i^2\right) \Big(-M_f^2 \,m_c^4+6\, M_f^4 \,m_c^2+\Big(m_c^4-3\,
   M_f^2\, m_c^2
   \nonumber\\
&&   \qquad \,+\,3 \,M_f^4\Big) M_i^2\Big)\, m_b^2
   +\left(-2\, m_c^4+M_f^2\, m_c^2+M_f^4\right)\left(M_f^3-M_f M_i^2\right)^2
   \nonumber\\
&&   \qquad \,-\,3 \,m_a^2\, M_f^2 \Big(\left(3\, M_f^2+M_i^2\right) m_b^4
   +\left(m_c^2 \,M_f^2-\left(m_c^2+4\,M_f^2\right) M_i^2\right) m_b^2
 \nonumber\\
&&   \qquad \,  +\,M_f^2 \left(m_c^2+M_f^2\right) \left(M_f^2-M_i^2\right)\Big)\Big]\,\bar I_{bc}
\nonumber\\
&&\,+\,12\, M_f^2\, M_i^4\,
   \Big[\left(M_i^2-M_f^2\right)^3 m_b^4+\left(m_b^2-m_a^2\right) \left(M_f^2-M_i^2\right)^2 m_b^4
 \nonumber\\
&&   \qquad \, +\,\left(m_b^2-m_a^2\right)^3 M_i^2 \,m_b^2+\left(m_b^3-m_a^2 \,m_b\right)^2 \left(m^2_a-m^2_c\right)
   \left(M_i^2-M_f^2\right)
    \nonumber\\
&&   \qquad \,+\,\left(m_a^2-m_b^2+M_f^2-M_i^2\right) \left(\left(m_a^2-m_b \,m_c\right) M_f^2
   -m_b \left(m_b-m_c\right)
   M_i^2\right)
 \nonumber\\
&&   \qquad \,\times
   \left(\left(m_a^2+m_b\, m_c\right) M_f^2-m_b \left(m_b+m_c\right) M_i^2\right)\Big]\,J_{abc}\,,
\end{eqnarray}
and
\begin{eqnarray}
&& \frac{24\,M_i^4}{(p\cdot q)^3}\,
P^{(1)}_{\bar \alpha \bar \beta, \mu, \alpha \beta }\,\Bigg(\tilde C^{\bar \alpha \bar \beta, \mu, \alpha \beta }_{abc}
 -\tilde C^{\bar \alpha \bar \beta, \mu, \alpha \beta }_{\chi, abc} \Bigg)
=    \Big[ M_i^8-2 \left(m_a^2-2 m_c^2\right) M_i^6
\nonumber\\
&&\qquad \,+\,\left(m_a^4+4\, m_c^2\, m_a^2-5\, m_c^4-\left(m_a^2+29\, m_c^2\right) M_f^2\right) M_i^4
\nonumber\\
&&\qquad \, +\,2 \left(m_a^4-3\, m_c^2\, m_a^2+2\, m_c^4\right) M_f^2\, M_i^2-\left(m_a^2-m_c^2\right)^3 M_f^2
\Big]\,I_{ac} \,,
\nonumber\\ \nonumber\\
    && \frac{24\,M_i^2}{(p\cdot q)^3}\,
P^{(2)}_{\bar \alpha \bar \beta, \mu, \alpha \beta }\,\Bigg(\tilde C^{\bar \alpha \bar \beta, \mu, \alpha \beta }_{abc}
-\tilde C^{\bar \alpha \bar \beta, \mu, \alpha \beta }_{\chi, abc} \Bigg)
=  \Big[-\left(m_a^2+23\, m_c^2+2\, M_f^2\right) M_i^4
\nonumber\\
&&\qquad \,+\,3 \,M_i^6-\left(m_a^4-6\, m_c^2 \,m_a^2+5 \,m_c^4 +2 \left(m_a^2+m_c^2\right)
   M_f^2\right) M_i^2
   \nonumber\\
&&\qquad \,-\,\left(m_a^2-m_c^2\right)^2 \left(m_a^2-m_c^2-4\, M_f^2\right)\Big]\,I_{ac} \,,
\end{eqnarray}

and
\begin{eqnarray}
&&48\,\frac{M_f^2\,M_i^2\,m_b^2}{(p \cdot q)^2}\,P^{(1)}_{\bar \alpha \bar \beta, \mu, \alpha \beta }\,
D^{\bar \alpha \bar \beta, \mu, \alpha \beta }_{ab} =
M_f^2 \,\Big[\left(6\, m_b^2-11\, M_f^2\right) M_i^6
  \nonumber\\
&&   \qquad \,+\,\left(6\, m_b^4-4\, M_f^2 \,m_b^2+11\, M_f^4
   +m_a^2 \left(7\, M_f^2-6\,m_b^2\right)\right) M_i^4
     \nonumber\\
&&   \qquad \,-\,M_f^2 \left(-4\, m_a^4+\left(4\, m_b^2+7\, M_f^2\right) m_a^2
   -8\, m_b^4+m_b^2\, M_f^2\right)
   M_i^2
     \nonumber\\
&&   \qquad \,+\,\left(m_a^2-m_b^2\right)^2 M_f^2 \left(m_b^2-4 M_f^2\right)\Big]\, I_{ab}
\nonumber\\
&&\,+\,  M_i^2 \Big[-3\, M_f^8+\left(3\, m_a^2+m_b^2\right)
   M_f^6-m_b^2 \left(5 \,m_a^2+11\, m_b^2\right) M_f^4
   \nonumber\\
&&   \qquad \,+\,m_b^2 \left(m_a^2-m_b^2\right)^2 M_f^2
   -\Big(-3\, M_f^6+\left(3 \,m_a^2+2\,m_b^2\right) M_f^4
  \nonumber\\
&&   \qquad \, -\,m_b^2 \left(7\, m_a^2+m_b^2\right) M_f^2+2 \left(m_b^3-m_a^2 \,m_b\right)^2\Big)\, M_i^2 \Big]\,
   \bar I_{ab}
\nonumber\\
&&\,+\, 6 \,m_b^2\, M_f^2\, M_i^2
\left(M_f^2-M_i^2\right) \Big[ m_b^4-m_a^2\, m_b^2+M_f^4+\left(m_b^2-M_f^2\right) M_i^2\Big]\,\bar J_{abb}\,,
\nonumber\\ \nonumber\\
&&48\,\frac{M_f^2\,M_i^2\,m_b^2}{(p \cdot q)^2}\,P^{(2)}_{\bar \alpha \bar \beta, \mu, \alpha \beta }\,
D^{\bar \alpha \bar \beta, \mu, \alpha \beta }_{ab} =
  2\, M_f^2 \,\Big[\left(3 \,m_a^2+4\, m_b^2+2\, M_f^2\right) M_i^6
\nonumber\\
&&\qquad\,-\,\left(-M_f^4+4 \left(m_a^2+m_b^2\right) M_f^2+2\, m_b^2
   \left(m_a^2+m_b^2\right)\right) M_i^4
 \nonumber\\
&&\qquad\,+\,\Big(\left(m_a^2+m_b^2\right) M_f^4+2 \left(m_a^2-m_b^2\right)^2
   M_f^2+\left(m_b^3-m_a^2\, m_b\right)^2\Big)\, M_i^2
 \nonumber\\
&&\qquad\,-\,3\, M_i^8 -2 \left(m_a^2-m_b^2\right)^2 M_f^4\Big]\,I_{ab}
\nonumber\\
&&\,-\,2\, m_b^2
   \left(\left(m_a-m_b\right)^2-M_f^2\right) \left(\left(m_a+m_b\right)^2-M_f^2\right) M_i^4 \,\bar I_{ab}\,,
\end{eqnarray}
and

\begin{eqnarray}
&&24\,\frac{M_f^2\,M_i^2\,m_b^2}{(p \cdot q)^2}\,P^{(1)}_{\bar \alpha \bar \beta, \mu, \alpha \beta }\,
 D^{\bar \alpha \bar \beta, \mu, \alpha \beta }_{\chi, ab} =
M_f^2 \Big[6 \,M_i^6-\left(6\, m_a^2-6\, m_b^2+7\, M_f^2\right) M_i^4
\nonumber\\
&&   \qquad \,+\,8 \left(m_a^2-2 \,m_b^2\right) M_f^2
   M_i^2-\left(m_a^2-m_b^2\right)^2 M_f^2\Big]\, I_{ab}
\nonumber\\
&& \,+\, \Big[\Big(5\, M_f^4-\left(7 \,m_a^2+m_b^2\right) M_f^2+2
   \left(m_a^2-m_b^2\right)^2\Big)\, M_i^4
\nonumber\\
&&   \qquad \, -\,M_f^2 \Big(4 \,M_f^4-\left(5 \,m_a^2+11\, m_b^2\right)
   M_f^2+\left(m_a^2-m_b^2\right)^2\Big)\, M_i^2 \Big]\,\bar I_{ab}
\nonumber\\
&& \,+\, 6\, m_b^2\, M_f^2\, M_i^2 \left(-m_a^2+m_b^2+M_i^2\right)
   \left(M_i^2-M_f^2\right) \bar J_{abb}\,,
\nonumber\\ \nonumber\\
&&12\,\frac{M_f^2\,m_b^2}{(p \cdot q)^2}\,P^{(2)}_{\bar \alpha \bar \beta, \mu, \alpha \beta }\,
 D^{\bar \alpha \bar \beta, \mu, \alpha \beta }_{\chi, ab} =
-M_f^2 \,\Big[2 \left(2\, m_a^2-4\, m_b^2+3\, M_f^2\right) M_i^2
\nonumber\\
&&   \qquad \,-\,5\, M_i^4+ \left(m_a^2-m_b^2\right) \left(m_a^2-m_b^2-6
   M_f^2\right)\Big]\,I_{ab}
\nonumber\\
&&\,+\,  \left(\left(m_a-m_b\right)^2-M_f^2\right) \left(\left(m_a+m_b\right)^2-M_f^2\right) M_i^2 \,\bar I_{ab}\,,
\end{eqnarray}

and

\begin{eqnarray}
&&48\,\frac{M_f^2\,M_i^4}{(p \cdot q)^2}\,P^{(1)}_{\bar \alpha \bar \beta, \mu, \alpha \beta }\,
\tilde D^{\bar \alpha \bar \beta, \mu, \alpha \beta }_{ab} =
M_f^2 \,\Big[-M_i^{10}+\left(2\, m_a^2-4 \,m_b^2+M_f^2\right) M_i^8
\nonumber\\
&&   \qquad \, -\,\left(m_a^4+\left(4 \,m_b^2+M_f^2\right) m_a^2-5\,
   m_b^4-M_f^4-10\, m_b^2 \,M_f^2\right) M_i^6
       \nonumber\\
&&   \qquad \, -\,M_f^2 \left(m_a^4+\left(3\, M_f^2-8\, m_b^2\right) m_a^2-m_b^2 \left(m_b^2+5\,
   M_f^2\right)\right) M_i^4
       \nonumber\\
&&   \qquad \, +\,\left(m_a^2-m_b^2\right) M_f^2 \left(m_a^4-\left(m_b^2-3 \,M_f^2\right) m_a^2-5\, m_b^2 \,M_f^2\right)
   M_i^2
       \nonumber\\
&&   \qquad \, -\,\left(m_a^2-m_b^2\right)^3 M_f^4\Big]\,I_{ab}
\nonumber\\
&&   \,+\,    \left(m_b^2+M_f^2\right) M_i^4 \,\Big[-2\, M_f^6+\left(m_a^2-17\, m_b^2\right)
   M_f^4+\left(m_a^2-m_b^2\right)^2 M_f^2
       \nonumber\\
&&   \qquad \, -\,\Big(-M_f^4-\left(m_a^2+7 \,m_b^2\right) M_f^2
   +2 \left(m_a^2-m_b^2\right)^2\Big)
   M_i^2\Big]\,\bar I_{ab}
\nonumber\\
&&   \,+\,6 \,m_b^2\, M_f^2 \left(m_b^2+M_f^2\right) M_i^4 \left(M_f^2-M_i^2\right) \Big[-m_a^2+m_b^2+2\,
   M_f^2-M_i^2\Big]\, \bar J_{abb} \,,
 \nonumber\\ \\
&&48\,\frac{M_f^2\,M_i^2}{(p \cdot q)^2}\,P^{(2)}_{\bar \alpha \bar \beta, \mu, \alpha \beta }\,
\tilde D^{\bar \alpha \bar \beta, \mu, \alpha \beta }_{ab} =
M_f^2\, \Big(-M_i^4+\left(m_a^2+m_b^2+3 \,M_f^2\right) M_i^2
          \nonumber\\
&&   \qquad\,-\,\left(m_a^2-m_b^2\right) M_f^2\Big) \,\Big[m_a^4-2\,
   \left(m_b^2+M_i^2\right) m_a^2+\left(m_b^2-M_i^2\right)^2\Big]\,I_{ab}
\nonumber\\
&&   \, +\,\Big[-2 \left(\left(m_a-m_b\right)^2-M_f^2\right)
   \left(\left(m_a+m_b\right)^2-M_f^2\right) \left(m_b^2+M_f^2\right) M_i^4 \,\Big]\, \bar I_{ab}\,,
   \nonumber
\end{eqnarray}
and
\begin{eqnarray}
&& \frac{24\,M_i^4}{(p\cdot q)^3}\,
P^{(1)}_{\bar \alpha \bar \beta, \mu, \alpha \beta }\,\Bigg(\tilde D^{\bar \alpha \bar \beta, \mu, \alpha \beta }_{ab}
+\frac{m_b^2+M_f^2}{2}\,\tilde D^{\bar \alpha \bar \beta, \mu, \alpha \beta }_{\chi, ab} \Bigg)
\nonumber\\
&& \,=\,\Big[ -M_i^8+2 \left(m_a^2-2 m_b^2\right) M_i^6-\left(m_a^2+5\, m_b^2\right)
\left(m_a^2-m_b^2-M_f^2\right) M_i^4
\nonumber\\
&& \qquad \,-\,2 \left(m_a^4-3\,m_b^2\, m_a^2+2\, m_b^4\right) M_f^2\, M_i^2
+\left(m_a^2-m_b^2\right)^3 M_f^2 \Big]\,I_{ab}\,,
\nonumber\\ \\
&& \frac{24\,M_i^4}{(p\cdot q)^3}\,
P^{(2)}_{\bar \alpha \bar \beta, \mu, \alpha \beta }\,\Bigg(\tilde D^{\bar \alpha \bar \beta, \mu, \alpha \beta }_{ab}
+\frac{m_b^2+M_f^2}{2}\,\tilde D^{\bar \alpha \bar \beta, \mu, \alpha \beta }_{\chi, ab} \Bigg)
\nonumber\\
&& \,=\,M_i^2\left((m_a-m_b)^2-M^2_i\right) \left((m_a+m_b)^2-M^2_i\right)
   \left(m_a^2-m_b^2-M_i^2\right)\, I_{ab}\,, \nonumber
\end{eqnarray}

and
\begin{eqnarray}
&&\frac{48}{(p \cdot q)}\, P^{(1)}_{\bar \alpha \bar \beta, \mu,
\alpha \beta }\,E^{\bar \alpha \bar \beta, \mu, \alpha \beta
}_{abc} = 3\, m_c^2\, M_f^2 \Big[M_i^4-\left(m_a^2-2\,
m_b^2+m_c^2+M_f^2\right) M_i^2
\nonumber\\
&& \qquad \,+\,\left(m_a^2-m_c^2\right)    M_f^2\Big]\,I_{ac}
\nonumber\\
&& \,+\Big[-\left(M_f^2-M_i^2\right)^2 m_b^4+ 2\,m_b^2\left(m_a^2+M_f^2\right)
\left(M_f^2-M_i^2\right)^2
\nonumber\\
&& \qquad \,-\,3\,m_b^2\,m_c^2\, M_f^2 \left(M_f^2+M_i^2\right)
   +6\, m_c^4\, M_f^4-\left(m_a^2-M_f^2\right)^2 \left(M_f^2-M_i^2\right)^2
   \nonumber\\
&& \qquad \,+\,3\, m_c^2\, M_f^2
   \left(M_f^2-m_a^2\right) \left(M_f^2-M_i^2\right) \Big]\,\bar I_{ab}
\nonumber\\
&& \,-\,6\, m_c^2\, M_f^2 \,\Big[\left(m_b^2-m_c^2\right) \left(M_f^2-M_i^2\right)
   m_a^2
   \nonumber\\
&& \qquad \,+\,\left(m^2_c \,M^2_f-m^2_b\, M^2_i\right) \left(-m_b^2+m_c^2+M_f^2-M_i^2\right)\Big]\, \bar J_{abc} \,,
\nonumber\\ \nonumber\\
&&\frac{48\,M_f^2}{(p \cdot q)}\,
P^{(2)}_{\bar \alpha \bar \beta, \mu, \alpha \beta }\,E^{\bar \alpha \bar \beta, \mu, \alpha \beta }_{abc} =
-3\, m_c^2 \,M_f^2 \Big[\left(M_f^4-4 \,M_i^2\, M_f^2-M_i^4\right) m_c^2
\nonumber\\
&& \qquad\,-\,\left(M_f^2-M_i^2\right)^2 \left(m_a^2+M_i^2\right)+2\,
   m_b^2\, M_i^2 \left(M_f^2+M_i^2\right)\Big]\,I_{ac}
\nonumber\\
&&\,- M_i^2 \Big[  \left(M_f^2-M_i^2\right)^2 m_b^4
-2\,\Big(6\, m_c^2\,M_f^4+\left(m_a^2+M_f^2\right) \left(M_f^2-M_i^2\right)^2\Big)\, m_b^2
   \nonumber\\
&& \qquad\,+12 \,m_c^4\, M_f^4+
   \left(m_a^2-M_f^2\right)^2 \left(M_f^2-M_i^2\right)^2\Big]\, \bar I_{ab}
\nonumber\\
&& \,+\, 6\, m_c^2 \left(m_c^2-m_b^2\right) M_f^2\, M_i^2
   \Big[M_f^4-\left(m_a^2+m_b^2-2\, m_c^2\right) M_f^2
\nonumber\\
&& \qquad\,-\,\left(-m_a^2+m_b^2+M_f^2\right) M_i^2\Big] \bar J_{abc}\,,
\end{eqnarray}
and

\begin{eqnarray}
&& \frac{12\,M_i^2}{(p\cdot q)^3}\,
P^{(1)}_{\bar \alpha \bar \beta, \mu, \alpha \beta }\,\Bigg(E^{\bar \alpha \bar \beta, \mu, \alpha \beta }_{ab}
-\bar E^{\bar \alpha \bar \beta, \mu, \alpha \beta }_{ab} \Bigg)
\nonumber\\
\,= \, && \frac{12\,M_f^2}{(p\cdot q)^3}\,
P^{(2)}_{\bar \alpha \bar \beta, \mu, \alpha \beta }\,\Bigg(E^{\bar \alpha \bar \beta, \mu, \alpha \beta }_{ab}
-\bar E^{\bar \alpha \bar \beta, \mu, \alpha \beta }_{ab} \Bigg)
\nonumber\\
\nonumber\\
\,= \, && -  M_i^2 \Big[m_a^4-2 \left(m_b^2+M_f^2\right) m_a^2+\left(m_b^2-M_f^2\right)^2\Big]
\,\bar I_{ab} \,,
\end{eqnarray}
and

\begin{eqnarray}
&&\frac{12\,M_f^2}{(p \cdot q)^3}\, P^{(1)}_{\bar \alpha \bar
\beta, \mu, \alpha \beta }\,F^{\bar \alpha \bar \beta, \mu, \alpha
\beta }_{ab} = \frac{12\,M_f^2}{(p \cdot q)^3}\, P^{(2)}_{\bar
\alpha \bar \beta, \mu, \alpha \beta }\,F^{\bar \alpha \bar \beta,
\mu, \alpha \beta }_{ab}
\nonumber\\
&& \qquad\,= \,\Big[ m_a^4
+\left(4\, m_b^2-2\, M_i^2\right) m_a^2-5\, m_b^4+M_i^4+4\, m_b^2 \,M_i^2 \Big]\,I_{ab}\,,
\nonumber\\
&&\frac{M_f^2}{(p \cdot q)^3}\,
P^{(1)}_{\bar \alpha \bar \beta, \mu, \alpha \beta }\,F^{\bar \alpha \bar \beta, \mu, \alpha \beta }_{\chi,ab} =
\frac{M_f^2}{(p \cdot q)^3}\,
P^{(2)}_{\bar \alpha \bar \beta, \mu, \alpha \beta }\,F^{\bar \alpha \bar \beta, \mu, \alpha \beta }_{\chi,ab}
\nonumber\\
&& \qquad\,= \,\Big[-m_a^2+m_b^2+M_i^2
\Big]\,I_{ab}\,.
\end{eqnarray}

\newpage

\section*{Appendix E}
\label{appendix:D}

In the following we present the radiative decay amplitudes not given in section \ref{sec:rad-decay}.
We establish
\begin{eqnarray}
\lefteqn{i\,M^{\mu, \alpha \beta}_{b_1^0 \to \, \gamma \,\eta}
=-\sqrt{3}\,\frac{e\,h_P\,m_V}{4\,f^2}\,h^{(b_1)}_{K\,\bar K^*}\,\Big\{A^{\mu, \alpha \beta}_{K K^*}(\bar p,p)
-\,\hat B^{\mu, \alpha \beta}_{K K^*,0}(\bar p,p)\Big\}   }
\nonumber\\
&& {} - \sqrt{3}\,\frac{h_A}{96\,f^2\,m_V}\,\eAcptn
\,h^{(b_1)}_{K\,\bar K^*}\,C^{\mu, \alpha \beta}_{K K^* K^*}(\bar p,p)
\nonumber\\
&& {} + \sqrt{3}\,\frac{\eA \,h_A}{48\,f^2\,m_V}\,
\Big\{h^{(b_1)}_{\pi\,\omega}\,C^{\mu, \alpha \beta}_{\pi \omega \omega}(\bar p,p)
+ \frac{1}{\sqrt{3}}\, h^{(b_1)}_{\eta\,\rho}\,C^{\mu, \alpha \beta}_{\eta \rho \rho}(\bar p,p) \Big\}
\nonumber\\
&& {} - \sqrt{3}\,\frac{h_A}{96\,f^2\,m_V}\,\eAcptnbA
\,h^{(b_1)}_{K\,\bar K^*}\,\bar C^{\mu, \alpha \beta}_{K K^* K^*}(\bar p,p)
\nonumber\\
&& {} + \sqrt{3}\,\frac{\eAbA \,h_A}{48\,f^2\,m_V}\,
\Big\{ h^{(b_1)}_{\pi\,\omega}\,\bar C^{\mu, \alpha \beta}_{\pi \omega \omega}(\bar p,p)
+ \frac{1}{\sqrt{3}}\, h^{(b_1)}_{\eta\,\rho}\,\bar C^{\mu, \alpha \beta}_{\eta \rho \rho}(\bar p,p) \Big\}
\nonumber\\
&& {} - \sqrt{3}\,\frac{b_A}{24\,f^2\,m_V}\,\Big(4\,m_K^2 -3\,m_\pi^2 \Big)\,\eAcptn
\,h^{(b_1)}_{K\,\bar K^*}\,\Cchi^{\mu, \alpha \beta}_{K K^* K^*}(\bar p,p)
\nonumber\\
&& {} + \sqrt{3}\,\frac{\eA \,b_A}{12\,f^2\,m_V}\, m_\pi^2 \,\Big\{ h^{(b_1)}_{\pi\,\omega}\,\Cchi^{\mu, \alpha \beta}_{\pi \omega \omega}(\bar p,p)
+ \frac{1}{\sqrt{3}}\, h^{(b_1)}_{\eta\,\rho}\,\Cchi^{\mu, \alpha \beta}_{\eta \rho \rho}(\bar p,p) \Big\}
\nonumber\\
&& {} - \sqrt{3}\,\frac{b_A}{24\,f^2\,m_V}\,\Big(4\,m_K^2 -3\,m_\pi^2\Big)\,\eAcptnbA
\,h^{(b_1)}_{K\,\bar K^*}\,\barCchi^{\mu, \alpha \beta}_{K K^* K^*}(\bar p,p)
\nonumber\\
&& {} + \sqrt{3}\,\frac{\eAbA \,b_A}{12\,f^2\,m_V}\,m_\pi^2 \, \Big\{h^{(b_1)}_{\pi\,\omega}\,\barCchi^{\mu, \alpha \beta}_{\pi \omega \omega}(\bar p,p)
+ \frac{1}{\sqrt{3}}\, h^{(b_1)}_{\eta\,\rho}\,\barCchi^{\mu, \alpha \beta}_{\eta \rho \rho}(\bar p,p) \Big\}
\nonumber\\
&& {} - \sqrt{3}\,\frac{e_V\,m_V}{32\,f^2}\, h^{(b_1)}_{K\,\bar K^*}\, E^{\mu, \alpha \beta}_{K K^*}(\bar p,p)
 + \frac{e_V}{16\,f^2\,m_V}\,\sqrt{3}\,h^{(b_1)}_{K\,\bar K^*}\, F^{\mu, \alpha \beta}_{K K^*}(\bar p,p)
\nonumber\\
&& {} -\sqrt{3}\,\frac{e_V\, g_D}{12\,f^2\,m_V}  \,
\Big\{h^{(b_1)}_{\pi\,\omega}\, G^{\mu, \alpha \beta}_{\pi \omega}(\bar p,p)
+\frac{1}{\sqrt{3}} \, h^{(b_1)}_{\eta\,\rho}\, G^{\mu, \alpha \beta}_{\eta \rho}(\bar p,p) \Big\}
\nonumber\\
&& {} -\sqrt{3}\,\frac{e_V}{48\,f^2\,m_V} \, \Big(g_D - 3 \, g_F\Big) \,
h^{(b_1)}_{K\,\bar K^*}\, G^{\mu, \alpha \beta}_{K K^*}(\bar p,p)
\nonumber\\
&& {} -\sqrt{3}\,\frac{e_V\,b_D}{24\,f^2\,m_V}\,m_\pi^2 \,\Big\{
h^{(b_1)}_{\pi\,\omega}\, E^{\mu, \alpha \beta}_{\pi \omega}(\bar p,p)
 + \frac{1}{\sqrt{3}} \,h^{(b_1)}_{\eta\,\rho}\, E^{\mu, \alpha \beta}_{\eta \rho}(\bar p,p) \Big\}
\nonumber\\
&& {} +\sqrt{3}\,\frac{e_V\,b_D}{96\,f^2\,m_V}\,(5\,m_K^2 - 3\,m_\pi^2) \, h^{(b_1)}_{K\,\bar K^*}\, E^{\mu, \alpha \beta}_{K K^*}(\bar p,p)
\nonumber\\
&& {} +\sqrt{3} \, \frac{ h_A \, b_A \, e_V}{48 \, f^2 \, m_V} \, m_\pi^2 \, \Big\{
h^{(b_1)}_{\pi\,\omega}\, H^{\mu, \alpha \beta}_{\pi \rho \omega}(\bar p,p)
+ \frac{1}{\sqrt{3}} \, h^{(b_1)}_{\eta\,\rho}\,
H^{\mu, \alpha \beta}_{\eta \rho \rho}(\bar p,p)
\nonumber\\
&& \hspace{7em} {}+ h^{(b_1)}_{K\,\bar K^*}\, H^{\mu, \alpha \beta}_{K \rho K^*}(\bar p,p) \Big\}
\nonumber\\
&& {} + \sqrt{3} \, \frac{(b_A)^2 \, e_V}{12 \, f^2 \, m_V} \, m_\pi^2 \, \Big\{
m_\pi^2 \, h^{(b_1)}_{\pi\,\omega}\,
   \bar H^{\mu, \alpha \beta}_{\pi \rho \omega}(\bar p,p)
+ \frac{1}{\sqrt{3}} \, m_\pi^2 \, h^{(b_1)}_{\eta\,\rho}\,
   \bar H^{\mu, \alpha \beta}_{\eta \rho \rho}(\bar p,p)
\nonumber\\
&& \hspace{7em} {}
+ m_K^2 \, h^{(b_1)}_{K\,\bar K^*}\,
   \bar H^{\mu, \alpha \beta}_{K \rho K^*}(\bar p,p) \Big\}\,,
\label{b1:gamma:eta}
\end{eqnarray}
and

\begin{eqnarray}
\lefteqn{M^{\bar \alpha \bar \beta, \mu, \alpha \beta}_{b_1^+ \to \, \gamma \,\rho^+}=
-\frac{e\,h_A}{8\,f}\,h^{(b_1)}_{K\,\bar K^*}\,\Big\{ \hat A^{\bar \alpha \bar \beta, \mu, \alpha \beta}_{K K^*,0}(\bar p,p)
 +\hat B^{\bar \alpha \bar \beta, \mu, \alpha \beta}_{K K^*,0}(\bar p,p)\Big\}  }
\nonumber\\
&& {} -\frac{e\,b_A}{2\,f}\,m_K^2 \,h^{(b_1)}_{K\,\bar K^*}\,\Big\{ \bar A^{\bar \alpha \bar \beta, \mu, \alpha \beta}_{K K^*,0}(\bar p,p)
 +\bar B^{\bar \alpha \bar \beta, \mu, \alpha \beta}_{K K^*,0}(\bar p,p)\Big\}
\nonumber\\
&& {}-\frac{e\,h_A}{4\,f}\,\Big\{
h^{(b_1)}_{\pi\,\omega}\, \hat A^{\bar \alpha \bar \beta, \mu, \alpha \beta}_{\pi \omega,0}(\bar p,p)
 +\frac{1}{\sqrt{3}}\,h^{(b_1)}_{\eta\,\rho}\,\hat B^{\bar \alpha \bar \beta, \mu, \alpha \beta}_{\eta \rho,0}(\bar p,p)\Big\}
\nonumber\\
&& {}-\frac{e\,b_A}{f}\,m_\pi^2 \,\Big\{
h^{(b_1)}_{\pi\,\omega}\, \bar A^{\bar \alpha \bar \beta, \mu, \alpha \beta}_{\pi \omega,0}(\bar p,p)
 +\frac{1}{\sqrt{3}}\,h^{(b_1)}_{\eta\,\rho}\,\bar B^{\bar \alpha \bar \beta, \mu, \alpha \beta}_{\eta \rho,0}(\bar p,p)\Big\}
\nonumber\\
&& {} + \frac{\tilde h_V}{8\,f\,m_V^2}\,  \eAcptn
\, h^{(b_1)}_{K\,\bar K^*}\,\hat C^{\bar \alpha \bar \beta, \mu, \alpha \beta}_{K K^* K^*}(\bar p,p)
\nonumber\\
&& {} + \frac{\tilde h_V}{8\,f\,m_V^2}\,  \eAcptnbA
\, h^{(b_1)}_{K\,\bar K^*}\,\bar C^{\bar \alpha \bar \beta, \mu, \alpha \beta}_{K K^* K^*}(\bar p,p)
\nonumber\\
&& {} + \sqrt{3}\,\frac{\eA\,\tilde h_V}{12\,f\,m_V^2}\,h^{(b_1)}_{\eta\,\rho}\,
\hat C^{\bar \alpha \bar \beta, \mu, \alpha \beta}_{\eta \rho \rho}(\bar p,p)
+ \sqrt{3}\,\frac{\eAbA\,\tilde h_V}{12\,f\,m_V^2}\,h^{(b_1)}_{\eta\,\rho}\,
\bar C^{\bar \alpha \bar \beta, \mu, \alpha \beta}_{\eta \rho \rho}(\bar p,p)
\nonumber\\
&& {} + \frac{\eA\,h_P}{8\,f^3}\,h^{(b_1)}_{\pi\,\omega}\, E^{\bar \alpha \bar \beta, \mu, \alpha \beta}_{\pi \pi \omega}(\bar p,p)
+ \frac{\eAbA\,h_P}{8\,f^3}\,h^{(b_1)}_{\pi\,\omega}\, \bar E^{\bar \alpha \bar \beta, \mu, \alpha \beta}_{\pi \pi \omega}(\bar p,p)
\nonumber\\
&& {} + \frac{h_P}{16\,f^3}\, \eAcptn
\, h^{(b_1)}_{K\,\bar K^*}\, E^{\bar \alpha \bar \beta, \mu, \alpha \beta}_{K K K^*}(\bar p,p)
\nonumber\\
&& {} + \frac{h_P}{16\,f^3}\, \eAcptnbA
\, h^{(b_1)}_{K\,\bar K^*}\, \bar E^{\bar \alpha \bar \beta, \mu, \alpha \beta}_{K K K^*}(\bar p,p)  \,,
\label{b1:gamma:rho}
\end{eqnarray}
and
\begin{eqnarray}
\lefteqn{i\,M^{\mu, \alpha \beta}_{h_1 \to \, \gamma \,\pi^0}
=- \frac{e\,h_P\,m_V}{4\,f^2}\,h^{(h_1)}_{K\,\bar K^*}\,
\Big\{ A^{\mu, \alpha \beta}_{K K^*}(\bar p,p)
 -\hat B^{\mu, \alpha \beta}_{K K^*,0}(\bar p,p)\Big\}   }
\nonumber\\
&& {} - \sqrt{3}\,\frac{e\,h_P\,m_V}{3\,f^2}\,h^{(h_1)}_{\pi\,\rho}\,
\Big\{ A^{\mu, \alpha \beta}_{\pi \rho}(\bar p,p)
 -\hat B^{\mu, \alpha \beta}_{\pi \rho,0}(\bar p,p)\Big\}
\nonumber\\
&& {} + \frac{h_A}{32\,f^2\,m_V}\,  \eAcptn
\,h^{(h_1)}_{K\,\bar K^*}\,C^{\mu, \alpha \beta}_{K K^* K^*}(\bar p,p)
\nonumber\\
&& {} +\sqrt{3}\,\frac{\eA\,h_A}{48\,f^2\,m_V}\,
\Big\{ h^{(h_1)}_{\pi\,\rho}\,C^{\mu, \alpha \beta}_{\pi \omega \rho}(\bar p,p)
+ h^{(h_1)}_{\eta\,\omega}\,C^{\mu, \alpha \beta}_{\eta \rho \omega}(\bar p,p)
 \Big\}
\nonumber\\
&& {} + \frac{h_A}{32\,f^2\,m_V}\,  \eAcptnbA
\,h^{(h_1)}_{K\,\bar K^*}\,\bar C^{\mu, \alpha \beta}_{K K^* K^*}(\bar p,p)
\nonumber\\
&& {} +\sqrt{3}\,\frac{\eAbA\,h_A}{48\,f^2\,m_V}\,
\Big\{ h^{(h_1)}_{\pi\,\rho}\,\bar C^{\mu, \alpha \beta}_{\pi \omega \rho}(\bar p,p)
+ h^{(h_1)}_{\eta\,\omega}\,\bar C^{\mu, \alpha \beta}_{\eta \rho \omega}(\bar p,p)
 \Big\}
\nonumber\\
&& {} + \frac{b_A}{8\,f^2\,m_V}\, m_\pi^2  \,\eAcptn
\,h^{(h_1)}_{K\,\bar K^*}\,\Cchi^{\mu, \alpha \beta}_{K K^* K^*}(\bar p,p)
\nonumber\\
&& {} +\sqrt{3}\,\frac{\eA\,b_A}{12\,f^2\,m_V}\, m_\pi^2  \,
\Big\{ h^{(h_1)}_{\pi\,\rho}\,\Cchi^{\mu, \alpha \beta}_{\pi \omega \rho}(\bar p,p)
+ h^{(h_1)}_{\eta\,\omega}\,\Cchi^{\mu, \alpha \beta}_{\eta \rho \omega}(\bar p,p)
 \Big\}
\nonumber\\
&& {} + \frac{b_A}{8\,f^2\,m_V}\,  m_\pi^2  \, \eAcptnbA
\,h^{(h_1)}_{K\,\bar K^*}\,\barCchi^{\mu, \alpha \beta}_{K K^* K^*}(\bar p,p)
\nonumber\\
&& {} +\sqrt{3}\,\frac{\eAbA\,b_A}{12\,f^2\,m_V}\, m_\pi^2 \,
\Big\{ h^{(h_1)}_{\pi\,\rho}\,\barCchi^{\mu, \alpha \beta}_{\pi \omega \rho}(\bar p,p)
+ h^{(h_1)}_{\eta\,\omega}\,\barCchi^{\mu, \alpha \beta}_{\eta \rho \omega}(\bar p,p)
 \Big\}
\nonumber\\
&& {} - \frac{e_V\,m_V}{32\,f^2}\,h^{(h_1)}_{K\,\bar K^*}\,E^{\mu, \alpha \beta}_{K K^*}(\bar p,p)
- \sqrt{3}\,\frac{e_V\,m_V}{24\,f^2}\,h^{(h_1)}_{\pi\,\rho}\,E^{\mu, \alpha \beta}_{\pi \rho}(\bar p,p)
\nonumber\\
&& {} + \frac{e_V}{16\,f^2 \, m_V}\,h^{(h_1)}_{K\,\bar K^*}\,F^{\mu, \alpha \beta}_{K K^*}(\bar p,p)
+ \sqrt{3}\,\frac{e_V}{12\,f^2\,m_V}\,h^{(h_1)}_{\pi\,\rho}\,F^{\mu, \alpha \beta}_{\pi \rho}(\bar p,p)
\nonumber\\
&& {} - \frac{e_V}{16\,f^2\,m_V}\,\Big(3\,g_D - g_F\Big)\,h^{(h_1)}_{K\,\bar K^*}\,G^{\mu, \alpha \beta}_{K K^*}(\bar p,p)
- \sqrt{3}\,\frac{e_V\,g_D}{12\,f^2\,m_V}\,h^{(h_1)}_{\eta\,\omega}\,G^{\mu, \alpha \beta}_{\eta \omega}(\bar p,p)
\nonumber\\
&& {} - \sqrt{3}\,\frac{e_V}{12\,f^2\,m_V}\,\Big(2\,g_D - g_F \Big)\, h^{(h_1)}_{\pi\,\rho}\,G^{\mu, \alpha \beta}_{\pi \rho}(\bar p,p)
\nonumber\\
&& {} - \frac{e_V\,b_D}{32\,f^2\,m_V}\, (m_\pi^2 + m_K^2)\,h^{(h_1)}_{K\,\bar K^*}\,E^{\mu, \alpha \beta}_{K K^*}(\bar p,p)
\nonumber\\
&& {} - \sqrt{3}\,\frac{e_V\,b_D}{24\,f^2\,m_V}\,m_\pi^2\,
\Big\{ h^{(h_1)}_{\pi\,\rho}\,E^{\mu, \alpha \beta}_{\pi \rho}(\bar p,p)
+h^{(h_1)}_{\eta\,\omega}\,E^{\mu, \alpha \beta}_{\eta \omega}(\bar p,p) \Big\}
\nonumber\\
&& {} + \frac{h_A \, b_A \, e_V}{16 \, f^2 \, m_V} \, m_\pi^2 \, \Big\{
\sqrt{3}\, h^{(h_1)}_{\pi\,\rho}\, H^{\mu, \alpha \beta}_{\pi \omega \rho}(\bar p,p)
+\frac{1}{\sqrt{3}} \, h^{(h_1)}_{\eta\,\omega}\,
H^{\mu, \alpha \beta}_{\eta \omega \omega}(\bar p,p)
\nonumber\\
&& \hspace{6em}  {}+ h^{(h_1)}_{K\,\bar K^*}\, H^{\mu, \alpha \beta}_{K \omega K^*}(\bar p,p)  \Big\}
\nonumber\\
&& {} + \frac{(b_A)^2 \, e_V}{4 \, f^2 \, m_V} \, m_\pi^2 \, \Big\{
\sqrt{3}\, m_\pi^2 \, h^{(h_1)}_{\pi\,\rho}\,
   \bar H^{\mu, \alpha \beta}_{\pi \omega \rho}(\bar p,p)
+\frac{1}{\sqrt{3}} \, m_\pi^2 \, h^{(h_1)}_{\eta\,\omega}\,
   \bar H^{\mu, \alpha \beta}_{\eta \omega \omega}(\bar p,p)
\nonumber\\
&& \hspace{6em}  {}
+ m_K^2 \, h^{(h_1)}_{K\,\bar K^*}\,
   \bar H^{\mu, \alpha \beta}_{K \omega K^*}(\bar p,p)  \Big\}\,,
\label{h1:gamma:pi}
\end{eqnarray}
and
\begin{eqnarray}
\lefteqn{i\,M^{\mu, \alpha \beta}_{h_1 \to \, \gamma \,\eta}
= -\sqrt{3}\,\frac{e\,h_P\,m_V}{4\,f^2}\,h^{(h_1)}_{K\,\bar K^*}\,\Big\{A^{\mu, \alpha \beta}_{K K^*}(\bar p,p)
-\,\hat B^{\mu, \alpha \beta}_{K K^*,2}(\bar p,p)\Big\}   }
\nonumber\\
&& {} + \sqrt{3}\,\frac{h_A}{288\,f^2\,m_V}\, \eAcmtn
\,h^{(h_1)}_{K\,\bar K^*}\,C^{\mu, \alpha \beta}_{K K^* K^*}(\bar p,p)
\nonumber\\
&& {} + \frac{\eA\,h_A}{48\,f^2\,m_V}\,\Big\{h^{(h_1)}_{\pi\,\rho}\,C^{\mu, \alpha \beta}_{\pi \rho \rho}(\bar p,p)
+ \frac{1}{3}\,h^{(h_1)}_{\eta\,\omega}\,C^{\mu, \alpha \beta}_{\eta \omega \omega}(\bar p,p) \Big\}
\nonumber\\
&& {}- \sqrt{2}\,\frac{h_A}{36\,f^2\,m_V}\, \eAphi
\, h^{(h_1)}_{\eta\,\phi}\,C^{\mu, \alpha \beta}_{\eta \phi \phi}(\bar p,p)
\nonumber\\
&& {} + \sqrt{3}\,\frac{h_A}{288\,f^2\,m_V}\, \eAcmtnbA
\,h^{(h_1)}_{K\,\bar K^*}\,\bar C^{\mu, \alpha \beta}_{K K^* K^*}(\bar p,p)
\nonumber\\
&& {} + \frac{\eAbA\,h_A}{48\,f^2\,m_V}\,\Big\{h^{(h_1)}_{\pi\,\rho}\,\bar C^{\mu, \alpha \beta}_{\pi \rho \rho}(\bar p,p)
+ \frac{1}{3}\,h^{(h_1)}_{\eta\,\omega}\,\bar C^{\mu, \alpha \beta}_{\eta \omega \omega}(\bar p,p) \Big\}
\nonumber\\
&& {}- \sqrt{2}\,\frac{h_A}{36\,f^2\,m_V}\, \eAphibA
\, h^{(h_1)}_{\eta\,\phi}\,\bar C^{\mu, \alpha \beta}_{\eta \phi \phi}(\bar p,p)
\nonumber\\
&& {} + \sqrt{3}\,\frac{b_A}{72\,f^2\,m_V}\,(4\,m_K^2-3\,m_\pi^2)\, \eAcmtn
\,h^{(h_1)}_{K\,\bar K^*}\,\Cchi^{\mu, \alpha \beta}_{K K^* K^*}(\bar p,p)
\nonumber\\
&& {} + \frac{\eA\,b_A}{12\,f^2\,m_V}\,m_\pi^2 \,\Big\{h^{(h_1)}_{\pi\,\rho}\,\Cchi^{\mu, \alpha \beta}_{\pi \rho \rho}(\bar p,p)
+ \frac{1}{3}\,h^{(h_1)}_{\eta\,\omega}\,\Cchi^{\mu, \alpha \beta}_{\eta \omega \omega}(\bar p,p) \Big\}
\nonumber\\
&& {}- \sqrt{2}\,\frac{b_A}{9\,f^2\,m_V}\,\Big(2\,m_K^2-m_\pi^2 \Big)\, \eAphi
\, h^{(h_1)}_{\eta\,\phi}\,\Cchi^{\mu, \alpha \beta}_{\eta \phi \phi}(\bar p,p)
\nonumber\\
&& {} + \sqrt{3}\,\frac{b_A}{72\,f^2\,m_V}\,(4\,m_K^2-3\,m_\pi^2)\, \eAcmtnbA
\,h^{(h_1)}_{K\,\bar K^*}\,\barCchi^{\mu, \alpha \beta}_{K K^* K^*}(\bar p,p)
\nonumber\\
&& {} + \frac{\eAbA\,b_A}{12\,f^2\,m_V}\,m_\pi^2 \,\Big\{h^{(h_1)}_{\pi\,\rho}\,\barCchi^{\mu, \alpha \beta}_{\pi \rho \rho}(\bar p,p)
+ \frac{1}{3}\,h^{(h_1)}_{\eta\,\omega}\,\barCchi^{\mu, \alpha \beta}_{\eta \omega \omega}(\bar p,p) \Big\}
\nonumber\\
&& {}- \sqrt{2}\,\frac{b_A}{9\,f^2\,m_V}\,\Big(2\,m_K^2-m_\pi^2 \Big)\, \eAphibA
\, h^{(h_1)}_{\eta\,\phi}\,\barCchi^{\mu, \alpha \beta}_{\eta \phi \phi}(\bar p,p)
\nonumber\\
&& {} - \sqrt{3}\,\frac{e_V\,m_V}{32\,f^2}\,h^{(h_1)}_{K\,\bar K^*}\,E^{\mu, \alpha \beta}_{K K^*}(\bar p,p)
+\sqrt{3}\,\frac{e_V}{48\,f^2\,m_V}\,
\left(2\,\ratVphi+1 \right) h^{(h_1)}_{K\,\bar K^*}\, F^{\mu, \alpha \beta}_{K K^*}(\bar p,p)
\nonumber\\
&& {} - \frac{e_V\,g_D}{12\,f^2\,m_V}\,\Big\{ h^{(h_1)}_{\pi\,\rho}\,G^{\mu, \alpha \beta}_{\pi \rho}(\bar p,p)
+ \frac{1}{3}\,h^{(h_1)}_{\eta\,\omega}\,G^{\mu, \alpha \beta}_{\eta \omega}(\bar p,p) \Big\}
\nonumber\\
&& {} + \sqrt{2}\,\frac{e_V\, \ratVphi\,g_D}{9\,f^2\,m_V}\,h^{(h_1)}_{\eta\,\phi}\,G^{\mu, \alpha \beta}_{\eta \phi}(\bar p,p)
\nonumber\\
&& {} - \sqrt{3}\,\frac{e_V}{144\,f^2\,m_V} \,\Big[
  \left( 1 + 10\,\ratVphi \right)\,g_D - 3\,\left( 1 + 2 \, \ratVphi \right) g_F \Big]\,
h^{(h_1)}_{K\,\bar K^*}\,G^{\mu, \alpha \beta}_{K K^*}(\bar p,p)
\nonumber\\
&& {} - \frac{e_V\,b_D}{24\,f^2\,m_V}\,m_\pi^2\,\Big\{ h^{(h_1)}_{\pi\,\rho}\,E^{\mu, \alpha \beta}_{\pi \rho}(\bar p,p)
+ \frac{1}{3}\, h^{(h_1)}_{\eta\,\omega}\,E^{\mu, \alpha \beta}_{\eta \omega}(\bar p,p) \Big\}
\nonumber\\
&& {} + \sqrt{2}\,\frac{e_V\, \ratVphi\,b_D}{18\,f^2\,m_V} \,
(2\,m_K^2-m_\pi^2) \, h^{(h_1)}_{\eta\,\phi}\, E^{\mu, \alpha \beta}_{\eta \phi}(\bar p,p)
\nonumber\\
&& {} + \sqrt{3}\,\frac{e_V\,b_D}{288\,f^2\,m_V} \,(5\,m_K^2-3\,m_\pi^2) \,
\left(1-2 \, \ratVphi \right)\,h^{(h_1)}_{K\,\bar K^*}\, E^{\mu, \alpha \beta}_{K K^*}(\bar p,p)
\nonumber\\
&& {} + \frac{h_A \, b_A \, e_V}{48 \, f^2 \, m_V} \, m_\pi^2 \, \Big\{
h^{(h_1)}_{\pi\,\rho}\, H^{\mu, \alpha \beta}_{\pi \omega \rho}(\bar p,p)
+\frac{1}{3} \, h^{(h_1)}_{\eta\,\omega}\,
H^{\mu, \alpha \beta}_{\eta \omega \omega}(\bar p,p)
+ \frac{1}{\sqrt{3}} \, h^{(h_1)}_{K\,\bar K^*}\,
H^{\mu, \alpha \beta}_{K \omega K^*}(\bar p,p)  \Big\}
\nonumber\\
&& {} + \frac{ \sqrt{2}\, h_A \, b_A \, e_V }{72 \, f^2 \, m_V} \, \ratVphi \,
(2\, m_K^2 - m_\pi^2) \, \Big\{
\sqrt{6} \, h^{(h_1)}_{K\,\bar K^*}\, H^{\mu, \alpha \beta}_{K \phi K^*}(\bar p,p)
- 2 \, h^{(h_1)}_{\eta\,\phi}\,
H^{\mu, \alpha \beta}_{\eta \phi \phi}(\bar p,p)  \Big\}
\nonumber\\
&& {} + \frac{(b_A)^2 \, e_V}{12 \, f^2 \, m_V} \, m_\pi^2 \, \Big\{
m_\pi^2 \, h^{(h_1)}_{\pi\,\rho}\,
   \bar H^{\mu, \alpha \beta}_{\pi \omega \rho}(\bar p,p)
+\frac{1}{3} \, m_\pi^2 \, h^{(h_1)}_{\eta\,\omega}\,
   \bar H^{\mu, \alpha \beta}_{\eta \omega \omega}(\bar p,p)
\nonumber\\
&& \hspace{7em}  {}
+ \frac{1}{\sqrt{3}} \, m_K^2 \, h^{(h_1)}_{K\,\bar K^*}\,
   \bar H^{\mu, \alpha \beta}_{K \omega K^*}(\bar p,p)  \Big\}
\nonumber\\
&& {} + \frac{ \sqrt{2}\, (b_A)^2 \, e_V }{18 \, f^2 \, m_V} \, \ratVphi \,
(2\, m_K^2 - m_\pi^2) \, \Big\{
\sqrt{6} \, m_K^2 \, h^{(h_1)}_{K\,\bar K^*}\,
   \bar H^{\mu, \alpha \beta}_{K \phi K^*}(\bar p,p)
\nonumber\\
&& \hspace{13em}  {}
- 2 \, (2 m_K^2 - m_\pi^2) \, h^{(h_1)}_{\eta\,\phi}\,
   \bar H^{\mu, \alpha \beta}_{\eta \phi \phi}(\bar p,p)  \Big\}
\,, \nonumber \\
\label{h1:gamma:eta}
\end{eqnarray}

and

\begin{eqnarray}
\lefteqn{i\,M^{\mu, \alpha \beta}_{a_1^+ \to \, \gamma \,\pi^+}=
\frac{e\,h_P\,m_V}{4\,f^2}\,h^{(a_1)}_{K\,\bar K^*}\,\Big\{ A^{\mu, \alpha \beta}_{K K^*}(\bar p,p)
 +\hat B^{\mu, \alpha \beta}_{K K^*,0}(\bar p,p)\Big\}  }
\nonumber\\
&& {} + \sqrt{2}\,\frac{e\,h_P\,m_V}{4\,f^2}\,h^{(a_1)}_{\pi\,\rho}\,
\Big\{ A^{\mu, \alpha \beta}_{\pi \rho}(\bar p,p)
 +\hat B^{\mu, \alpha \beta}_{\pi \rho,0}(\bar p,p)\Big\}
\nonumber\\
&& {} - \frac{h_A}{32\,f^2\,m_V}\, \eAcptn
\, h^{(a_1)}_{K\,\bar K^*}\,C^{\mu, \alpha \beta}_{K K^* K^*}(\bar p,p)
\nonumber\\
&& {} - \sqrt{2}\,\frac{\eA\,h_A}{32\,f^2\,m_V}\,h^{(a_1)}_{\pi\,\rho}\,C^{\mu, \alpha \beta}_{\pi \omega \rho}(\bar p,p)
\nonumber\\
&& {} - \frac{h_A}{32\,f^2\,m_V}\, \eAcptnbA
\, h^{(a_1)}_{K\,\bar K^*}\,\bar C^{\mu, \alpha \beta}_{K K^* K^*}(\bar p,p)
\nonumber\\
&& {} - \sqrt{2}\,\frac{\eAbA\,h_A}{32\,f^2\,m_V}\,h^{(a_1)}_{\pi\,\rho}\,\bar C^{\mu, \alpha \beta}_{\pi \omega \rho}(\bar p,p)
\nonumber\\
&& {} - \frac{b_A}{8\,f^2\,m_V}\, m_\pi^2\,\eAcptn
\, h^{(a_1)}_{K\,\bar K^*}\,\Cchi^{\mu, \alpha \beta}_{K K^* K^*}(\bar p,p)
\nonumber\\
&& {} - \sqrt{2}\,\frac{\eA\,b_A}{8\,f^2\,m_V}\,m_\pi^2\,h^{(a_1)}_{\pi\,\rho}\,\Cchi^{\mu, \alpha \beta}_{\pi \omega \rho}(\bar p,p)
\nonumber\\
&& {} - \frac{b_A}{8\,f^2\,m_V}\,m_\pi^2 \, \eAcptnbA
\, h^{(a_1)}_{K\,\bar K^*}\,\barCchi^{\mu, \alpha \beta}_{K K^* K^*}(\bar p,p)
\nonumber\\
&& {} - \sqrt{2}\,\frac{\eAbA\,b_A}{8\,f^2\,m_V}\,m_\pi^2 \,h^{(a_1)}_{\pi\,\rho}\,\barCchi^{\mu, \alpha \beta}_{\pi \omega \rho}(\bar p,p)
\nonumber\\
&& {} + \frac{3\,e_V\,m_V}{32\,f^2}\,\Big\{ h^{(a_1)}_{K\,\bar K^*}\, E^{\mu, \alpha \beta}_{K K^*}(\bar p,p)
+ \sqrt{2}\,h^{(a_1)}_{\pi\,\rho}\, E^{\mu, \alpha \beta}_{\pi \rho}(\bar p,p) \Big\}
\nonumber\\
&& {} + \frac{e_V}{16\,f^2 m_V}\,\Big\{h^{(a_1)}_{K\,\bar K^*}\, F^{\mu, \alpha \beta}_{K K^*}(\bar p,p)
+ \sqrt{2}\,h^{(a_1)}_{\pi\,\rho}\, F^{\mu, \alpha \beta}_{\pi \rho}(\bar p,p) \Big\}
 \nonumber\\
&& {} + \frac{e_V}{16\,f^2\,m_V}\,\Big(g_D-3\, g_F\Big)\,
\Big\{ h^{(a_1)}_{K\,\bar K^*}\, G^{\mu, \alpha \beta}_{K K^*}(\bar p,p)
+\sqrt{2}\,h^{(a_1)}_{\pi\,\rho}\, G^{\mu, \alpha \beta}_{\pi \rho}(\bar p,p) \Big\}
\nonumber\\
&& {} - \sqrt{2}\,\frac{e_V\,b_D}{16 \,f^2\,m_V}\,m_\pi^2\, h^{(a_1)}_{\pi\,\rho}\, E^{\mu, \alpha \beta}_{\pi \rho}(\bar p,p)
\nonumber\\
&& {} - \frac{e_V\,b_D}{32\,f^2\,m_V} \, (m_\pi^2 + m_K^2)\, h^{(a_1)}_{K\,\bar K^*}\, E^{\mu, \alpha \beta}_{K K^*}(\bar p,p)
\,,
\label{a1:gamma:pi}
\end{eqnarray}
and
\begin{eqnarray}
\lefteqn{
M^{\bar \alpha \bar \beta, \mu, \alpha \beta}_{a_1^+ \to \, \gamma\,\rho^+}=
M^{\bar \alpha \bar \beta, \mu, \alpha \beta}_{a_1^0 \to \, \gamma\,\rho^0}=
\frac{e\,h_A}{8\,f}\,h^{(a_1)}_{K\,\bar K^*}\,\Big\{ \hat A^{\bar \alpha \bar \beta, \mu, \alpha \beta}_{K K^*,2}(\bar p,p)
 -\hat B^{\bar \alpha \bar \beta, \mu, \alpha \beta}_{K K^*,2}(\bar p,p)\Big\}  }
\nonumber\\
&& {} + \frac{e\,b_A}{2\,f}\,m_K^2\,h^{(a_1)}_{K\,\bar K^*}\,\Big\{ \bar A^{\bar \alpha \bar \beta, \mu, \alpha \beta}_{K K^*,2}(\bar p,p)
 -\bar B^{\bar \alpha \bar \beta, \mu, \alpha \beta}_{K K^*,2}(\bar p,p)\Big\}
\nonumber\\
&& {} + \frac{\tilde h_V}{24\,f\,m_V^2}\, \eAcmtn
\,h^{(a_1)}_{K\,\bar K^*}\,\hat C^{\bar \alpha \bar \beta, \mu, \alpha \beta}_{K K^* K^*}(\bar p,p)
\nonumber\\
&& {} + \frac{\tilde h_V}{24\,f\,m_V^2}\, \eAcmtnbA
\,h^{(a_1)}_{K\,\bar K^*}\,\bar C^{\bar \alpha \bar \beta, \mu, \alpha \beta}_{K K^* K^*}(\bar p,p)
\nonumber\\
&& {} - \sqrt{2}\,\frac{\eA\,\tilde h_V}{12\,f\,m_V^2}\,h^{(a_1)}_{\pi\,\rho}\,\hat C^{\bar \alpha \bar \beta, \mu, \alpha \beta}_{\pi \rho \rho}(\bar p,p)
 - \sqrt{2}\,\frac{\eAbA\,\tilde h_V}{12\,f\,m_V^2}\,h^{(a_1)}_{\pi\,\rho}\,\bar C^{\bar \alpha \bar \beta, \mu, \alpha \beta}_{\pi \rho \rho}(\bar p,p)
\nonumber\\
&& {}+ \sqrt{2}\,\frac{\eA\,h_P}{24 \,f^3}\,h^{(a_1)}_{\pi\,\rho}\, E^{\bar \alpha \bar \beta, \mu, \alpha \beta}_{\pi \pi \rho}(\bar p,p)
+ \sqrt{2}\,\frac{\eAbA\,h_P}{24 \,f^3}\,h^{(a_1)}_{\pi\,\rho}\, \bar E^{\bar \alpha \bar \beta, \mu, \alpha \beta}_{\pi \pi \rho}(\bar p,p)
\nonumber\\
&& {} - \frac{h_P}{48\,f^3}\, \eAcmtn
\, h^{(a_1)}_{K\,\bar K^*}\, E^{\bar \alpha \bar \beta, \mu, \alpha \beta}_{K K K^*}(\bar p,p)
\nonumber\\
&& {} - \frac{h_P}{48\,f^3}\, \eAcmtnbA
\, h^{(a_1)}_{K\,\bar K^*}\, \bar E^{\bar \alpha \bar \beta, \mu, \alpha \beta}_{K K K^*}(\bar p,p)  \,,
\label{a1:gamma:rho}
\end{eqnarray}
and
\begin{eqnarray}
\lefteqn{M^{\bar \alpha \bar \beta, \mu, \alpha \beta}_{a_1^0 \to \, \gamma \,\omega}
=\frac{e\,h_A}{8\,f}\,h^{(a_1)}_{K\,\bar K^*}\,\Big\{ \hat A^{\bar \alpha \bar \beta, \mu, \alpha \beta}_{K K^*,0}(\bar p,p)
 -\hat B^{\bar \alpha \bar \beta, \mu, \alpha \beta}_{K K^*,0}(\bar p,p)\Big\}  }
\nonumber\\
&& {} + \frac{e\,b_A}{2\,f}\,m_K^2\,h^{(a_1)}_{K\,\bar K^*}\,\Big\{ \bar A^{\bar \alpha \bar \beta, \mu, \alpha \beta}_{K K^*,0}(\bar p,p)
 -\bar B^{\bar \alpha \bar \beta, \mu, \alpha \beta}_{K K^*,0}(\bar p,p)\Big\}
\nonumber\\
&& {} +\sqrt{2}\,\frac{e\,h_A}{4\,f}\,
h^{(a_1)}_{\pi\,\rho}\,\Big\{\hat A^{\bar \alpha \bar \beta, \mu, \alpha \beta}_{\pi \rho,0}(\bar p,p)
- \hat B^{\bar \alpha \bar \beta, \mu, \alpha \beta}_{\pi \rho,0}(\bar p,p)\Big\}
\nonumber\\
&& {} +\sqrt{2}\,\frac{e\,b_A}{f}\,m_\pi^2\,
h^{(a_1)}_{\pi\,\rho}\,\Big\{\bar A^{\bar \alpha \bar \beta, \mu, \alpha \beta}_{\pi \rho,0}(\bar p,p)
- \bar B^{\bar \alpha \bar \beta, \mu, \alpha \beta}_{\pi \rho,0}(\bar p,p)\Big\}
\nonumber\\
&& {} - \frac{\tilde h_V}{8\,f\,m_V^2}\, \eAcptn \,h^{(a_1)}_{K\,\bar K^*}\,\hat C^{\bar \alpha \bar \beta, \mu, \alpha \beta}_{K K^* K^*}(\bar p,p)
\nonumber\\
&& {} - \frac{\tilde h_V}{8\,f\,m_V^2}\, \eAcptnbA \,h^{(a_1)}_{K\,\bar K^*}\,\bar C^{\bar \alpha \bar \beta, \mu, \alpha \beta}_{K K^* K^*}(\bar p,p)
\nonumber\\
&& {} + \frac{h_P}{16\,f^3}\, \eAcptn
\,h^{(a_1)}_{K\,\bar K^*}\, E^{\bar \alpha \bar \beta, \mu, \alpha \beta}_{K K K^*}(\bar p,p)
\nonumber\\
&& {} + \frac{h_P}{16\,f^3}\, \eAcptnbA
\,h^{(a_1)}_{K\,\bar K^*}\, \bar E^{\bar \alpha \bar \beta, \mu, \alpha \beta}_{K K K^*}(\bar p,p) \,,
\label{a1:gamma:omega}
\end{eqnarray}
and
\begin{eqnarray}
\lefteqn{M^{\bar \alpha \bar \beta, \mu, \alpha \beta}_{a_1^0 \to \, \gamma\,\phi}
= \sqrt{2}\,\frac{e\,h_A}{8\,f}\,h^{(a_1)}_{K\,\bar K^*}\,
\Big\{ \hat A^{\bar \alpha \bar \beta, \mu, \alpha \beta}_{K K^*,0}(\bar p,p)
 -\hat B^{\bar \alpha \bar \beta, \mu, \alpha \beta}_{K K^*,0}(\bar p,p)\Big\}  }
\nonumber\\
&& {} + \sqrt{2}\,\frac{e\,b_A}{2\,f}\,m_K^2\,h^{(a_1)}_{K\,\bar K^*}\,
\Big\{ \bar A^{\bar \alpha \bar \beta, \mu, \alpha \beta}_{K K^*,0}(\bar p,p)
 -\bar B^{\bar \alpha \bar \beta, \mu, \alpha \beta}_{K K^*,0}(\bar p,p)\Big\}
\nonumber\\
&& {} + \sqrt{2}\,\frac{\tilde h_V}{8\,f\,m_V^2}\, \eAcptn
\, h^{(a_1)}_{K\,\bar K^*}\,
\hat C^{\bar \alpha \bar \beta, \mu, \alpha \beta}_{K K^* K^*}(\bar p,p)
\nonumber\\
&& {} + \sqrt{2}\,\frac{\tilde h_V}{8\,f\,m_V^2}\, \eAcptnbA
\, h^{(a_1)}_{K\,\bar K^*}\,
\bar C^{\bar \alpha \bar \beta, \mu, \alpha \beta}_{K K^* K^*}(\bar p,p)
\nonumber\\
&& {} -\sqrt{2}\,\frac{h_P}{16\,f^3}\, \eAcptn \,h^{(a_1)}_{K\,\bar K^*}\,
E^{\bar \alpha \bar \beta, \mu, \alpha \beta}_{K K K^*}(\bar p,p)
\nonumber\\
&& {} -\sqrt{2}\,\frac{h_P}{16\,f^3}\, \eAcptnbA \,h^{(a_1)}_{K\,\bar K^*}\,
\bar E^{\bar \alpha \bar \beta, \mu, \alpha \beta}_{K K K^*}(\bar p,p) \,,
\label{a1:gamma:phi}
\end{eqnarray}
and

\begin{eqnarray}
\lefteqn{i\,M^{\mu, \alpha \beta}_{K_1^+ \to \, \gamma \,K^+} =
\frac{e\,h_P\,m_V}{4\,f^2}\,\Big\{ \frac{1}{\sqrt{3}}\,h^{(K_1)}_{K\,\rho}\,
\left(A^{\mu, \alpha \beta}_{K \rho}(\bar p,p) + 2\,\hat B^{\mu, \alpha \beta}_{K \rho,0}(\bar p,p)  \right)  }
\nonumber\\
&& \hspace*{5em} {} + h^{(K_1)}_{K\,\omega}\,A^{\mu, \alpha \beta}_{K \omega}(\bar p,p)
- \sqrt{2}\,h^{(K_1)}_{K\,\phi}\,A^{\mu, \alpha \beta}_{K \phi}(\bar p,p)
\nonumber\\
&& \hspace*{5em} {} -\sqrt{3}\,h^{(K_1)}_{\eta\,K^*}\,\hat B^{\mu, \alpha \beta}_{\eta K^*,1/2}(\bar p,p)
 \nonumber\\
&& \hspace*{5em} {} - \frac{1}{\sqrt{3}}\,h^{(K_1)}_{\pi\,K^*}\,
\left(2\,A^{\mu, \alpha \beta}_{\pi K^*}(\bar p,p)+\hat B^{\mu, \alpha \beta}_{\pi K^*,\infty}(\bar p,p) \right)
\Big\}
\nonumber\\
&& {} + \frac{h_A}{96\,f^2\,m_V}\, \Big\{
\eAc \, h^{(K_1)}_{K\,\omega}\,C^{\mu, \alpha \beta}_{K K^* \omega}(\bar p,p)
\nonumber\\
&& \hspace*{5em} {} - \eAcmfn \, \sqrt{3}\,h^{(K_1)}_{K\,\rho}\,C^{\mu, \alpha \beta}_{K K^* \rho}(\bar p,p)
\nonumber\\
&& \hspace*{5em} {} + \sqrt{2}\,\eAc \, h^{(K_1)}_{K\,\phi}\,C^{\mu, \alpha \beta}_{K K^* \phi}(\bar p,p)
\nonumber\\
&& \hspace*{5em} {} + \sqrt{3}\,\eA \, h^{(K_1)}_{\pi\,K^*}\,C^{\mu, \alpha \beta}_{\pi \omega K^*}(\bar p,p)
+\sqrt{3} \, \eA \,h^{(K_1)}_{\pi\,K^*}\,C^{\mu, \alpha \beta}_{\pi \rho K^*}(\bar p,p)
\nonumber\\
&& \hspace*{5em} {} + \frac{1}{\sqrt{3}}\,\eA \,h^{(K_1)}_{\eta\,K^*}\,C^{\mu, \alpha \beta}_{\eta \omega K^*}(\bar p,p)
+ \sqrt{3}\,\eA \, h^{(K_1)}_{\eta\,K^*}\,C^{\mu, \alpha \beta}_{\eta \rho K^*}(\bar p,p)
\nonumber\\
&& \hspace*{5em} {} +\frac{4}{\sqrt{3}}\,\eAphi \, h^{(K_1)}_{\eta\,K^*}\,C^{\mu, \alpha \beta}_{\eta \phi K^*}(\bar p,p)
\Big\}
\nonumber\\
&& {} + \frac{h_A}{96\,f^2\,m_V}\, \Big\{
\eAcbA \, h^{(K_1)}_{K\,\omega}\,\bar C^{\mu, \alpha \beta}_{K K^* \omega}(\bar p,p)
\nonumber\\
&& \hspace*{5em} {} - \eAcmfnbA \, \sqrt{3}\,h^{(K_1)}_{K\,\rho}\,\bar C^{\mu, \alpha \beta}_{K K^* \rho}(\bar p,p)
\nonumber\\
&& \hspace*{5em} {} + \sqrt{2}\,\eAcbA \, h^{(K_1)}_{K\,\phi}\,\bar C^{\mu, \alpha \beta}_{K K^* \phi}(\bar p,p)
\nonumber\\
&& \hspace*{5em} {} + \sqrt{3}\,\eAbA \, h^{(K_1)}_{\pi\,K^*}\,\bar C^{\mu, \alpha \beta}_{\pi \omega K^*}(\bar p,p)
+\sqrt{3} \, \eAbA \,h^{(K_1)}_{\pi\,K^*}\,\bar C^{\mu, \alpha \beta}_{\pi \rho K^*}(\bar p,p)
\nonumber\\
&& \hspace*{5em} {} + \frac{1}{\sqrt{3}}\,\eAbA \,h^{(K_1)}_{\eta\,K^*}\,\bar C^{\mu, \alpha \beta}_{\eta \omega K^*}(\bar p,p)
+ \sqrt{3}\,\eAbA \, h^{(K_1)}_{\eta\,K^*}\,\bar C^{\mu, \alpha \beta}_{\eta \rho K^*}(\bar p,p)
\nonumber\\
&& \hspace*{5em} {} +\frac{4}{\sqrt{3}}\,\eAphibA \, h^{(K_1)}_{\eta\,K^*}\,\bar C^{\mu, \alpha \beta}_{\eta \phi K^*}(\bar p,p)
\Big\}
\nonumber\\
&& {} + \frac{b_A\,m_K^2}{24\,f^2\,m_V}\, \Big\{
\eAc \, h^{(K_1)}_{K\,\omega}\,\Cchi^{\mu, \alpha \beta}_{K K^* \omega}(\bar p,p)
\nonumber\\
&& \hspace*{5em} {} - \eAcmfn \, \sqrt{3}\,h^{(K_1)}_{K\,\rho}\,\Cchi^{\mu, \alpha \beta}_{K K^* \rho}(\bar p,p)
\nonumber\\
&& \hspace*{5em} {} + \sqrt{2}\,\eAc \, h^{(K_1)}_{K\,\phi}\,\Cchi^{\mu, \alpha \beta}_{K K^* \phi}(\bar p,p)
\nonumber\\
&& \hspace*{5em} {} + \sqrt{3}\,\eA \, h^{(K_1)}_{\pi\,K^*}\,\Cchi^{\mu, \alpha \beta}_{\pi \omega K^*}(\bar p,p)
+\sqrt{3} \, \eA \,h^{(K_1)}_{\pi\,K^*}\,\Cchi^{\mu, \alpha \beta}_{\pi \rho K^*}(\bar p,p)
\nonumber\\
&& \hspace*{5em} {} + \frac{1}{\sqrt{3}}\,\eA \,h^{(K_1)}_{\eta\,K^*}\,\Cchi^{\mu, \alpha \beta}_{\eta \omega K^*}(\bar p,p)
+ \sqrt{3}\,\eA \, h^{(K_1)}_{\eta\,K^*}\,\Cchi^{\mu, \alpha \beta}_{\eta \rho K^*}(\bar p,p)
\nonumber\\
&& \hspace*{5em} {} +\frac{4}{\sqrt{3}}\,\eAphi \, h^{(K_1)}_{\eta\,K^*}\,\Cchi^{\mu, \alpha \beta}_{\eta \phi K^*}(\bar p,p)
\Big\}
\nonumber\\
&& {} + \frac{b_A\,m_K^2}{24\,f^2\,m_V}\, \Big\{
\eAcbA \, h^{(K_1)}_{K\,\omega}\,\barCchi^{\mu, \alpha \beta}_{K K^* \omega}(\bar p,p)
\nonumber\\
&& \hspace*{5em} {} - \eAcmfnbA \, \sqrt{3}\,h^{(K_1)}_{K\,\rho}\,\barCchi^{\mu, \alpha \beta}_{K K^* \rho}(\bar p,p)
\nonumber\\
&& \hspace*{5em} {} + \sqrt{2}\,\eAcbA \, h^{(K_1)}_{K\,\phi}\,\barCchi^{\mu, \alpha \beta}_{K K^* \phi}(\bar p,p)
\nonumber\\
&& \hspace*{5em} {} + \sqrt{3}\,\eAbA \, h^{(K_1)}_{\pi\,K^*}\,\barCchi^{\mu, \alpha \beta}_{\pi \omega K^*}(\bar p,p)
+\sqrt{3} \, \eAbA \,h^{(K_1)}_{\pi\,K^*}\,\barCchi^{\mu, \alpha \beta}_{\pi \rho K^*}(\bar p,p)
\nonumber\\
&& \hspace*{5em} {} + \frac{1}{\sqrt{3}}\,\eAbA \,h^{(K_1)}_{\eta\,K^*}\,\barCchi^{\mu, \alpha \beta}_{\eta \omega K^*}(\bar p,p)
+ \sqrt{3}\,\eAbA \, h^{(K_1)}_{\eta\,K^*}\,\barCchi^{\mu, \alpha \beta}_{\eta \rho K^*}(\bar p,p)
\nonumber\\
&& \hspace*{5em} {} +\frac{4}{\sqrt{3}}\,\eAphibA \, h^{(K_1)}_{\eta\,K^*}\,\barCchi^{\mu, \alpha \beta}_{\eta \phi K^*}(\bar p,p)
\Big\}
\nonumber\\
&& {} + \frac{e_V\,m_V}{8\,f^2}\,\Big\{
\frac{1}{\sqrt{3}}\,h^{(K_1)}_{K\,\rho}\,E^{\mu, \alpha \beta}_{K \rho}(\bar p,p)
- \frac{5}{\sqrt{3}\,4}\,h^{(K_1)}_{\pi\,K^*}\,E^{\mu, \alpha \beta}_{\pi K^*}(\bar p,p)
\nonumber\\
&& \hspace*{5em} {} +\frac{1}{2} \,h^{(K_1)}_{K\,\omega}\,E^{\mu, \alpha \beta}_{K \omega}(\bar p,p)
- \frac{\sqrt{3}}{4}\,h^{(K_1)}_{\eta\,K^*}\,E^{\mu, \alpha \beta}_{\eta K^*}(\bar p,p)
\nonumber\\
&& \hspace*{5em} {} -\frac{1}{\sqrt{2}}\,h^{(K_1)}_{K\,\phi}\,E^{\mu, \alpha \beta}_{K \phi}(\bar p,p) \Big\}
\nonumber\\
&& {} -\frac{e_V}{16 \,f^2\,m_V}\,\Big\{-\frac{2}{\sqrt{3}}\,h^{(K_1)}_{K\,\rho}\,
F^{\mu, \alpha \beta}_{K \rho}(\bar p,p)
+ \frac{1}{\sqrt{3}}\,h^{(K_1)}_{\pi\,K^*}\,\ratVphi\, F^{\mu, \alpha \beta}_{\pi K^*}(\bar p,p)
\nonumber\\
&& \hspace*{5em} {} +  \,\frac{2+\ratVphi }{\sqrt{3}} \,
h^{(K_1)}_{\eta\,K^*}\,F^{\mu, \alpha \beta}_{\eta K^*}(\bar p,p)  \Big\}
\nonumber\\
&& {} -\frac{e_V}{16 \,f^2\,m_V}\,\Big\{
\frac{1}{\sqrt{3}}\,\Big[
  \left(4 - \ratVphi \right) g_D - \left(4 + \ratVphi \right) g_F \Big]\,h^{(K_1)}_{\pi\,K^*}\,
G^{\mu, \alpha \beta}_{\pi K^*}(\bar p,p)
\nonumber\\
&& \hspace*{5em} {} + \frac{2}{\sqrt{3}}\,\Big[
  \left(1 - \ratVphi \right) g_D + \left(1 + \ratVphi \right) g_F
\Big]\,h^{(K_1)}_{K\,\rho}\,G^{\mu, \alpha \beta}_{K \rho}(\bar p,p)
\nonumber\\
&& \hspace*{5em} {} +\frac{2}{3}\,\Big[
  \left(2 - \ratVphi \right) g_D + \left(2 + \ratVphi \right) g_F \Big]\,h^{(K_1)}_{K\,\omega}\,G^{\mu, \alpha \beta}_{K \omega}(\bar p,p)
\nonumber\\
&& \hspace*{5em} {} + \frac{1}{3\sqrt{3}}\,\Big[
  \left(2 +5 \, \ratVphi \right) g_D - 3\, \left(2 + \ratVphi \right) g_F
\Big]\,h^{(K_1)}_{\eta\,K^*}\,G^{\mu, \alpha \beta}_{\eta K^*}(\bar p,p)
\nonumber\\
&& \hspace*{5em} {}+\frac{2\,\sqrt{2}}{3}\,\Big[
  \left(2 - \ratVphi \right) g_D - \left(2 + \ratVphi \right) g_F  \Big]\,
h^{(K_1)}_{K\,\phi}\,G^{\mu, \alpha \beta}_{K \phi}(\bar p,p)\Big\}
\nonumber\\
&& {} -\frac{e_V\, b_D}{96 \,f^2\,m_V}\,\Big\{
-\sqrt{3} \, \ratVphi \, (m_\pi^2 + m_K^2)
\, h^{(K_1)}_{\pi\,K^*}\,E^{\mu, \alpha \beta}_{\pi K^*}(\bar p,p)
\nonumber\\
&& \hspace*{5em} {}+ 4\,\sqrt{3}\,m_K^2\,h^{(K_1)}_{K\,\rho}\,E^{\mu, \alpha \beta}_{K \rho}(\bar p,p)
\nonumber\\
&& \hspace*{5em} {} + 8\,m_K^2  \,h^{(K_1)}_{K\,\omega}\,E^{\mu, \alpha \beta}_{K \omega}(\bar p,p)
\nonumber\\
&& \hspace*{5em} {} + \sqrt{3} \left(m_\pi^2 - \frac{5}{3} \, m_K^2 \right)\left(2-\ratVphi \right)\,h^{(K_1)}_{\eta\,K^*}\,
E^{\mu, \alpha \beta}_{\eta K^*}(\bar p,p)
\nonumber\\
&& \hspace*{5em} {}- 4\,\sqrt{2} \,\ratVphi \, m_K^2 \, h^{(K_1)}_{K\,\phi}\,E^{\mu, \alpha \beta}_{K \phi}(\bar p,p)  \Big\}
\nonumber\\
&& {}+ \frac{h_A \, b_A \, e_V}{96 \, f^2 \, m_V} \, (2-\ratVphi) \, m_K^2 \,
\Big\{
\sqrt{3} \, h^{(K_1)}_{\pi\,K^*}\, H^{\mu, \alpha \beta}_{\pi K^* K^*}(\bar p,p)
+ \sqrt{3} \, h^{(K_1)}_{K\,\rho}\, H^{\mu, \alpha \beta}_{K K^* \rho}(\bar p,p)
\nonumber\\
&& \hspace*{10em} {}
+ h^{(K_1)}_{K\,\omega}\, H^{\mu, \alpha \beta}_{K K^* \omega}(\bar p,p)
- \frac{1}{\sqrt{3}} \, h^{(K_1)}_{\eta\,K^*}\,
   H^{\mu, \alpha \beta}_{\eta K^* K^*}(\bar p,p)
\nonumber\\
&& \hspace*{10em} {}
+ \sqrt{2} \, h^{(K_1)}_{K\,\phi}\, H^{\mu, \alpha \beta}_{K K^* \phi}(\bar p,p)
\Big\}
\nonumber\\
&& {}+ \frac{(b_A)^2 \, e_V}{24 \, f^2 \, m_V} \, (2-\ratVphi) \, m_K^2 \,
\Big\{
\sqrt{3} \, m_\pi^2 \, h^{(K_1)}_{\pi\,K^*}\,
   \bar H^{\mu, \alpha \beta}_{\pi K^* K^*}(\bar p,p)
\nonumber\\
&& \hspace*{10em} {}
+ \sqrt{3} \, m_K^2 \, h^{(K_1)}_{K\,\rho}\,
   \bar H^{\mu, \alpha \beta}_{K K^* \rho}(\bar p,p)
\nonumber\\
&& \hspace*{10em} {}
+ m_K^2 \, h^{(K_1)}_{K\,\omega}\,
   \bar H^{\mu, \alpha \beta}_{K K^* \omega}(\bar p,p)
\nonumber\\
&& \hspace*{10em} {}
- \frac{1}{\sqrt{3}} \, (4 \, m_K^2 - 3\, m_\pi^2) \, h^{(K_1)}_{\eta\,K^*}\,
   \bar H^{\mu, \alpha \beta}_{\eta K^* K^*}(\bar p,p)
\nonumber\\
&& \hspace*{10em} {}
+ \sqrt{2} \, m_K^2 \, h^{(K_1)}_{K\,\phi}\,
   \bar H^{\mu, \alpha \beta}_{K K^* \phi}(\bar p,p)
\Big\}\,,
\label{K1:gamma:Kp}
\end{eqnarray}

and
\begin{eqnarray}
\lefteqn{M^{\bar \alpha \bar \beta, \mu, \alpha \beta}_{K_1^+ \to \, \gamma \,K^+_*}
= -\frac{e\,h_A}{8\,f}\,\Big\{ h^{(K_1)}_{K\,\omega}\,
\hat A^{\bar \alpha \bar \beta, \mu, \alpha \beta}_{K \omega,1/2}(\bar p,p)
+\frac{1}{\sqrt{3}}\,h^{(K_1)}_{K\,\rho}\,
\hat A^{\bar \alpha \bar \beta, \mu, \alpha \beta}_{K \rho, \infty }(\bar p,p)     }
\nonumber\\
&& \hspace*{5em} {} + \sqrt{2}\,h^{(K_1)}_{K\,\phi}\,\hat A^{\bar \alpha \bar \beta, \mu, \alpha \beta}_{K \phi,1/2}(\bar p,p)\Big\}
\nonumber\\
&& {}-\frac{e\,b_A}{2\,f}\,m_K^2 \,\Big\{ h^{(K_1)}_{K\,\omega}\,
\bar A^{\bar \alpha \bar \beta, \mu, \alpha \beta}_{K \omega,1/2}(\bar p,p)
+  \sqrt{2}\,h^{(K_1)}_{K\,\phi}\,\bar A^{\bar \alpha \bar \beta, \mu, \alpha \beta}_{K \phi,1/2}(\bar p,p)
\nonumber\\
&& \hspace*{5em} {} + \frac{1}{\sqrt{3}}\,h^{(K_1)}_{K\,\rho}\, \Big(
\bar A^{\bar \alpha \bar \beta, \mu, \alpha \beta}_{K \rho, \infty }(\bar p,p)
+2\,\bar B^{\bar \alpha \bar \beta, \mu, \alpha \beta}_{K \rho,0}(\bar p,p) \Big)
\Big\}
 \nonumber\\
&& {} +\frac{e\,h_A}{8\,\sqrt{3}\,f}\,\Big\{ h^{(K_1)}_{\eta\,K^*}\,
\hat B^{\bar \alpha \bar \beta, \mu, \alpha \beta}_{\eta K^*,1/2}(\bar p,p)
-2\,h^{(K_1)}_{K\,\rho}\,\hat B^{\bar \alpha \bar \beta, \mu, \alpha \beta}_{K \rho,0}(\bar p,p)
 \nonumber\\
&& \hspace*{5em} {}- h^{(K_1)}_{\pi\,K^*}\,
\left(2\,\hat A^{\bar \alpha \bar \beta, \mu, \alpha \beta}_{\pi K^*,0}(\bar p,p)
+\hat B^{\bar \alpha \bar \beta, \mu, \alpha \beta}_{\pi K^*,\infty}(\bar p,p)
\right)
\Big\}
 \nonumber\\
&& {} +\frac{e\,b_A}{2\,\sqrt{3}\,f}\,\Big\{ (4\,m_K^2-3\,m_\pi^2)\,h^{(K_1)}_{\eta\,K^*}\,
\bar B^{\bar \alpha \bar \beta, \mu, \alpha \beta}_{\eta K^*,1/2}(\bar p,p)
 \nonumber\\
&& \hspace*{5em} {}- m_\pi^2\,h^{(K_1)}_{\pi\,K^*}\,
\left(2\,\bar A^{\bar \alpha \bar \beta, \mu, \alpha \beta}_{\pi K^*,0}(\bar p,p)
+\bar B^{\bar \alpha \bar \beta, \mu, \alpha \beta}_{\pi K^*,\infty}(\bar p,p)
\right)
\Big\}
\nonumber\\
&& {}+ \frac{\tilde h_V}{8\,f\,m_V^2}\, \Big\{ -\frac{1}{3}\,\eAc \, h^{(K_1)}_{K\,\omega}\,
\hat C^{\bar \alpha \bar \beta, \mu, \alpha \beta}_{K K^* \omega}(\bar p,p)
\nonumber\\
&& \hspace*{5em} {}+\frac{1}{\sqrt{3}}\,\eAcmfn \,h^{(K_1)}_{K\,\rho}\,
\hat C^{\bar \alpha \bar \beta, \mu, \alpha \beta}_{K K^* \rho}(\bar p,p)
\nonumber\\
&& \hspace*{5em} {}+ \frac{\sqrt{2}}{3}\,\eAc \, h^{(K_1)}_{K\,\phi}\,
\hat C^{\bar \alpha \bar \beta, \mu, \alpha \beta}_{K K^* \phi}(\bar p,p)
\nonumber\\
&& \hspace*{5em} {}+ \frac{1}{\sqrt{3}}\,\eA \, h^{(K_1)}_{\pi\,K^*}\,
\hat C^{\bar \alpha \bar \beta, \mu, \alpha \beta}_{\pi \omega K^*}(\bar p,p)
+ \frac{1}{\sqrt{3}}\,\eA \, h^{(K_1)}_{\pi\,K^*}\,\hat C^{\bar \alpha \bar \beta, \mu, \alpha \beta}_{\pi \rho K^*}(\bar p,p)
\nonumber\\
&& \hspace*{5em}
{}+\frac{1}{\sqrt{3}\,3}\,\eA \, h^{(K_1)}_{\eta\,K^*}\,\hat C^{\bar \alpha \bar \beta, \mu, \alpha \beta}_{\eta \omega K^*}(\bar p,p)
+ \frac{1}{\sqrt{3}}\,\eA \, h^{(K_1)}_{\eta\,K^*}\,\hat C^{\bar \alpha \bar \beta, \mu, \alpha \beta}_{\eta \rho K^*}(\bar p,p)
\nonumber\\
&& \hspace*{5em} {}-\frac{4}{\sqrt{3}\,3}\,\eAphi \, h^{(K_1)}_{\eta\,K^*}\,
\hat C^{\bar \alpha \bar \beta, \mu, \alpha \beta}_{\eta \phi K^*}(\bar p,p)
\Big\}
\nonumber\\
&& {}+ \frac{\tilde h_V}{8\,f\,m_V^2}\, \Big\{ -\frac{1}{3}\,\eAcbA \, h^{(K_1)}_{K\,\omega}\,
\bar C^{\bar \alpha \bar \beta, \mu, \alpha \beta}_{K K^* \omega}(\bar p,p)
\nonumber\\
&& \hspace*{5em} {}+\frac{1}{\sqrt{3}}\,\eAcmfnbA \,h^{(K_1)}_{K\,\rho}\,
\bar C^{\bar \alpha \bar \beta, \mu, \alpha \beta}_{K K^* \rho}(\bar p,p)
\nonumber\\
&& \hspace*{5em} {}+ \frac{\sqrt{2}}{3}\,\eAcbA \, h^{(K_1)}_{K\,\phi}\,
\bar C^{\bar \alpha \bar \beta, \mu, \alpha \beta}_{K K^* \phi}(\bar p,p)
\nonumber\\
&& \hspace*{5em} {}+ \frac{1}{\sqrt{3}}\,\eAbA \, h^{(K_1)}_{\pi\,K^*}\,
\bar C^{\bar \alpha \bar \beta, \mu, \alpha \beta}_{\pi \omega K^*}(\bar p,p)
+ \frac{1}{\sqrt{3}}\,\eAbA \, h^{(K_1)}_{\pi\,K^*}\,\bar C^{\bar \alpha \bar \beta, \mu, \alpha \beta}_{\pi \rho K^*}(\bar p,p)
\nonumber\\
&& \hspace*{5em}
{}+\frac{1}{\sqrt{3}\,3}\,\eAbA \, h^{(K_1)}_{\eta\,K^*}\,\bar C^{\bar \alpha \bar \beta, \mu, \alpha \beta}_{\eta \omega K^*}(\bar p,p)
+ \frac{1}{\sqrt{3}}\,\eAbA \, h^{(K_1)}_{\eta\,K^*}\,\bar C^{\bar \alpha \bar \beta, \mu, \alpha \beta}_{\eta \rho K^*}(\bar p,p)
\nonumber\\
&& \hspace*{5em} {}-\frac{4}{\sqrt{3}\,3}\,\eAphibA \, h^{(K_1)}_{\eta\,K^*}\,
\bar C^{\bar \alpha \bar \beta, \mu, \alpha \beta}_{\eta \phi K^*}(\bar p,p)
\Big\}
\nonumber\\
&& {}+\frac{h_P}{16\,f^3}\,\Big\{ \eA \, h^{(K_1)}_{K\,\omega}\,
E^{\bar \alpha \bar \beta, \mu, \alpha \beta}_{K  \pi\omega}(\bar p,p)
+\frac{1}{3}\,\eA \, h^{(K_1)}_{K\,\omega}\,E^{\bar \alpha \bar \beta, \mu, \alpha \beta}_{K \eta \omega}(\bar p,p)
\nonumber\\
&& \hspace*{5em} {}+\frac{1}{\sqrt{3}}\,\eA \, h^{(K_1)}_{K\,\rho}\,
E^{\bar \alpha \bar \beta, \mu, \alpha \beta}_{K \pi \rho}(\bar p,p)
+\frac{1}{\sqrt{3}}\,\eA \, h^{(K_1)}_{K\,\rho}\,
E^{\bar \alpha \bar \beta, \mu, \alpha \beta}_{K \eta \rho}(\bar p,p)
\nonumber\\
&& \hspace*{5em} {}+\frac{\sqrt{2}\,2}{3}\,\eAphi \, h^{(K_1)}_{K\,\phi}\,
E^{\bar \alpha \bar \beta, \mu, \alpha \beta}_{K \eta \phi}(\bar p,p)
\nonumber\\
&& \hspace*{5em} {}+\frac{1}{\sqrt{3}}\,\eAcmfn \, h^{(K_1)}_{\pi\,K^*}\,E^{\bar \alpha \bar \beta, \mu, \alpha \beta}_{\pi K K^*}(\bar p,p)
\nonumber\\
&& \hspace*{5em} {}-\frac{1}{\sqrt{3}}\,\eAc \, h^{(K_1)}_{\eta\,K^*}\,E^{\bar \alpha \bar \beta, \mu, \alpha \beta}_{\eta K K^*}(\bar p,p)
\Big\}
\nonumber\\
&& {}+\frac{h_P}{16\,f^3}\,\Big\{ \eAbA \, h^{(K_1)}_{K\,\omega}\,
\bar E^{\bar \alpha \bar \beta, \mu, \alpha \beta}_{K  \pi\omega}(\bar p,p)
+\frac{1}{3}\,\eAbA \, h^{(K_1)}_{K\,\omega}\,\bar E^{\bar \alpha \bar \beta, \mu, \alpha \beta}_{K \eta \omega}(\bar p,p)
\nonumber\\
&& \hspace*{5em} {}+\frac{1}{\sqrt{3}}\,\eAbA \, h^{(K_1)}_{K\,\rho}\,
\bar E^{\bar \alpha \bar \beta, \mu, \alpha \beta}_{K \pi \rho}(\bar p,p)
+\frac{1}{\sqrt{3}}\,\eAbA \, h^{(K_1)}_{K\,\rho}\,
\bar E^{\bar \alpha \bar \beta, \mu, \alpha \beta}_{K \eta \rho}(\bar p,p)
\nonumber\\
&& \hspace*{5em} {}+\frac{\sqrt{2}\,2}{3}\,\eAphibA \, h^{(K_1)}_{K\,\phi}\,
\bar E^{\bar \alpha \bar \beta, \mu, \alpha \beta}_{K \eta \phi}(\bar p,p)
\nonumber\\
&& \hspace*{5em} {}+\frac{1}{\sqrt{3}}\,\eAcmfnbA \, h^{(K_1)}_{\pi\,K^*}\,\bar E^{\bar \alpha \bar \beta, \mu, \alpha \beta}_{\pi K K^*}(\bar p,p)
\nonumber\\
&& \hspace*{5em} {}-\frac{1}{\sqrt{3}}\,\eAcbA \, h^{(K_1)}_{\eta\,K^*}\,\bar E^{\bar \alpha \bar \beta, \mu, \alpha \beta}_{\eta K K^*}(\bar p,p)
\Big\}
\,,
\label{K1:gamma:Kmup}
\end{eqnarray}
and
\begin{eqnarray}
\lefteqn{i\,M^{\mu, \alpha \beta}_{K_1^0 \to \, \gamma \,K^0} =
\frac{e\,h_P\,m_V}{\sqrt{3}\,2\,f^2}\,\Big\{ h^{(K_1)}_{K\,\rho}\,
\Big( A^{\mu, \alpha \beta}_{K \rho}(\bar p,p) - \hat B^{\mu, \alpha \beta}_{K \rho,0}(\bar p,p)
\Big)  }
\nonumber\\
&& \hspace*{5em} {} +h^{(K_1)}_{\pi\,K^*}\,
\Big(A^{\mu, \alpha \beta}_{\pi K^*}(\bar p,p)
      - \hat B^{\mu, \alpha \beta}_{\pi K^*,1}(\bar p,p) \Big) \Big\}
\nonumber\\
&& {} + \frac{h_A}{96\,f^2\,m_V}\, \Big\{ -2\, \eAn \, h^{(K_1)}_{K\,\omega}\,C^{\mu, \alpha \beta}_{K K^* \omega}(\bar p,p)
\nonumber\\
&& \hspace*{5em} {} + \frac{2}{\sqrt{3}} \eAcmn \, h^{(K_1)}_{K\,\rho}\,C^{\mu, \alpha \beta}_{K K^* \rho}(\bar p,p)
\nonumber\\
&& \hspace*{5em} {} - 2 \, \sqrt{2}\,\eAn \, h^{(K_1)}_{K\,\phi}\,C^{\mu, \alpha \beta}_{K K^* \phi}(\bar p,p)
\nonumber\\
&& \hspace*{5em} {} -\sqrt{3}\,\eA \, h^{(K_1)}_{\pi\,K^*}\,C^{\mu, \alpha \beta}_{\pi \omega K^*}(\bar p,p)
+ \sqrt{3}\,\eA \, h^{(K_1)}_{\pi\,K^*}\,C^{\mu, \alpha \beta}_{\pi \rho K^*}(\bar p,p)
\nonumber\\
&& \hspace*{5em} {} +\frac{1}{\sqrt{3}} \,\eA \,h^{(K_1)}_{\eta\,K^*}\,C^{\mu, \alpha \beta}_{\eta \omega K^*}(\bar p,p)
- \sqrt{3}\,\eA \, h^{(K_1)}_{\eta\,K^*}\,C^{\mu, \alpha \beta}_{\eta \rho K^*}(\bar p,p)
\nonumber\\
&& \hspace*{5em} {} +\frac{4}{\sqrt{3}}\,\eAphi \, h^{(K_1)}_{\eta\,K^*}\,C^{\mu, \alpha \beta}_{\eta \phi K^*}(\bar p,p)
\Big\}
\nonumber\\
&& {} + \frac{h_A}{96\,f^2\,m_V}\, \Big\{ -2\, \eAnbA \, h^{(K_1)}_{K\,\omega}\,\bar C^{\mu, \alpha \beta}_{K K^* \omega}(\bar p,p)
\nonumber\\
&& \hspace*{5em} {} + \frac{2}{\sqrt{3}} \eAcmnbA \, h^{(K_1)}_{K\,\rho}\,\bar C^{\mu, \alpha \beta}_{K K^* \rho}(\bar p,p)
\nonumber\\
&& \hspace*{5em} {} - 2 \, \sqrt{2}\,\eAnbA \, h^{(K_1)}_{K\,\phi}\,\bar C^{\mu, \alpha \beta}_{K K^* \phi}(\bar p,p)
\nonumber\\
&& \hspace*{5em} {} -\sqrt{3}\,\eAbA \, h^{(K_1)}_{\pi\,K^*}\,\bar C^{\mu, \alpha \beta}_{\pi \omega K^*}(\bar p,p)
+ \sqrt{3}\,\eAbA \, h^{(K_1)}_{\pi\,K^*}\, \bar C^{\mu, \alpha \beta}_{\pi \rho K^*}(\bar p,p)
\nonumber\\
&& \hspace*{5em} {} +\frac{1}{\sqrt{3}} \,\eAbA \,h^{(K_1)}_{\eta\,K^*}\,\bar C^{\mu, \alpha \beta}_{\eta \omega K^*}(\bar p,p)
- \sqrt{3}\,\eAbA \, h^{(K_1)}_{\eta\,K^*}\,\bar C^{\mu, \alpha \beta}_{\eta \rho K^*}(\bar p,p)
\nonumber\\
&& \hspace*{5em} {} +\frac{4}{\sqrt{3}}\,\eAphibA \, h^{(K_1)}_{\eta\,K^*}\,\bar C^{\mu, \alpha \beta}_{\eta \phi K^*}(\bar p,p)
\Big\}
\nonumber\\
&& {} + \frac{b_A\,m_K^2}{24\,f^2\,m_V}\, \Big\{ -2\, \eAn \, h^{(K_1)}_{K\,\omega}\,\Cchi^{\mu, \alpha \beta}_{K K^* \omega}(\bar p,p)
\nonumber\\
&& \hspace*{5em} {} + \frac{2}{\sqrt{3}} \eAcmn \, h^{(K_1)}_{K\,\rho}\,\Cchi^{\mu, \alpha \beta}_{K K^* \rho}(\bar p,p)
\nonumber\\
&& \hspace*{5em} {} - 2 \, \sqrt{2}\,\eAn \, h^{(K_1)}_{K\,\phi}\,\Cchi^{\mu, \alpha \beta}_{K K^* \phi}(\bar p,p)
\nonumber\\
&& \hspace*{5em} {} -\sqrt{3}\,\eA \, h^{(K_1)}_{\pi\,K^*}\,\Cchi^{\mu, \alpha \beta}_{\pi \omega K^*}(\bar p,p)
+ \sqrt{3}\,\eA \, h^{(K_1)}_{\pi\,K^*}\,\Cchi^{\mu, \alpha \beta}_{\pi \rho K^*}(\bar p,p)
\nonumber\\
&& \hspace*{5em} {} +\frac{1}{\sqrt{3}} \,\eA \,h^{(K_1)}_{\eta\,K^*}\,\Cchi^{\mu, \alpha \beta}_{\eta \omega K^*}(\bar p,p)
- \sqrt{3}\,\eA \, h^{(K_1)}_{\eta\,K^*}\,\Cchi^{\mu, \alpha \beta}_{\eta \rho K^*}(\bar p,p)
\nonumber\\
&& \hspace*{5em} {} +\frac{4}{\sqrt{3}}\,\eAphi \, h^{(K_1)}_{\eta\,K^*}\,\Cchi^{\mu, \alpha \beta}_{\eta \phi K^*}(\bar p,p)
\Big\}
\nonumber\\
&& {} + \frac{b_A\,m_K^2}{24\,f^2\,m_V}\, \Big\{ -2\, \eAnbA \, h^{(K_1)}_{K\,\omega}\,\barCchi^{\mu, \alpha \beta}_{K K^* \omega}(\bar p,p)
\nonumber\\
&& \hspace*{5em} {} + \frac{2}{\sqrt{3}} \eAcmnbA \, h^{(K_1)}_{K\,\rho}\,\barCchi^{\mu, \alpha \beta}_{K K^* \rho}(\bar p,p)
\nonumber\\
&& \hspace*{5em} {} - 2 \, \sqrt{2}\,\eAnbA \, h^{(K_1)}_{K\,\phi}\,\barCchi^{\mu, \alpha \beta}_{K K^* \phi}(\bar p,p)
\nonumber\\
&& \hspace*{5em} {} -\sqrt{3}\,\eAbA \, h^{(K_1)}_{\pi\,K^*}\,\barCchi^{\mu, \alpha \beta}_{\pi \omega K^*}(\bar p,p)
+ \sqrt{3}\,\eAbA \, h^{(K_1)}_{\pi\,K^*}\, \barCchi^{\mu, \alpha \beta}_{\pi \rho K^*}(\bar p,p)
\nonumber\\
&& \hspace*{5em} {} +\frac{1}{\sqrt{3}} \,\eAbA \,h^{(K_1)}_{\eta\,K^*}\,\barCchi^{\mu, \alpha \beta}_{\eta \omega K^*}(\bar p,p)
- \sqrt{3}\,\eAbA \, h^{(K_1)}_{\eta\,K^*}\,\barCchi^{\mu, \alpha \beta}_{\eta \rho K^*}(\bar p,p)
\nonumber\\
&& \hspace*{5em} {} +\frac{4}{\sqrt{3}}\,\eAphibA \, h^{(K_1)}_{\eta\,K^*}\,\barCchi^{\mu, \alpha \beta}_{\eta \phi K^*}(\bar p,p)
\Big\}
\nonumber\\
&&  {} + \frac{(1-r)\,e_V\,h_P}{16\,\sqrt{3}\,f^2\,m_V}\,h^{(K_1)}_{\eta\,K^*}\,
\Big\{3\,m_V^2\,h_V\, D^{\mu, \alpha \beta}_{\eta K^*}(\bar p,p) +\tilde h_V\, \tilde D^{\mu, \alpha \beta}_{\eta K^*}(\bar p,p)
\Big\}
\nonumber\\
&& {} +\frac{e_V\,m_V}{16\,f^2}\,\Big\{ \frac{1}{\sqrt{3}}\,h^{(K_1)}_{K\,\rho}\,E^{\mu, \alpha \beta}_{K \rho}(\bar p,p)
+ \frac{1}{\sqrt{3}}\,h^{(K_1)}_{\pi\,K^*}\,E^{\mu, \alpha \beta}_{\pi K^*}(\bar p,p) \Big\}
\nonumber\\
&& {} - \frac{e_V}{8\,\sqrt{3} \, f^2\,m_V}\,\Big\{h^{(K_1)}_{K\,\rho}\,
F^{\mu, \alpha \beta}_{K \rho}(\bar p,p) + \frac{1}{2} \, \left(\ratVphi +1 \right)
h^{(K_1)}_{\pi\,K^*}\,F^{\mu, \alpha \beta}_{\pi K^*}(\bar p,p)
\nonumber\\
&& \hspace*{5em} {}+\frac{1}{2} \, \left(\ratVphi -1 \right) h_{\eta K^*}^{(K_1)}\,F_{\eta K^*}^{\mu, \alpha\beta}(\bar p,p) \Big\}
\nonumber\\
&& {} -\frac{e_V}{16 \,f^2\,m_V}\,\Big\{ -\frac{1}{\sqrt{3}}\,\Big[
  \left(1 + \ratVphi \right) g_D - \left(3- \ratVphi \right) g_F
\Big]\,h^{(K_1)}_{\pi\,K^*}\, G^{\mu, \alpha \beta}_{\pi K^*}(\bar p,p)
\nonumber\\
&& \hspace*{5em} {} - \frac{2}{\sqrt{3}}\,\Big[ \ratVphi\,g_D - \ratVphi\,g_F  \Big]\,
h^{(K_1)}_{K\,\rho}\,G^{\mu, \alpha \beta}_{K \rho}(\bar p,p)
\nonumber\\
&& \hspace*{5em} {} -\frac{2}{3}\,\Big[
  \left(1+ \ratVphi \right) g_D + \left(1- \ratVphi \right) g_F  \Big]
h^{(K_1)}_{K\,\omega}\,G^{\mu, \alpha \beta}_{K \omega}(\bar p,p)
\nonumber\\
&& \hspace*{5em} {} - \frac{1}{3\,\sqrt{3}}\,\Big[
  \left(1-5 \, \ratVphi \right) g_D - 3\, \left(1- \ratVphi \right) g_F
\Big]\,h^{(K_1)}_{\eta\,K^*}\, G^{\mu, \alpha \beta}_{\eta K^*}(\bar p,p)
\nonumber\\
&& \hspace*{5em} {}-\frac{2\,\sqrt{2}}{3}\,\Big[
  \left(1+ \ratVphi \right) g_D - \left(1- \ratVphi \right) g_F  \Big]\,
h^{(K_1)}_{K\,\phi}\,G^{\mu, \alpha \beta}_{K \phi}(\bar p,p) \Big\}
\nonumber\\
&& {} -\frac{e_V\, b_D}{96\,f^2\,m_V}\,\Big\{
\sqrt{3}\,\left(1 - \ratVphi \right) (m_\pi^2 + m_K^2) \, h^{(K_1)}_{\pi\,K^*}\,E^{\mu, \alpha \beta}_{\pi K^*}(\bar p,p)
\nonumber\\
&& \hspace*{5em} {} - 4\,m_K^2\, h^{(K_1)}_{K\,\omega}\,E^{\mu, \alpha \beta}_{K \omega}(\bar p,p)
\nonumber\\
&& \hspace*{5em} {} + \frac{1}{\sqrt{3}} \left( 5\,m_K^2-3\,m_\pi^2  \right)
\left(1+\ratVphi \right) \, h^{(K_1)}_{\eta\,K^*}\,E^{\mu, \alpha \beta}_{\eta K^*}(\bar p,p)
\nonumber\\
&& \hspace*{5em} {}- 4\,\sqrt{2}\,\ratVphi \,m_K^2\, h^{(K_1)}_{K\,\phi}\,E^{\mu, \alpha \beta}_{K \phi}(\bar p,p) \Big\}
\nonumber\\
&& {}- \frac{h_A \, b_A \, e_V}{96 \, f^2 \, m_V} \, (1+\ratVphi) \, m_K^2 \,
\Big\{ \sqrt{3} \, h^{(K_1)}_{\pi\,K^*}\, H^{\mu, \alpha \beta}_{\pi K^* K^*}(\bar p,p)
+ \sqrt{3} \, h^{(K_1)}_{K\,\rho}\, H^{\mu, \alpha \beta}_{K K^* \rho}(\bar p,p)
\nonumber\\
&& \hspace*{10em} {}
+ h^{(K_1)}_{K\,\omega}\, H^{\mu, \alpha \beta}_{K K^* \omega}(\bar p,p)
- \frac{1}{\sqrt{3}} \, h^{(K_1)}_{\eta\,K^*}\,
   H^{\mu, \alpha \beta}_{\eta K^* K^*}(\bar p,p)
\nonumber\\
&& \hspace*{10em} {}
+ \sqrt{2} \, h^{(K_1)}_{K\,\phi}\, H^{\mu, \alpha \beta}_{K K^* \phi}(\bar p,p)
\Big\}
\nonumber\\
&& {}- \frac{(b_A)^2 \, e_V}{24 \, f^2 \, m_V} \, (1+\ratVphi) \, m_K^2 \,
\Big\{
\sqrt{3} \, m_\pi^2 \, h^{(K_1)}_{\pi\,K^*}\,
   \bar H^{\mu, \alpha \beta}_{\pi K^* K^*}(\bar p,p)
\nonumber\\
&& \hspace*{10em} {}
+ \sqrt{3} \, m_K^2 \, h^{(K_1)}_{K\,\rho}\,
   \bar H^{\mu, \alpha \beta}_{K K^* \rho}(\bar p,p)
\nonumber\\
&& \hspace*{10em} {}
+ m_K^2 \, h^{(K_1)}_{K\,\omega}\,
    \bar H^{\mu, \alpha \beta}_{K K^* \omega}(\bar p,p)
\nonumber\\
&& \hspace*{10em} {}
- \frac{1}{\sqrt{3}} \, (4 \, m_K^2 - 3 \, m_\pi^2) \, h^{(K_1)}_{\eta\,K^*}\,
    \bar H^{\mu, \alpha \beta}_{\eta K^* K^*}(\bar p,p)
\nonumber\\
&& \hspace*{10em} {}
+ \sqrt{2} \, m_K^2 \, h^{(K_1)}_{K\,\phi}\,
     \bar H^{\mu, \alpha \beta}_{K K^* \phi}(\bar p,p)
\Big\}\,,
\label{K1:gamma:K0}
\end{eqnarray}
and
\begin{eqnarray}
\lefteqn{M^{\bar \alpha \bar \beta, \mu, \alpha \beta}_{K_1^0 \to \, \gamma \,K^0_*}
=  -\frac{1}{\sqrt{3}}\,\frac{e\,h_A}{4\,f}\,h^{(K_1)}_{K\,\rho}\,
\Big\{ \hat A^{\bar \alpha \bar \beta, \mu, \alpha \beta}_{K \rho,1}(\bar p,p)
-\hat B^{\bar \alpha \bar \beta, \mu, \alpha \beta}_{K \rho,0}(\bar p,p)
 \Big\}  }
\nonumber\\
&& {} -\frac{1}{\sqrt{3}}\,\frac{e\,b_A}{f}\,m_K^2\,h^{(K_1)}_{K\,\rho}\,
\Big\{ \bar A^{\bar \alpha \bar \beta, \mu, \alpha \beta}_{K \rho,1}(\bar p,p)
-\bar  B^{\bar \alpha \bar \beta, \mu, \alpha \beta}_{K \rho,0}(\bar p,p)
 \Big\}
\nonumber\\
&& {} +\frac{1}{\sqrt{3}}\, \frac{e\,h_A}{4\,f}\, h^{(K_1)}_{\pi\,K^*}\,
\Big\{ \hat A^{\bar \alpha \bar \beta, \mu, \alpha \beta}_{\pi K^*,0}(\bar p,p)
-\hat B^{\bar \alpha \bar \beta, \mu, \alpha \beta}_{\pi K^*,1}(\bar p,p)
\Big\}
\nonumber\\
&& {} + \frac{1}{\sqrt{3}}\,\frac{e\,b_A}{f}\,m_\pi^2 \, h^{(K_1)}_{\pi\,K^*}\,
\Big\{ \bar A^{\bar \alpha \bar \beta, \mu, \alpha \beta}_{\pi K^*,0}(\bar p,p)
-\bar B^{\bar \alpha \bar \beta, \mu, \alpha \beta}_{\pi K^*,1}(\bar p,p)
\Big\}
\nonumber\\
&& {} - \frac{(1-r)\,e_V\,h_A}{96\,\sqrt{3}\,f\,m_V^2}\,h^{(K_1)}_{\eta\,K^*}\,\Big\{
3\,m_V^2\,h_V\,D^{\bar \alpha \bar \beta, \mu, \alpha \beta}_{\eta K^*}(\bar p,p)
+\tilde h_V\, \tilde D^{\bar \alpha \bar \beta, \mu, \alpha \beta}_{\eta K^*}(\bar p,p) \Big\}
\nonumber\\
&& {} - \frac{(1-r)\,e_V\,b_A}{24\,\sqrt{3}\,f\,m_V^2}\,(4\,m_K^2-3\,m_\pi^2)\,h^{(K_1)}_{\eta\,K^*}\,\Big\{
3\,m_V^2\,h_V\,D^{\bar \alpha \bar \beta, \mu, \alpha \beta}_{\chi, \eta K^*}(\bar p,p)
\nonumber\\
&& \hspace*{5em} {} +\tilde h_V\, \tilde D^{\bar \alpha \bar \beta, \mu, \alpha \beta}_{\chi,\eta K^*}(\bar p,p)\Big\}
\nonumber\\
&& {} + \frac{\tilde h_V}{8\,f\,m_V^2}\, \Big\{\frac{2}{3}\, \eAn \, h^{(K_1)}_{K\,\omega}\,
\hat C^{\bar \alpha \bar \beta, \mu, \alpha \beta}_{K K^* \omega}(\bar p,p)
\nonumber\\
&& \hspace*{5em} {} - \frac{2}{3 \sqrt{3}}\,\eAcmn \, h^{(K_1)}_{K\,\rho}\,
\hat C^{\bar \alpha \bar \beta, \mu, \alpha \beta}_{K K^* \rho}(\bar p,p)
\nonumber\\
&& \hspace*{5em} {} -\frac{2\,\sqrt{2}}{3}\,\eAn \, h^{(K_1)}_{K\,\phi}\,
\hat C^{\bar \alpha \bar \beta, \mu, \alpha \beta}_{K K^* \phi}(\bar p,p)
\nonumber\\
&& \hspace*{5em} {} - \frac{1}{\sqrt{3}}\,\eA \, h^{(K_1)}_{\pi\,K^*}\,
\hat C^{\bar \alpha \bar \beta, \mu, \alpha \beta}_{\pi \omega K^*}(\bar p,p)
+\,\frac{1}{\sqrt{3}}\,\eA \, h^{(K_1)}_{\pi\,K^*}\,\hat C^{\bar \alpha \bar \beta, \mu, \alpha \beta}_{\pi \rho K^*}(\bar p,p)
\nonumber\\
&& \hspace*{5em} {} +\frac{1}{\sqrt{3}\,3}\,\eA \, h^{(K_1)}_{\eta\,K^*}\,
\hat C^{\bar \alpha \bar \beta, \mu, \alpha \beta}_{\eta \omega K^*}(\bar p,p)
- \frac{1}{\sqrt{3}}\,\eA \, h^{(K_1)}_{\eta\,K^*}\,\hat C^{\bar \alpha \bar \beta, \mu, \alpha \beta}_{\eta \rho K^*}(\bar p,p)
\nonumber\\
&& \hspace*{5em} {} -\frac{4}{\sqrt{3}\,3}\,\eAphi \, h^{(K_1)}_{\eta\,K^*}\,\hat C^{\bar \alpha \bar \beta, \mu, \alpha \beta}_{\eta \phi K^*}(\bar p,p)
\Big\}
\nonumber\\
&& {} + \frac{\tilde h_V}{8\,f\,m_V^2}\, \Big\{\frac{2}{3}\, \eAnbA \, h^{(K_1)}_{K\,\omega}\,
\bar C^{\bar \alpha \bar \beta, \mu, \alpha \beta}_{K K^* \omega}(\bar p,p)
\nonumber\\
&& \hspace*{5em} {} - \frac{2}{3 \sqrt{3}}\,\eAcmnbA \, h^{(K_1)}_{K\,\rho}\,
\bar C^{\bar \alpha \bar \beta, \mu, \alpha \beta}_{K K^* \rho}(\bar p,p)
\nonumber\\
&& \hspace*{5em} {} -\frac{2\,\sqrt{2}}{3}\,\eAnbA \, h^{(K_1)}_{K\,\phi}\,
\bar C^{\bar \alpha \bar \beta, \mu, \alpha \beta}_{K K^* \phi}(\bar p,p)
\nonumber\\
&& \hspace*{5em} {} - \frac{1}{\sqrt{3}}\,\eAbA \, h^{(K_1)}_{\pi\,K^*}\,
\bar C^{\bar \alpha \bar \beta, \mu, \alpha \beta}_{\pi \omega K^*}(\bar p,p)
+\,\frac{1}{\sqrt{3}}\,\eAbA \, h^{(K_1)}_{\pi\,K^*}\,\bar C^{\bar \alpha \bar \beta, \mu, \alpha \beta}_{\pi \rho K^*}(\bar p,p)
\nonumber\\
&& \hspace*{5em} {} +\frac{1}{\sqrt{3}\,3}\,\eAbA \, h^{(K_1)}_{\eta\,K^*}\,
\bar C^{\bar \alpha \bar \beta, \mu, \alpha \beta}_{\eta \omega K^*}(\bar p,p)
- \frac{1}{\sqrt{3}}\,\eAbA \, h^{(K_1)}_{\eta\,K^*}\,\bar C^{\bar \alpha \bar \beta, \mu, \alpha \beta}_{\eta \rho K^*}(\bar p,p)
\nonumber\\
&& \hspace*{5em} {} -\frac{4}{\sqrt{3}\,3}\,\eAphibA \, h^{(K_1)}_{\eta\,K^*}\,\bar C^{\bar \alpha \bar \beta, \mu, \alpha \beta}_{\eta \phi K^*}(\bar p,p)
\Big\}
\nonumber\\
&& {} + \frac{h_P}{16\,f^3}\,\Big\{
-\eA \, h^{(K_1)}_{K\,\omega}\,E^{\bar \alpha \bar \beta, \mu, \alpha \beta}_{K \pi \omega}(\bar p,p)
+\frac{1}{3}\,\eA \, h^{(K_1)}_{K\,\omega}\,E^{\bar \alpha \bar \beta, \mu, \alpha \beta}_{K \eta \omega}(\bar p,p)
\nonumber\\
&& \hspace*{5em} {} +\frac{1}{\sqrt{3}}\,\eA \, h^{(K_1)}_{K\,\rho}\,E^{\bar \alpha \bar \beta, \mu, \alpha \beta}_{K \pi \rho}(\bar p,p)
-\frac{1}{\sqrt{3}}\,\eA \, h^{(K_1)}_{K\,\rho}\,E^{\bar \alpha \bar \beta, \mu, \alpha \beta}_{K \eta \rho}(\bar p,p)
\nonumber\\
&& \hspace*{5em} {} +\frac{\sqrt{2}\,2}{3}\,\eAphi \, h^{(K_1)}_{K\,\phi}\,E^{\bar \alpha \bar \beta, \mu, \alpha \beta}_{K \eta \phi}(\bar p,p)
\nonumber\\
&& \hspace*{5em} {} +\frac{2}{\sqrt{3}}\,\eAn \, h^{(K_1)}_{\eta\,K^*}\,E^{\bar \alpha \bar \beta, \mu, \alpha \beta}_{\eta K K^*}(\bar p,p)
\nonumber\\
&& \hspace*{5em} {} -\frac{2}{3\,\sqrt{3}}\,\eAcmn \, h^{(K_1)}_{\pi\,K^*}\,E^{\bar \alpha \bar \beta, \mu, \alpha \beta}_{\pi K K^*}(\bar p,p)
\Big\}
\nonumber\\
&& {} + \frac{h_P}{16\,f^3}\,\Big\{
-\eAbA \, h^{(K_1)}_{K\,\omega}\,\bar E^{\bar \alpha \bar \beta, \mu, \alpha \beta}_{K \pi \omega}(\bar p,p)
+\frac{1}{3}\,\eAbA \, h^{(K_1)}_{K\,\omega}\,\bar E^{\bar \alpha \bar \beta, \mu, \alpha \beta}_{K \eta \omega}(\bar p,p)
\nonumber\\
&& \hspace*{5em} {} +\frac{1}{\sqrt{3}}\,\eAbA \, h^{(K_1)}_{K\,\rho}\,\bar E^{\bar \alpha \bar \beta, \mu, \alpha \beta}_{K \pi \rho}(\bar p,p)
-\frac{1}{\sqrt{3}}\,\eAbA \, h^{(K_1)}_{K\,\rho}\,\bar E^{\bar \alpha \bar \beta, \mu, \alpha \beta}_{K \eta \rho}(\bar p,p)
\nonumber\\
&& \hspace*{5em} {} +\frac{\sqrt{2}\,2}{3}\,\eAphibA \, h^{(K_1)}_{K\,\phi}\,\bar E^{\bar \alpha \bar \beta, \mu, \alpha \beta}_{K \eta \phi}(\bar p,p)
\nonumber\\
&& \hspace*{5em} {} +\frac{2}{\sqrt{3}}\,\eAnbA \, h^{(K_1)}_{\eta\,K^*}\,\bar E^{\bar \alpha \bar \beta, \mu, \alpha \beta}_{\eta K K^*}(\bar p,p)
\nonumber\\
&& \hspace*{5em} {} -\frac{2}{3\,\sqrt{3}}\,\eAcmnbA \, h^{(K_1)}_{\pi\,K^*}\,\bar E^{\bar \alpha \bar \beta, \mu, \alpha \beta}_{\pi K K^*}(\bar p,p)
\Big\}
\nonumber\\
&& {} + \frac{(1-r)\,e_V\,h_V}{32\,f}\,h^{(K_1)}_{K\,\omega}\,\Big\{ h_A\,F^{\bar \alpha \bar \beta, \mu, \alpha \beta}_{K \omega}(\bar p,p)
+ 4\,m_K^2\,b_A\,F^{\bar \alpha \bar \beta, \mu, \alpha \beta}_{\chi, K \omega}(\bar p,p)\Big\}
\nonumber\\
&& {} - \frac{(1-r)\,e_V\,h_V}{32\,\sqrt{3}\,f}\,h^{(K_1)}_{\eta\,K^*}\,\Big\{ h_A\,F^{\bar \alpha \bar \beta, \mu, \alpha \beta}_{\eta K^*}(\bar p,p)
\nonumber\\
&& \hspace*{5em} {} + 4\,(4\,m_K^2-3\,m_\pi^2)\,b_A\,F^{\bar \alpha \bar \beta, \mu, \alpha \beta}_{\chi, \eta K^*}(\bar p,p)\Big\}
\nonumber\\
&& {} + \frac{(1-r)\,e_V\,h_V}{16\,\sqrt{2}\,f}\,h^{(K_1)}_{K\,\phi}\,\Big\{ h_A\,F^{\bar \alpha \bar \beta, \mu, \alpha \beta}_{K \phi}(\bar p,p)
+ 4\,m_K^2\,b_A\,F^{\bar \alpha \bar \beta, \mu, \alpha \beta}_{\chi, K \phi}(\bar p,p)\Big\}
\,,
\label{K1:gamma:Kmu0}
\end{eqnarray}
where
\begin{eqnarray}
&& \hat e_A^{(K^*_0)\,}  =  e_A + \frac18 \,\left(1+\frac{m_V^2}{m_\phi^2} \right)  e_V \, h_A \,,
\nonumber \\
&& \hat e_A^{(K^*_+)}  =  e_A + \frac14 \, \left(2-\frac{m_V^2}{m_\phi^2} \right) e_V \,  h_A \,,
  \label{eq:hateA:Appendix}
\end{eqnarray}
and
\begin{eqnarray}
&& \bar e_A^{(K^*_0)\,}  =  - \,  \left(1+\frac{m_V^2}{m_\phi^2} \right) \left(
\frac{m_K^2}{m_{K^*}^2} \right)e_V \, b_A \,,
\nonumber \\
&& \bar e_A^{(K^*_+)}  =   - 2  \, \left(2-\frac{m_V^2}{m_\phi^2} \right) \left(
 \frac{m_K^2}{m_{K^*}^2} \right)\, e_V \, b_A \,,
  \label{eq:bareA:Appendix-strange}
\end{eqnarray}
and
\begin{eqnarray}
&& \bar e_A = - 2\,\left( \frac{m_\pi^2}{m_V^2}\right) \,e_V\,b_A\,, \qquad \quad
\bar e_A^{(\phi)} \;\; =  - 2  \,  \left(\frac{m_V^2}{m_\phi^2}\right) \left(
  \frac{2 \,m_K^2 -m_\pi^2}{m_\phi^2}  \right)\,e_V \,b_A \,,
\nonumber\\
&&\hat e_A^{}  =  e_A + \frac14 \,  e_V  \, h_A  \,, \qquad
\hat e_A^{(\phi)} \;\; =  e_A + \frac14 \,\left(\frac{m_V^2}{m_\phi^2}\right)   e_V  \, h_A  \,.
\label{eq:bareA:Appendix-nonstr}
\end{eqnarray}
We identify $m_V=776 $ MeV with the average of the $\rho$ and $\omega $ mass. Contributions proportional
to $m_\rho-m_\omega$ are neglected.


\newpage
\begin{thebibliography}{99}

\bibitem{Godfrey:1985xj}
S.~Godfrey and N.~Isgur,
Phys.\ Rev.\  {\bf D 32} (1985) 189.

\bibitem{Ishida:Yamada:Oda:1989}
S. Ishida, K. Yamada and M. Oda, Phys. Rev. {\bf D 40} (1989) 1497.

\bibitem{Lutz:habil}
M.F.M. Lutz, GSI-Habil-2002-1.

\bibitem{Lutz-Kolomeitsev:2004}
M.F.M. Lutz and E.E. Kolomeitsev, Nucl. Phys. {\bf A 730} (2004) 392.

\bibitem{Wagner:Leupold:2007}
M. Wagner and S. Leupold,  arXiv:0708.2223 [hep-ph].

\bibitem{Wagner:2008gz}
  M.~Wagner and S.~Leupold,
  arXiv:0801.0814 [hep-ph].

\bibitem{Kolomeitsev-Lutz:2004}
E.E. Kolomeitsev and M.F.M. Lutz, Phys. Lett. {\bf B 582} (2004) 39.

\bibitem{Hofmann-Lutz:2004}
J. Hofmann and M.F.M. Lutz, Nucl. Phys. {\bf A 733} (2004) 142.

\bibitem{Lutz-Soyeur:2007}
M.F.M. Lutz and M. Soyeur, arXiv:0710.1545[hep-ph].


\bibitem{Roca:Hosaka:Oset:2007}
L. Roca, A. Hosaka and E. Oset, Phys.\ Lett.\  {\bf B 658} (2007) 17.

\bibitem{Bando}
M. Bando, T. Kugo, S. Uehara, K. Yamawaki and T.
Yanagida, Phys. Rev. Lett. {\bf 54} (1985) 1215;
M. Bando, T. Kugo and K. Yamawaki, Phys. Rep. {\bf 164} (1988) 217.

\bibitem{Birse:1996}
M. Birse, Z. Phys. {\bf A 355} (1996) 231.

\bibitem{Rosner:1981}
J.L. Rosner, Phys. Rev. {\bf D 23} (1981) 1127.

\bibitem{Roca:Palomar:Oset:2004}
L. Roca, J.E. Palomar and E. Oset, Phys. Rev. {\bf D 70} (2004) 094006.

\bibitem{Aznaurian:Oganesian:1988}
I.G. Aznaurian and  K.A. Oganesian, Sov. J. Nucl. Phys. {\bf 47} (1988) 1097.


\bibitem{PDG:2006}
W.-M.Yao et al. (Particle Data Group), J. Phys. {\bf G 33} (2006) 1.

\bibitem{Kyriakopoulos:1971}
E. Kyriakopoulos, Phys. Rev. {\bf D 4} (1971) 2002.

\bibitem{Kyriakopoulos:1972}
E. Kyriakopoulos, Phys. Rev. {\bf D 6} (1972) 2207.

\bibitem{Ecker:1989}
G. Ecker et al., Nucl. Phys. {\bf B 321} (1989) 311.

\bibitem{Krause:1990}
A.~Krause, Helv. Phys. Acta {\bf 63} (1990) 3.

\bibitem{Lutz:Kolomeitsev:2002}
M.F.M. Lutz and E. E. Kolomeitsev, Nucl. Phys. {\bf A 700} (2002) 193.

\bibitem{Jenkins:1995vb}
E.~E.~Jenkins, A.~V.~Manohar and M.~B.~Wise,
Phys.\ Rev.\ Lett.\  {\bf 75} (1995) 2272.

\bibitem{Scherer:1}
D. Djukanovic et al.,  Phys. Rev. Lett. {\bf 93} (2004) 122002.

\bibitem{Scherer:2}
M.R. Schindler, J. Gegelia and S. Scherer, Eur. Phys. J. {\bf A 26} (2005) 1.

\bibitem{Cirigliano:2003yq}
V.~Cirigliano, G.~Ecker, H.~Neufeld and A.~Pich,
JHEP {\bf 0306} (2003) 012.

\bibitem{large-N_c-reference}
G. 't Hooft, Nucl. Phys. {\bf B 72} (1964) 461.


\bibitem{Roca:Oset:Singh:2005}
L. Roca, E. Oset and J. Singh, Phys. Rev. {\bf D 72} (2005) 014002.


\bibitem{Harada:2003jx}
M.~Harada and K.~Yamawaki,
Phys.\ Rept.\  {\bf 381} (2003) 1.


\bibitem{Fearing:1994ga}
H.~W.~Fearing and S.~Scherer,
Phys.\ Rev.\  {\bf D 53} (1996) 315.

\bibitem{RuizFemenia:2003hm}
  P.D.~Ruiz-Femenia, A.~Pich and J.~Portoles,
  JHEP {\bf 0307} (2003) 003.

\bibitem{Jones:1962}
H.F. Jones, Nuovo Cimento {\bf 26} (1962) 790.

\bibitem{Barua:1977gc}
  D.~Barua and S.N.~Gupta,
  Phys.\ Rev.\  {\bf D 15} (1977) 509.

\bibitem{Brodsky:Hiller:1992}
S.J. Brodsky and J.R. Hiller, Phys. Rev. {\bf D 46} (1992) 2141.

\bibitem{lattice:Hedlich:2007}
J.N. Hedditch et al., Phys. Rev. {\bf D 75} (2007) 094504.

\bibitem{Samsonov:2003hs}
  A.~Samsonov,
  JHEP {\bf 0312} (2003) 061.

\bibitem{Aliev:2003ba}
  T.~M.~Aliev, I.~Kanik and M.~Savci,
  Phys.\ Rev.\  {\bf D 68} (2003) 056002.

\bibitem{Choi:Ji:2004}
H.-M. Coi and Ch.-R. Ji, Phys. Rev. {\bf D 70} (2004) 053015.

\bibitem{Leupold:2005ep}
  S.~Leupold,
  Phys.\ Rev.\  {\bf D 73} (2006) 085013.

\bibitem{Passarino:Veltman:1979}
G. Passarino and M. Veltman, Nucl. Phys. {\bf B 160} (1979) 151.



\bibitem{Amelin:1994ii}
  D.~V.~Amelin {\it et al.},
  Z.\ Phys.\  {\bf C 66} (1995) 71.

\bibitem{Collick:1984}
B. Collick et al., Phys. Rev. Lett. {\bf 53} (1984) 2374.

\bibitem{Zielinski:1984}
M. Zielinski et al., Phys. Rev. Lett. {\bf 52} (1984) 1195.

\bibitem{KTeV:2002}
KTeV Collaboration, Phys. Rev. Lett. {\bf 89} (2002) 072001.


\bibitem{Ecker:1989yg}
  G.~Ecker, J.~Gasser, H.~Leutwyler, A.~Pich and E.~de Rafael,
  Phys.\ Lett.\  B {\bf 223} (1989) 425.

\end{thebibliography}
\end{document}